\documentclass[10pt]{article}

\pdfoutput=1 

\usepackage{jheppub} 

\usepackage[T1]{fontenc} 
\usepackage{enumitem}

\usepackage{xcolor}
\usepackage{xspace}
\usepackage{amssymb,amsthm,cancel,hyperref,graphicx,xcolor}
\usepackage{mathrsfs,wasysym}
\usepackage{booktabs}
\usepackage{subcaption}
\usepackage{tikz}
\usetikzlibrary{calc}
\usepackage{placeins}
\usepackage{tabu}
\usepackage{pdflscape}
\usepackage{afterpage}
\usepackage{graphicx}
\usepackage{makecell}

\DeclareMathOperator{\Tr}{Tr}   
\newcommand{\gsim}{\gtrsim}

\newcommand{\as}{\alpha_s}

\let\originalleft\left
\let\originalright\right
\renewcommand{\left}{\mathopen{}\mathclose\bgroup\originalleft}
\renewcommand{\right}{\aftergroup\egroup\originalright}

\newcommand{\zc}{z_{\rm cut}}

\newcommand{\la}{\lambda_\alpha}

\newcommand{\sherpa}{S\protect\scalebox{0.8}{HERPA}\xspace}
\newcommand{\pythia}{P\protect\scalebox{0.8}{YTHIA}\xspace}
\newcommand{\herwig}{H\protect\scalebox{0.8}{ERWIG}\xspace}

\newcommand{\comix}{C\protect\scalebox{0.8}{OMIX}\xspace}

\newcommand{\Caesar}{C\protect\scalebox{0.8}{AESAR}\xspace}
\newcommand{\fastjet}{F\protect\scalebox{0.8}{AST}J\protect\scalebox{0.8}{ET}\xspace}
\newcommand{\rivet}{R\protect\scalebox{0.8}{IVET}\xspace}
\newcommand{\recola}{R\protect\scalebox{0.8}{ECOLA}\xspace}
\newcommand{\collier}{C\protect\scalebox{0.8}{OLLIER}\xspace}
\newcommand{\OpenLoops}{O\protect\scalebox{0.8}{PEN}L\protect\scalebox{0.8}{OOPS}\xspace}
\newcommand{\mcfm}{M\protect\scalebox{0.8}{CFM}\xspace}
\newcommand{\powheg}{P\protect\scalebox{0.8}{OWHEG}\xspace}
\newcommand{\mgamcnlo}{M\protect\scalebox{0.8}{ADGRAPH}{5\_aMC@NLO}\xspace}

\newcommand{\softdrop}{\textsf{SoftDrop }}

\newcommand{\muR}{\ensuremath{\mu_{\text{R}}}}
\newcommand{\muF}{\ensuremath{\mu_{\text{F}}}}

\newcommand{\zcut}{\ensuremath{z_{\text{cut}}}}
\newcommand{\alphaS}{\alpha_\text{s}\xspace}

\newcommand{\NLO}{\text{NLO}\xspace}    
\newcommand{\NLL}{\text{NLL}\xspace}
\newcommand{\NLLp}{\ensuremath{\text{NLL}^\prime}\xspace}
\newcommand{\NLOpNLL}{\ensuremath{\NLO+\NLL}\xspace}
    
\newcommand{\NLOpNLLp}{\ensuremath{\NLOpNLL^\prime}\xspace}

\def\beq{\begin{equation}}  
\def\eeq{\end{equation}}
\def\({\left(}
\def\){\right)}
\def\[{\left[}
\def\]{\right]}

\usepackage{color}
\definecolor{darkblue}{rgb}{0,0,0.5}
\definecolor{darkred}{rgb}{0.5,0,0}
\definecolor{darkgreen}{rgb}{0,0.5,0}

\newcommand{\NPPS}{$\left(d\sigma^{\rm HL} / d\lambda\right) / \left(d\sigma^{\rm PL} / d\lambda\right)$~}

\preprint{\newline MCNET-21-06}

\title{\boldmath Jet Angularities in Z+jet production at the LHC}

\author[a]{Simone~Caletti,}
\author[a]{Oleh~Fedkevych,} 
\author[a]{Simone~Marzani,}
\author[b]{Daniel~Reichelt,}
\author[b]{Steffen~Schumann,}
\author[d]{Gregory~Soyez,}
\author[b,d]{Vincent~Theeuwes}

\affiliation[a]{Dipartimento di Fisica, Universit\`a di Genova and INFN, Sezione di Genova,\\ Via Dodecaneso 33, 16146, Genoa, Italy}
\affiliation[b]{Institut f{\"u}r Theoretische Physik, Georg-August-Universit{\"a}t G\"ottingen,\\ Friedrich-Hund-Platz 1, 37077 G\"ottingen, Germany}
\affiliation[d]{Institut de Physique Th\'eorique, Paris Saclay University, CEA, CNRS, F-91191 Gif-sur-Yvette France}
\emailAdd{simone.caletti@ge.infn.it}
\emailAdd{simone.marzani@ge.infn.it}
\emailAdd{daniel.reichelt@uni-goettingen.de}
\emailAdd{steffen.schumann@phys.uni-goettingen.de}
\emailAdd{gregory.soyez@ipht.fr}
\emailAdd{vincent.theeuwes@uni-goettingen.de}

\abstract
{
We present a phenomenological study of angularities measured on the
highest transverse-momentum jet in LHC events that feature the
associate production of a $Z$ boson and one or more jets. In
particular, we study angularity distributions that are measured on
jets with and without the \softdrop grooming procedure. We begin our
analysis exploiting state-of-the-art Monte Carlo parton shower
simulations and we quantitatively assess the impact of next-to-leading
order (NLO) matching and merging procedures. We then move to analytic
resummation and arrive at an all-order expression that features the
resummation of large logarithms at next-to-leading logarithmic
accuracy (NLL) and is matched to the exact NLO result. Our predictions
include the effect of soft emissions at large angles, treated as a power
expansion in the jet radius, and non-global logarithms. Furthermore, matching
to fixed-order is performed in such a way to ensure what is usually referred
to as NLL$'$ accuracy. Our results account for realistic experimental cuts
and can be easily compared to upcoming measurements of jet angularities
from the LHC collaborations. \\[1cm]
}

\begin{document}
\maketitle

\section{Introduction}\label{sec:intro}
During the first two runs of the CERN Large Hadron Collider (LHC) the experimental collaborations
have accumulated a vast amount of high-quality data of proton--proton collisions at energies as high as $13$~TeV. In the
current shutdown period, the collaborations refine their search and measurement strategies, in order to
fully exploit the physics potential of the acquired data, and to prepare for the upcoming third run of the LHC.
However, no significant increase in collision energy is planned for the foreseeable future. Accordingly,
the focus of the theory and experimental communities should be on devising new effective and robust
analysis techniques to interrogate the data, so that no stone is left unturned. In this context, the use of
deep-learning algorithms to augment performance is now becoming standard practice
(see \textit{e.g.}~\citep{Larkoski:2017jix,Kasieczka:2019dbj,Benato:2020sbi} for recent reviews).
Complementary to this effort, measurement campaigns that deliver unfolded data that can be compared to
state-of-the-art theoretical calculations have been, and will further be, pursued in order to stress-test
our understanding of the Standard Model. 

The physics of jets and their structure plays a special role in this effort. First of all, jets,
\textit{i.e.}\ collimated sprays of particles, are ubiquitous objects in hadronic-collision events. Furthermore, in
experimental analyses, jets can be used in the context of searches for new physics, but also as a probe of
strong-interaction phenomena and dynamics. 
Measurements of jet cross sections provide stringent tests of our understanding of the strong interaction,
ranging over several orders of magnitude in the relevant energy scales. High-quality jet data are used in fits
for the strong coupling constant, see  \textit{e.g.}~\citep{Britzger:2017maj,Chatrchyan:2013txa,Aaboud:2017fml,ATLAS:2015yaa,CMS:2014mna}
and form an important input for constraining parton distribution functions (PDFs)~\citep{Aad:2021zrv,Aad:2013lpa,Khachatryan:2014waa,Khachatryan:2016mlc,AbdulKhalek:2020jut,Harland-Lang:2017ytb,Pumplin:2009nk,Watt:2013oha}. 

An accurate description of jets and their structure poses numerous theoretical challenges. Despite the fact that
$2 \to 2$ processes involving either two jets or a jet and an electroweak gauge boson are known to next-to-next-to-leading
order (NNLO) accuracy in the strong coupling expansion, the presence of multiple scales, such as the jet transverse momentum, the
jet radius and the substructure variables we would like to consider, \textit{e.g.}\ the jet mass, renders fixed-order
predictions in QCD insufficient. In order to achieve higher precision, fixed-order results need to be combined with
resummed calculations. 
However, all-order evaluations of jet substructure observables are far from trivial due to the presence of hard
boundaries in phase space, which give rise to non-global effects~\citep{Dasgupta:2001sh,Dasgupta:2002bw}, and because of the algorithmic
nature of jet definitions that make all-order factorisations difficult to achieve. These challenges have been tackled by the
theory community over the past decade and nowadays, thanks to numerous QCD studies, \textit{e.g.}~\citep{Dasgupta:2013ihk,Dasgupta:2013via,Larkoski:2014wba, Dasgupta:2015lxh,Salam:2016yht, Dasgupta:2016ktv,Dasgupta:2015yua,Larkoski:2014gra,Larkoski:2015kga,Larkoski:2020wgx, Larkoski:2013eya,Kang:2019prh,Cal:2019gxa,Cal:2020flh}, a deeper understanding of jet substructure has been reached. 
One of the lessons we have learnt from these investigations is that so-called grooming techniques, \textit{i.e.}\
jet substructure algorithms that aim to clean up a jet by removing from its constituents the ones that originate from soft
physics, actually improve our ability to use perturbation theory to describe jet physics. They reduce the impact of
non-perturbative effects due to hadronisation and the underlying event (UE), for a recent review, see \textit{e.g.}~\citep{Marzani:2019hun}.
Furthermore, the all-order structure of the perturbative result is also simplified because groomers, such as the
\textsf{modified-MassDrop}/\softdrop algorithm~\citep{Dasgupta:2013ihk,Larkoski:2014wba}, can eliminate the logarithmic enhancement due to soft gluons at wide angles,
including the intricate structure of non-global logarithms by turning logarithms of the observable under consideration into logarithms of an external parameter, such as $\zcut$ in the case of \softdrop. 

In this study we concentrate on high-energy proton--proton collisions that result in the production of an electroweak
$Z$ boson (decaying into a pair of muons) in association with one or more jets.  
For the jet with the highest transverse momentum, we are going to consider jet angularity observables~\citep{Larkoski:2014pca},
which are variables that probe the energy flow within a jet. Prior to the angularity evaluation we optionally perform the \softdrop
procedure on the candidate jet. The considered jet angularities form an interesting testbed for our theoretical
  understanding of intra-jet QCD dynamics. Furthermore, they over great application potential, \emph{e.g.}\ in distinguishing
  quark-like from gluon-like jets, in a theoretically well-defined way~\citep{Badger:2016bpw,Gras:2017jty,Amoroso:2020lgh}, or
  in extractions of the strong coupling constant~\citep{Bendavid:2018nar}.
Our analysis begins with a detailed phenomenological study of jet angularities, which is performed exploiting state-of-the-art
Monte Carlo (MC) parton shower simulations. Beside assessing the relative importance of the different contributions that affect the
angularity distributions, such as perturbative radiation, hadronisation and the UE, our studies aim to quantify
the role of modern NLO merging techniques on jet substructure distributions. We then move to the second part of the paper,
where we present first-principle theoretical predictions for the observables of interest. Our resummed calculation is performed
at next-to-leading logarithmic accuracy (NLL) matched to fixed-order predictions evaluated at NLO in $\alpha_s$.
Furthermore, by keeping track of the jet flavour in our matching procedure, we are able to obtain what is usually referred to
as \NLLp accuracy. We stress that such accuracy is achieved for both the groomed and the ungroomed distributions, which implies
that we account for single-logarithmic corrections originating from soft emissions at wide angle, that we treat as an expansion
in the jet radius parameter, and from non-global logarithms, in the limit of large number of colours ($N_c$).\footnote{The counting
  is different in the case of \softdrop jets with angular exponent $\beta=0$ because only collinear radiation is kept by the groomer
  and the resulting distributions are single-logarithmic. In this case, our calculations correctly capture the leading-logarithmic
  contributions with non-vanishing coefficients.}
Furthermore, our initial detailed study of MC predictions allows us to supplement our perturbative \NLOpNLLp predictions
with non-perturbative corrections extracted from hadron-level MC simulations.
Our theoretical predictions are fully differential in the kinematics of the jets and of the leptonic decay products
of the $Z$ boson, so that we can impose realistic fiducial cuts and compare them to the distributions obtained from MC
event generators. Our predictions are obtained with the resummation plugin of the \sherpa generator framework~\citep{Gerwick:2014gya}
and they can be directly compared to unfolded measurements from the LHC collaborations, once these become available.\footnote{We note that similar studies have been performed also in the context of Soft-Collinear Effective Theory (SCET), see \textit{e.g.}~\cite{Ellis:2010rwa,Hornig:2016ahz,Kang:2018qra,Kang:2018vgn}. To the best of our knowledge, those calculations, when performed for proton--proton collisions, are limited to the leading contribution in the small jet-radius limit.}

The paper is organised as follows: in Section~\ref{sec:def} we introduce the class of angularity observables to consider
and detail our event selection cuts. In Section~\ref{sec:def-mc} we present and compare hadron-level predictions for groomed and ungroomed
jet angularities from the \herwig and \pythia event generators, based on the leading-order
matrix elements, and \sherpa, using multijet merging at leading
and next-to-leading order. Section~\ref{sec:resummation} is devoted to the all-orders evaluation of jet angularities at \NLOpNLLp accuracy
using the \sherpa resummation framework. In Section~\ref{sec:mc-comp} we compare our resummed predictions with the parton-level results of
our MC simulations. We extract non-perturbative corrections from the full particle-level simulations and apply them to our
\NLOpNLLp predictions. Our conclusions are presented in Section~\ref{sec:conclusions}. Additional results are collected in Appendix~\ref{app:appendix}.

\section{Observable definition}\label{sec:def}
In this study we concentrate on the production of a leptonically decaying $Z$ boson associated with one or more jets at the LHC.
Jets are defined using the anti-$k_t$ clustering algorithm~\citep{Cacciari:2008gp} with radius parameter $R_0$ and standard $E$-scheme
recombination, \emph{i.e.}\ the momenta of objects that are paired together are simply added so that the resulting jet momentum is given by
the sum of its constituents' momenta. 

On the hardest jet, \textit{i.e.}\ the jet with the largest transverse
momentum $p_{T, \text{jet}}$, we measure the
angularities~\citep{Berger:2003iw,Almeida:2008yp,Larkoski:2014pca}
\begin{equation}\label{eq:ang-def}
\la^\kappa= \sum_{i \in \text{jet}}\left(\frac{p_{T,i}}{\sum_{j \in \text{jet}} p_{T,j}}\right)^\kappa\left(\frac{\Delta_i}{R_0} \right)^\alpha\,,
\end{equation}
where 
\begin{equation}\label{eq:dist-def}
\Delta_i=\sqrt{(y_i-y_\text{jet})^2+(\phi_i-\phi_\text{jet})^2}\,, 
\end{equation}
is the Euclidean azimuth-rapidity distance of particle $i$ from the jet axis. 
Our study is limited to a subset of the above defined general angularities,
namely the ones that exhibit infrared and collinear (IRC) safety. This poses
the restrictions $\kappa=1$ and $\alpha>0$. 
Furthermore, it is well-known~\citep{Banfi:2004yd,Larkoski:2013eya}
that IRC safe angularities with $\alpha \le 1$ are sensitive to recoil
against soft emissions, leading to a rather complicated resummation structure. To circumvent these additional complications,
we compute for these angularities the distance measure in Eq.~\eqref{eq:dist-def} with respect to the jet axis
obtained by reclustering the jet constitutents with the anti-$k_t$ algorithm but using the Winner-Take-All (WTA) recombination
scheme~\citep{Larkoski:2014uqa}.
We are also interested in computing angularities on groomed jets. In this case,
we recluster the jet with the Cambridge--Aachen (C/A) algorithm and consider
the \softdrop grooming algorithm with parameters $\zc$ and $\beta$~\citep{Larkoski:2014wba}. The angularity is then
computed on the resulting \softdrop jet, \textit{i.e.}\ both sums in Eq.~\eqref{eq:ang-def} are restricted to the particles
that survived the grooming. 
For groomed jets we also adopt the WTA prescription for angularities with $\alpha \le 1$. 

In order to highlight the behaviour of the angularity distributions in different energy regimes, standard and groomed
angularities are considered in different jet transverse momentum bins. This will allow us, for instance, to better study the
impact of non-perturbative corrections, \emph{e.g.}\ from the parton-to-hadron transition.
We note that, in order to avoid issues related to bin-migration, which greatly complicates the structure of the resummation
of  the \softdrop $\beta=0$ (\textsf{modified MassDrop}) angularities~\citep{Marzani:2017mva}, we shall always choose the reference transverse momentum $p_{T,\text{jet}}$
to be the one \emph{before} grooming. 
Our theoretical predictions, both the ones obtained with matrix-element improved parton showers and with all-order resummation
matched to fixed-order calculations, account for fiducial cuts and therefore can directly be compared to unfolded measurements. 

\subsection*{Event selection cuts}

We close this section by detailing the fiducial volume for the jets
and leptons in our subsequent studies.
Our choices follow the selection of a recent (preliminary) CMS measurement~\cite{CMS-PAS-SMP-20-010}.
We consider the inclusive production of a pair of oppositely charged muons in proton--proton collisions at
$13~\text{TeV}$ centre-of-mass energy. We require all final state particles to
have pseudo-rapidity $|\eta|<5$. For both muon candidates we require
\begin{equation}
  p_{T,\mu} > 26~\mathrm{GeV}\,,\;\;\text{and}\;\;|\eta_\mu|<2.4\,.
\end{equation}
The lepton pair has to pass the additional conditions
\begin{equation}
\quad70~\mathrm{GeV}<m_{\mu^+\mu^-}<110~\mathrm{GeV}\,,\;\;\text{and}\;\; p_{T,\mu^+\mu^-}>30\;\text{GeV}\,.
\end{equation}
In what follows we will refer to the lepton pair as $Z$ boson, where
nevertheless we imply the inclusion of off-shell effects. We then select events
that exhibit at least one anti-$k_t$ jet \citep{Cacciari:2008gp} with 
\begin{equation}
  |y_{\text{jet}}| < 1.7\,,\quad\text{and}\quad R_0=0.8~\footnote{Results for $R_0=0.4$ are collected in Appendix~\ref{app:NLL_NP_R4}.},
\end{equation}
and consider several bins in $p_{T,\text{jet}}$ starting at $50~\text{GeV}$
following~\cite{CMS-PAS-SMP-20-010}. We compiled
results for all bins used there\footnote{Our framework is, of course,
  general and theoretical predictions for different jet radii, angularity exponents and \softdrop
  parameters, as well as different selection cuts and $p_{T,\text{jet}}$ bins,
  can be easily obtained. They can be provided upon request.}, but will in the
following limit the discussion to $p_{T,\text{jet}}\in [120,150]~\text{GeV}$ and
$p_{T,\text{jet}}\in [408,1500]~\text{GeV}$ that we find representative for the
lower and higher jet scale selections.
In order to gain better control over NLO QCD corrections for the $Zj$ production process,
that become large when the $Z$-boson transverse momentum is significantly smaller than
$p_{T,\text{jet}}$~\citep{Rubin:2010xp}, we require the transverse momenta of the lepton pair
and the leading jet to be largely balanced. To this end we impose the constraint
\begin{equation}\label{eq:imbalance_cut}
\Delta^{p_T}_{Z, {\rm jet}}  \equiv \left| \frac{p_{T,\rm jet} - p_{T, \mu^+\mu^-} }{ p_{T,\rm jet} + p_{T, \mu^+\mu^-}}  \right| < 0.3\,.
\end{equation}
Finally, we require the $Z$-boson and the leading jet to be well separated in azimuthal angle, \textit{i.e.}\
\begin{equation}
  \Delta^\phi_{Z, {\rm jet}}  \equiv \left|\phi_Z - \phi_{\rm jet}\right| > 2\,.
\end{equation}

\noindent
In our following studies we will consider the jet angularities $\la^\kappa$
with $\left(\kappa, \alpha\right) = (1, 1/2), \, (1, 1)$ and  $(1, 2)$. In previous studies~\citep{Larkoski:2014pca,Badger:2016bpw}
these have been dubbed Les Houches Angularity (LHA) $\lambda^1_{1/2}$, Width $\lambda^1_1$, and
Thrust $\lambda^1_2$, respectively. We adopt this naming convention here. For \softdrop grooming we
use $\beta = 0$, $\zcut = 0.1$ throughout.

\section{State-of-the-art Monte Carlo study of jet angularities}\label{sec:def-mc}

The Born-level contributions to $Zj$ production are given by the two channels
\mbox{$pp \rightarrow \mu^+\mu^- + q/g$}, where $q$ represents a massless quark or
anti-quark.
Jet angularities become non-trivial upon inclusion of additional radiation, which in
general purpose MC generators is accounted for by parton shower simulations~\citep{Buckley:2011ms}.
However, to properly include non-logarithmic corrections,
\textit{e.g.}\ from hard real emissions, these should be matched to
exact higher-order matrix elements, as obtained, for example using
event generators such as
\mcfm~\cite{Campbell:1999ah,Campbell:2011bn,Boughezal:2016wmq},
\powheg~\cite{Nason:2004rx, Frixione:2007vw,Alioli:2010xd} or
\mgamcnlo~\cite{Frixione:2002ik,Alwall:2014hca}. 
Here we consider simulations including the complete set of NLO QCD corrections to the
$Zj$ and $Zjj$ production processes matched to the \sherpa parton shower in
the MEPS@NLO formalism~\citep{Hoeche:2012yf}. To address the sensitivity of the
angularities to non-perturbative corrections we also account for UE contributions to
the collision final states and model the parton-to-hadron
transition. In this section we describe the calculational  setups for our
Monte Carlo simulations and then present corresponding predictions for the
groomed and ungroomed LHA, Width, and Thrust angularities.

\subsection{Monte Carlo generator setup}\label{sec:mc-studies}

We compile MC predictions for the jet angularities with the
\sherpa~\citep{Gleisberg:2008ta,Bothmann:2019yzt} event generator, version 2.2.10,
using the NNPDF-3.0 NNLO PDF set \citep{Ball:2014uwa}.
We consider the inclusive $Zj$ production process in the MEPS@NLO
multijet merging formalism~\citep{Hoeche:2012yf}, thereby combining
the NLO QCD matrix elements for $\mu^+\mu^-j$ and $\mu^+\mu^-jj$ production,
matched with the \sherpa Catani--Seymour dipole shower~\citep{Schumann:2007mg}.
We set  the merging-scale parameter to
$Q_{\text{cut}}=30\,\text{GeV}$ (which is of the order of the jet
$p_T$ cut used in our event selection).
We obtain the QCD one-loop amplitudes for the one- and two-jet processes
from \recola~\citep{Actis:2016mpe,Biedermann:2017yoi}, using
the \collier library~\citep{Denner:2016kdg} for the evaluation of tensor and
scalar integrals. To assess the impact of the QCD one-loop corrections, we
furthermore compile MEPS@LO~\citep{Hoeche:2009rj} predictions based on merging
the one- and two-jet leading-order matrix elements, using $Q_{\text{cut}}=30\,\text{GeV}$
as well.

The perturbative scales entering the calculation are defined according to
the CKKW-style scale setting prescription~\citep{Catani:2001cc,Hoeche:2009rj}.
In this procedure the hard-process partons are clustered into a Born-like
$2\to 2$ configuration that defines the \emph{core process} with an associated
scale $\mu_{\text{core}}$. For the production channels considered here the event-wise
determined core process, of order $\alphaS^{n_{\text{core}}}$, can correspond
to $jj\to\mu^+\mu^-$ ($n_{\text{core}}=0$), $jj\to Zj$ ($n_{\text{core}}=1$),
or $jj\to jj$ ($n_{\text{core}}=2$). For the three possible cluster configurations
the corresponding core-process scale is given by
\begin{equation}
  \mu_{\text{core}}=\left\{\begin{array}{lcl}
  m_{\mu^+\mu^-} &:& n_{\text{core}}=0\\
  \sqrt{m^2_{\mu^+\mu^-}+p^2_{T,\mu^+\mu^-}} &:& n_{\text{core}}=1\;.\\
  p_{T,j} &:& n_{\text{core}}=2
  \end{array}\right.
\end{equation}

The core-process scale is then used to define
the factorisation scale and the parton shower starting scale of the core process, \textit{i.e.}\
\begin{equation}
  \mu_{\text{F}}=\mu_{\text{core}}\,,\quad \mu_{\text{Q}}=R_0\mu_{\text{core}}\,.
\end{equation}
Our choice of $\mu_{\text{Q}}$ here is motivated by the corresponding scale in
the resummed calculation, \emph{c.f.} Sec.~\ref{sec:resummation}. The effective
renormalisation scale, $\mu_{\text{CKKW}}$, of the $n$-parton hard matrix
elements corresponds to 
\begin{equation}
\alphaS^n(\mu^2_{\text{CKKW}})=\alphaS^{n_{\text{core}}}(\mu^2_{\text{core}})\prod\limits_{i=1}^{n-n_{\text{core}}}\alphaS(t_{i})\,,
\end{equation}
with $t_i$ the reconstructed shower-branching scales. 
To estimate the perturbative uncertainties of our MC predictions,
we do on-the-fly~\citep{Bothmann:2016nao} $7$-point
variations~\citep{Cacciari:2003fi} of the factorisation and
renormalisation scales in the matrix elements and the parton shower. The uncertainty
bands given later on correspond to the envelope of the settings
$\{(\tfrac{1}{2}\muR, \tfrac{1}{2}\muF)$, $(\tfrac{1}{2}\muR,\muF)$,
$(\muR,\tfrac{1}{2}\muF)$, $(\muR,\muF)$, $(\muR,2\muF)$, $(2\muR,\muF)$, $(2\muR,2\muF)\}$.

To model the parton-to-hadron transition we use the \sherpa cluster fragmentation
model~\citep{Winter:2003tt}. The UE simulation relies on the \sherpa
implementation of the Sj\"ostrand--Zijl multiple-parton interaction
model~\citep{Sjostrand:1987su}. In both models the default set of tuning parameters
is used, see~\citep{Bothmann:2019yzt} for details.

To obtain further independent predictions, in particular for the modelling of
non-perturbative effects, we compile additional results from
\herwig~\citep{Bellm:2015jjp, Bellm:2019zci}, version 7.2.1, and \pythia~\citep{Sjostrand:2014zea},
version 8.303. For both generators we consider the
$\mu^+\mu^-j$-production process at leading order,\footnote{In the
  case of \herwig, we could have also included NLO-accurate matrix
  elements~\citep{Platzer:2011bc,Bellm:2017ktr}. However, since the baseline for our MC simulations are \sherpa MEPS@NLO predictions, we restrict ourselves to leading-order matrix elements for
  simplicity.}
invoking the respective default models for the QCD parton shower, hadronisation and
UE simulation.

For event selection and analysis we employ the \rivet analysis
package~\citep{Buckley:2010ar,Bierlich:2019rhm}. For jet reconstruction we use
the \fastjet~package~\citep{Cacciari:2011ma}. For the \softdrop grooming we
rely on the implementation in the \verb|RecursiveTools| class which is
a part of the \fastjet ~\verb|contrib| package. 

\subsection{Hadron-level Monte Carlo predictions}

To set the stage we begin by directly comparing the hadron-level predictions obtained from the
various event generators, \emph{i.e.}\ calculational setups, for the three considered jet angularities,
both, with and without \softdrop grooming. These will later serve as means
to extract non-perturbative corrections and estimates for related uncertainties for our resummed
predictions, see Section~\ref{sec:mc-comp}. 
%

\begin{figure}
  \centering
  \includegraphics[width=0.44\linewidth]{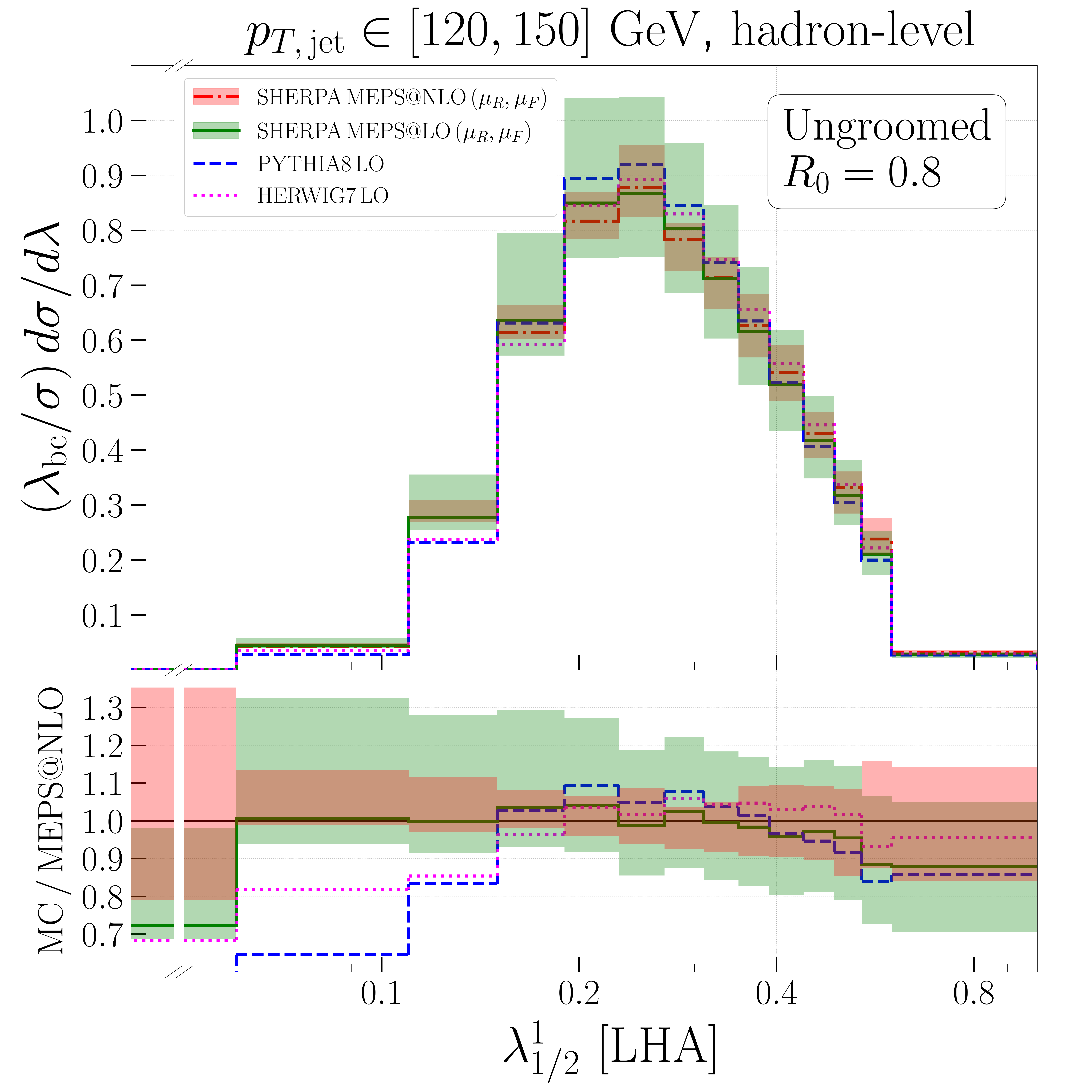}
  \hspace{1em}
  \includegraphics[width=0.44\linewidth]{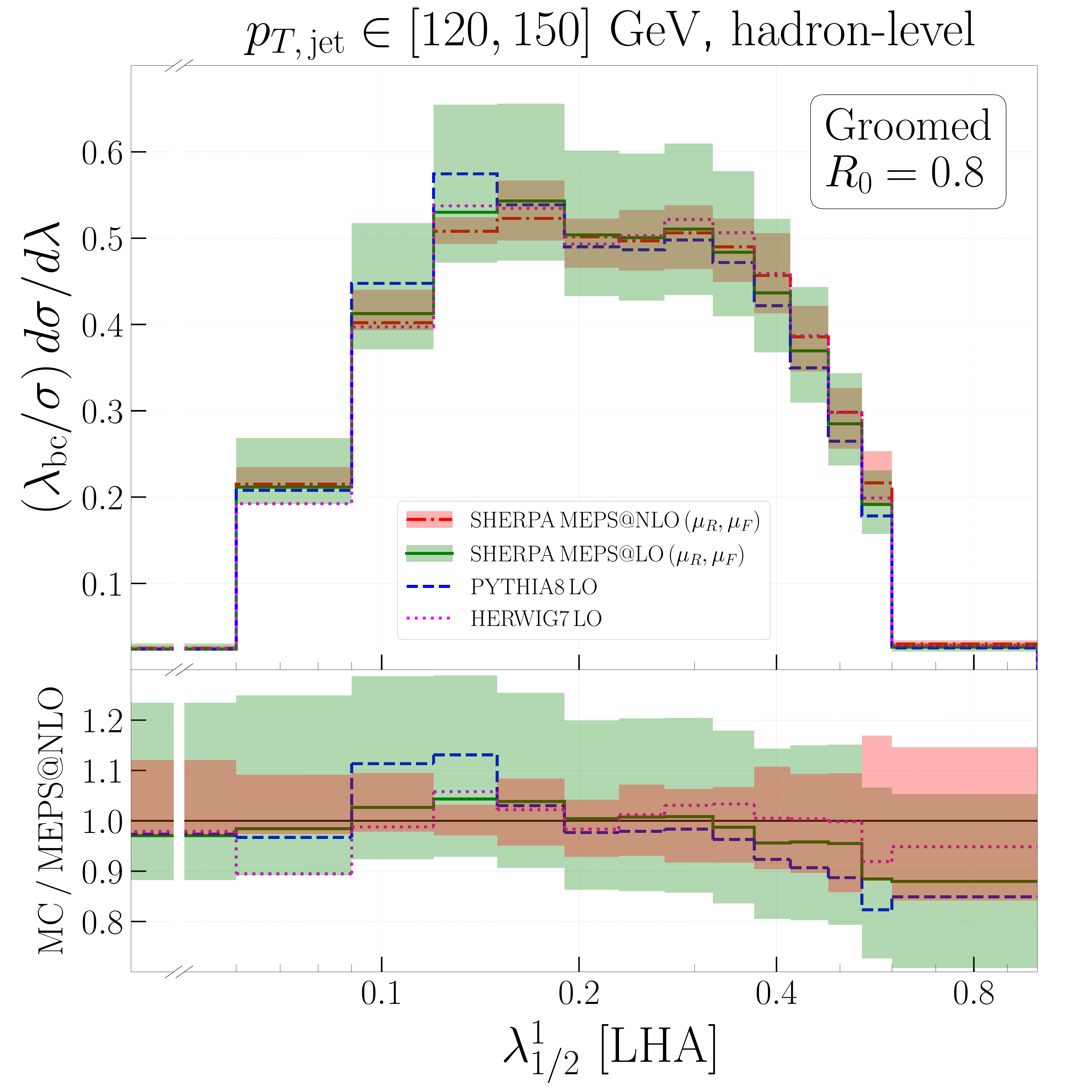}
  \centering
  \includegraphics[width=0.44\linewidth]{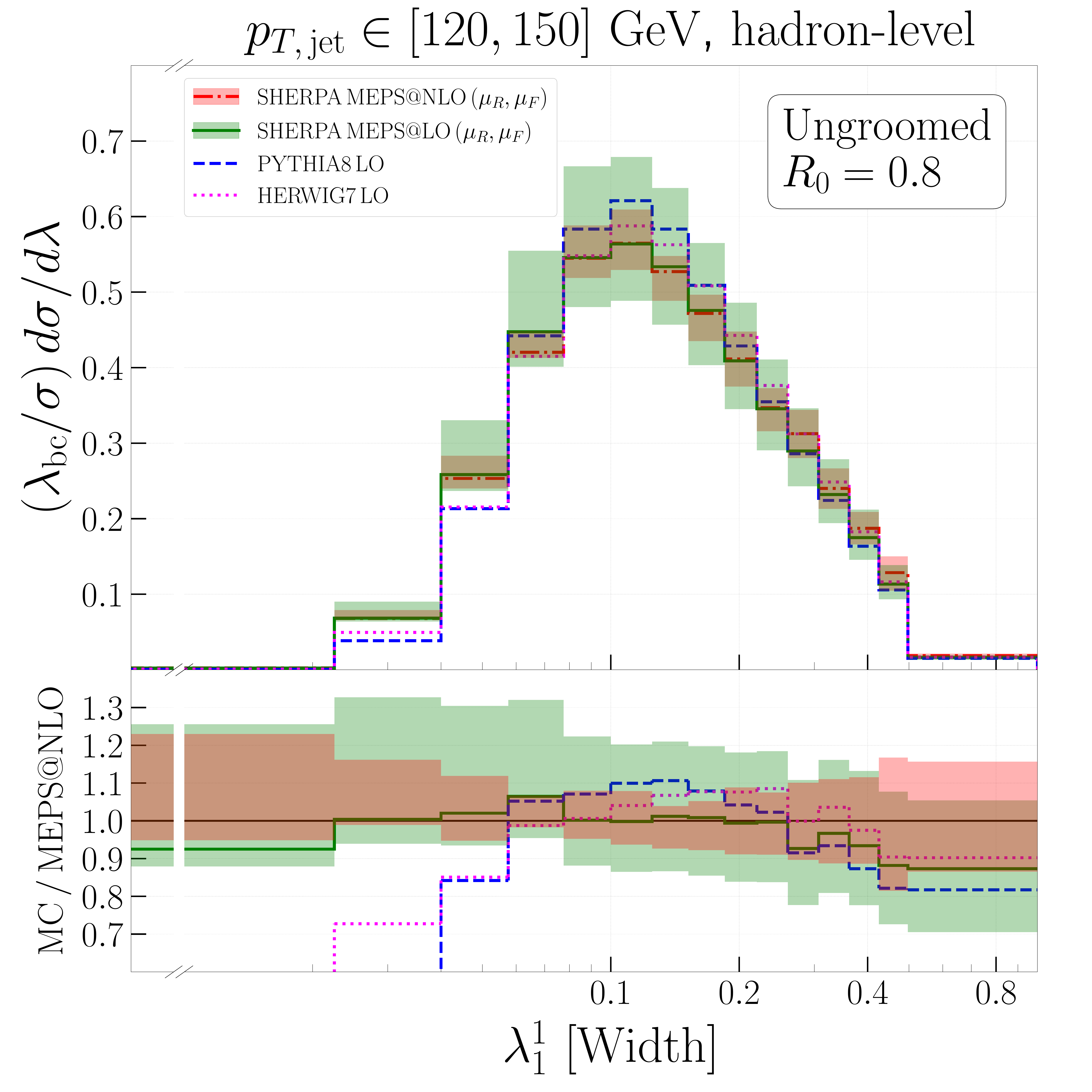}
  \hspace{1em}
  \includegraphics[width=0.44\linewidth]{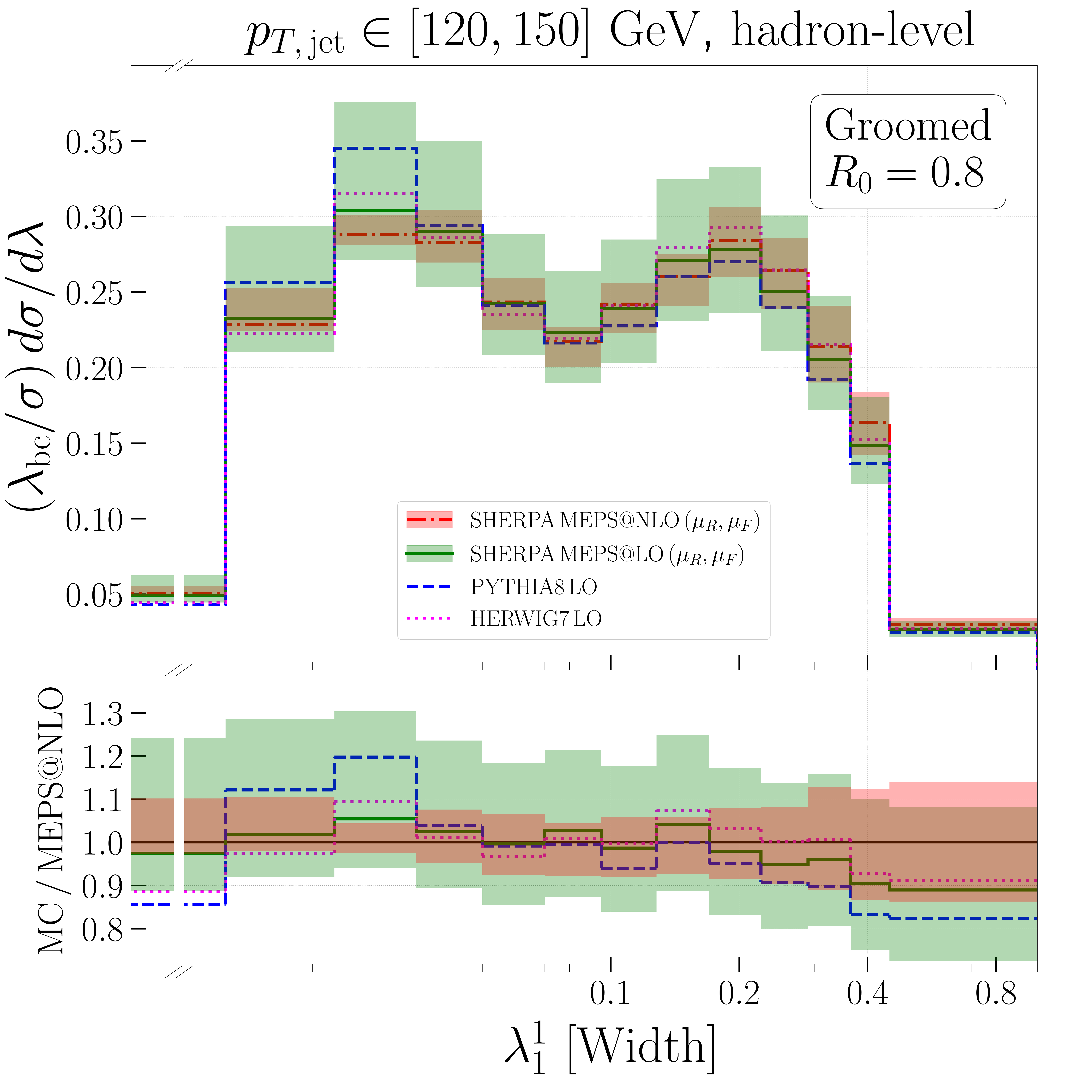}
  \centering
  \includegraphics[width=0.44\linewidth]{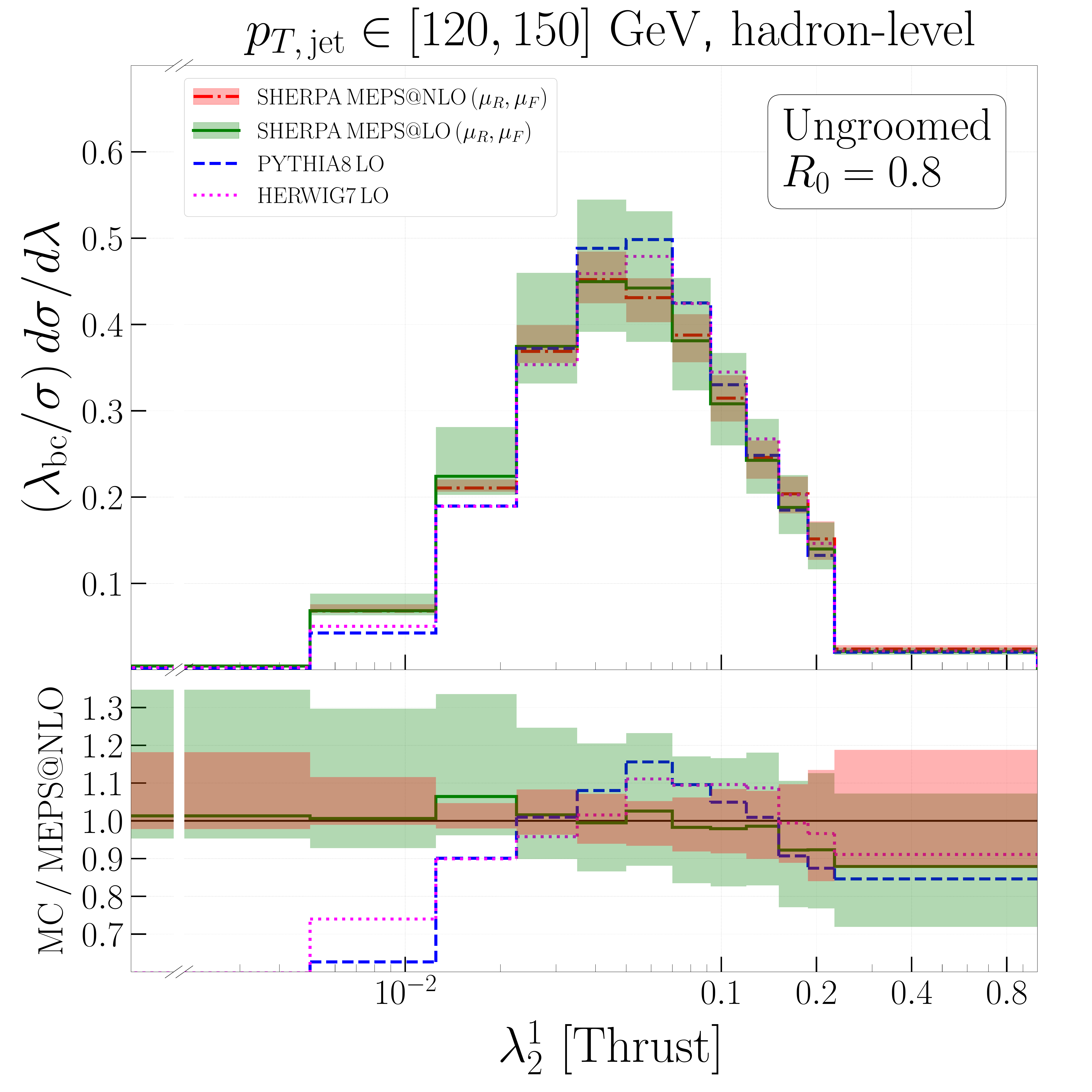}
  \hspace{1em}
  \includegraphics[width=0.44\linewidth]{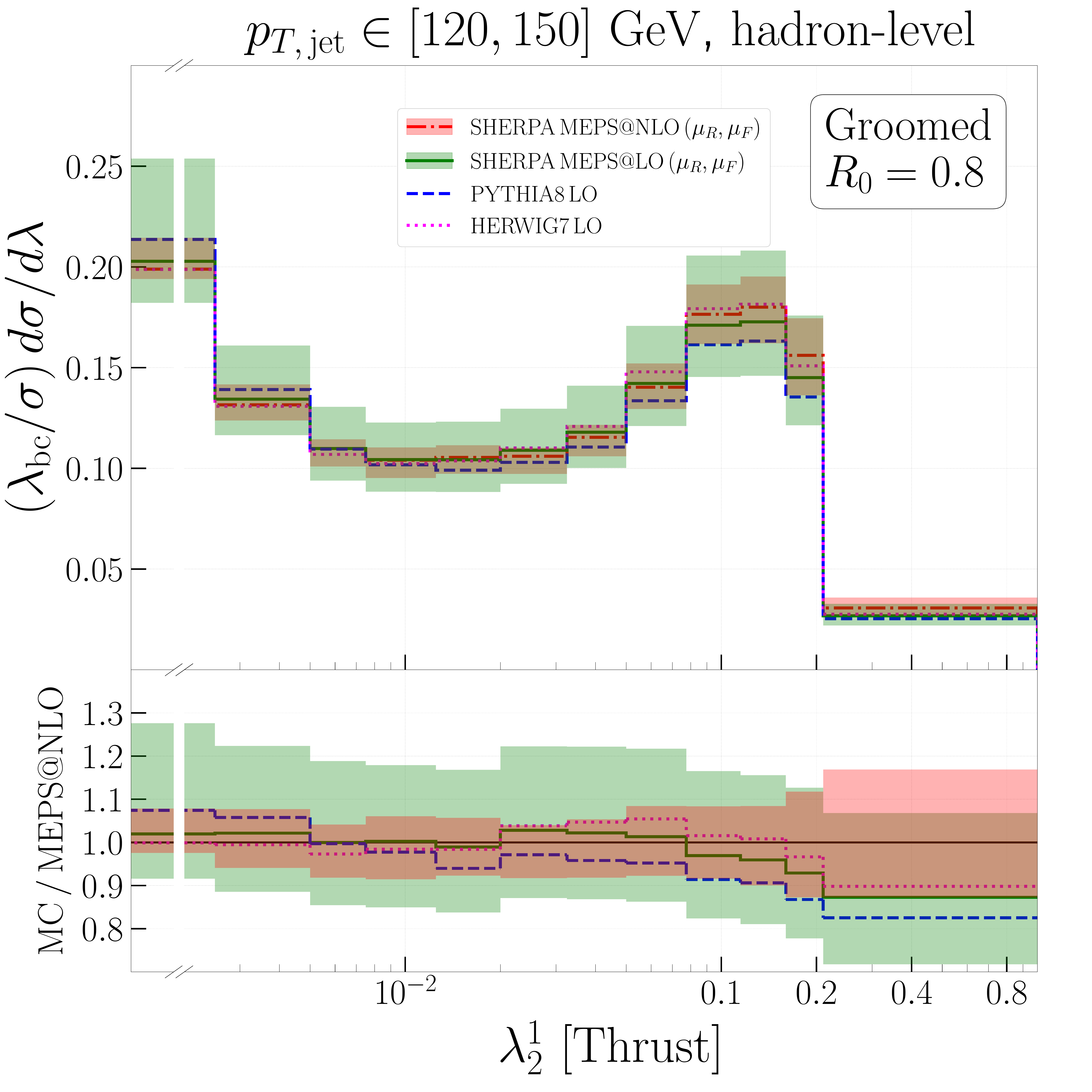}
  \caption{Comparison of hadron-level predictions for ungroomed and groomed jet-angularities in $Zj$ production from \pythia and \herwig (both based on the LO $Zj$ matrix element), and MEPS@LO as well as MEPS@NLO results from \sherpa. Here  $p_{T,\text{jet}}\in[120,150]\;\text{GeV}$ and $\lambda_\text{bc}$ stands for the bin centre.}
\label{fig:all_mc_np_log_pT120} 
\end{figure}

\begin{figure}
  \centering
  \includegraphics[width=0.44\linewidth]{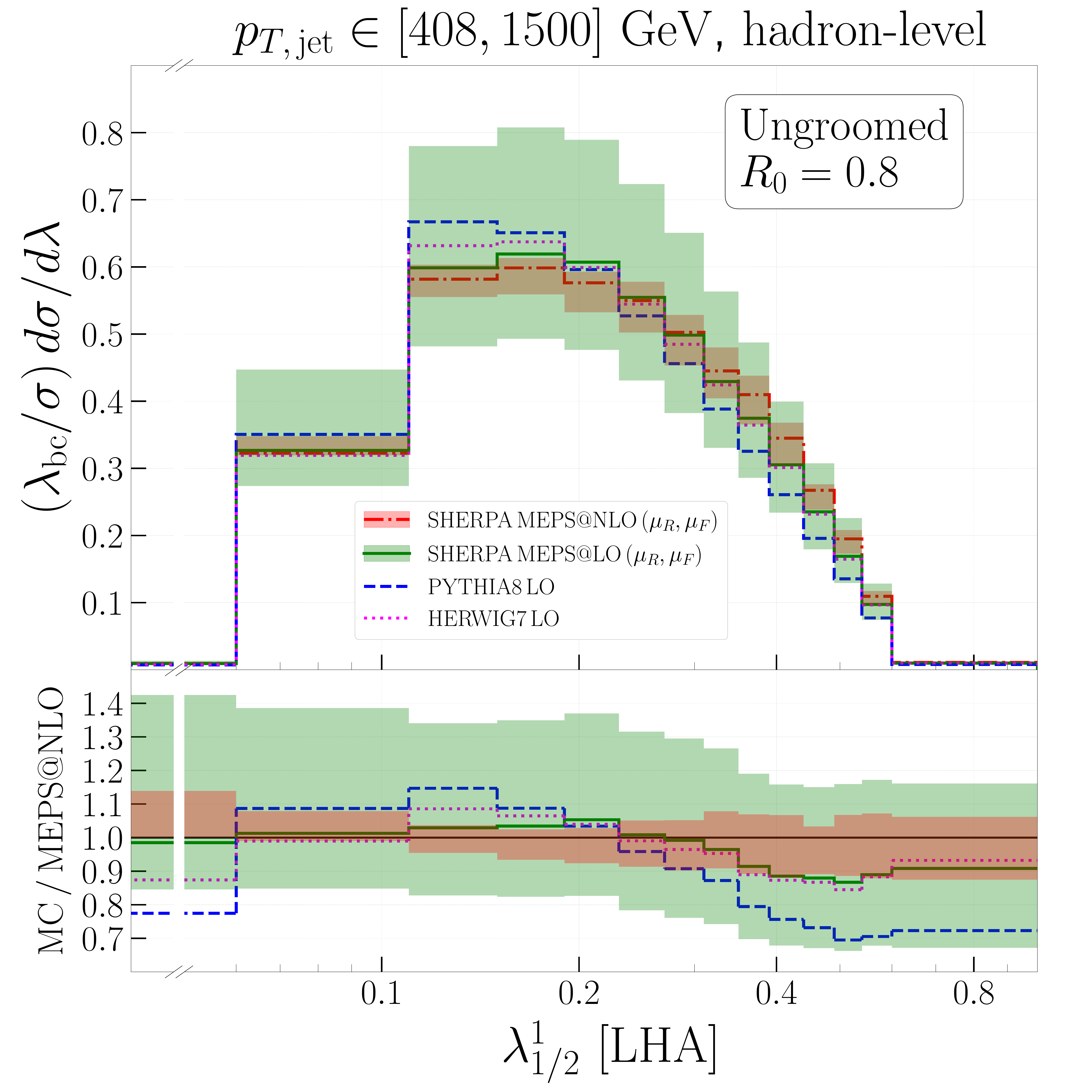}
  \hspace{1em}
  \includegraphics[width=0.44\linewidth]{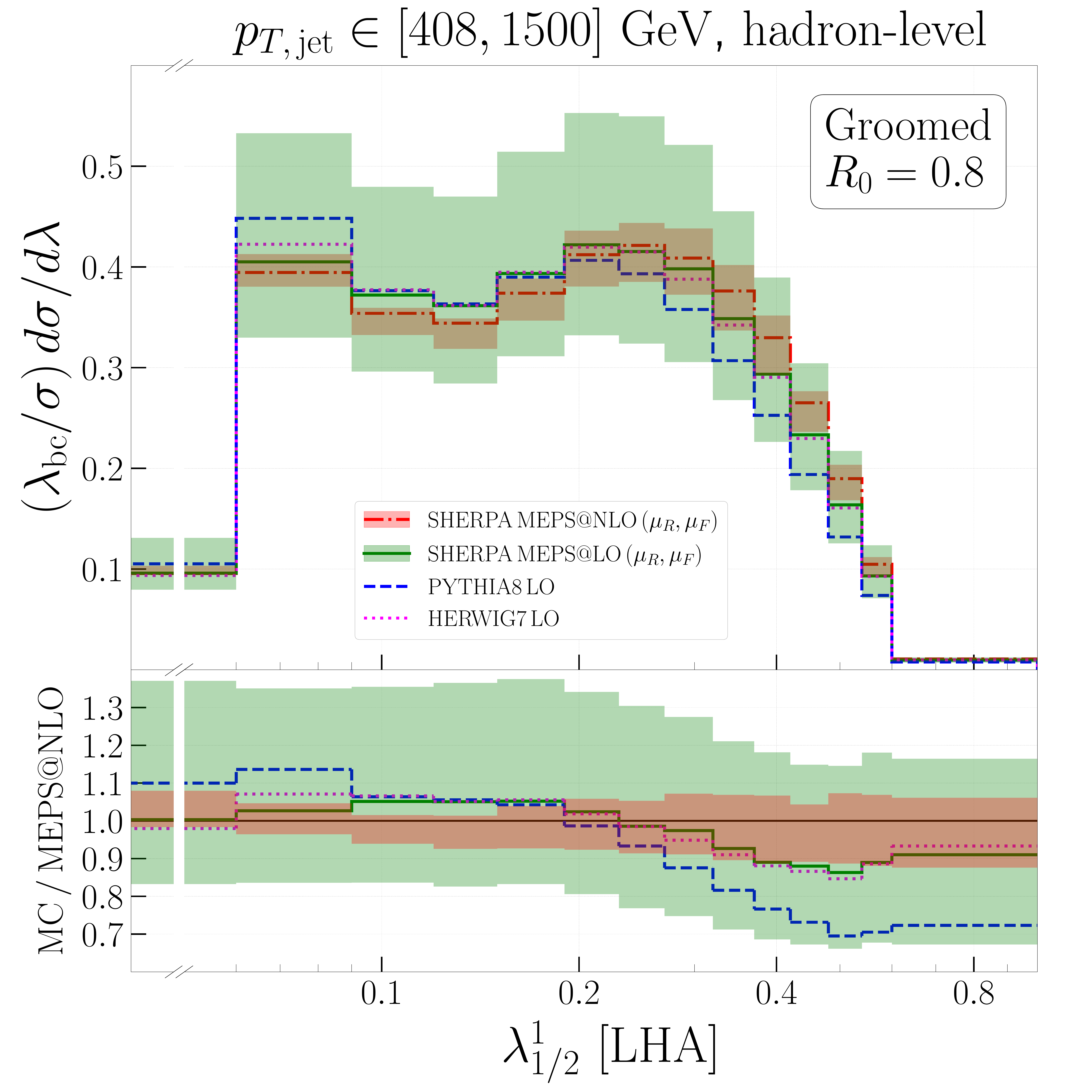}
  \centering
  \includegraphics[width=0.44\linewidth]{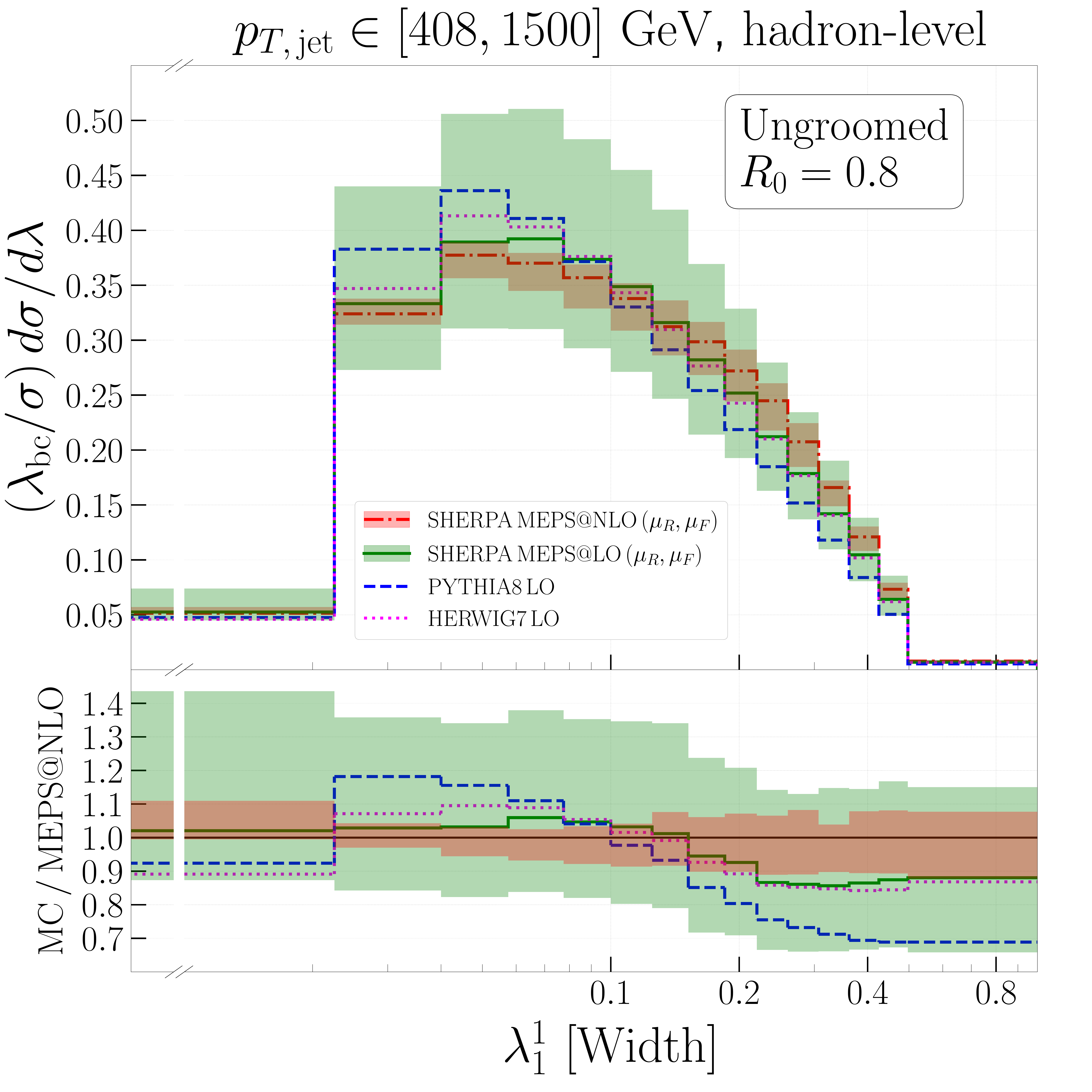}
  \hspace{1em}
  \includegraphics[width=0.44\linewidth]{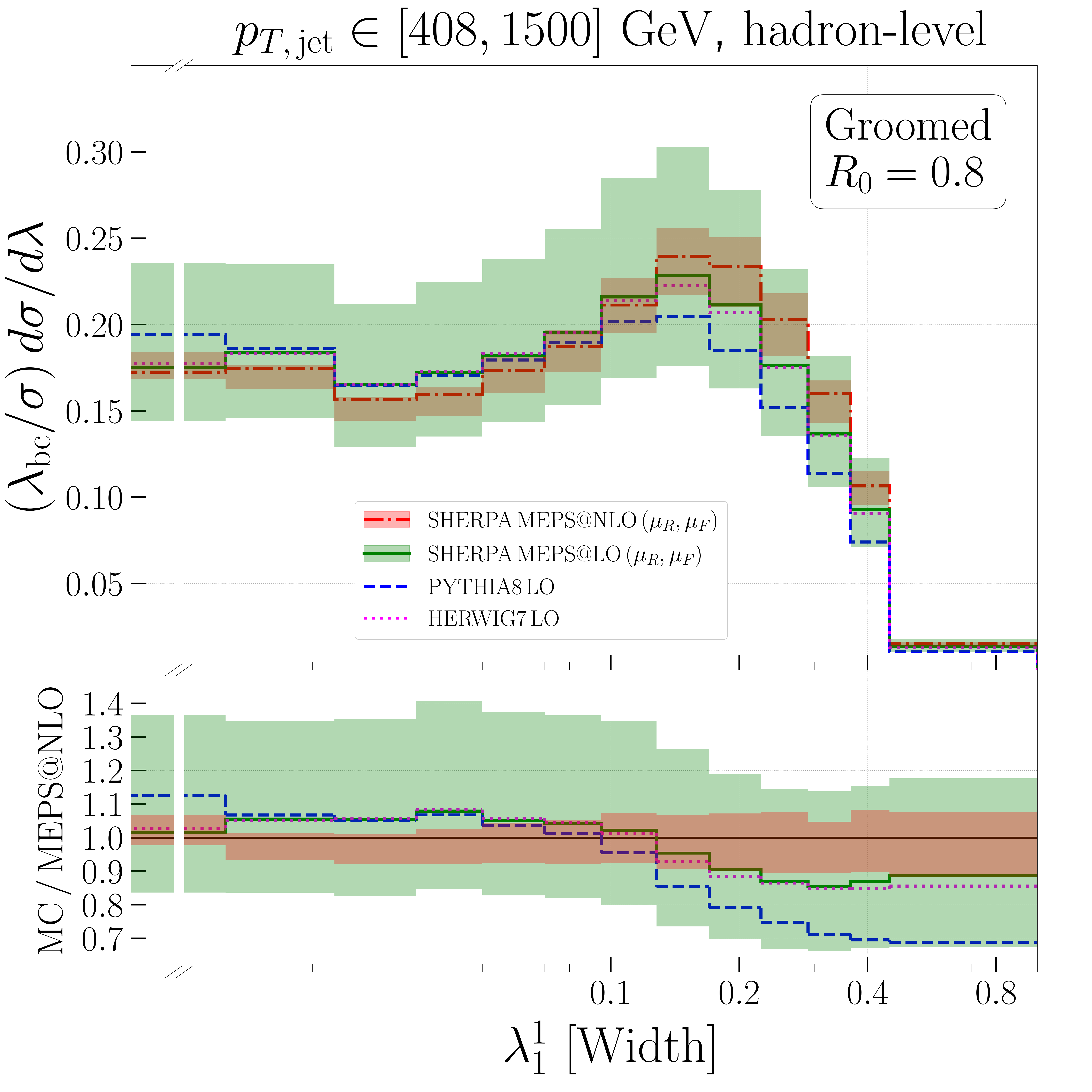}
  \centering
  \includegraphics[width=0.44\linewidth]{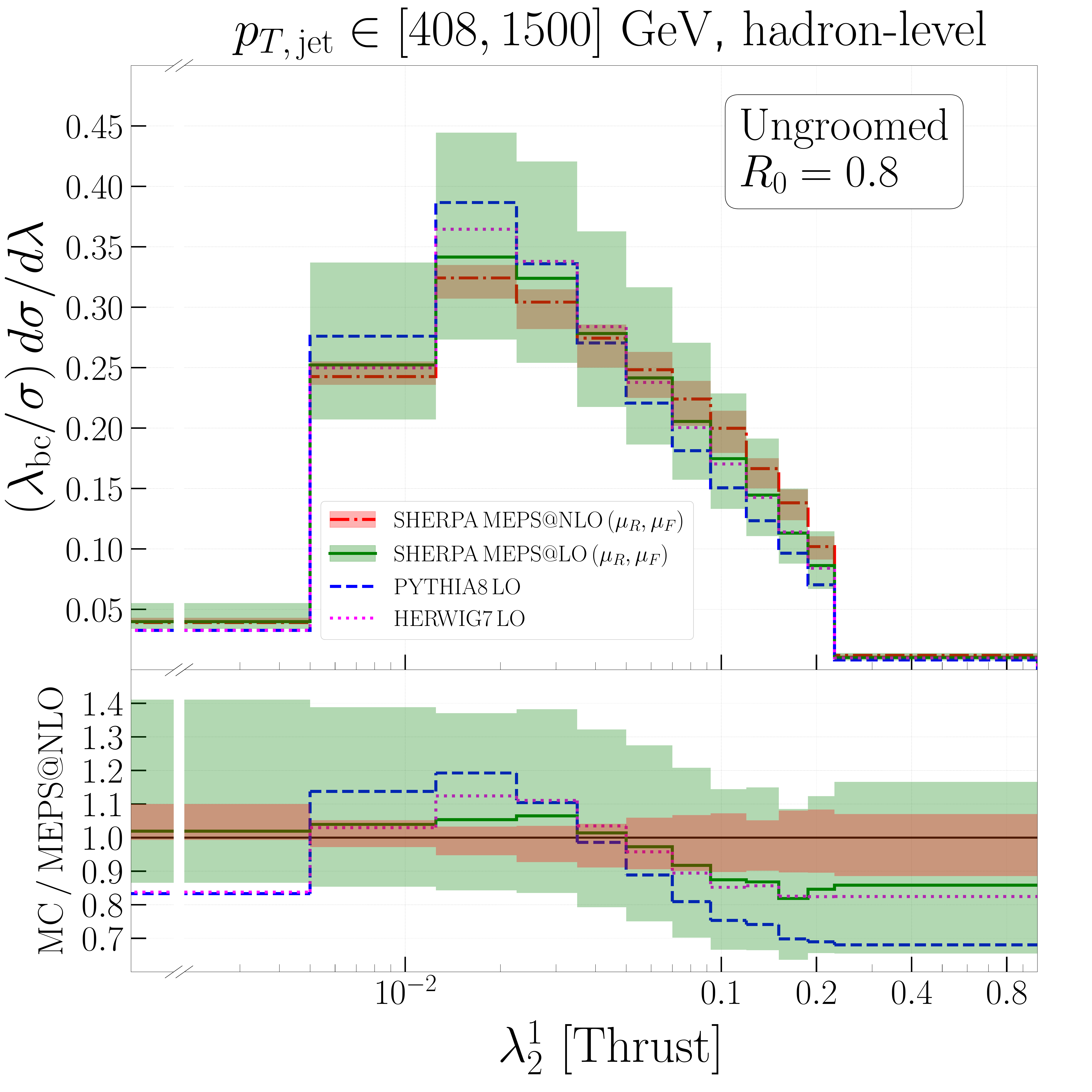}
  \hspace{1em}
  \includegraphics[width=0.44\linewidth]{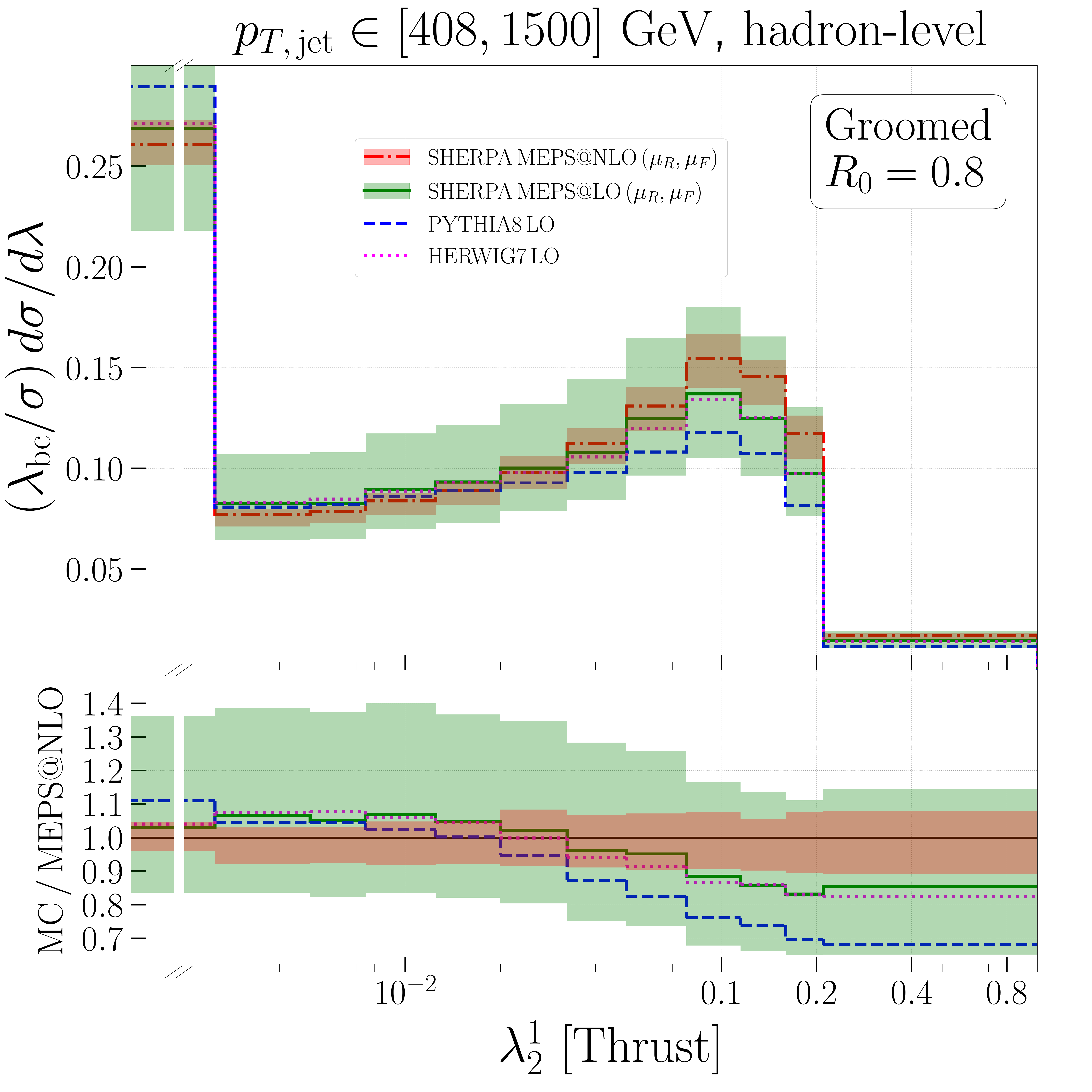}
  \caption{Comparison of hadron-level predictions for ungroomed and groomed jet-angularities in $Zj$ production from \pythia and \herwig (both based on the LO $Zj$ matrix element), and MEPS@LO as well as MEPS@NLO results from \sherpa. Here  $p_{T,\text{jet}}\in[408,1500]\;\text{GeV}$ and $\lambda_\text{bc}$ stands for the bin centre.}
\label{fig:all_mc_np_log_pT408} 
\end{figure}

In Fig.~\ref{fig:all_mc_np_log_pT120} and Fig.~\ref{fig:all_mc_np_log_pT408}  we collate the generators'
predictions for the lower and higher $p_{T,\text{jet}}$ slice, respectively. For each angularity we provide
results without (left column) and with (right column) \softdrop grooming. The broken $x$-axes indicate
the fact that the first bins start at zero and, hence, they appear unbounded on a logarithmic scale. 
For each individual panel we include a ratio plot, with the \sherpa MEPS@NLO result taken as reference.
For the \sherpa MEPS@LO and MEPS@NLO predictions we include uncertainty bands, green solid and red hatched,
respectively, that reflect the envelope of the 7-point scale variations in both the matrix-element and
parton-shower component. The inclusion of the exact NLO QCD one-loop corrections in the MEPS@NLO method
results in a significant reduction of scale uncertainties, for all angularities both in their groomed and
ungroomed variants.  

Considering the ungroomed angularities for the lower $p_{T,\text{jet}}$ window first, we observe
that with increasing $\alpha$ the distributions peak at lower observable values. While the
LHA angularity distribution exhibits a Sudakov peak around  $\lambda^1_{1/2}\approx 0.25$, for
the Thrust variable the maximum is at $\lambda^1_2\approx 0.05$. All generator predictions largely
agree on the peak position. However, for lower observable values quite sizeable deviations can be
observed, reaching and partially exceeding $50\%$. Notably, the two \sherpa predictions agree quite nicely
in this observable range, that in fact is significantly affected by non-perturbative corrections,
described by the same models in the MEPS@LO and MEPS@NLO simulations. However,
\herwig and \pythia use alternative models and parameter tunes for hadronisation and UE. Both generators
predict somewhat narrower distributions in comparison to \sherpa.
For large values of the angularities, \emph{i.e.}\ towards the
kinematical endpoints, the MEPS@NLO calculation predicts somewhat
larger event fractions than what is obtained using LO matrix elements,
with the largest deviation appearing in comparison to the \pythia LO
result. This systematic effect which barely exceeds 15\% is, however,
within the scale uncertainty band.
It is interesting to note that a similar pattern between LO and NLO
matrix elements was already observed
by the CMS collaboration~\cite{Sirunyan:2018xdh} when comparing their
measurement of the groomed jet mass to analytic QCD predictions.

Upon invoking \softdrop grooming of the jet constituents, the spread in the generator predictions
is sizeably reduced. In particular, in the region of very small observable values, dominated by
non-perturbative effects in the ungroomed case, the predictions now agree to within 10\%.
It is clearly visible that grooming decreases the jet-angularity values, and in fact
significantly sculpts the observable distributions. With increasing $\alpha$, a secondary
peak emerges below the grooming transition point that moves towards smaller values of $\lambda$.
For large observable values grooming has much smaller impact and the spread in the generator
predictions observed in the ungroomed case basically remains unaltered.

For higher jet transverse momenta, \emph{i.e.}\ $p_{T,\text{jet}}\in [408,1500]\;\text{GeV}$ presented in
Fig.~\ref{fig:all_mc_np_log_pT408}, the picture somewhat changes. Both for the groomed and ungroomed
angularities we observe a larger variation of the predictions for $\lambda\gsim 0.1$. The differences
between the \sherpa MEPS@NLO predictions and the \pythia LO results reach up to 30\%. This increase
with respect to the lower-$p_{T,\text{jet}}$ slice can be expected as the inclusive $K$-factors, with the fiducial cuts adopted here, 
are also found to be increasing functions of jet transverse momentum. 
 It is interesting to note that the \sherpa
MEPS@LO results agree very well with the \herwig predictions,
for all three considered angularities, with and without grooming. The inclusion of the full set of
one-loop virtual corrections for the one- and two-jet processes in the MEPS@NLO calculation
translates into corrections of ${\cal{O}}(10-15\%)$ for larger values of $\lambda$. Despite the
larger sensitivity to higher-order perturbative corrections, the relative impact of non-perturbative
effects, most relevant for small angularity values, is reduced for jets with high transverse momentum. 

We note that measurements of (groomed) jet angularities differential in $p_{T,\text{jet}}$ provide
means to probe the modelling of perturbative as well as non-perturbative corrections in
MC generators. In Sec.~\ref{sec:mc-comp} we will use the presented simulations
to extract non-perturbative corrections for the resummed calculations. There we will also present
a comparison of the generators' parton-level predictions, \emph{i.e.}\ without the inclusion of
UE and hadronisation, with the \NLOpNLLp results.

\clearpage

\section{All-order resummation at next-to-leading logarithmic accuracy}\label{sec:resummation}

To perform the NLL resummation of the jet angularities we use the
\sherpa implementation of the \Caesar resummation formalism~\citep{Banfi:2004yd},
first presented in~\citep{Gerwick:2014gya}. 
This framework has recently been employed to obtain resummed predictions for \softdrop
thrust~\citep{Marzani:2019evv} and multijet resolution
scales~\citep{Baberuxki:2019ifp} in electron--positron collisions,
as well as \NLOpNLLp predictions for \softdrop groomed hadronic
event shapes, in particular groomed transverse thrust~\citep{Baron:2020xoi}.

For a generic observable, the master formula for the all-order
cumulative cross section for observable values up to $v$, with
$L=-\ln(v)$, can be written as a sum over partonic channels $\delta$:
\begin{equation}\label{eq:CAESAR}
  \begin{split}
    \Sigma_\mathrm{res}(v) &= \sum_\delta \Sigma_\mathrm{res}^\delta(v)\,,\,\,\text{with} \\  
    \Sigma_\mathrm{res}^\delta(v) &= \int d\mathcal{B_\delta}
    \frac{\mathop{d\sigma_\delta}}{\mathop{d\mathcal{B_\delta}}} \exp\left[-\sum_{l\in\delta}
      R_l^\mathcal{B_\delta}(L)\right]\mathcal{P}^{\mathcal{B}_\delta}(L)\mathcal{S}^\mathcal{B_\delta}(L)\mathcal{F}^\mathcal{B_\delta}(L)\mathcal{H}^{\delta}(\mathcal{B_\delta})\,,
  \end{split}
\end{equation}
where $\frac{\mathop{d\sigma_\delta}}{\mathop{d\mathcal{B_\delta}}}$ is the fully differential Born cross section for the partonic channel $\delta$ and $\mathcal{H}$ implements the kinematic cuts applied to the Born
phase space $\mathcal{B}$.  $\mathcal{F}$ denotes the multiple
emission function which, for additive observables such as the
angularities considered in this paper, is simply given by
$\mathcal{F}(L) = e^{-\gamma_E R^\prime}/\Gamma(1+R^\prime)$,
with $R^\prime(L)=\partial R/\partial L$ and
$R(L)=\sum_{l\in \delta} R_l(L)$.  The ratio of
parton-distribution-functions (PDFs) $\mathcal{P}$ takes into account
the true initial-state collinear scale. The soft function
$\mathcal{S}$ implements the non-trivial aspects of colour evolution.
The collinear radiators $R_l$ for the hard legs $l$ were computed
in~\citep{Banfi:2004yd} for a general observable $V$ scaling for the
emission of a soft-gluon of relative transverse momentum $k_t^{(l)}$
and relative rapidity $\eta^{(l)}$ with respect to leg $l$ as
\begin{equation}\label{eq:CAESAR_param}
  V(k)=\left(\frac{k_{t}^{\left(l\right)}}{\mu_Q}\right)^{a}e^{-b_{l}\eta^{\left(l\right)}}d_{l}\left(\mu_Q\right)g_{l}\left(\phi\right)\,.
\end{equation}
The \Caesar resummation plugin to \sherpa hooks into the event generation framework, facilitating
the process management, and providing access to the \comix matrix-element generator~\citep{Gleisberg:2008fv},
as well as phase-space integration and event-analysis functionalities. The \sherpa framework is also
used to compile all the required higher-order tree-level and one-loop calculations. For the NLO QCD
computations we use the \sherpa implementation of the Catani--Seymour dipole subtraction~\citep{Gleisberg:2007md}
and the interfaces to the \recola~\citep{Actis:2016mpe,Biedermann:2017yoi} and \OpenLoops~\citep{Cascioli:2011va}
one-loop amplitude generators. The plugin implements the building blocks of the \Caesar master formula
Eq.~\eqref{eq:CAESAR}, along with the necessary expansion in $\alpha_s$ used in the matching with fixed-order
calculations. The building blocks are evaluated fully differentially for each Born-level
configuration $\mathcal{B}_\delta$ of a given flavour and momentum configuration.\footnote{Note, for the case of non-additive observables that
  feature a non-trivial multiple emission function, we pre-compute ${\cal{F}}$ numerically and tabulate
  its values on a grid for read-out during event generation.}

In Ref.~\citep{Baron:2020xoi} the \Caesar formalism and the corresponding implementation in the \sherpa
plugin were extended to include the phase-space constraints given by \softdrop grooming with general
parameters $\zcut$ and $\beta$.
Note
that, as already pointed out in~\citep{Larkoski:2014wba}, differential
distributions computed from Eq.~(\ref{eq:CAESAR}) at NLL are
discontinuous at $L=L_z=-\ln z_\text{cut}$. 
A way to cure this behaviour was proposed in Ref.~\citep{Marzani:2019evv}. However, as this discontinuity is a
subleading (NNLL) effect, we decided to leave it
untouched and checked that the two approaches were in agreement within our theoretical uncertainties.
Furthermore, we note that our resummation is strictly valid in the limit of small
  $\zcut$, \textit{i.e.} $\lambda_\alpha \ll \zcut \ll 1$. However, for $\beta=0$ \softdrop, finite-$\zcut$ corrections are
  already present at the (leading) single-logarithmic
  accuracy~\cite{Dasgupta:2013ihk}. For the specific choice of
  $\zcut=0.1$ adopted in this paper, these corrections have been
  studied in~\cite{Marzani:2017mva} in the context of the groomed jet
  mass in dijet processes, and found to have a negligible effect
  around one percent.

The treatment of the kinematic endpoint is implemented in the same way as in Ref.~\citep{Baron:2020xoi} by shifting the relevant logarithms and adding power-suppressed terms to achieve a cumulative distribution that approaches one
at the kinematic endpoint and has a smooth derivative approaching zero. In
particular, we introduce the additional parameters $p, x_L, v_\mathrm{max}$ and
modify all logarithms by power-suppressed terms according to 
\begin{equation}
   \ln \left(\frac{x_L}{v}\right) \to \frac{1}{p} \ln\left[\left(\frac{x_L}{
         v}\right)^p-\left(\frac{x_L}{v_\mathrm{max}}\right)^p+1\right]=L\,.
\end{equation}
Here we set $v_\mathrm{max}$ to the numerically determined endpoint of the NLO distribution,
and by default use $p=1$. 

We fix the renormalisation and factorisation scale to the transverse momentum of
the Z boson, $p_{T,\mu^+\mu^-}$, and the resummation scale to $\mu_Q =
p_{T,\mu^+\mu^-} R_0$. Note, for the Born configuration, on which the
resummation is performed, it holds $p_{T,\mu^+\mu^-}\equiv p_{T,\text{jet}}$. To evaluate 
the perturbative uncertainties  of our results, we vary $\mu_R$ and $\mu_F$
according to a 7-point variation, 
\emph{i.e.}\ $(\mu_R/p_{T,\mu^+\mu^-},\mu_F/p_{T,\mu^+\mu^-}) \in
\{(0.5,0.5),(0.5,1),(1,0.5),(1,1),(1,2),(2,1),(2,2)\}$, simultaneously in the
fixed-order calculation and the resummation. The argument of $\alphaS$ in the
resummation is always taken to be $\mu_R$, with a compensating term for the
LL dependence to remain \NLL accurate and ignoring the NNLL ambiguity this
introduces in pure \NLL terms. Additionally, we calculate the
resummed distribution with $x_L\in\{0.5,1,2\}$ for $\mu_R =\mu_F =
p_{T,\mu^+\mu^-}$. The total uncertainty is obtained by taking the envelope
of all those predictions with $(\mu_R/p_{T,\mu^+\mu^-},\mu_F/p_{T,\mu^+\mu^-},x_L) = (1,1,1)$ as central value.

The ingredients of the master formula Eq.~(\ref{eq:CAESAR}) are readily available
in the case of global event shapes, with or without \softdrop
grooming.  However, some adaptations are needed
if we want to apply this approach to the resummation of non-global jet
angularity distributions measured on the leading jet in $Z$+jet events.
These are described in Section~\ref{sec:sherpa4jets} below. We
then discuss aspects related to the matching to NLO fixed-order
calculations in Section~\ref{sec:matching} and present numerical results
in Section~\ref{sec:matched-res}.

\subsection{Jet angularities resummation within the \sherpa framework}\label{sec:sherpa4jets}
Compared to the formalism described above and used in \citep{Gerwick:2014gya,
  Baberuxki:2019ifp, Baron:2020xoi}, some adjustments are necessary to the
\sherpa resummation framework. In particular we have to take into
account the fact that
the angularities are sensitive to radiation within the hardest jet in the
event only. 
Let us label, for definiteness, the initial state legs as $l=1,2$ and the measured
final state leg as $l=3$. As our observable is not sensitive to radiation
collinear to the initial state legs, we are allowed to set $\mathcal{P}=1$
and $R_1=R_2=0$ in Eq.~\eqref{eq:CAESAR}.
Keeping in mind that the Born phase space is given by final states
consisting of a single parton plus the $Z$ boson, only one radiator is
left. In terms of Eq.~\eqref{eq:CAESAR_param}, the angularity
observables $\lambda^1_\alpha$ can be parametrised by
\begin{align}
  a &= 1\,, \\
  b_3 &= \alpha-1\,,\\
  g_3 d_3(\mu_Q) &= \left(2\frac{\cosh(y_3)}{R_0}\right)^{\alpha-1} \frac{\mu_Q}{p_{T,\text{jet}} R_0}\,,
\end{align}
where $y_3\equiv y_\text{jet}$ denotes the rapidity of the hard jet. Note that
for our choice of $\mu_Q = p_{T,\text{jet}} R_0$ the last factor in $g_3 d_3$ equals unity.

A number of complications arises when considering the presence of a jet boundary. 
First, when originally computing the
radiators, one conventionally divides the phase space for each dipole at
$\eta=0$.
An arbitrary rapidity condition with respect to leg $l$, $\eta_l >
\eta_{\text{min},l}$, leads to an additional contribution 
\begin{equation}
  \Delta R_l(L) = -2 C_l\left(\eta_{\text{min},l}-\ln 2 E_l/Q\right) t(L/a)\,,
\end{equation}
where we have introduced
\begin{equation}
  t(L) = \int^{\mu_Q}_{\mu_Qe^{-L}}
  \frac{dk_{t}}{k_{t}}\frac{\alphaS\left(k_{t}\right)}{\pi} = \frac{-\ln(1-2\alpha_s\beta_0L)}{2\pi\beta_0}\,,\label{eq:T}
\end{equation}
with $\beta_0=(11C_A-2n_f)/(12\pi)$. Note that the last equality holds with $\alphaS =
\alphaS(\mu_R)$ at \NLL accuracy.
We can use this to reflect the rapidity requirement given by the jet condition
by setting $\eta_{\text{min},3} = \ln\left(2\cosh(y_3)/R_0\right)$.
Here $Q$ is an arbitrary scale, which we choose to identify with the
partonic centre-of-mass energy, that ultimately cancels in the final expressions.
For global observables this cancellation happens with the soft function $\mathcal{S}$, however, the
restriction to radiation inside the jet additionally affects the
structure of soft emissions. The details were worked out in~\citep{Dasgupta:2012hg}
for the jet-mass observable. Those results apply to the general class of
angularities as well, as they depend on the scaling of the observable with $k_t$ only, \textit{i.e.}\ on the parameter $a$ in Eq.~\eqref{eq:CAESAR_param}. This implies (using $a=1$)
\begin{equation}
  \mathcal{S}^{\mathcal{B}_\delta}(L) = S_\text{global}^{\mathcal{B}_\delta}\left(t(L)\right)S_\text{non-global}^{\mathcal{B}_\delta}\left(t(L)\right)\;.
\end{equation}
In a general way, the global contribution can be written as
\begin{equation} \label{soft-gen}
  S_\text{global}^{\mathcal{B}_\delta}(t) = \Tr\left[H e^{-t\left(\mathbf{\Gamma}^{\mathcal{B}_\delta}\right)^\dagger}ce^{-t\mathbf{\Gamma}^{\mathcal{B}_\delta}}\right]/\Tr\left[cH\right]\;,
\end{equation}
where $c$ and $H$ are the colour metric and hard function, respectively, see
\citep{Gerwick:2014gya, Baberuxki:2019ifp} for details of the notation. We can
write the matrix $\mathbf{\Gamma}$ in colour space as a sum over dipoles
\begin{align}
  \mathbf{\Gamma}^{\mathcal{B}_\delta} &= \sum_{i>j\in\mathcal{B}_\delta} \mathbf{T}_i\cdot\mathbf{T}_j I_{ij}\,.
\end{align}
In our case only three dipoles, namely 12, 13 and 23, contribute to the soft function. As a consequence,
by exploiting colour conservation, all colour operators $ \mathbf{T}_i\cdot\mathbf{T}_j$ can be written
as linear combinations of the $SU(3)$ Casimir invariants $C_F$ and $C_A$ and, consequently, the matrix
structure of Eq.~(\ref{soft-gen}) disappears. 
The coefficients of the single-logarithmic function $t$, can be
  obtained by integrating the matrix elements for each dipole $ij$
  over the appropriate phase space. Because the collinear
  contributions have already been included in the radiators $R_l$ in
  Eq.~(\ref{eq:CAESAR}), the $I_{ij}$ coefficients are purely due to
  soft wide-angle radiation. Consequently, they are functions of the
  jet radius $R_0$ only, and are independent of the angularity
  exponent $\alpha$. In particular, they coincide with those computed in~\citep{Dasgupta:2012hg} for the jet mass:
\begin{align}
  I_{12} &= \frac{R^2_0}{4}\,,\\
  I_{13} = I_{23} &= \frac{R^2_0}{16} + \mathcal{O}(R^4_0)\,.
\end{align}
Note that by including the above dipoles we account for initial-state radiation at NLL accuracy, as an expansion in powers of the jet radius.
Higher order terms in the jet radius expansion can be computed, but in addition to the
suppression in $R^2_0$ the corresponding coefficient drops rapidly, and even for
$R_0=1$ the first terms are an appropriate approximation
\citep{Dasgupta:2012hg}.

\begin{figure}
  \centering
  \includegraphics[width=0.5\textwidth,page=2]{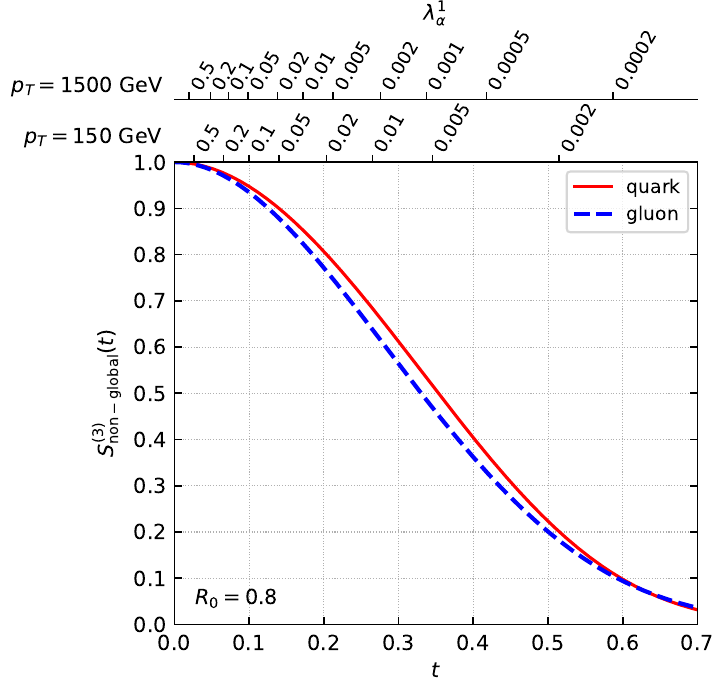}
  \caption{Sudakov suppression factor associated with the non-global
    logarithms, $S_\text{non-global}^{(3)}\left(t(L)\right)$. The two
    curves correspond to the case where the final-state parton is either a
    quark (solid red) or a gluon (dashed blue).
    The Sudakov is plotted as a function of the scaled variable $t(L)$
    (cf.\ Eq.~(\ref{eq:T})). For illustrative purpose, the top axes
    show the corresponding values of $\lambda_\alpha^1$ (irrespective
    of $\alpha$) for two representative jet transverse momenta.}
  \label{fig:ngls}
\end{figure}

The non-global part $S_\text{non-global}^{\mathcal{B}_\delta}(t)$ is
computed numerically, in the large-$N_c$ limit, following the algorithm
highlighted in~\citep{Dasgupta:2001sh}, as also done
in~\citep{Dasgupta:2012hg}.\footnote{Note that our definition
  of the scaled variable $t(L)$ differs from the one
  in~\citep{Dasgupta:2001sh} by a factor 2 and from the one in
  \emph{e.g.}~\cite{Banfi:2004yd,Gerwick:2014gya,Baberuxki:2019ifp,Baron:2020xoi}
  by factor of $1/2$.}
This can straightforwardly be applied to ungroomed angularities.
For \softdrop groomed distributions, the non-global factor
remains the same for $v\ge z_\text{cut}$ and saturates at that value,
\textit{i.e.}\
$S_\text{non-global}^\text{(groomed)}(v) =
S_\text{non-global}^\text{(ungroomed)}(\text{max}(v,z_\text{cut}))$.
We also note that the non-global contribution only depends on $v$ (and
the jet transverse momentum through $\alpha_s(\mu_R)$) but not on the angularity
parameter $\alpha$.
Similarly, for the groomed case, the contribution from non-global logarithms only depends on $z_\text{cut}$ and
not on $\beta$.
The main reason behind this is that, at single-logarithmic accuracy, non-global logarithms originate
from configurations that are strongly ordered in energy with angles
commensurate with the jet radius $R_0$.

In practice, $S_\text{non-global}(t)$ is computed separately for each
possible colour dipole, either incoming--incoming or incoming--final.
These contributions are independent of the jet rapidity meaning, in
particular, that the two possible incoming--final configurations are
equal.
One can then reconstruct the two Born channels $\delta$ from the
dipole contributions as done in~\citep{Dasgupta:2012hg}.
This procedure allows one to recover the full-$N_c$ result at least
for the first non-trivial term of the expansion in $\alpha_s$.
We finally note that the numerical procedure described
in~\citep{Dasgupta:2001sh} introduces an angular cut-off
$\theta_\text{min}$ to regulate the collinear divergence. We have
performed 4 independent runs with
$\theta_\text{min}=\{0.008,0.004,0.002,0.001\}$ and extrapolated the
result to $\theta_\text{min}\to 0$ using the same approach as
in~\citep{Dasgupta:2020fwr}.

Fig.~\ref{fig:ngls} shows the resulting non-global contributions
$S_\text{non-global}^{\mathcal{B}_\delta}(t)$ for both (Born-level)
quark and gluon jets.
The upper set of axes indicate what value of $t$ corresponds to the
observable value $\lambda^1_\alpha$ for fixed $p_{T,\text{jet}}$.
We have checked that they agree within $\sim 1\%$ with the results
obtained in~\citep{Dasgupta:2012hg} at
$t\lesssim 0.4$.
However, the results presented here extend to larger values of $t$, as
appropriate for the kinematic range investigated in this paper.

\begin{figure}
  \includegraphics[width=0.44\textwidth]{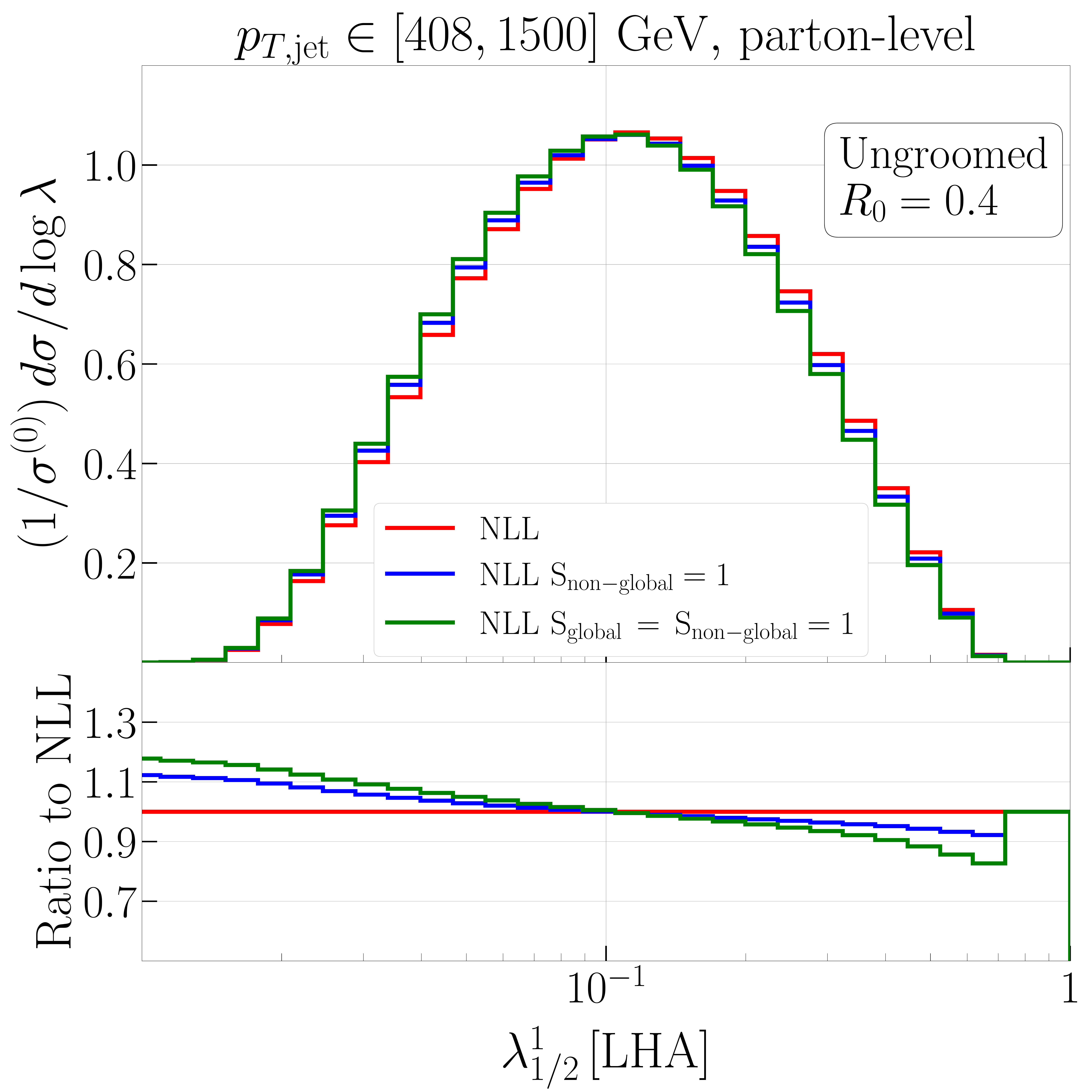}\hfill
  \includegraphics[width=0.44\textwidth]{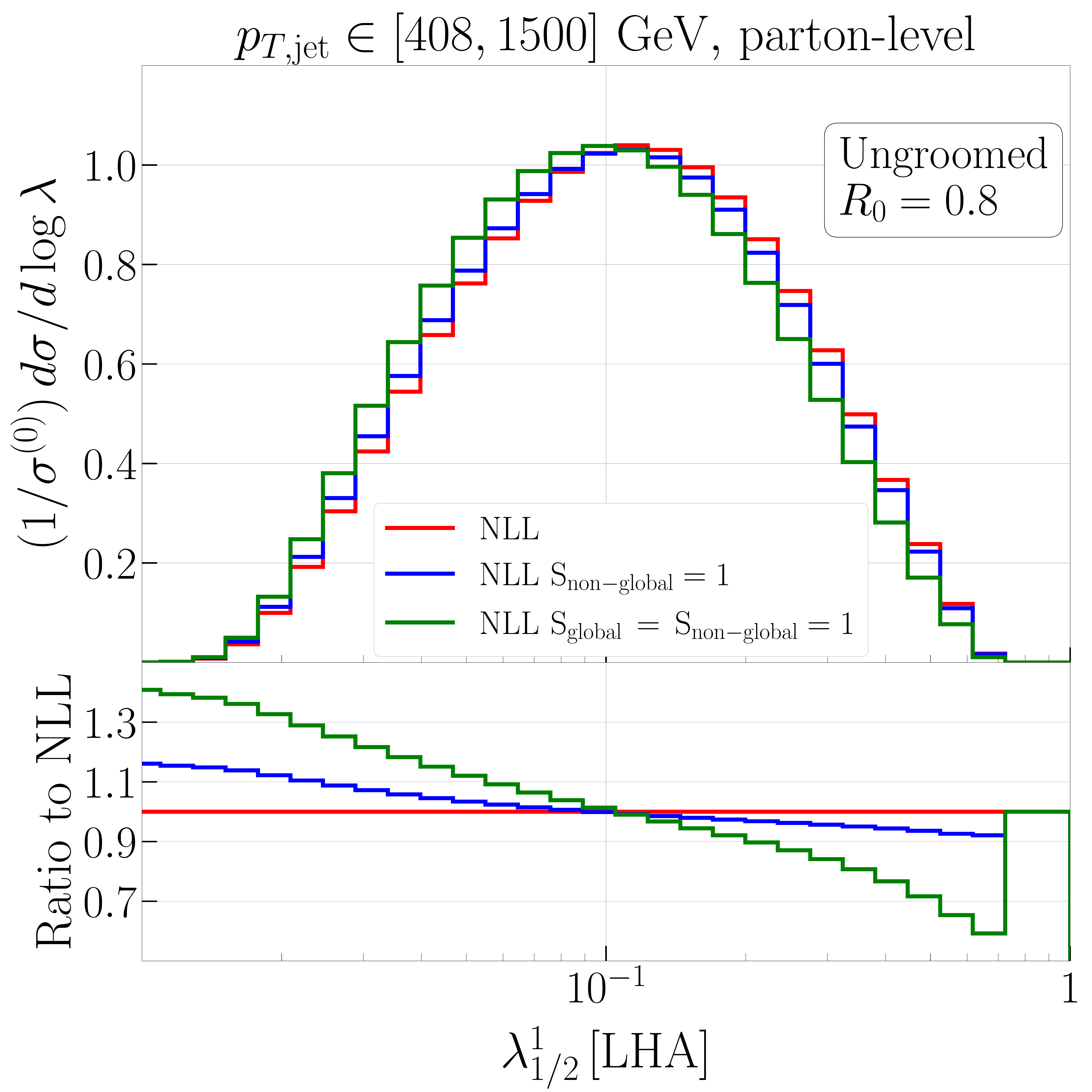}

  \includegraphics[width=0.44\textwidth]{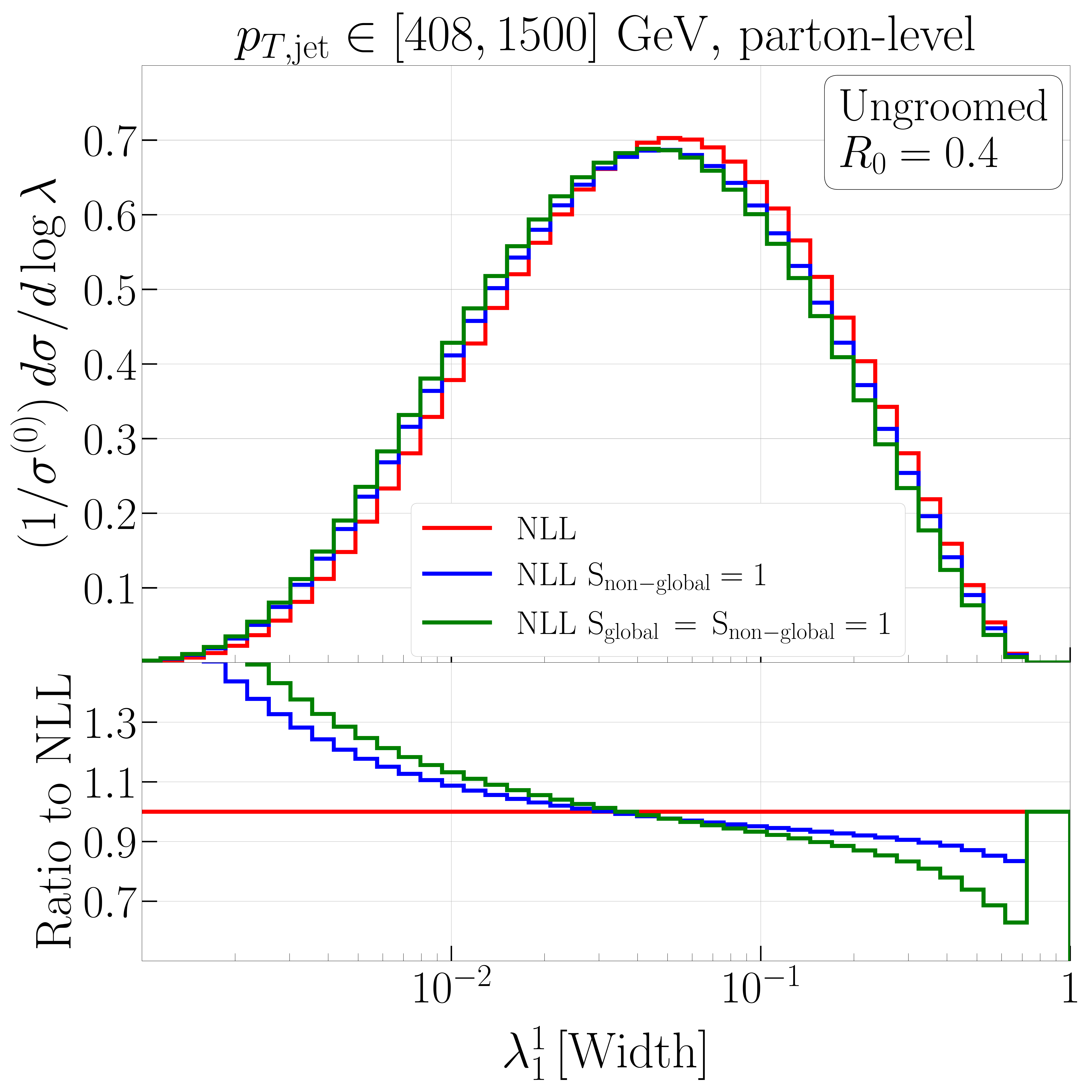}\hfill
  \includegraphics[width=0.44\textwidth]{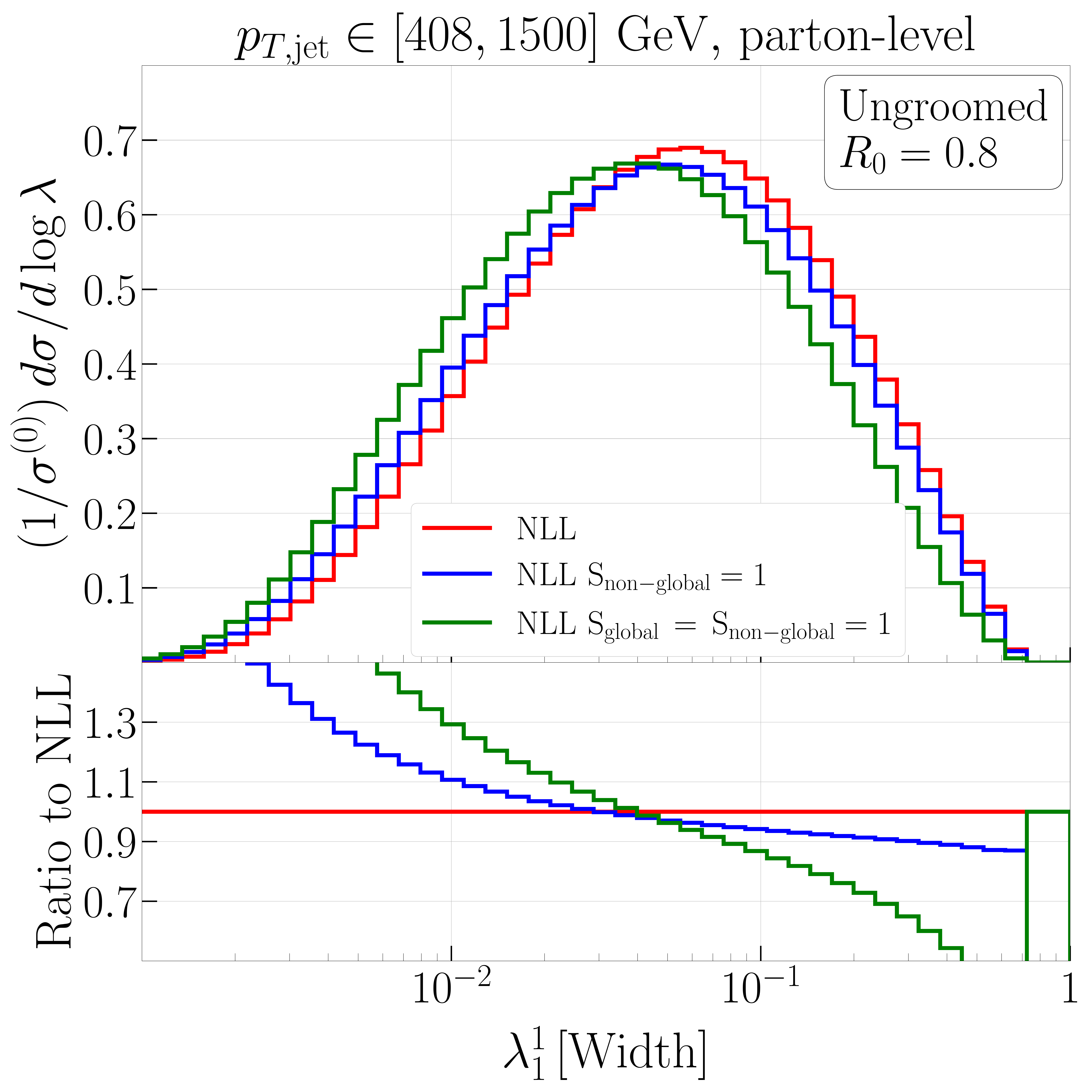}

  \includegraphics[width=0.44\textwidth]{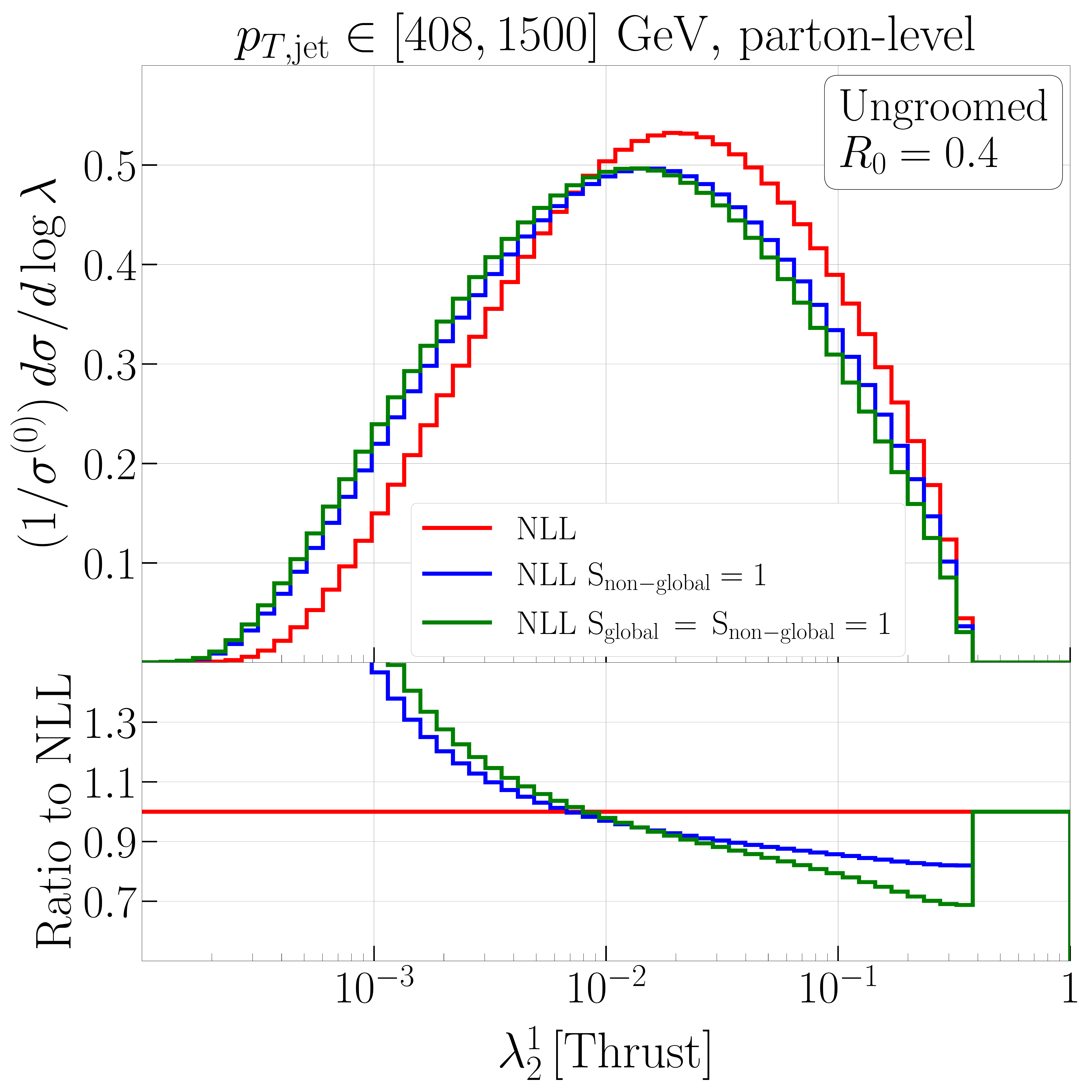}\hfill
  \includegraphics[width=0.44\textwidth]{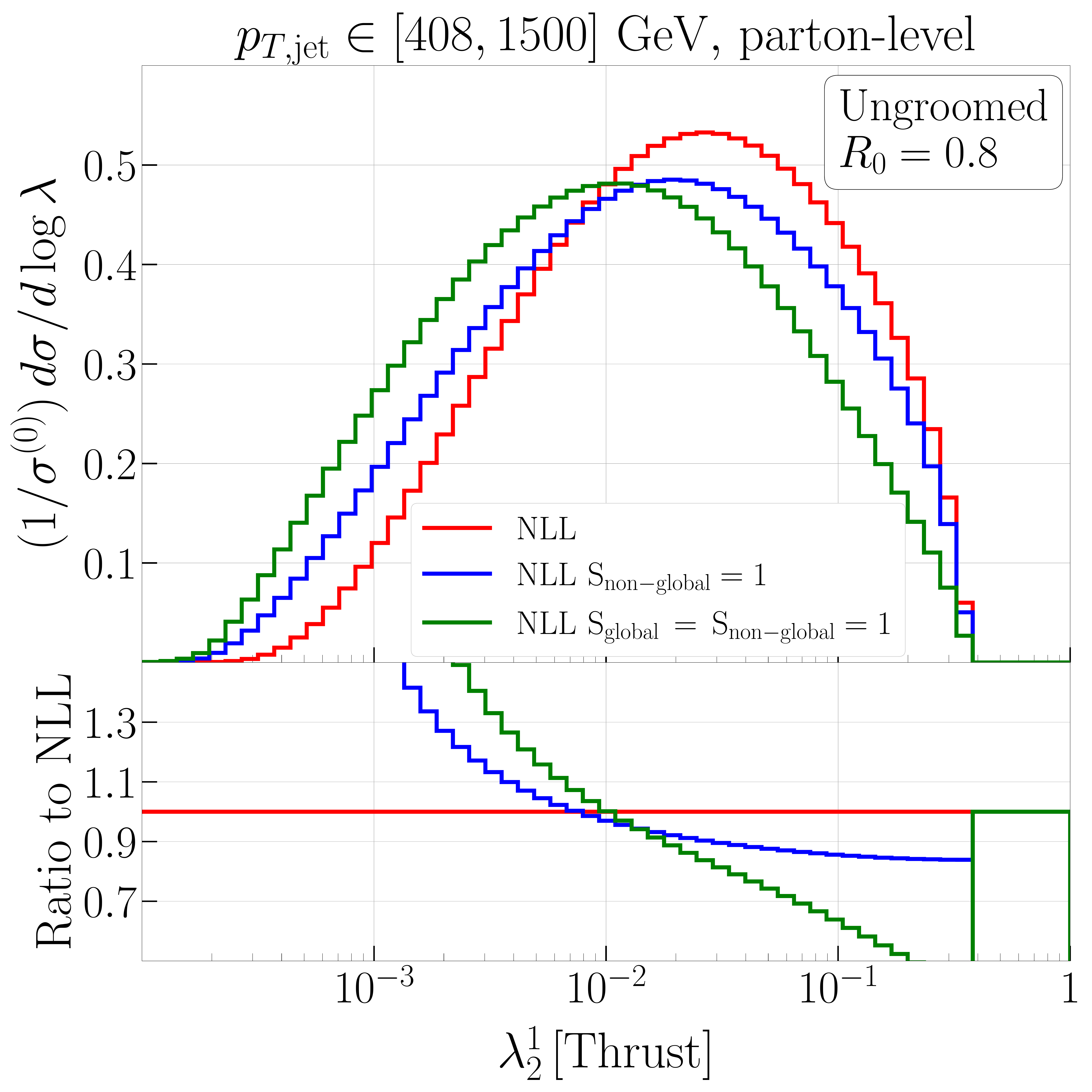}
  \caption{Anatomy of soft structure of the NLL predictions for angularities
    $\lambda^1_\alpha$, for $\alpha\in\{1/2,1,2\}$ (top to bottom), for jet radius
    $R_0=0.4$ (left) and $R_0=0.8$ (right), with $p_{T,\text{jet}}\in[408,1500]~\text{GeV}$.
    Besides the full NLL distribution (red) variants obtained by either excluding
    the non-global soft contribution ($S_\text{non-global}=1$, blue) or the entire soft
    contributions ($S_\text{global}=S_\text{non-global}=1$,
    green) are shown.}\label{fig:ngl_numerics}
\end{figure}

To gauge the numerical impact of the global and non-global soft function on our NLL
  predictions for the jet-angularity variables, we present in Fig.~\ref{fig:ngl_numerics}
  results for the ungroomed case. As both $\mathcal{S}$-function contributions contain terms
  expected to vanish in the $R_0 \to 0$ limit,\footnote{The global
    part of the soft function becomes unity in the $R_0\to 0$ limit while
    the non-global soft function remains non-trivial.} we here show results for the two
  considered jet radii, $R_0 = 0.4$ and $R_0 = 0.8$. Besides the complete
  NLL distribution (in red), we depict results for
  $S_\text{non-global} = 1$ (in blue), allowing us to
  quantify the numerical effect of non-global logarithms, and
  $S_\text{global}=S_\text{non-global} = 1$ (in green),
  effectively neglecting soft wide-angle emissions entirely.
  We observe that these contributions have a bigger impact for larger
  jet radii as theoretically anticipated.
  Furthermore, we observe a somewhat milder (relative) impact of
  $\mathcal{S}$ in the $\alpha=0.5$ case, when comparing to
  $\alpha=2$.
  For the latter a quite significant shift of the distribution is
  found. The case $\alpha=1$ lies in between the other two.
  This is understood from the fact that, for a given value of the
  angularity, the soft function is independent of $\alpha$ while the
  effect of the collinear contribution increases when decreasing
  $\alpha$.
  When invoking grooming, the soft function remains unchanged for
  observable values larger than the transition point. However, below
  the transition point the soft-function contributions become
  constant, given by $S_\text{non-global}(\zcut)$ and
  $S_\text{non-global}(\zcut) S_\text{global}(\zcut)$,
  respectively. This is illustrated and confirmed in
  Fig.~\ref{fig:ngl_numerics_groomed} in App.~\ref{app:ngl}. We have checked
  explicitly that our observations on the impact of the soft function both for
  the groomed and ungroomed case also hold after matching to NLO, as described in what follows.

We close this discussion of the NLL resummation, by noting, had we defined jets with algorithms
that differ from anti-$k_t$, for instance, C/A or $k_t$, a new class of so-called clustering
logarithms would have appeared at NLL, see
\textit{e.g.}~\citep{Banfi:2005gj,Delenda:2006nf,Banfi:2010pa,Delenda:2012mm,Lifson:2020gua}.

\FloatBarrier

\subsection{Matching to NLO and achieving \NLLp accuracy}\label{sec:matching}
In order to achieve a faithful description of the angularity distributions across their whole
spectrum, we need to match our all-order results with fixed-order
predictions, computed here at NLO
accuracy. Our matching procedure will keep track of the jet flavour, so that we can obtain what
is usually referred to as \NLOpNLLp accuracy.

The matching procedure is defined for the cumulative distribution (for
simplicity, we use $\lambda\equiv \lambda_\alpha^1$)
\begin{equation}
  \Sigma(\lambda) = \int_0^\lambda \mathop{d\sigma}.
\end{equation}
We also  introduce the notation 
$\Sigma^\delta_\mathrm{res,fo,match}$ (see also~\citep{Baberuxki:2019ifp,Baron:2020xoi}) in order to
denote the results for the cumulative distribution for the (family of) flavour channel(s)
$\delta$ computed, with resummation, at fixed order, or matched between the two, respectively.
In the following, the channel label is omitted in general
expressions and we use the shorthand $\sigma = \Sigma(1)$. We denote the expansion of any
cumulative distribution $\Sigma$ to order $\as^2$ relative to the $pp\to \mu^+\mu^-j$ Born process as
\begin{equation}
  \Sigma = \Sigma^{(0)} + \Sigma^{(1)} + \Sigma^{(2)} + {\cal{O}}(\as^{4})\,,
  \hspace{1cm} \Sigma^{(k)} \propto \alpha^2_\text{EW}\,\as^{1+k}\,.
\end{equation}
In practice, at least for $\Sigma^{(2)}_\mathrm{fo}$, we only calculate
\begin{equation}
  \overline{\Sigma}^{(2)}_\mathrm{fo}(\lambda) = \int_\lambda^1 \mathop{d\sigma^{(2)}}~, \hspace{1cm}   \Sigma^{(2)}_\mathrm{fo} = \sigma^{(2)}_\mathrm{fo} - \overline{\Sigma}^{(2)}_\mathrm{fo}\,.
\end{equation}
We employ a multiplicative matching scheme, in which, for every partonic channel $\delta$, we have
\begin{multline}\label{eq:mult_match}
  \Sigma_\text{match,mult}^{\delta}(\lambda) = 
  \Sigma_\text{res}^\delta(\lambda)
  \left[1+\frac{\Sigma_\mathrm{fo}^{\delta,(1)}(\lambda)-\Sigma_\mathrm{res}^{\delta,(1)}(\lambda)}{\sigma^{\delta,(0)}}\right.\\
    \left.+\frac{1}{\sigma^{\delta,(0)}}\left(-\overline{\Sigma}^{\delta,(2)}_\mathrm{fo}(\lambda)-\Sigma_\mathrm{res}^{\delta,(2)}(\lambda)-\Sigma_\mathrm{res}^{\delta,(1)}(\lambda)\frac{\Sigma_\mathrm{fo}^{\delta,(1)}(\lambda)-\Sigma_\mathrm{res}^{\delta,(1)}(\lambda)}{\sigma^{\delta,(0)}}\right)\right]\;.
\end{multline}
By this procedure we automatically include the correct coefficients
\begin{equation} \label{eq:c1}
\frac{\as}{2\pi}C^{\delta,(1)} \equiv
\lim\limits_{\lambda\to0}\frac{\Sigma_\mathrm{fo}^{\delta,(1)}(\lambda)-\Sigma_\mathrm{res}^{\delta,(1)}(\lambda)}{\sigma^{\delta,(0)}}\,,
\end{equation}
thus reaching \NLOpNLLp accuracy. The final result we present corresponds to
\begin{equation}
  \Sigma_\text{match,mult}(\lambda) = \sum_\delta \Sigma_\text{match,mult}^{\delta}(\lambda)\,.
\end{equation}
%

\begin{figure}
  \centering
  \includegraphics[width=0.22\linewidth,page=1]{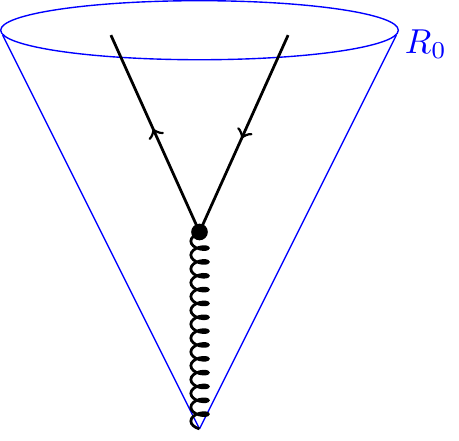}
  \includegraphics[width=0.24\linewidth,page=2]{figures/BSZ_jets.pdf}
  \includegraphics[width=0.24\linewidth,page=3]{figures/BSZ_jets.pdf}
  \includegraphics[width=0.24\linewidth,page=4]{figures/BSZ_jets.pdf}
  \centerline{(a)\hspace*{3.1cm}(b)\hspace*{3.4cm}(c)\hspace*{3.4cm}(d)\hspace*{0.8cm}}
  \caption{Examples for partonic input configurations to the flavour assignment
    algorithm. Configuration (a) needs to be identified as a gluon jet in the collinear limit,
    whereas for (b) we need to perform the LL resummation for a quark jet. IR safety
    requires that configurations (c) and (d) in the limit where $p_g\to0$ are identified
    as gluon and quark jet, respectively.}\label{fig:BSZ_examples} 
\end{figure}

We use the anti-$k_t$ algorithm with our choice of parameters for
defining jets, and employ the flavour-$k_t$ algorithm from
\citep{Banfi:2006hf} (BSZ) to assign the flavour channels. However, the BSZ algorithm is not immediately applicable here, due
  to the non-global structure of the angularities. In particular, at
  any perturbative order there exist contributions with only a single
  jet constituent, such that the angularity variable equals zero. In
  principle we need to perform the resummation around multi-jet
  configurations. However, to achieve our target accuracy, it is
  sufficient to guarantee that configurations occuring at LO are
  dressed with the proper LL Sudakov factor. An explicit example is
  given in Fig.~\ref{fig:BSZ_examples}(b).
  Our matching scheme achieves this by
  multiplying these $Z+jj$ events with the full NLL $\Sigma^\delta_\text{res}$
  corresponding to the channel $\delta$ obtained after
  dropping the second jet, which will contain the correct LL factor.   
  It is hence crucial that the flavour channel of those contributions is defined by
  the single object inside the leading jet alone. At the same time, in the
  example shown in Fig.~\ref{fig:BSZ_examples}(a), the quarks need to be
  clustered in the collinear limit and the jet identified as a gluon jet. While
  at this stage we could just use the combined flavour of all anti-$k_t$ jet
  constituents, NLO contributions would spoil the IR safety of this
  procedure. Examples of relevant configurations are given in
  Fig.~\ref{fig:BSZ_examples}(c) and~(d), where an IR-safe algorithm needs to
  ensure that the $g\to q\bar{q}$ splitting is undone first in the soft gluon
  limit. Here we propose and employ the following algorithm to assign flavour to
  the leading anti-$k_t$ jet:
\begin{enumerate}
\item[0.] Start with the list $\mathcal{O}$ of all coloured final-state objects,
  containing particle four-momenta and flavour labels, and the beams $B, \bar{B}$ with
  their respective flavours.
\item[1.] Run the standard anti-$k_t$ algorithm with radius parameter $R_0$ on
  $\mathcal{O}$, and obtain the objects in the leading, \emph{i.e.}\ highest $p_T$,
  jet $J\subset\mathcal{O}$.
\item[2.] If $J$ consists of only one object, $J=\{j\in\mathcal{O}\}$, terminate
  the algorithm. The flavour of $j$ defines the flavour channel $\delta$.
\item[3.] Otherwise, determine the pair $\{i,k\} \subset \mathcal{O}$ that
  minimises the BSZ measure $d_{ik}^\text{BSZ}$ and the objects $l,m$ that have
  minimal BSZ distances to the beams 
  $d_{lB}^\text{BSZ},d_{m\bar{B}}^\text{BSZ}$. Perform a cluster step according to
  $d^\text{BSZ}=\min(d_{ik}^\text{BSZ},d_{lB}^\text{BSZ},d_{m\bar{B}}^\text{BSZ})$:
  \begin{enumerate}
  \item If $d^\text{BSZ}=d_{ik}^\text{BSZ}$, update $\mathcal{O}$ by
    removing $i$ and $k$ and adding a new object with momentum
    $p_i+p_k$ and the combined flavour of objects $i$ and $k$.
  \item If $d^\text{BSZ}=d_{lB}^\text{BSZ}$
    ($d^\text{BSZ}=d_{m\bar{B}}^\text{BSZ}$), update $\mathcal{O}$ by removing
    $l$ ($m$) and assign the combined flavour of $l$ and
    $B$ ($m$ and $\bar{B}$) to the beam $B$ ($\bar{B}$).
  \end{enumerate}
  Go to step 1 and repeat.
\end{enumerate}
We finally can define \emph{bland} versions of this algorithm, vetoing clusterings that
would lead to jets with multiple flavours by setting the corresponding distance
measures to infinity. We use this \emph{bland} variant and identify every event as
having a leading jet that is either quark- or gluon-like.

Note that the only contribution
that matters are the cross terms between $C^{\delta,(1)}$ and the leading
logarithms, which only depend on this flavour assignment and not on details of
partons outside the jet. Therefore, we can sort any configuration with the
same flavour assignment of the leading jet into the same channel. With this
procedure we can ensure a correct assignment of the LO real corrections, while
maintaining an infrared-safe definition of jet flavour also at NLO. 

\subsection{\NLOpNLLp validation and predictions}\label{sec:matched-res}

\begin{figure}
  \centering
  \includegraphics[width=0.44\linewidth]{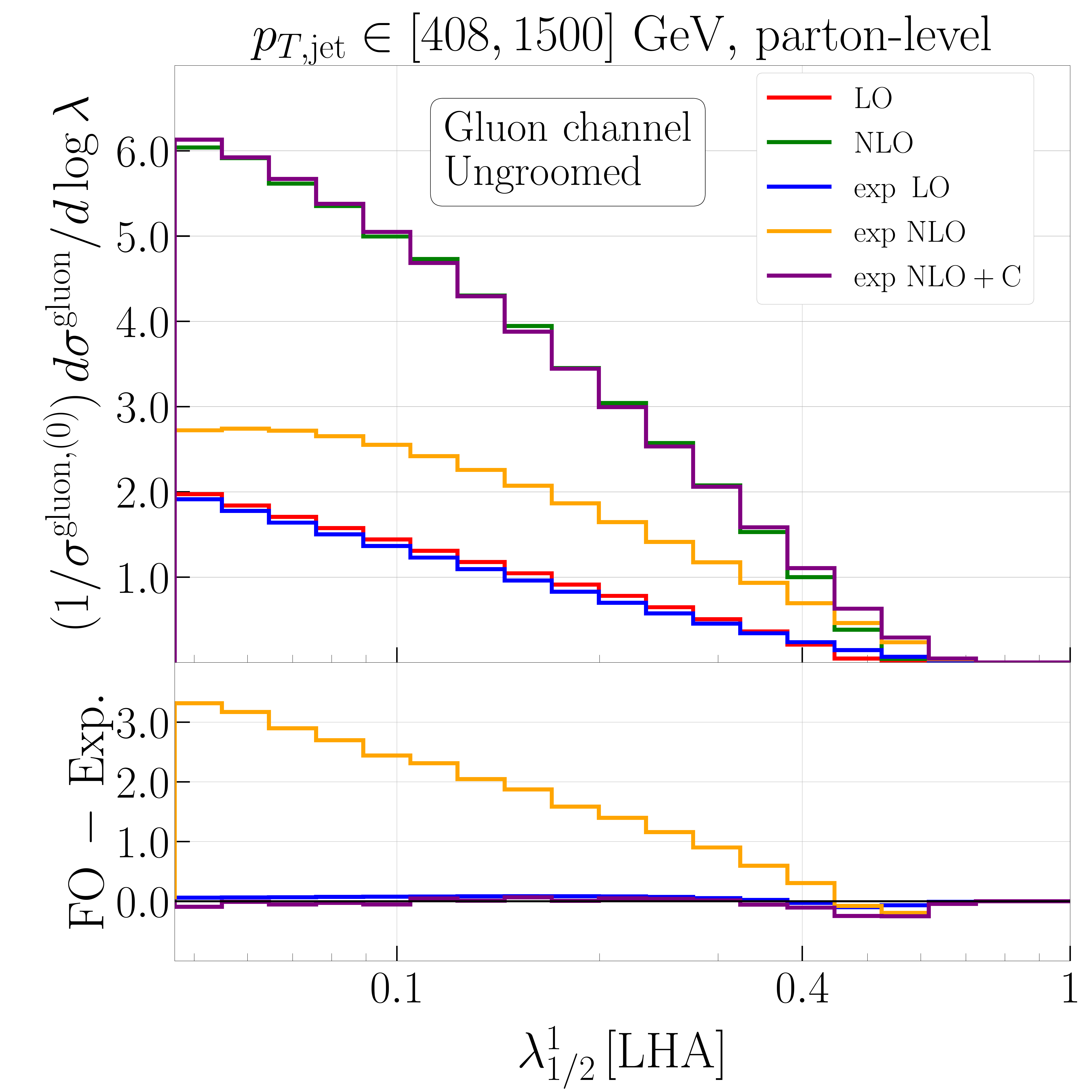}
  \hspace{1em}
  \includegraphics[width=0.44\linewidth]{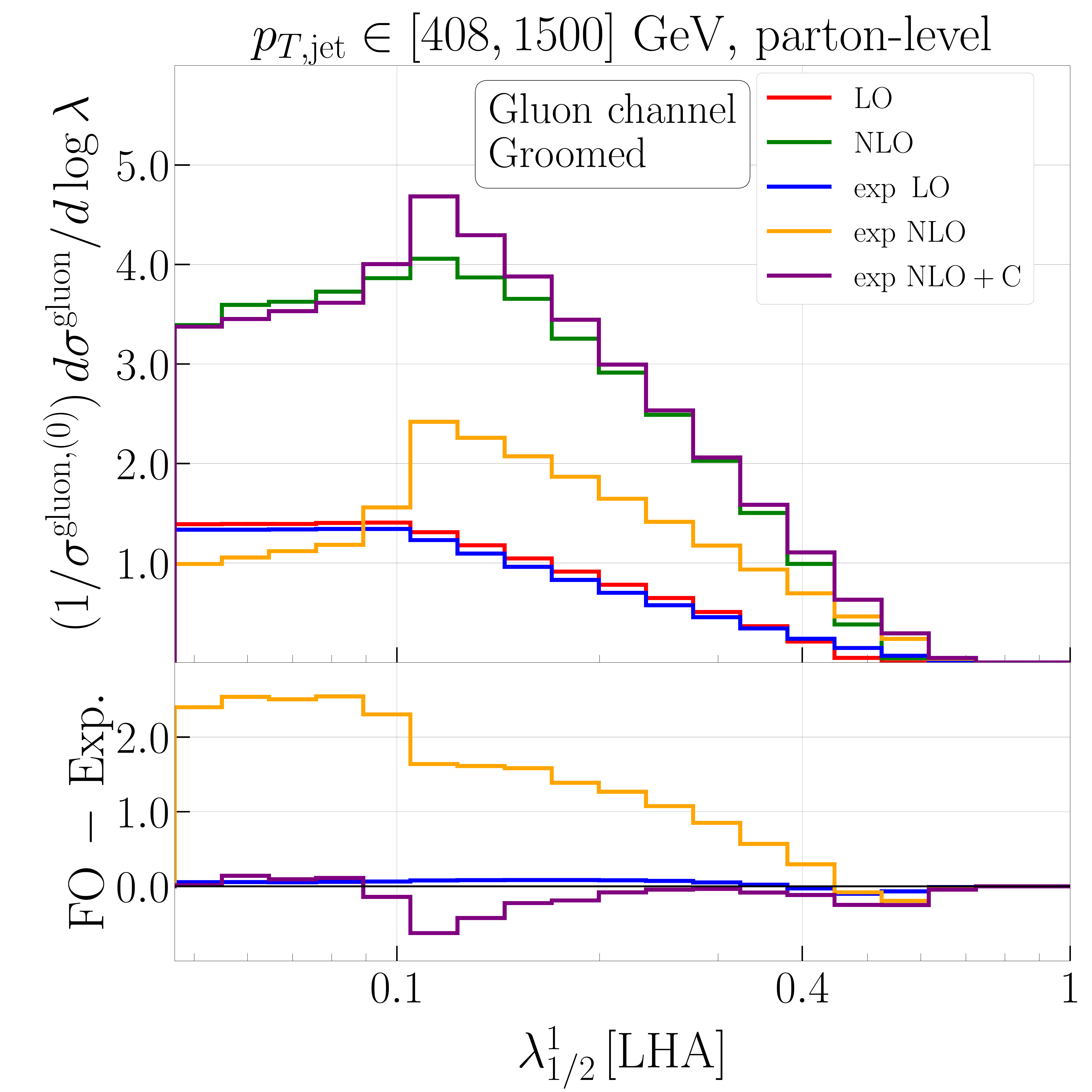}
  \centering
  \includegraphics[width=0.44\linewidth]{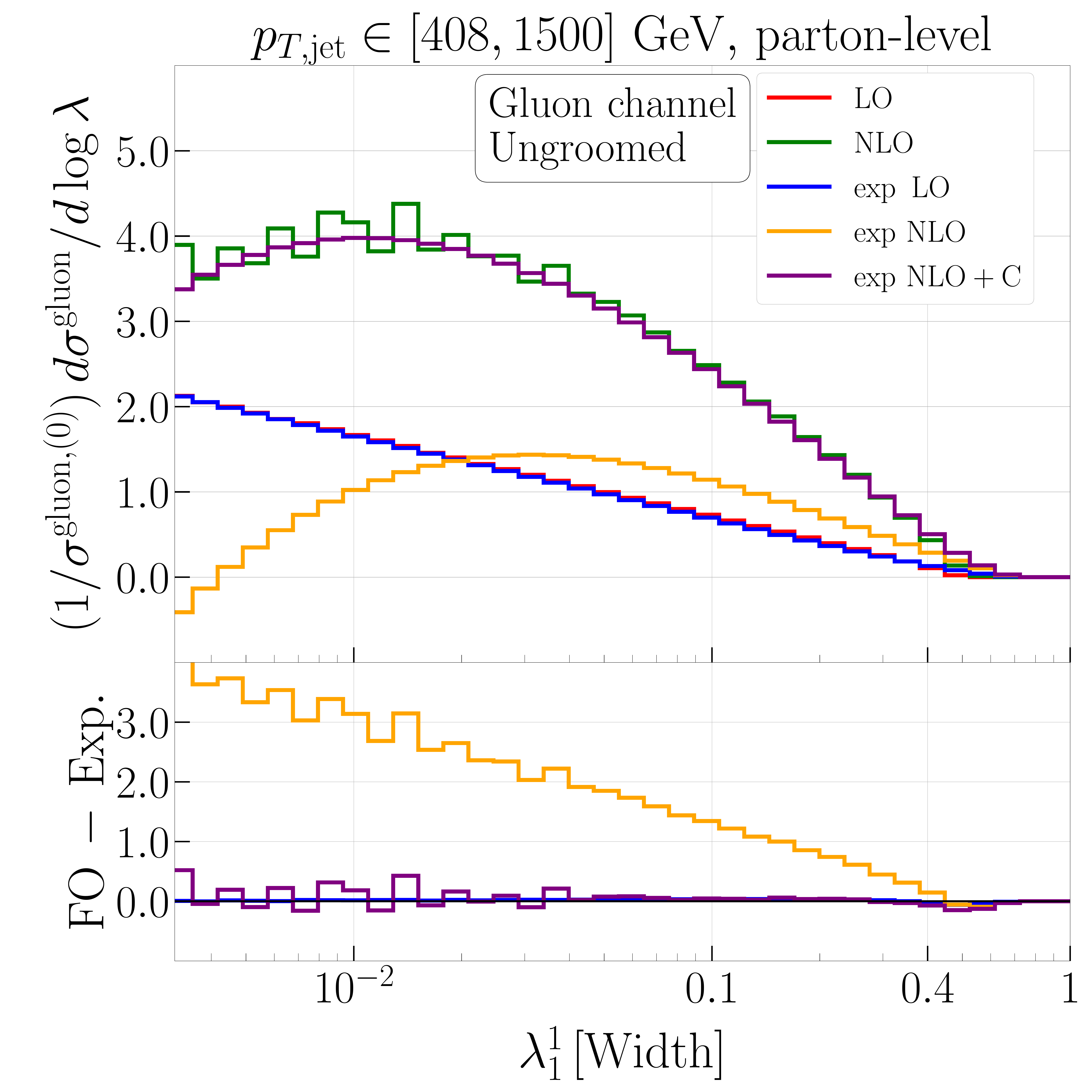}
  \hspace{1em}
  \includegraphics[width=0.44\linewidth]{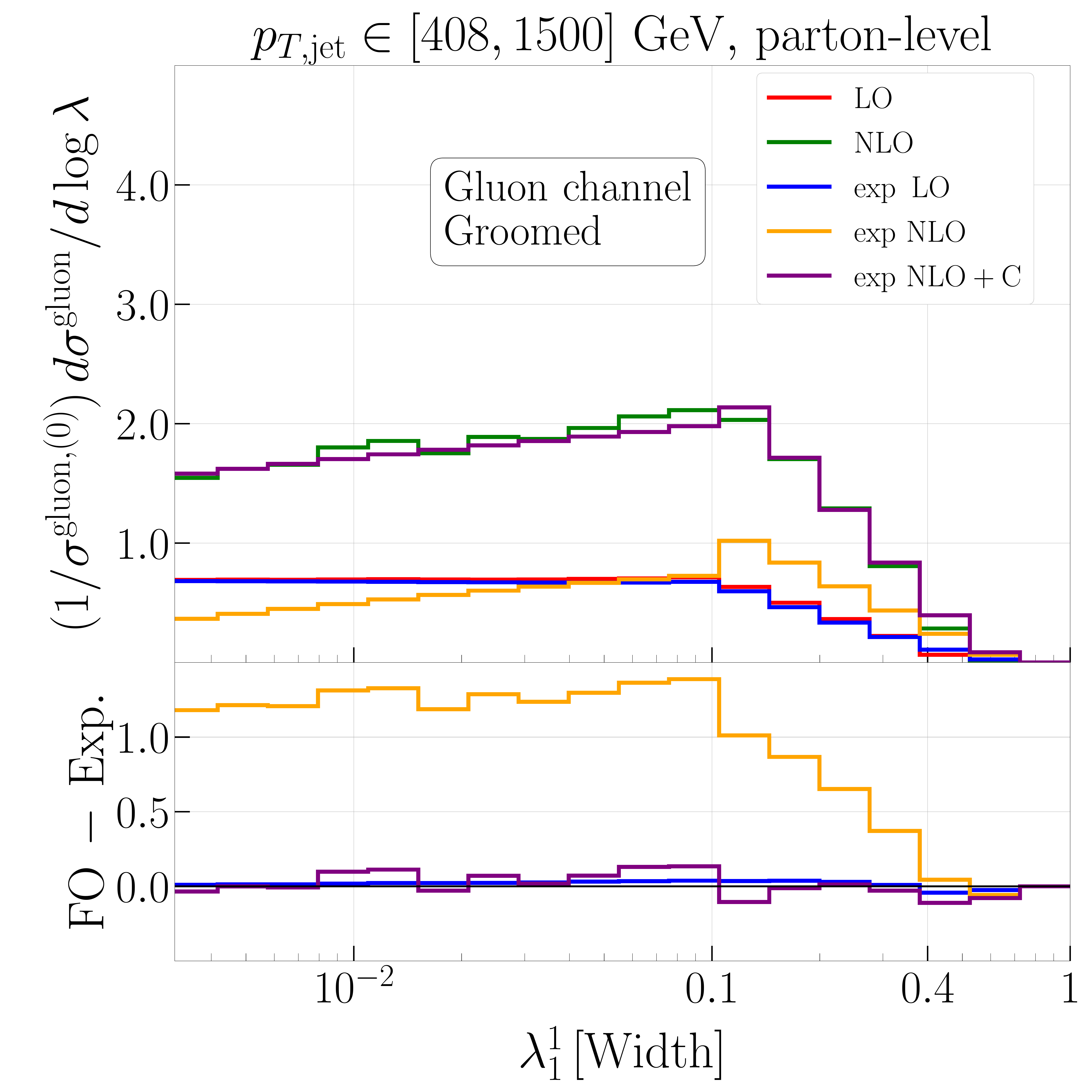}
  \centering
  \includegraphics[width=0.44\linewidth]{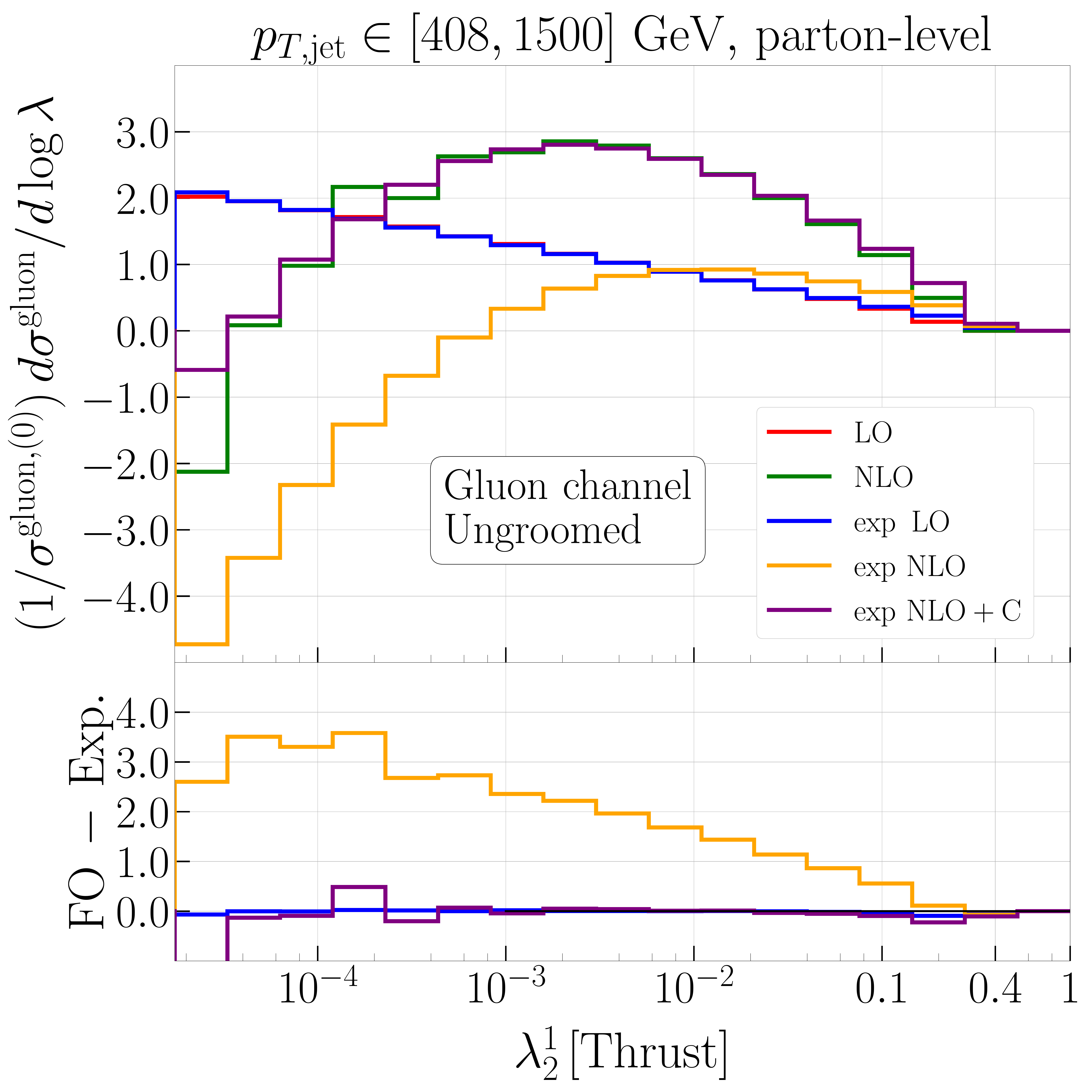}
  \hspace{1em}
  \includegraphics[width=0.44\linewidth]{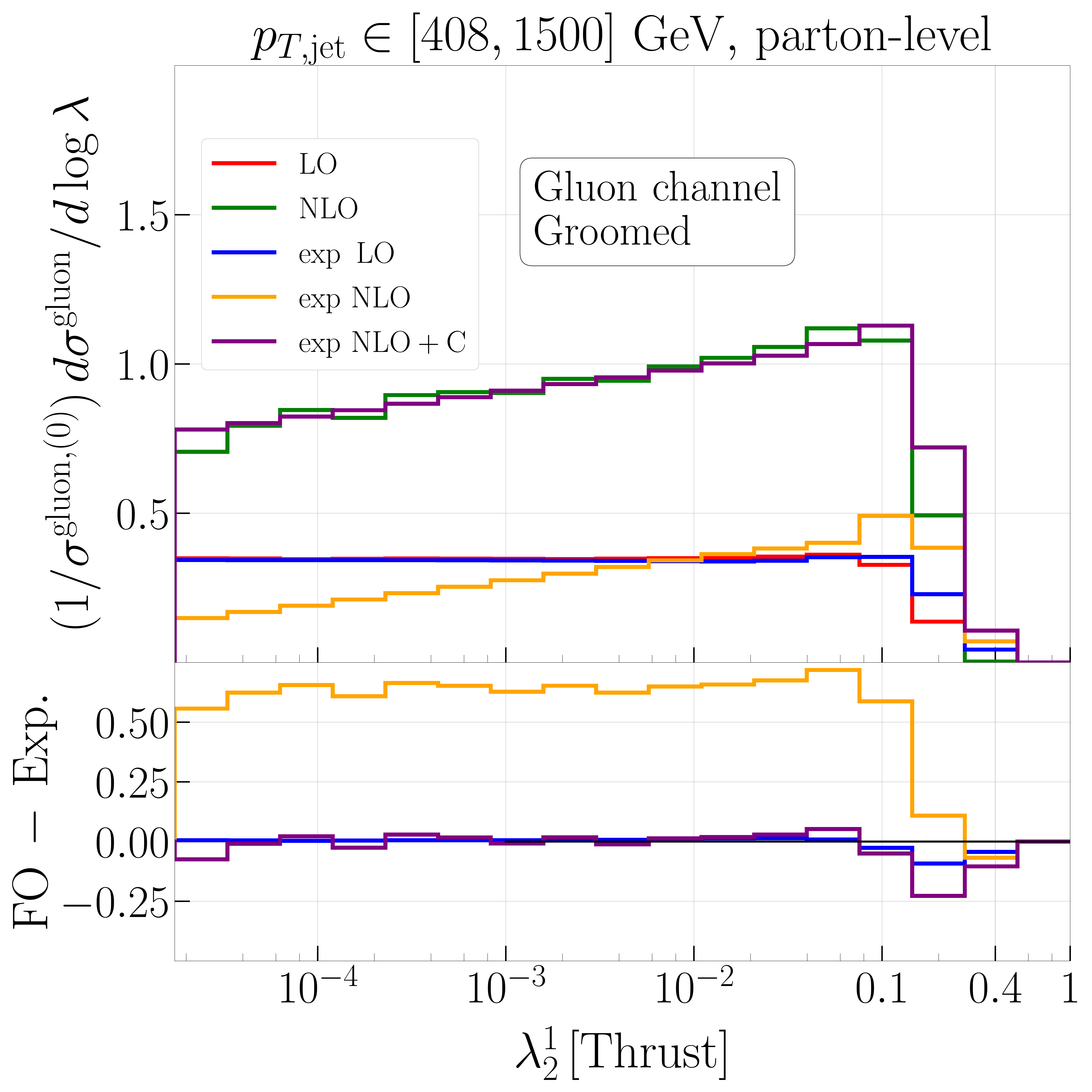}
   \caption{Fixed-order predictions for the gluon channel, identified with the BSZ algorithm, for ungroomed (left column) and groomed (right column) angularities $\lambda^1_\alpha$, for $\alpha\in\{1/2,1,2\}$, compared to the expansion of the resummation at the corresponding order of $\alphaS$, see text for details. The jet transverse momentum is constrained to $p_{T,\text{jet}}\in[408,1500]~\text{GeV}$.}\label{fig:FOvsExp_Gluon} 
\end{figure}

\begin{figure}
  \centering
  \includegraphics[width=0.44\linewidth]{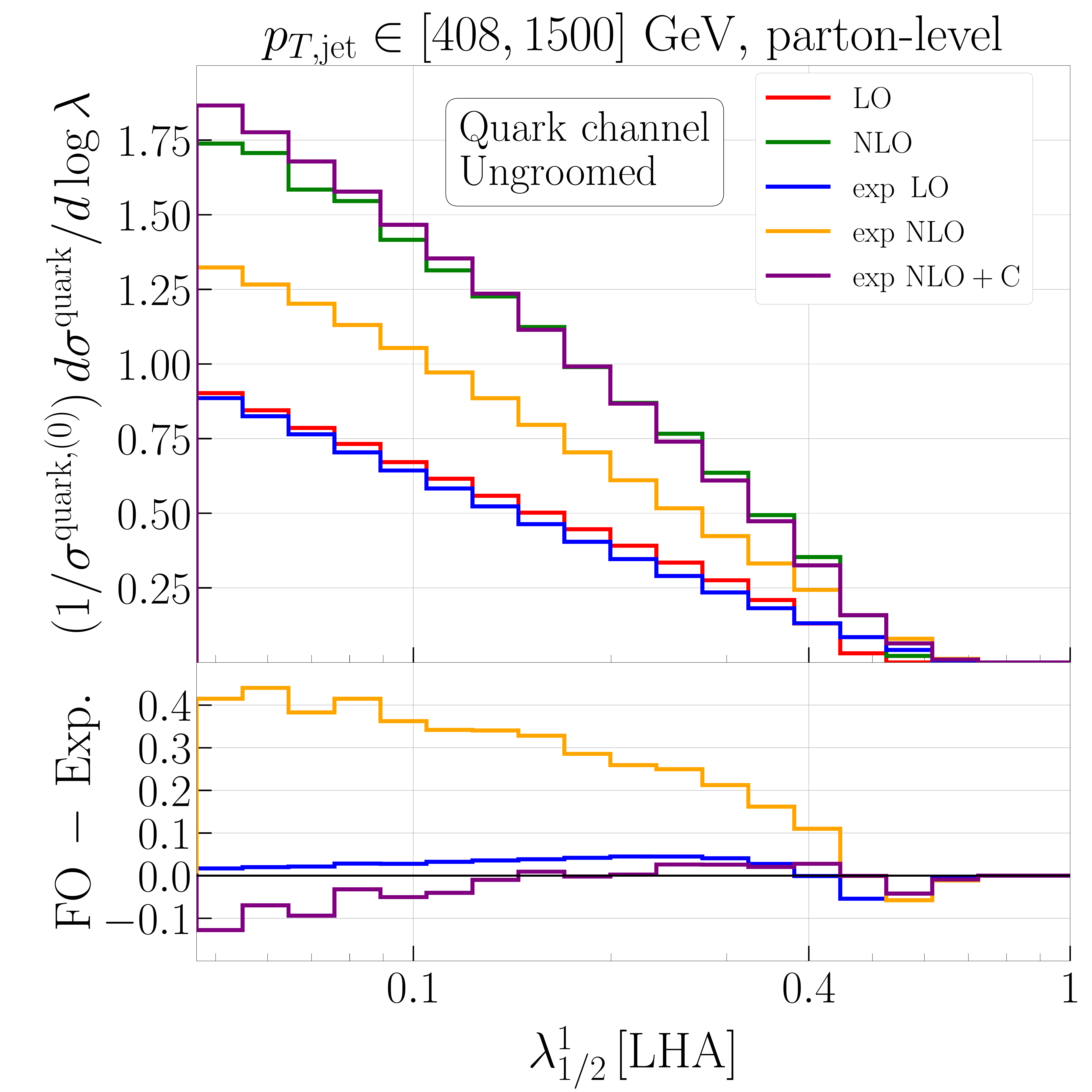}
  \hspace{1em}
  \includegraphics[width=0.44\linewidth]{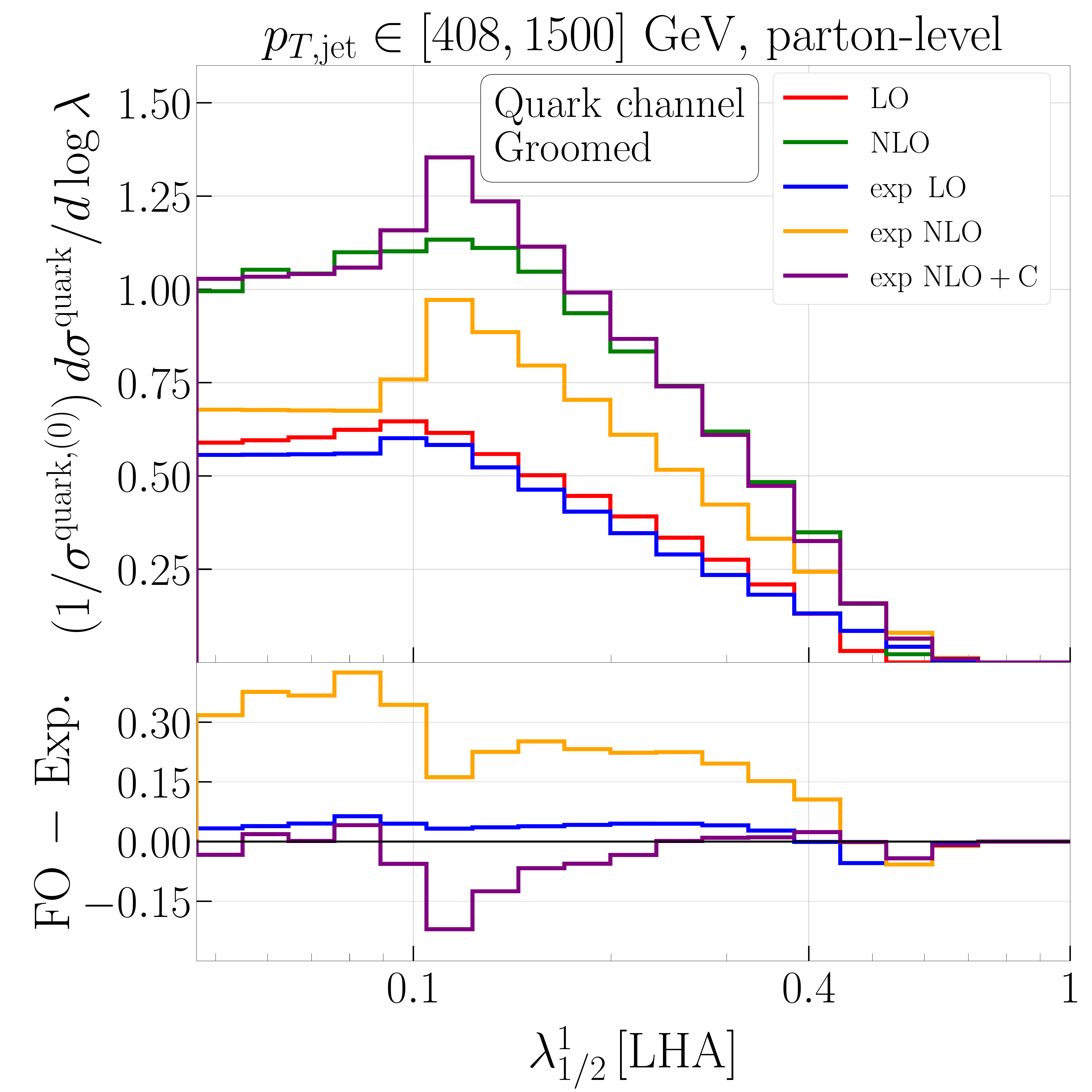}
  \centering
  \includegraphics[width=0.44\linewidth]{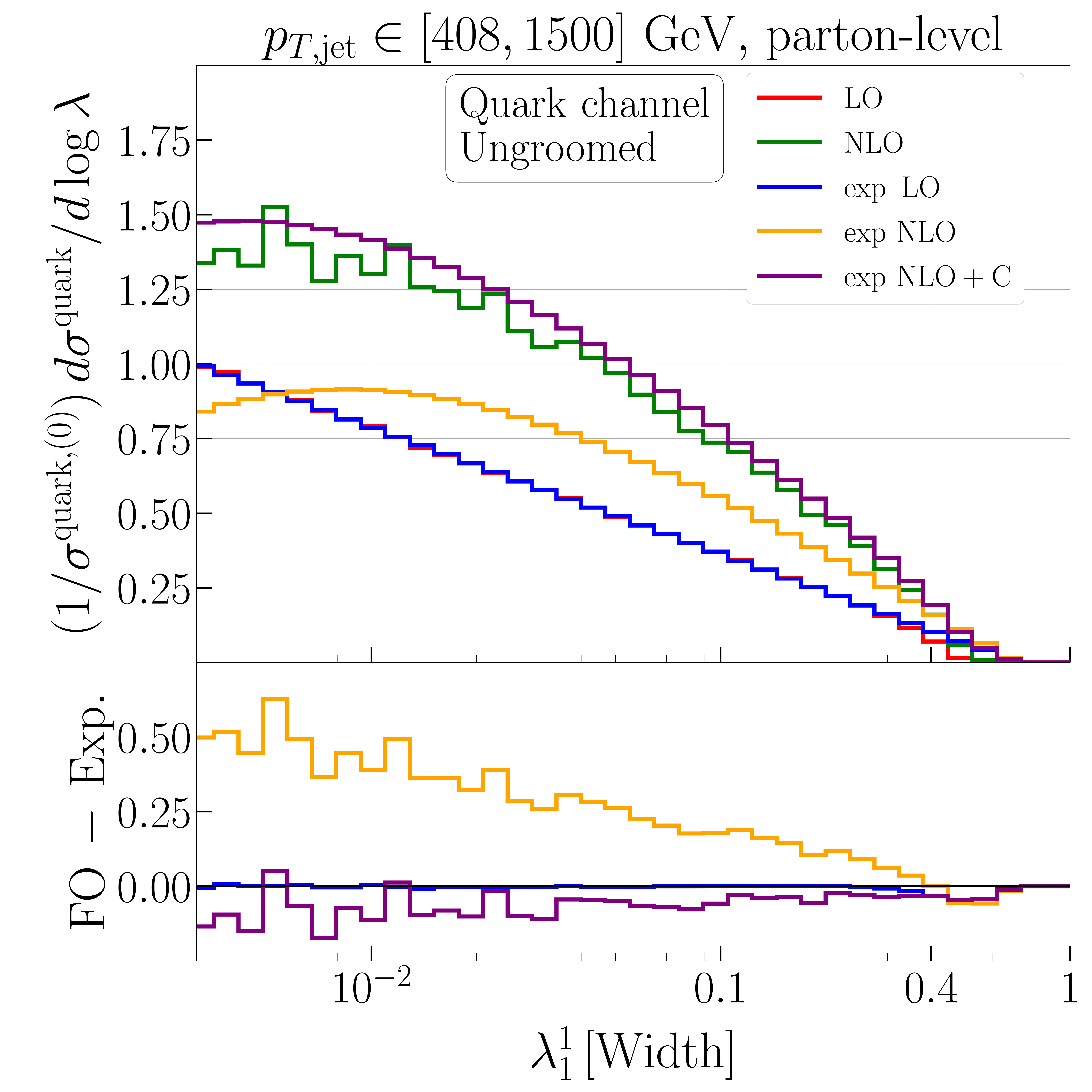}
  \hspace{1em}
  \includegraphics[width=0.44\linewidth]{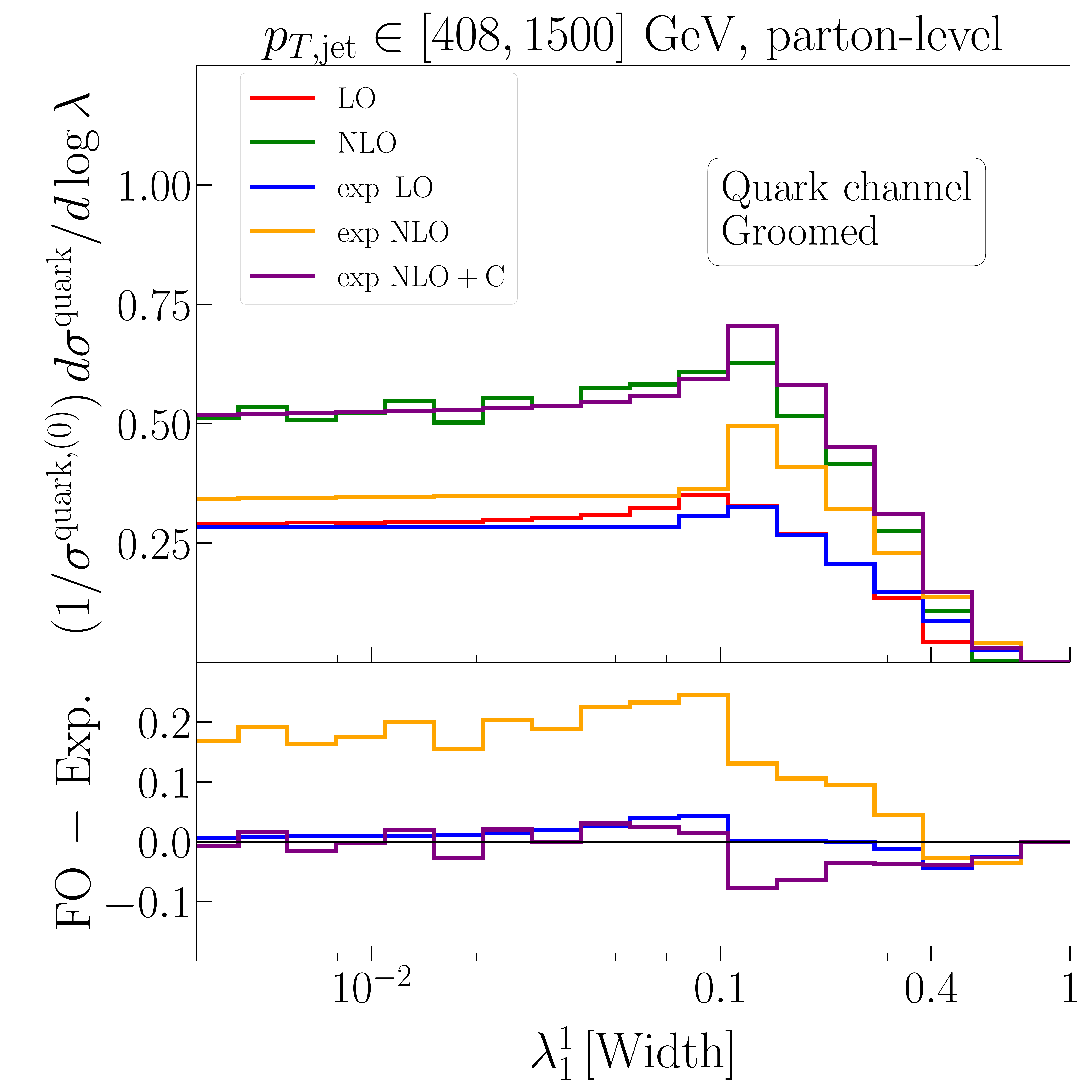}
  \centering
  \includegraphics[width=0.44\linewidth]{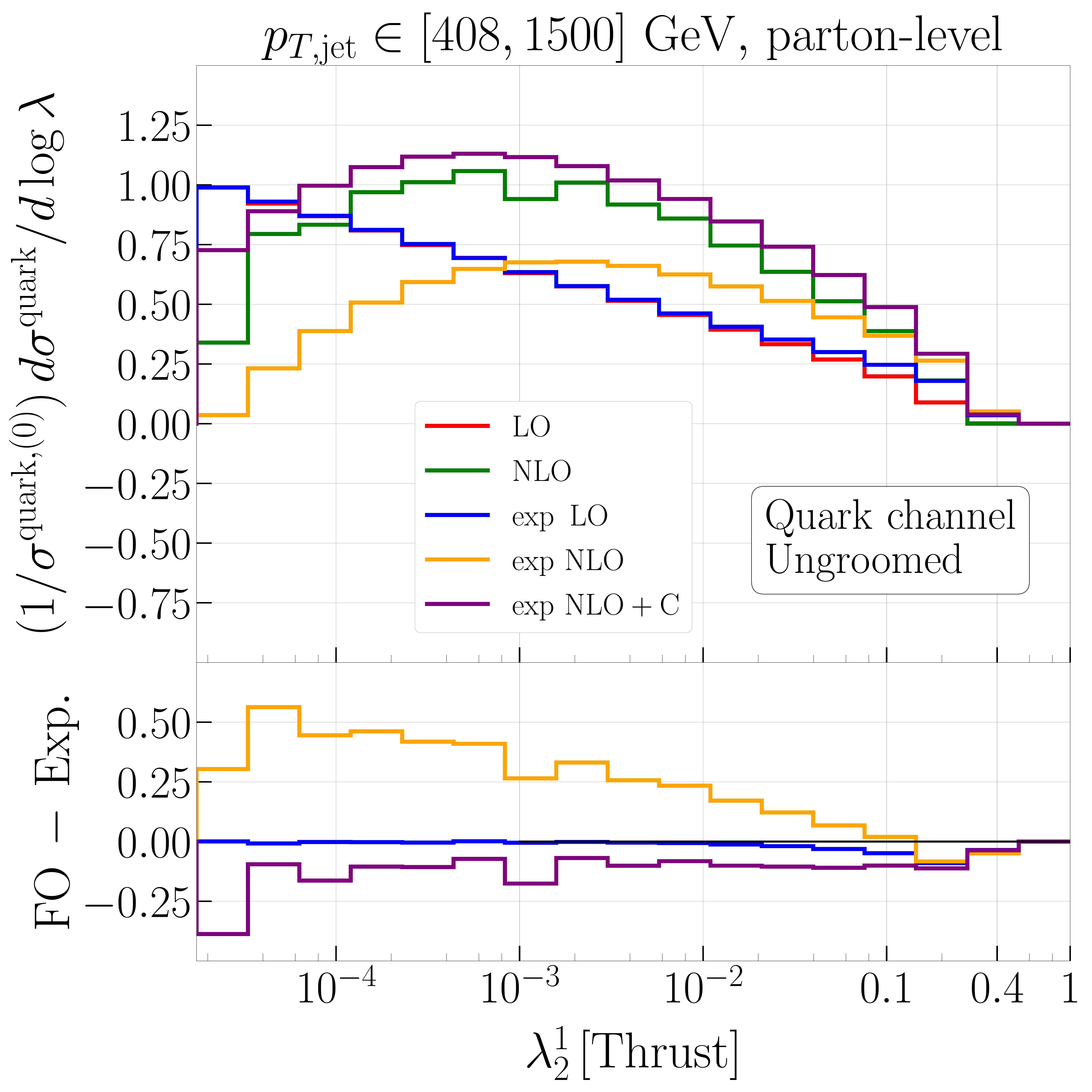}
  \hspace{1em}
  \includegraphics[width=0.44\linewidth]{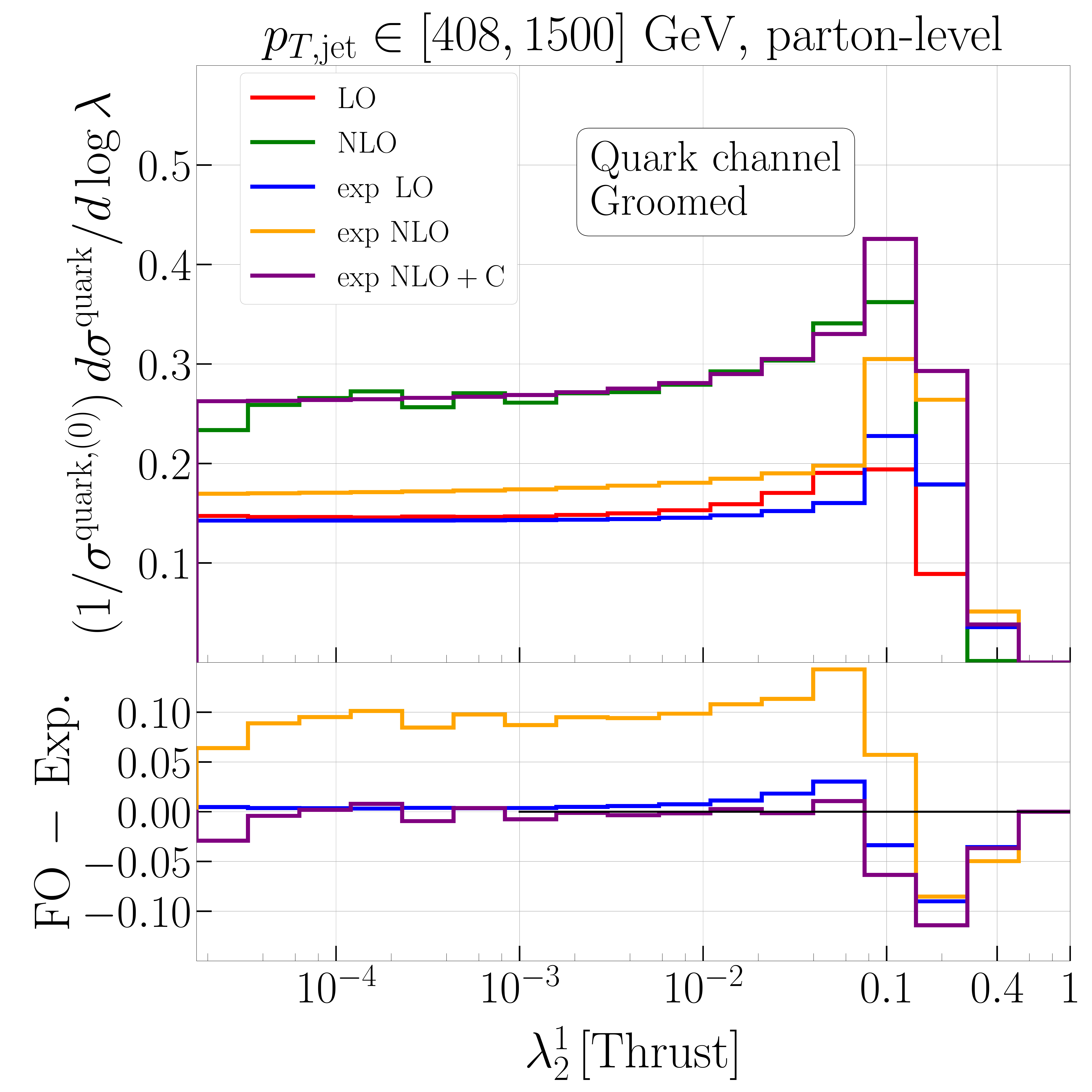}
   \caption{Fixed-order predictions for the quark channel, identified with the BSZ algorithm, for ungroomed (left column) and groomed (right column) angularities $\lambda^1_\alpha$, for $\alpha\in\{1/2,1,2\}$, compared to the expansion of the resummation at the corresponding order of $\alphaS$, see text for details. The jet transverse momentum is constrained to $p_{T,\text{jet}}\in[408,1500]~\text{GeV}$.}\label{fig:FOvsExp_Quark} 
\end{figure}

Splitting up the full fixed-order calculation into separate flavour channels in
an infrared safe way, enables us to validate our results, in particular the effective
inclusion of the $C^{\delta,(1)}$ coefficients, for each partonic channel $\delta$
individually. We hence compare the leading-order contributions
\begin{align}
  \Sigma_\text{LO}^{\delta}(\lambda) &= \sigma^{\delta,(0)} +
    \Sigma_\mathrm{fo}^{\delta,(1)}(\lambda)\,, \label{eq:SigmaFOLO}\\
  \Sigma_\text{exp LO}^{\delta}(\lambda) &= \sigma^{\delta,(0)} +
    \Sigma_\mathrm{res}^{\delta,(1)}(\lambda)\,,\label{eq:SigmaEXPLO}
\end{align}
obtained respectively for the exact matrix element and for the
expansion of the resummed calculation.
Similarly, at NLO, we compare
\begin{align}
  \Sigma_\text{NLO}^{\delta}(\lambda) &= \sigma^{\delta,(0)} +
    \Sigma_\mathrm{fo}^{\delta,(1)}(\lambda) -
    \overline{\Sigma}^{\delta,(2)}_\mathrm{fo}(\lambda)\,,\label{eq:SigmaFONLO}\\
  \Sigma_\text{exp NLO}^{\delta}(\lambda) &= \sigma^{\delta,(0)} +
  \Sigma_\mathrm{res}^{\delta,(1)}(\lambda) +
  \Sigma^{\delta,(2)}_\mathrm{res}(\lambda) \,,\label{eq:SigmaEXPNLO}\\
  \Sigma_\text{exp NLO+C}^{\delta}(\lambda) &= \sigma^{\delta,(0)} +
  \left(1+\frac{\Sigma_\mathrm{fo}^{\delta,(1)}(\lambda)-\Sigma_\mathrm{res}^{\delta,(1)}(\lambda)}{\sigma^{\delta,(0)}}\right)
  \Sigma_\mathrm{res}^{\delta,(1)}(\lambda) + \Sigma^{\delta,(2)}_\mathrm{res}\,,\label{eq:SigmaEXPCNLO}
\end{align}
where $\Sigma_\text{exp NLO}^{\delta}(\lambda)$ is the NLO expansion
of the all-order result, while $\Sigma_\text{exp
  NLO+C}^{\delta}(\lambda)$ also includes the contribution from the $C^{\delta,(1)}$ coefficient.
Note that $\Sigma_\text{NLO}(1) = \Sigma_\text{LO}(1) = \sum_\delta\left(
\sigma^{\delta,(0)} + \sigma^{\delta,(1)}_\mathrm{fo}\right)$. 

In Figs.~\ref{fig:FOvsExp_Gluon} and~\ref{fig:FOvsExp_Quark}
we present the differential distributions $d\sigma/d\log\lambda\equiv
d\Sigma(\lambda)/d\log\lambda$ in $\lambda$,
separately for the gluon (Fig.~\ref{fig:FOvsExp_Gluon})
and quark (Fig.~\ref{fig:FOvsExp_Quark}) channel, identified by the BSZ flavour-$k_t$ algorithm, for the
representative jet-$p_{T}$ slice $p_{T,\text{jet}}\in [408,1500]\;\text{GeV}$.
We have checked that analogous results are obtained for the other transverse momentum slices. 
In all these validations we assume a jet radius of $R_0=0.8$. As before, we consider the angularities
$\lambda^1_\alpha$ with $\alpha\in\{1/2,1,2\}$ either without grooming or with \softdrop grooming
using $\zcut=0.1,\,\beta=0$.
For each observable we include an additional panel that contains
the differences between the fixed-order result, \emph{i.e.}\ Eqs.~\eqref{eq:SigmaFOLO} and
\eqref{eq:SigmaFONLO}, and the expansions of the resummation up to order $\alphaS$ and $\alphaS^2$,
\emph{i.e.}\ Eqs.~\eqref{eq:SigmaEXPLO} and \eqref{eq:SigmaEXPNLO}, \eqref{eq:SigmaEXPCNLO}, respectively.

We first note that for all angularities --- groomed or ungroomed --- the difference
between the derivatives of $\Sigma_\text{LO}$ and $\Sigma_\text{exp LO}$
vanishes for $\lambda\ll 1$, as expected in the ungroomed case. For the groomed angularities
there are in principle, for $\beta = 0$, deviations of
$\mathcal{O}(\zcut)$. Those appear to be too small to observe numerically here,
as was also observed for example for event shapes in \cite{Baron:2020xoi}. In comparing  
$d\Sigma_\text{NLO}/dL$ and  
$d\Sigma_\text{exp NLO}/dL$ (with $L\equiv \log(1/\lambda)$) we
observe a linearly rising difference, indicating missing
terms of order $\alphaS^2 L^2$ in $\Sigma_\text{exp NLO}$. By including the
$C^{\delta,(1)}$ coefficient in $\Sigma_\text{exp NLO+C}$, the derivatives
only differ by a constant, confirming that missing terms in the
cumulative distribution are reduced to order $\alphaS^2 L$.

\begin{figure}
  \centering
  \includegraphics[width=0.44\linewidth]{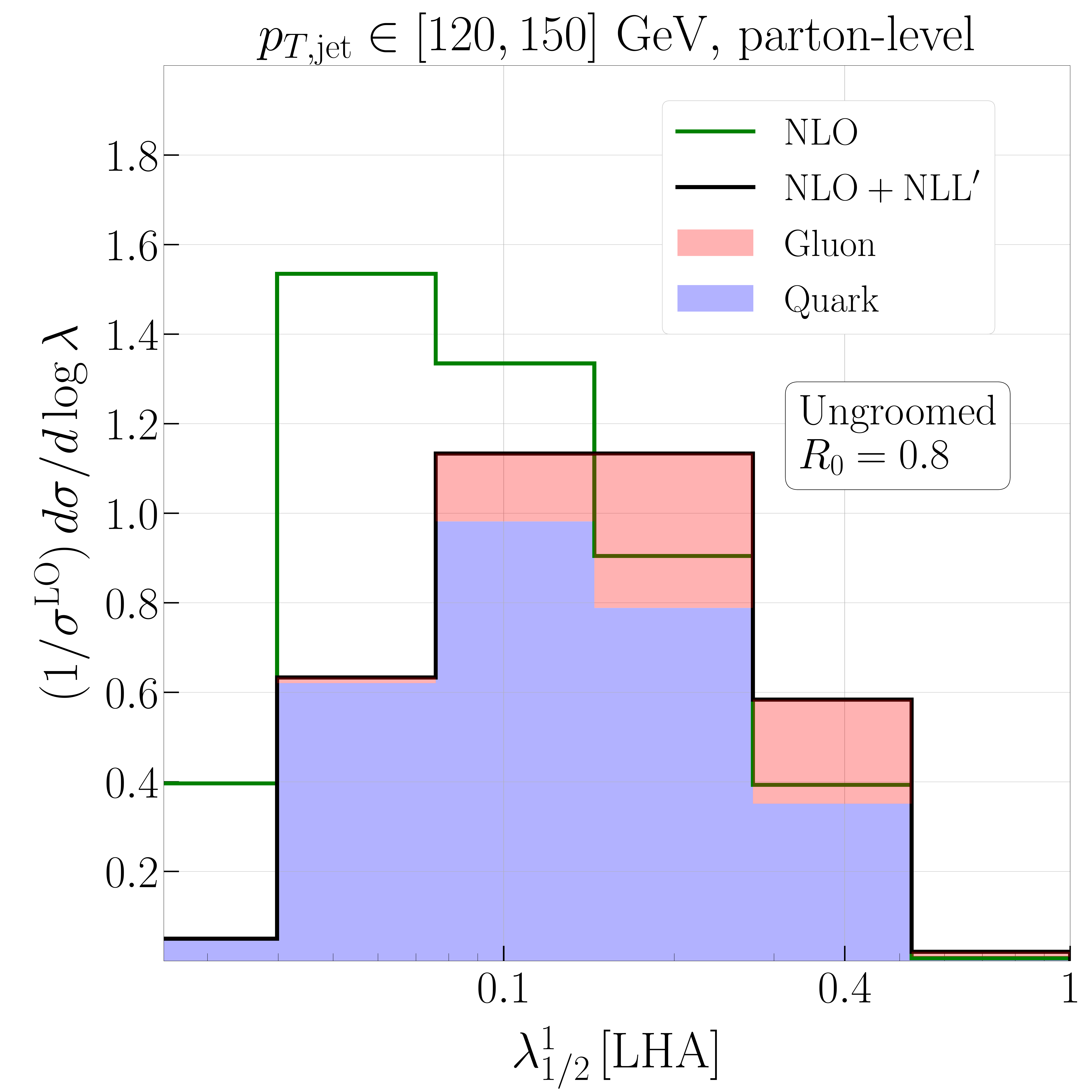}
  \hspace{1em}
  \includegraphics[width=0.44\linewidth]{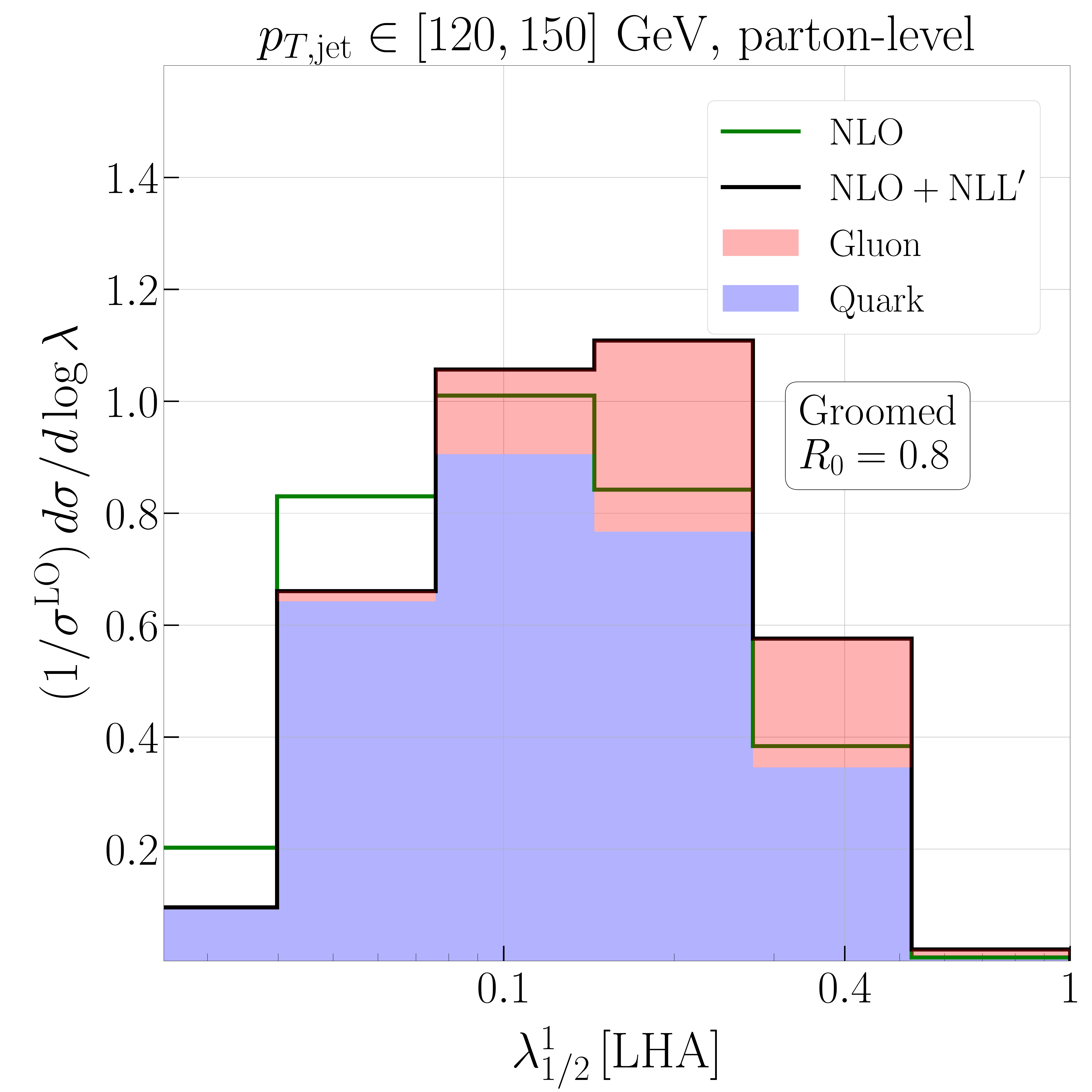}
  \centering
  \includegraphics[width=0.44\linewidth]{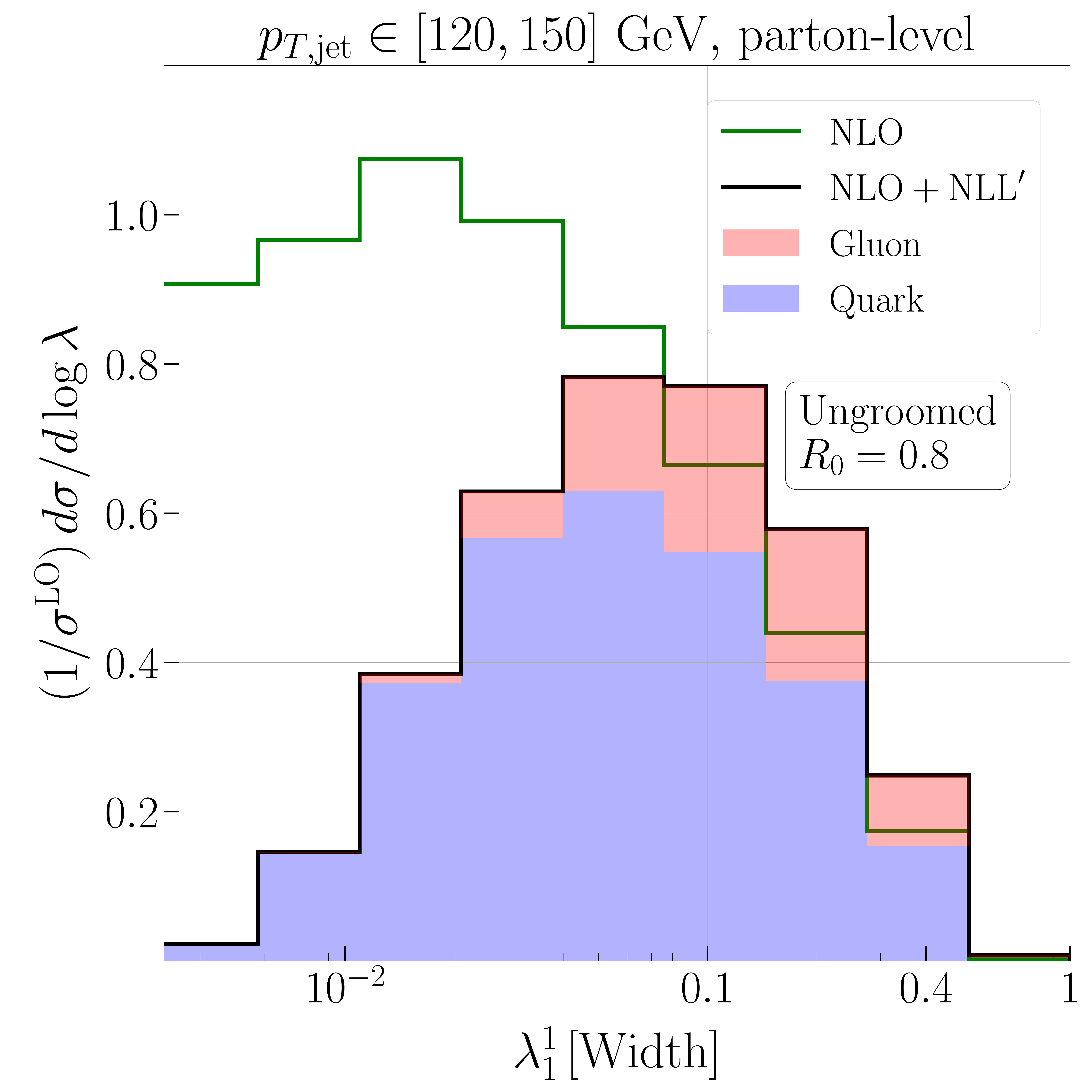}
  \hspace{1em}
  \includegraphics[width=0.44\linewidth]{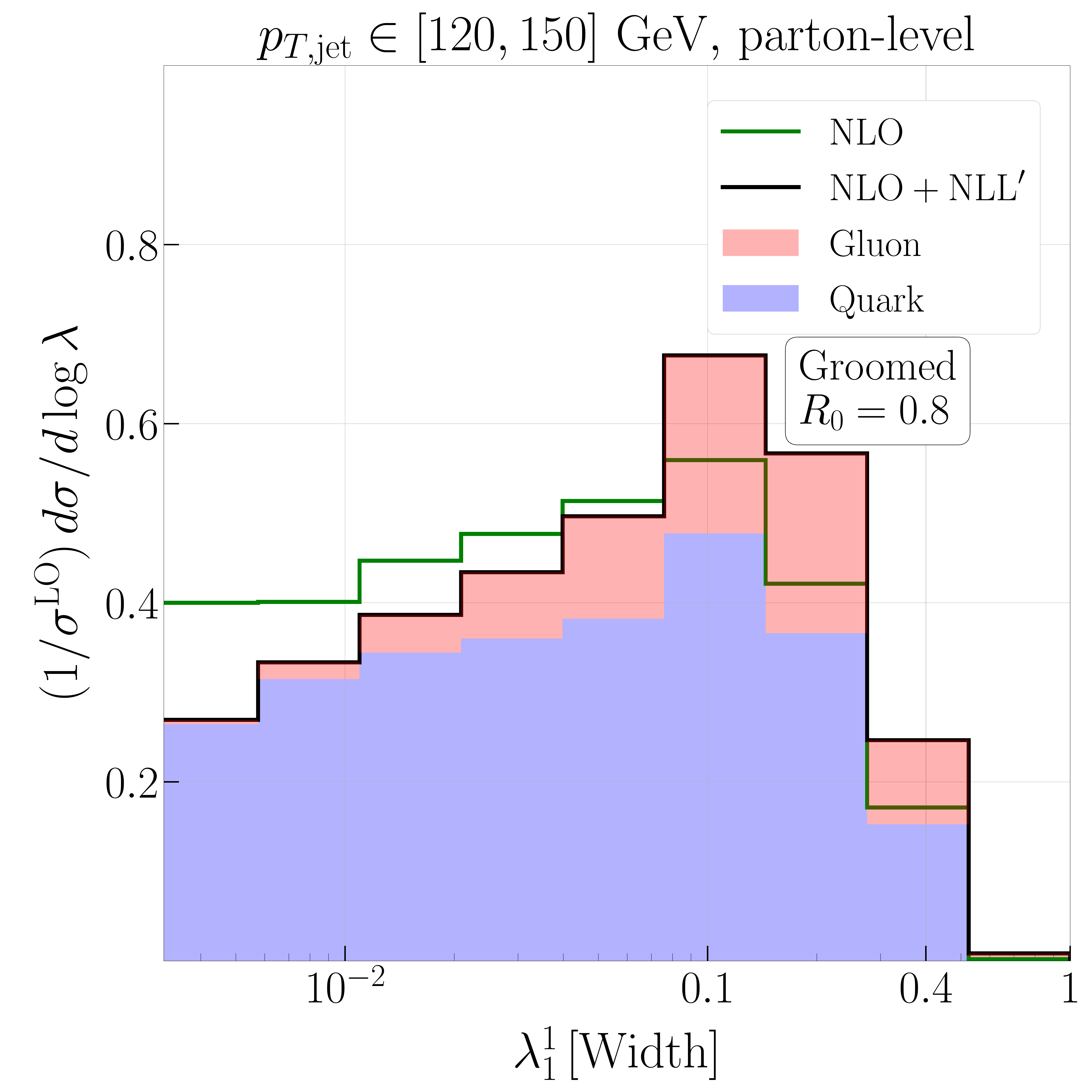}
  \centering
  \includegraphics[width=0.44\linewidth]{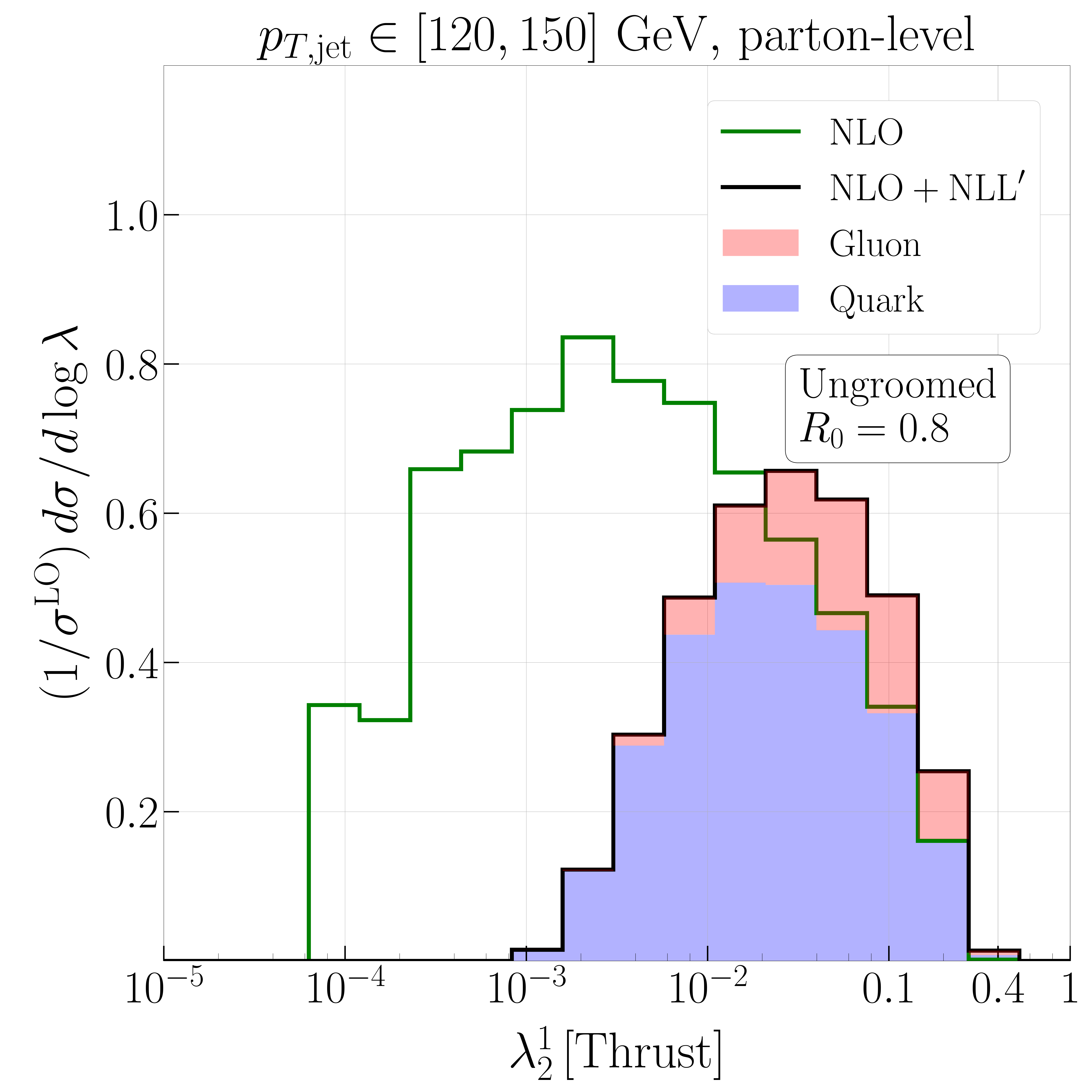}
  \hspace{1em}
  \includegraphics[width=0.44\linewidth]{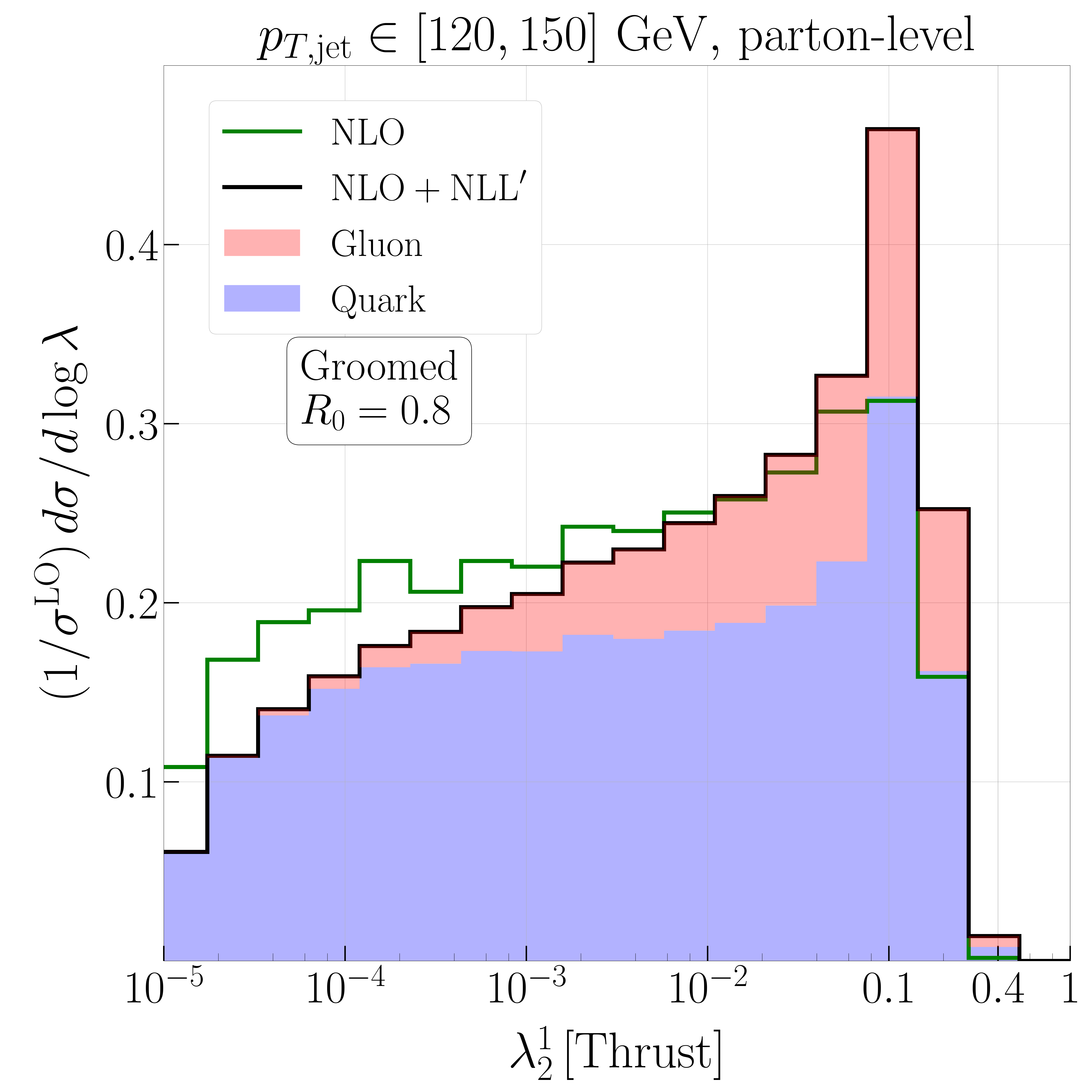}
  \caption{Fixed-order and matched predictions at \NLO and \NLOpNLLp accuracy,
    for ungroomed (left row) and groomed (right row) angularities, for
    $\alpha\in\{1/2,1,2\}$. The colour scheme indicates how the gluon (red) and
    quark (blue) channel as identified with the BSZ algorithm stack up to form
    the \NLOpNLLp prediction. The jet transverse momentum is
    constrained to $p_{T,\text{jet}}\in[120,150]~\text{GeV}$.}\label{fig:Matched_PT120}
\end{figure}

\begin{figure}
  \centering
  \includegraphics[width=0.44\linewidth]{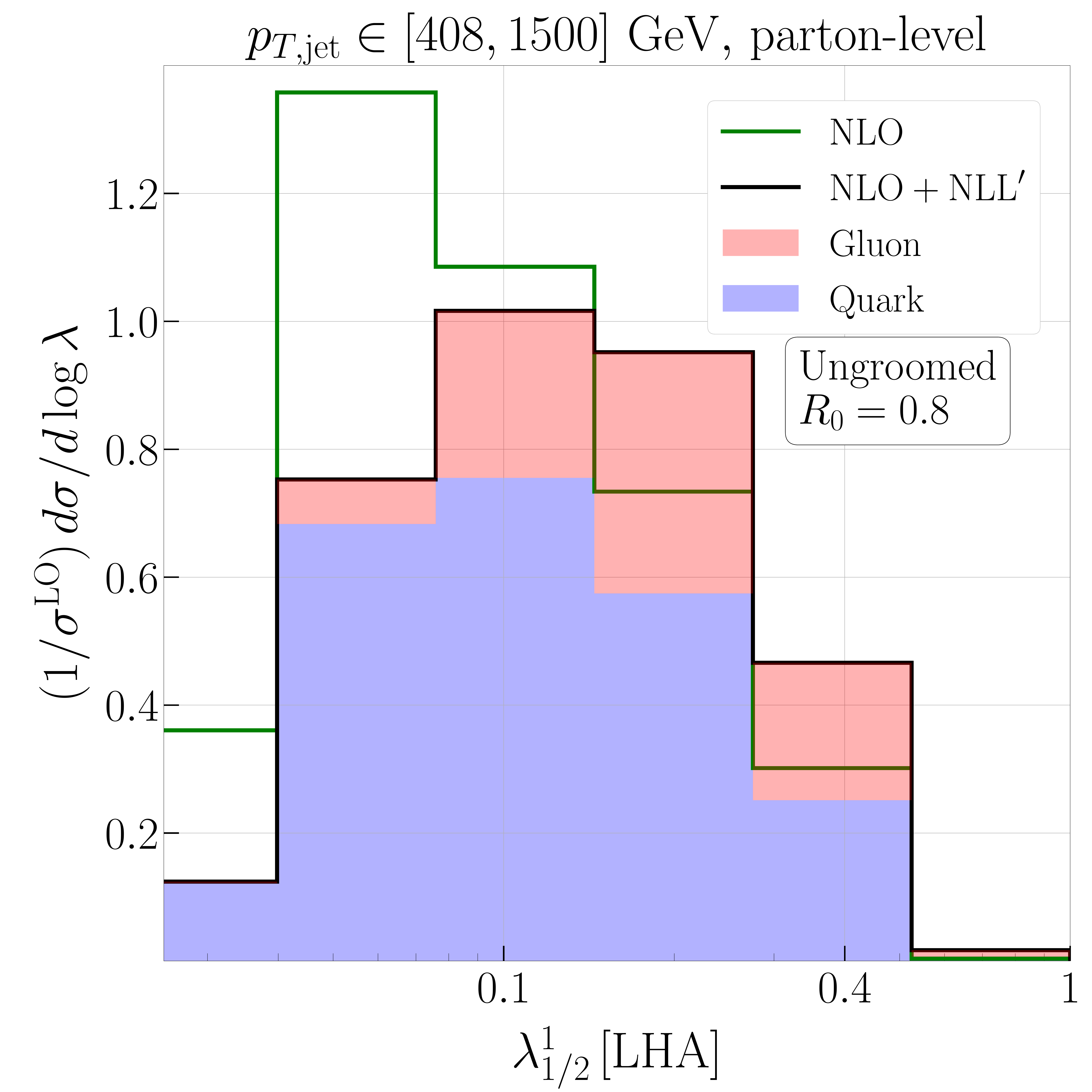}
  \hspace{1em}
  \includegraphics[width=0.44\linewidth]{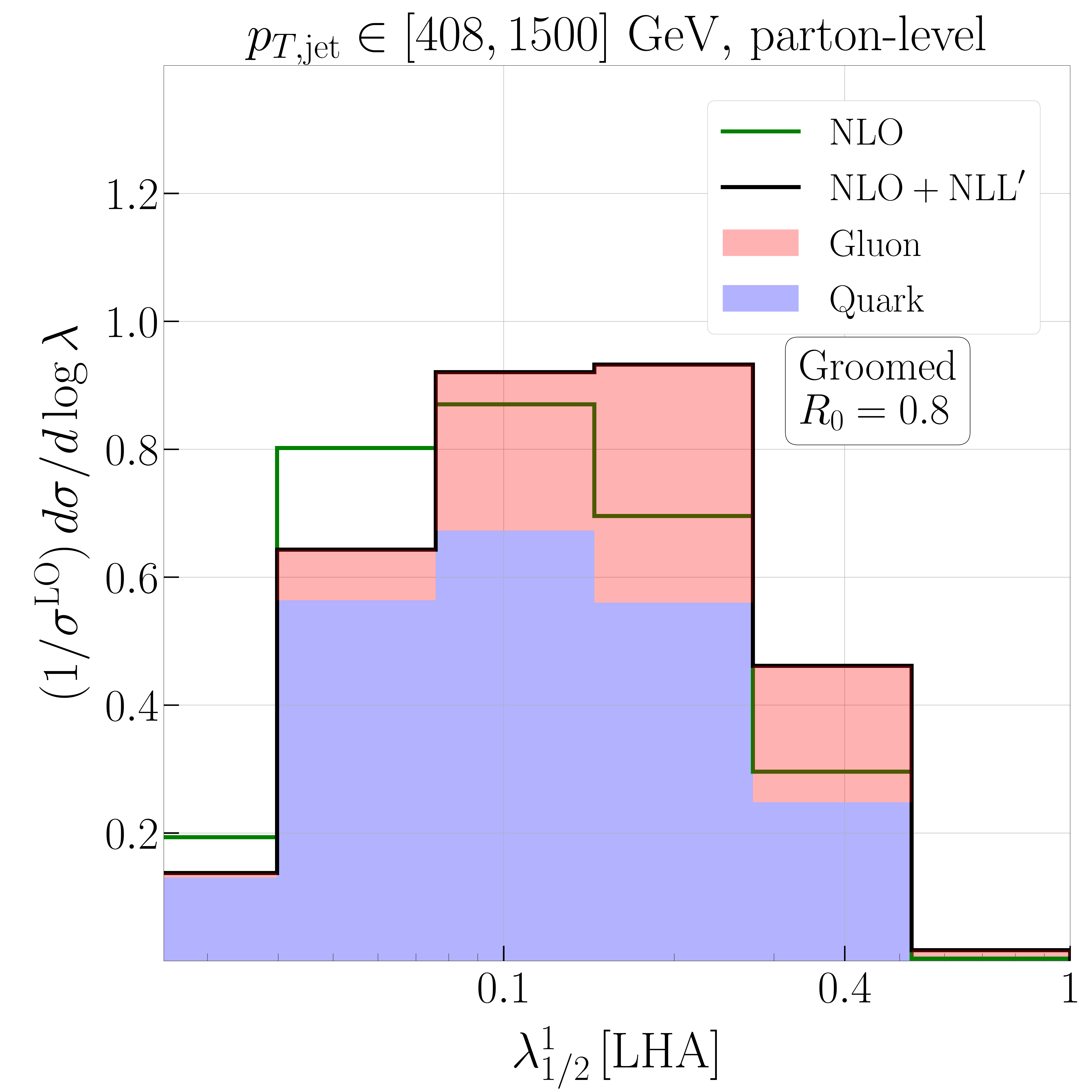}
  \centering
  \includegraphics[width=0.44\linewidth]{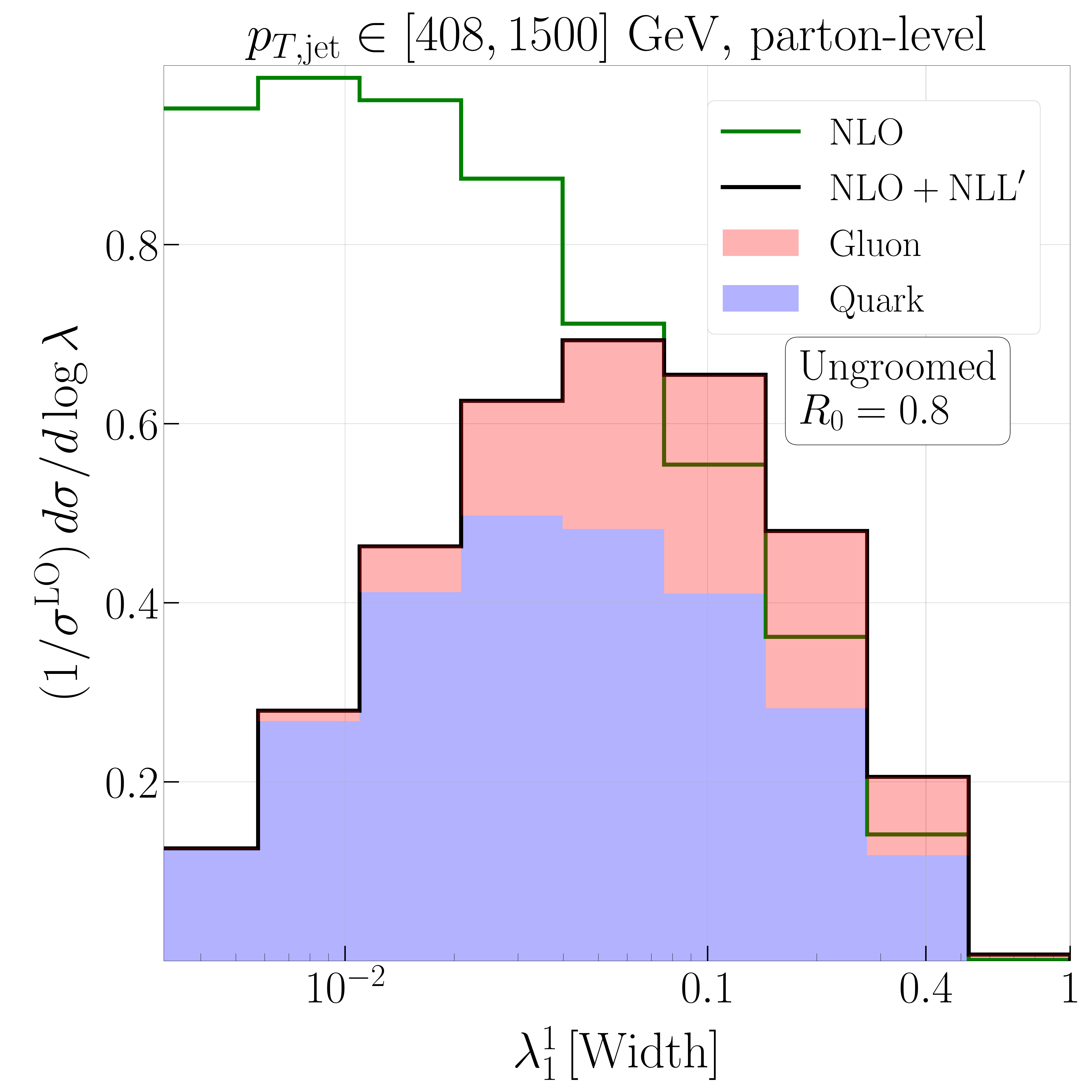}
  \hspace{1em}
  \includegraphics[width=0.44\linewidth]{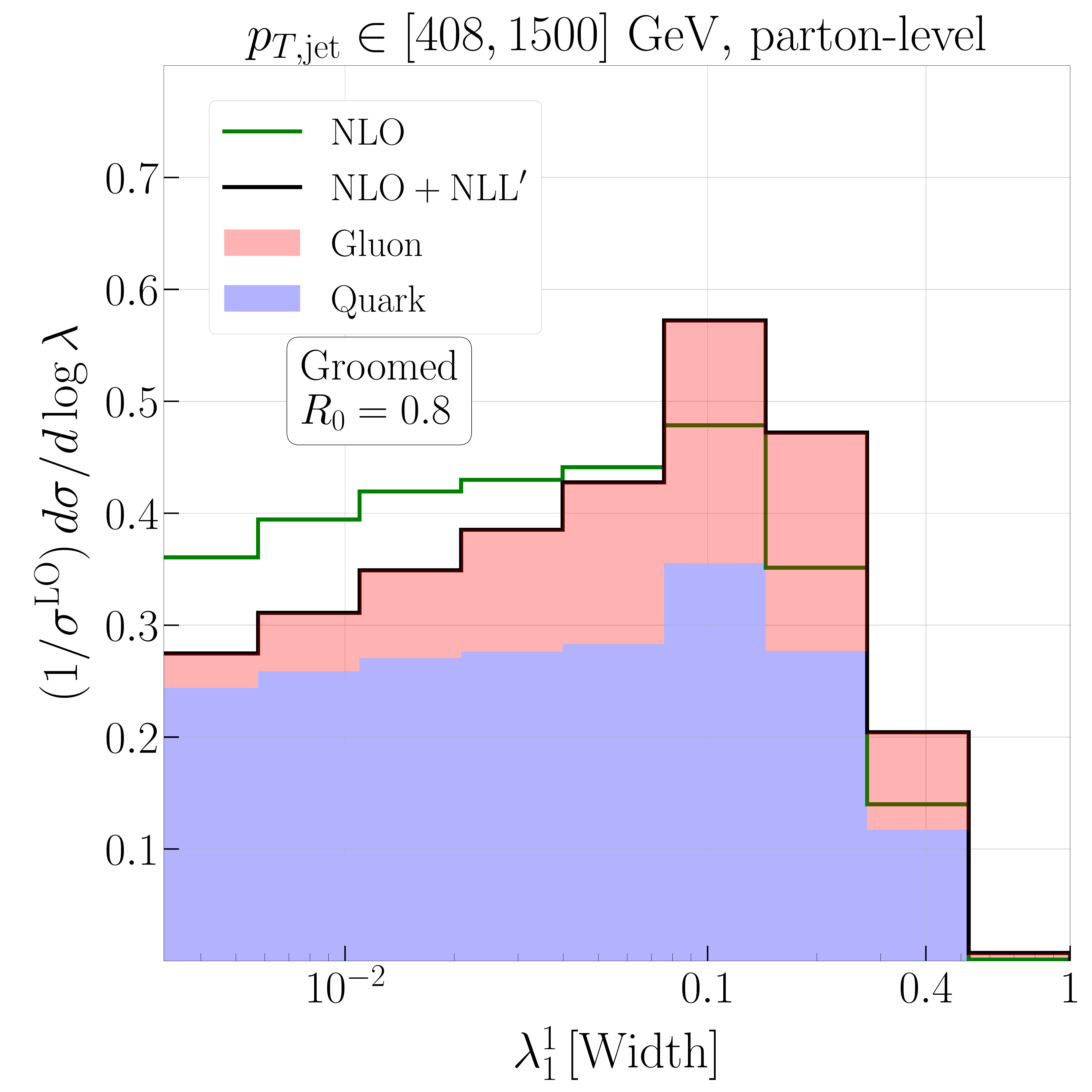}
  \centering
  \includegraphics[width=0.44\linewidth]{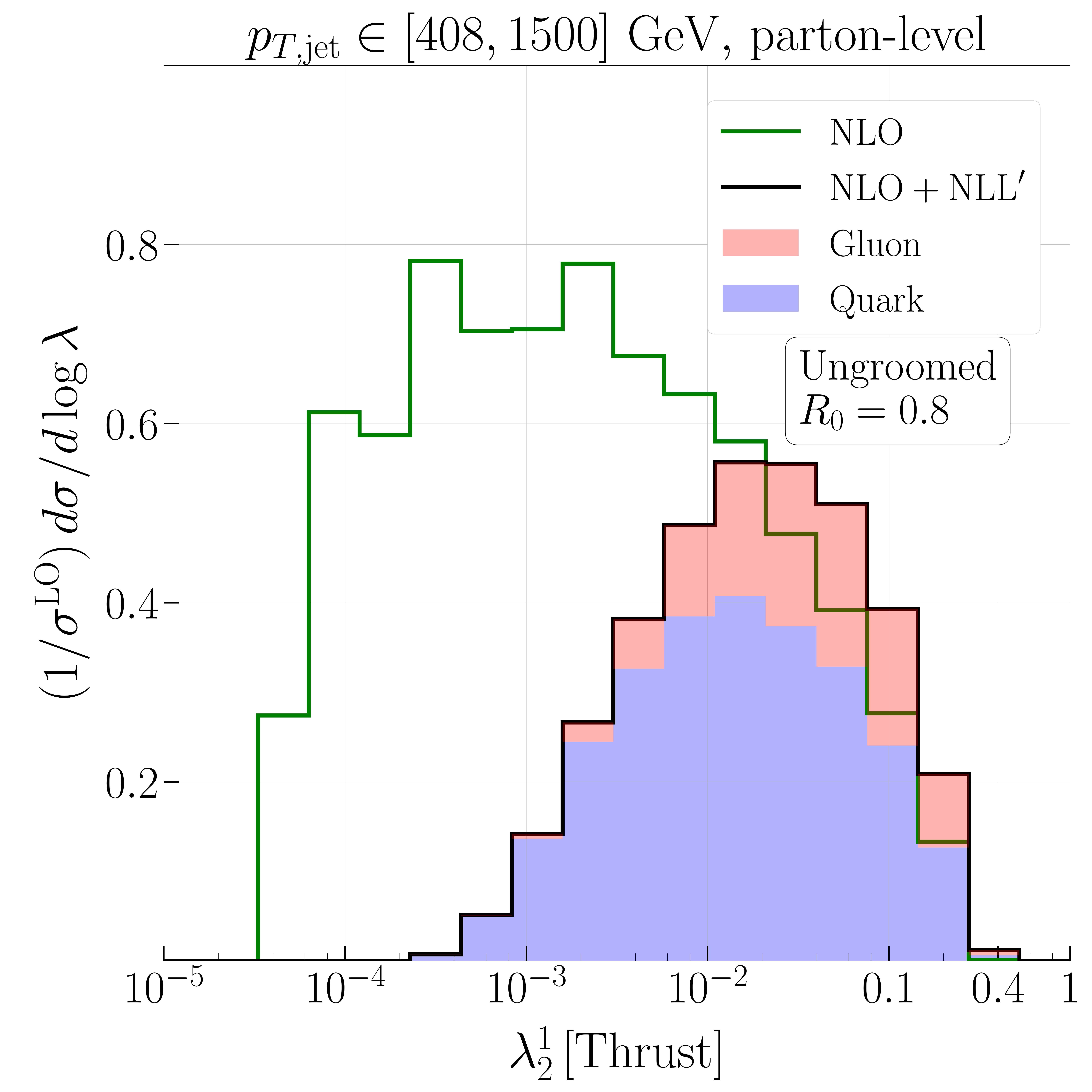}
  \hspace{1em}
  \includegraphics[width=0.44\linewidth]{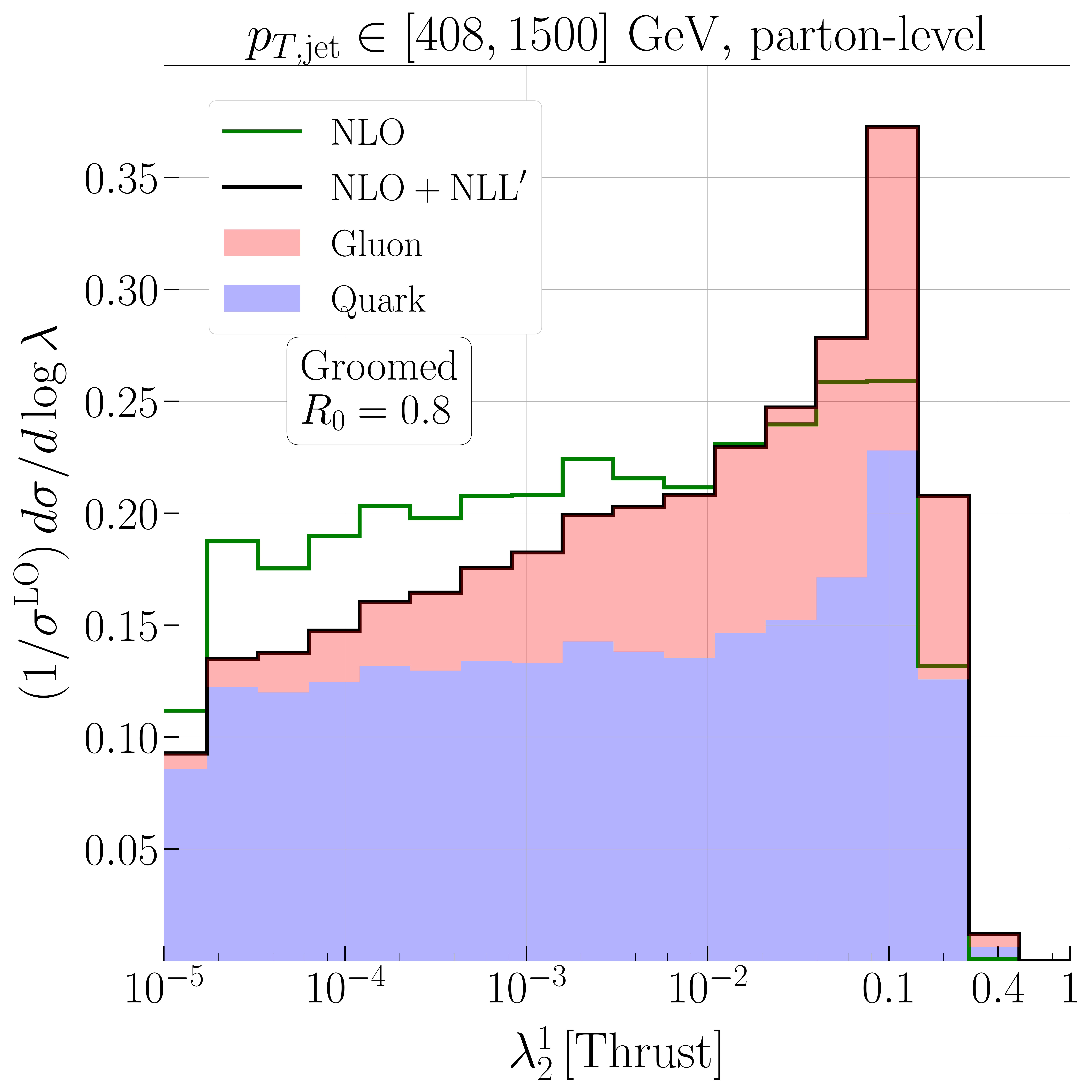}
  \caption{Fixed-order and matched predictions at \NLO and \NLOpNLLp accuracy,
    for ungroomed (left row) and groomed (right row) angularities, for
    $\alpha\in\{1/2,1,2\}$. The colour scheme indicates how the gluon (red) and
    quark (blue) channel as identified with the BSZ algorithm stack up to form
    the \NLOpNLLp prediction. The jet transverse momentum is
    constrained to $p_{T,\text{jet}}\in[408,1500]~\text{GeV}$.}\label{fig:Matched_PT408} 
\end{figure}

Having validated our resummation calculations by comparing their
expansions to the corresponding fixed-order predictions,
we finally present our matched resummed \NLOpNLLp results and compare them to their NLO counterparts. 
Figs.~\ref{fig:Matched_PT120} and~~\ref{fig:Matched_PT408} 
contain our predictions for the two considered transverse-momentum
slices, using the same set of observables and grooming parameters as
above.
These figures also illustrate how the full result is obtained as the
sum of the two flavour channels --- quarks and gluons ---  identified by our BSZ flavour-assignment
procedure.
We note that,  for all observables, the contribution from gluon jets
is increased  for the higher $p_{T,\text{jet}}$ slice  compared to the case $p_{T,\text{jet}} \in [120,150]~\text{GeV}$. However, for both slices we observe
a more significant contribution from gluon jets for larger values of the angularities, while the low-$\lambda^1_\alpha$
tails are entirely dominated by quark jets. 
This is a confirmation that a cut on the jet angularity can serve as a theoretically well-defined and IRC safe
quark--gluon discriminant, as pointed out, for instance, in Refs.~\citep{Gras:2017jty,Badger:2016bpw,Larkoski:2013eya,Larkoski:2014pca}.
We leave further investigation on this topic to future work. 
In addition to the matched resummation, we show the full fixed-order predictions at \NLO accuracy.
All-order effects turn out to be important essentially over the entire observable range. 
In particular, only for the highest values of the observables do \NLO and matched
predictions start to look similar. This appears to be the case for both ungroomed and groomed distributions.
After a detailed analysis, we have concluded that this effect, the size of which is rather surprising,
is driven by the large constant contribution of Eq.~(\ref{eq:c1}), which in turn originate from the large
perturbative corrections that characterise the $Z$+jets process. Similar observations have been
reported for the groomed jet mass in $Zj$ production in Ref.~\citep{Frye:2016aiz}.

\FloatBarrier

\section{Final predictions}
\label{sec:mc-comp}

In this section we provide our final theoretical predictions for the leading-jet
angularities. We start the
discussion with a comparison of our \NLOpNLLp  results with parton-level MC predictions from the \pythia, \herwig
and \sherpa event generators. We then discuss the size of non-perturbative corrections due to hadronisation and 
UE, and demonstrate how to account for these effects in our final \NLOpNLLp predictions.

\subsection{\NLOpNLLp at parton level}
\label{sec:nlo_nll_final_pl}

\begin{figure}
  \centering
  \includegraphics[width=0.44\linewidth]{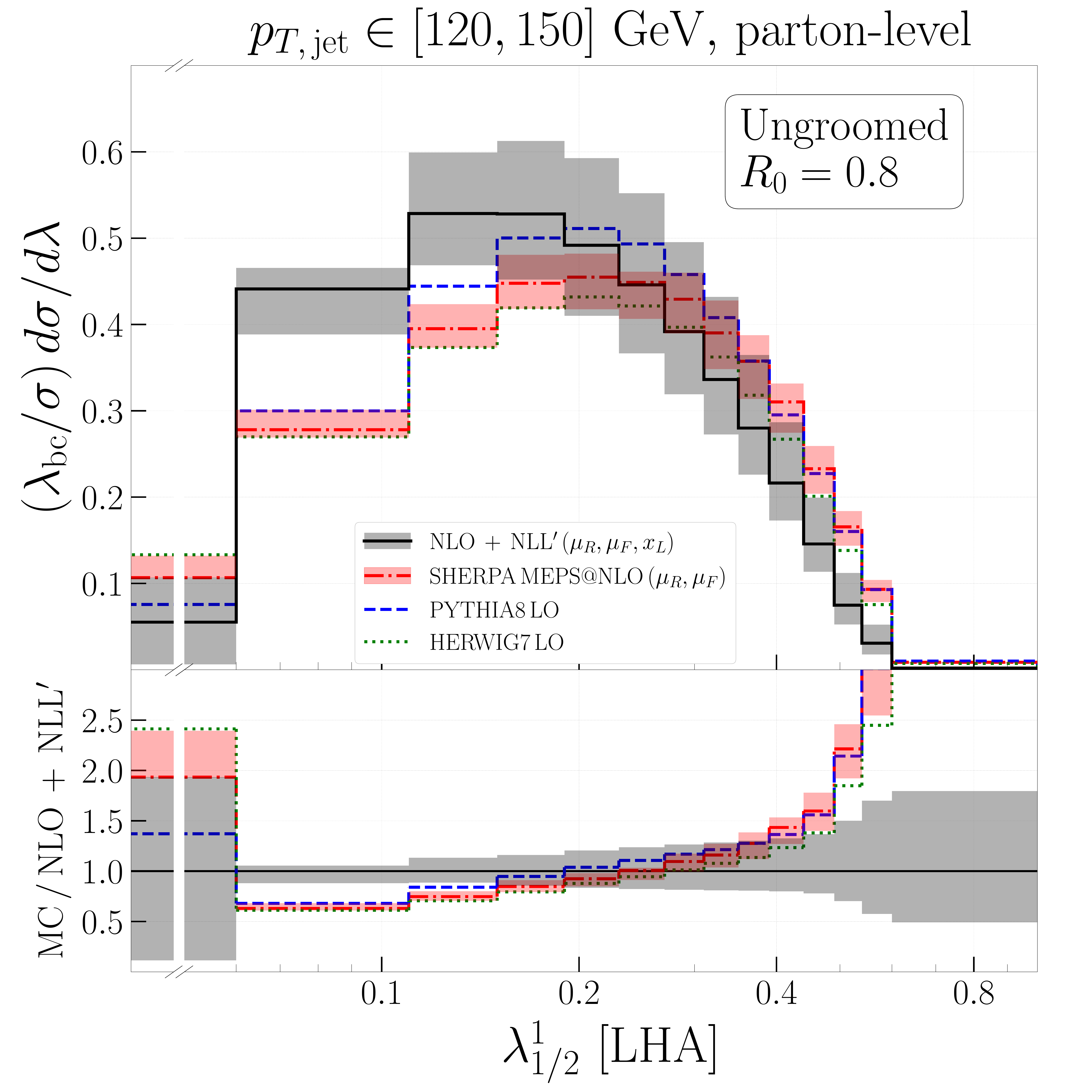}
  \hspace{1em}
  \includegraphics[width=0.44\linewidth]{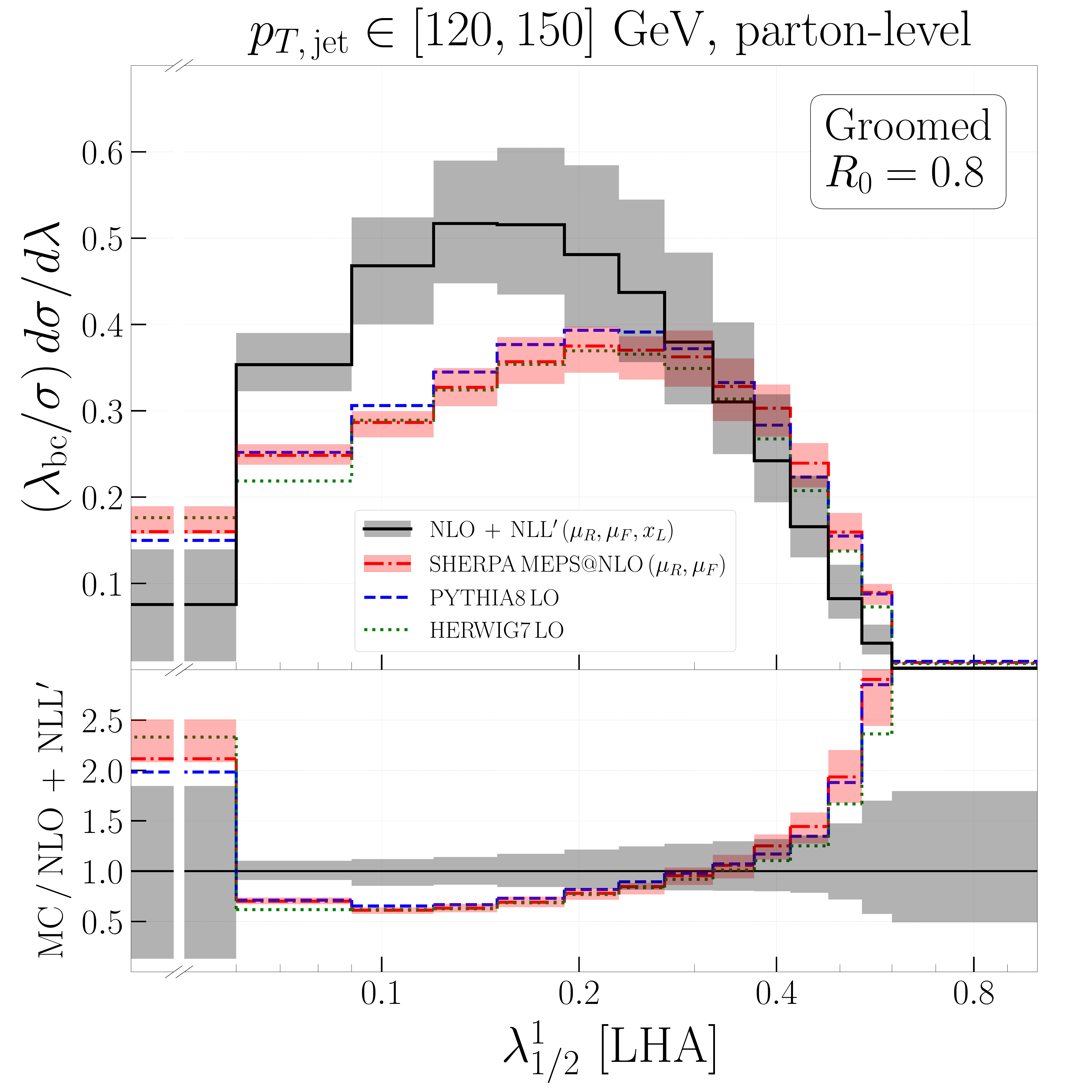}
  \centering
  \includegraphics[width=0.44\linewidth]{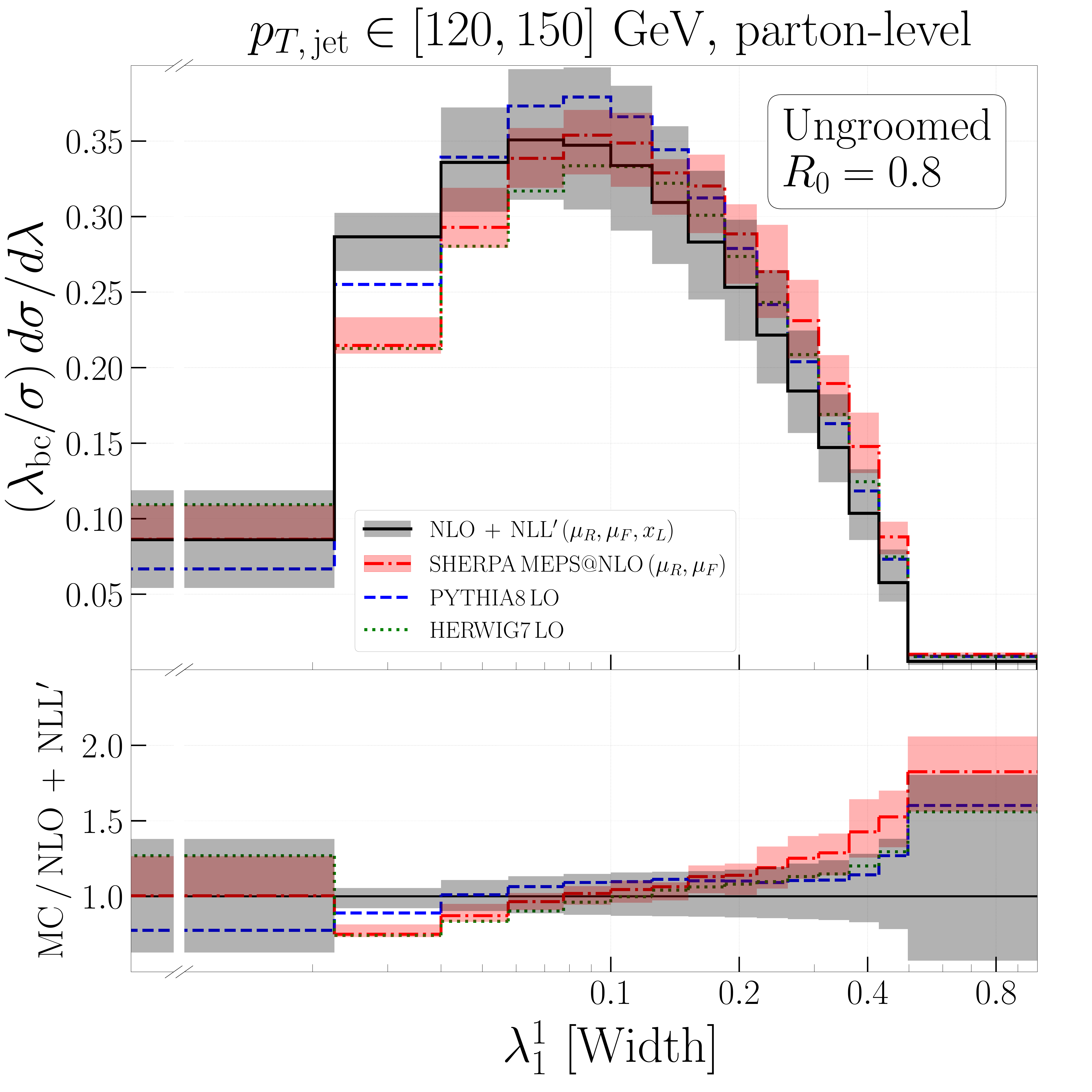}
  \hspace{1em}
  \includegraphics[width=0.44\linewidth]{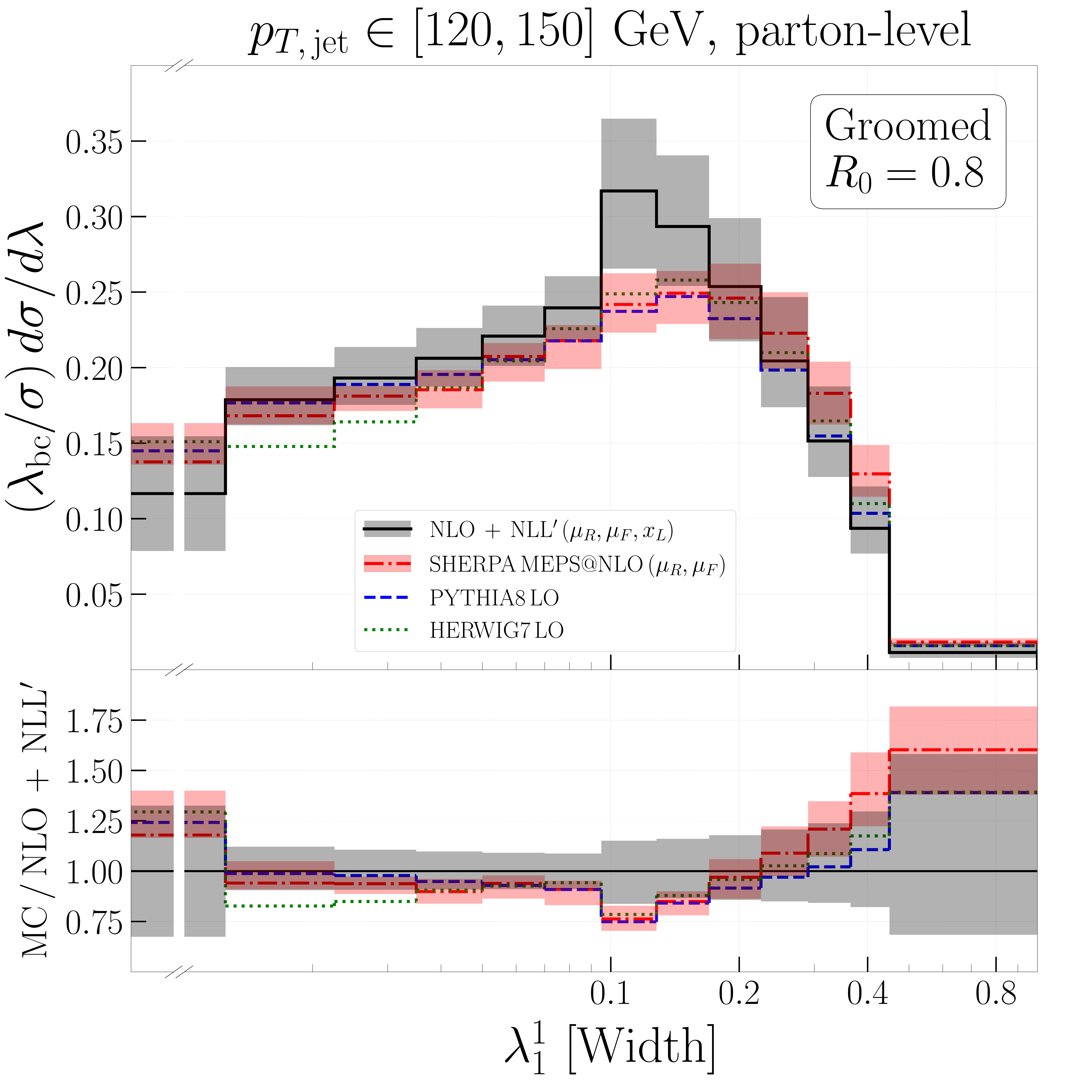}
  \centering
  \includegraphics[width=0.44\linewidth]{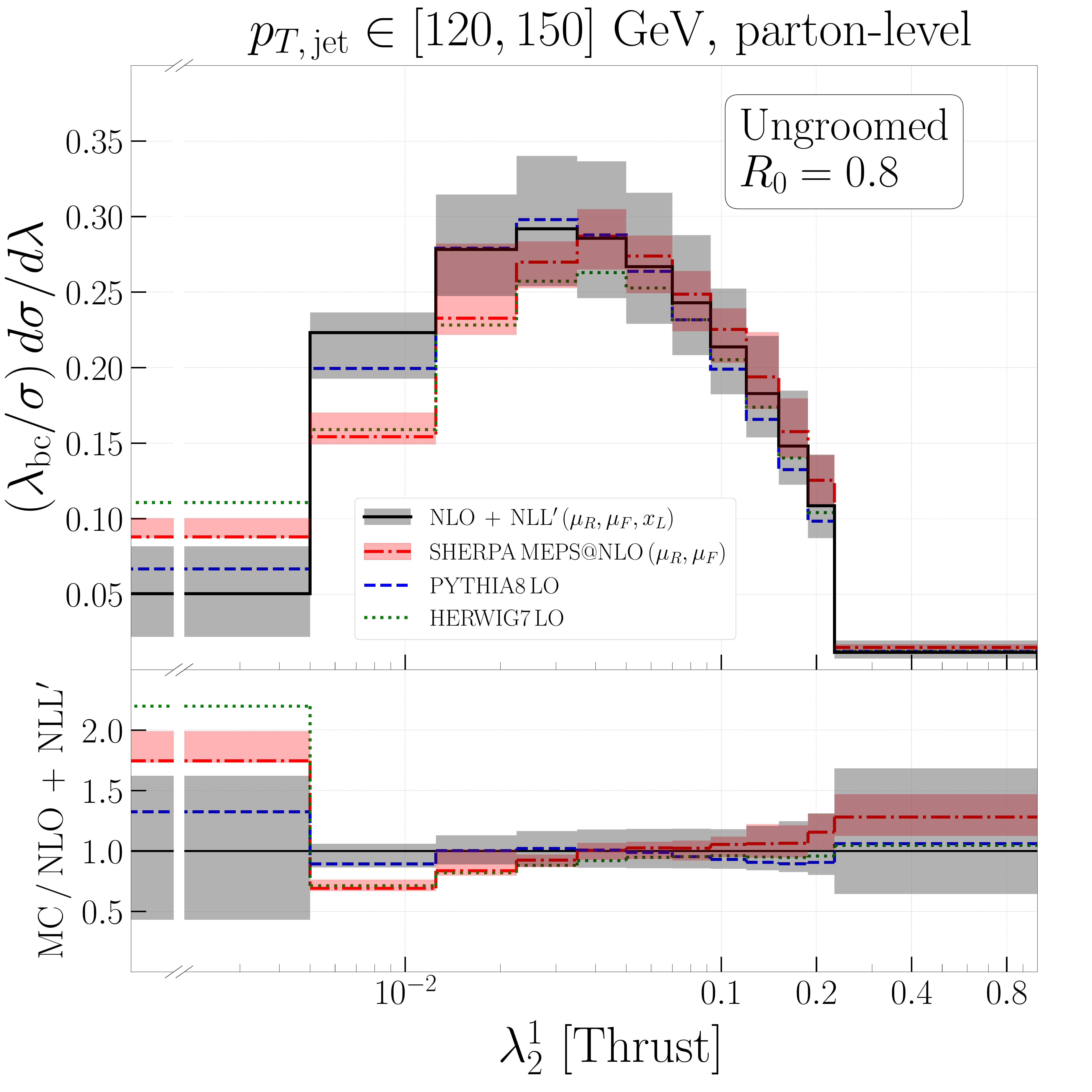}
  \hspace{1em}
  \includegraphics[width=0.44\linewidth]{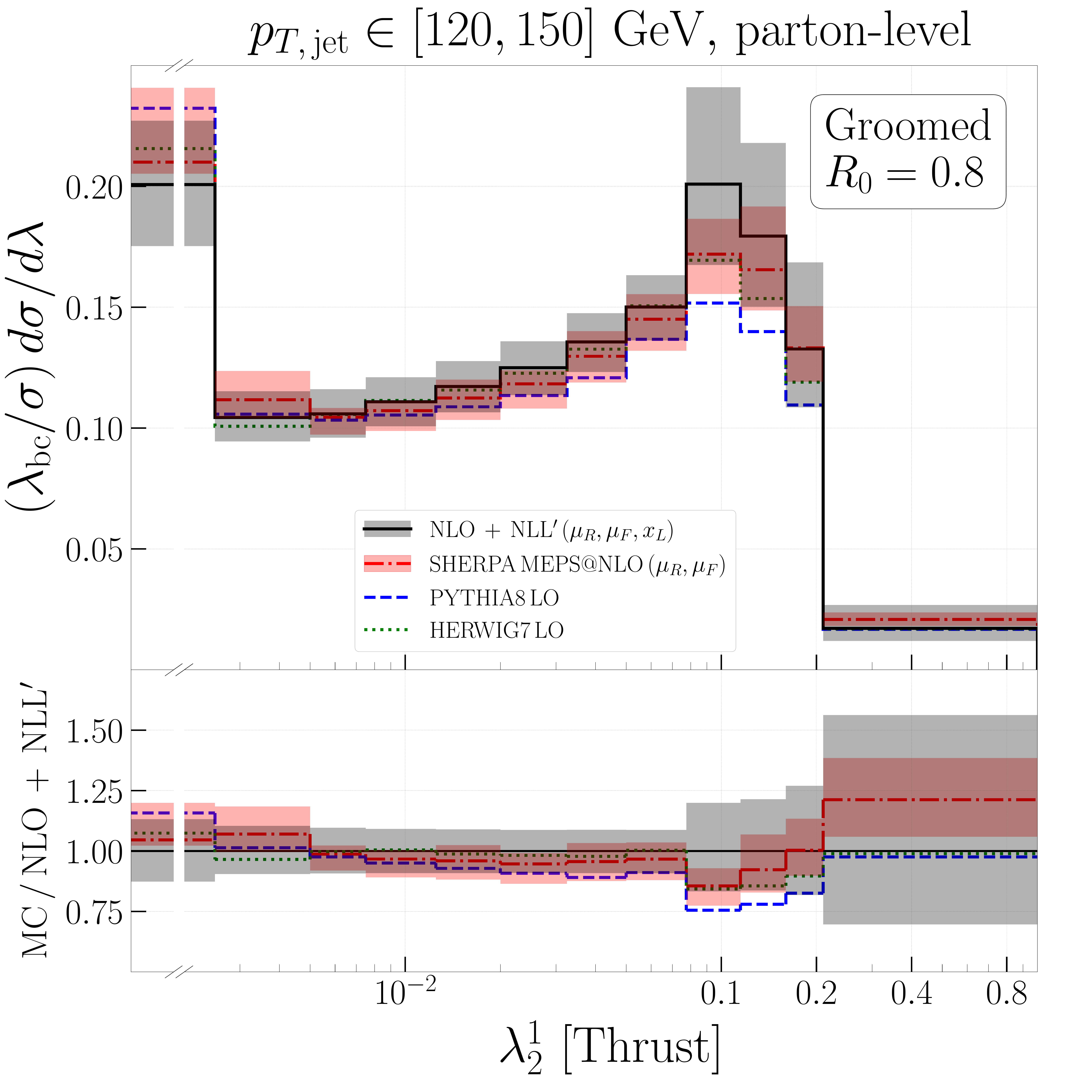}
  \caption{Comparison of parton-level predictions for ungroomed and groomed jet-angularities in $Zj$
    production from \pythia, \herwig (both LO), and \sherpa (MEPS@NLO) with \NLOpNLLp results for
    $p_{T,\text{jet}}\in[120,150]\;\text{GeV}$. Here $\lambda_\text{bc}$ stands for the bin centre. }
\label{fig:all_mc_vs_res_ps_pT120} 
\end{figure}

\begin{figure}
  \centering
  \includegraphics[width=0.44\linewidth]{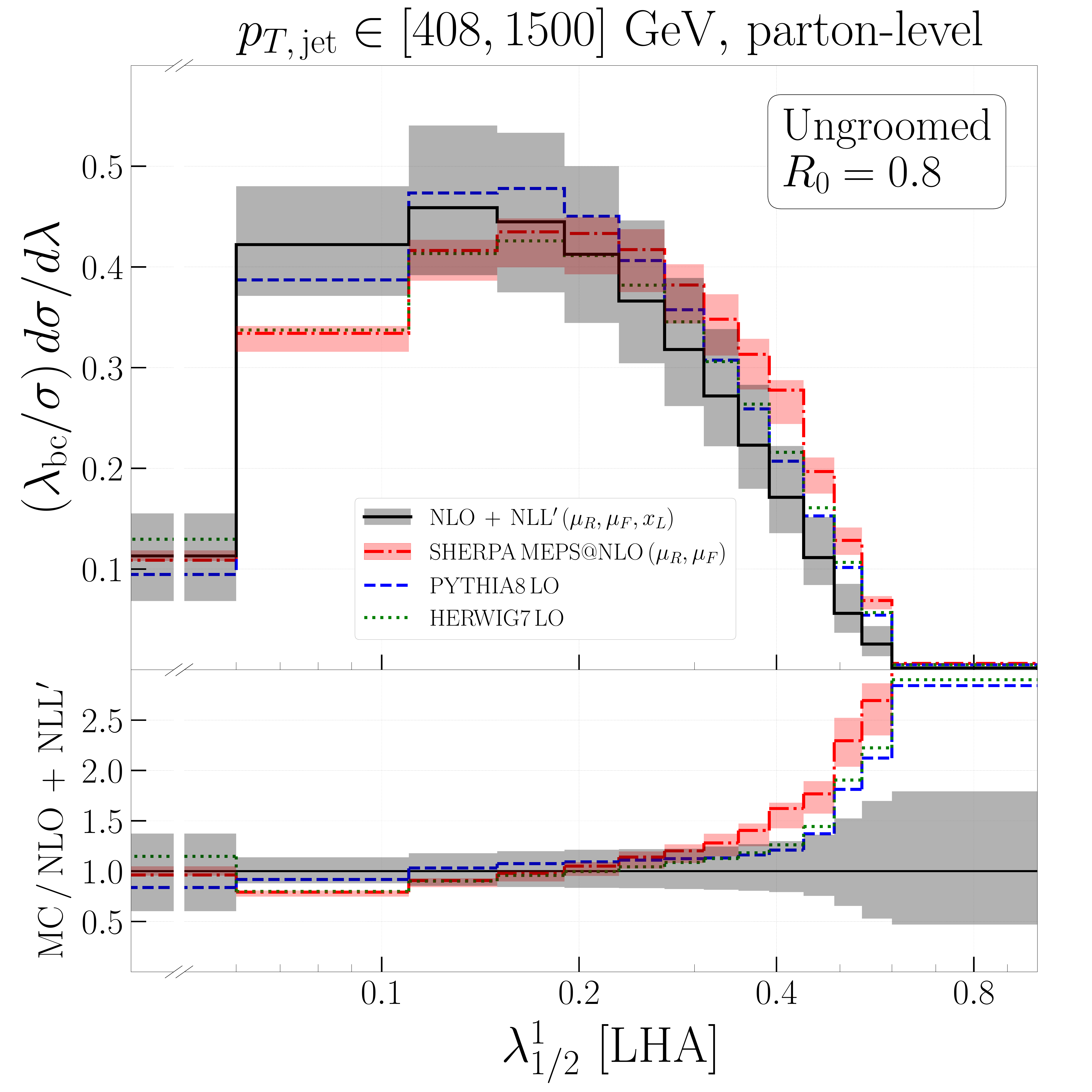}
  \hspace{1em}
  \includegraphics[width=0.44\linewidth]{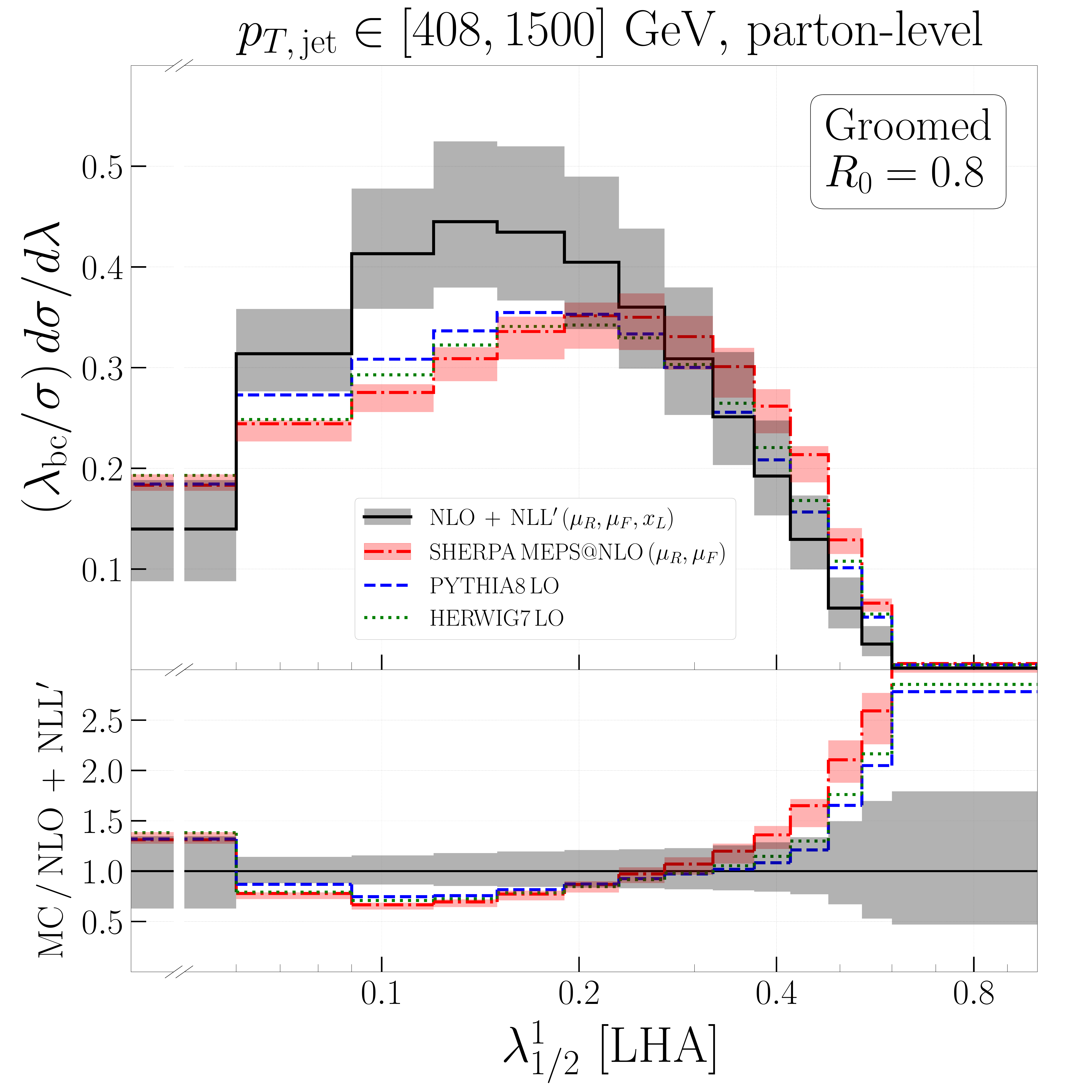}
  \centering
  \includegraphics[width=0.44\linewidth]{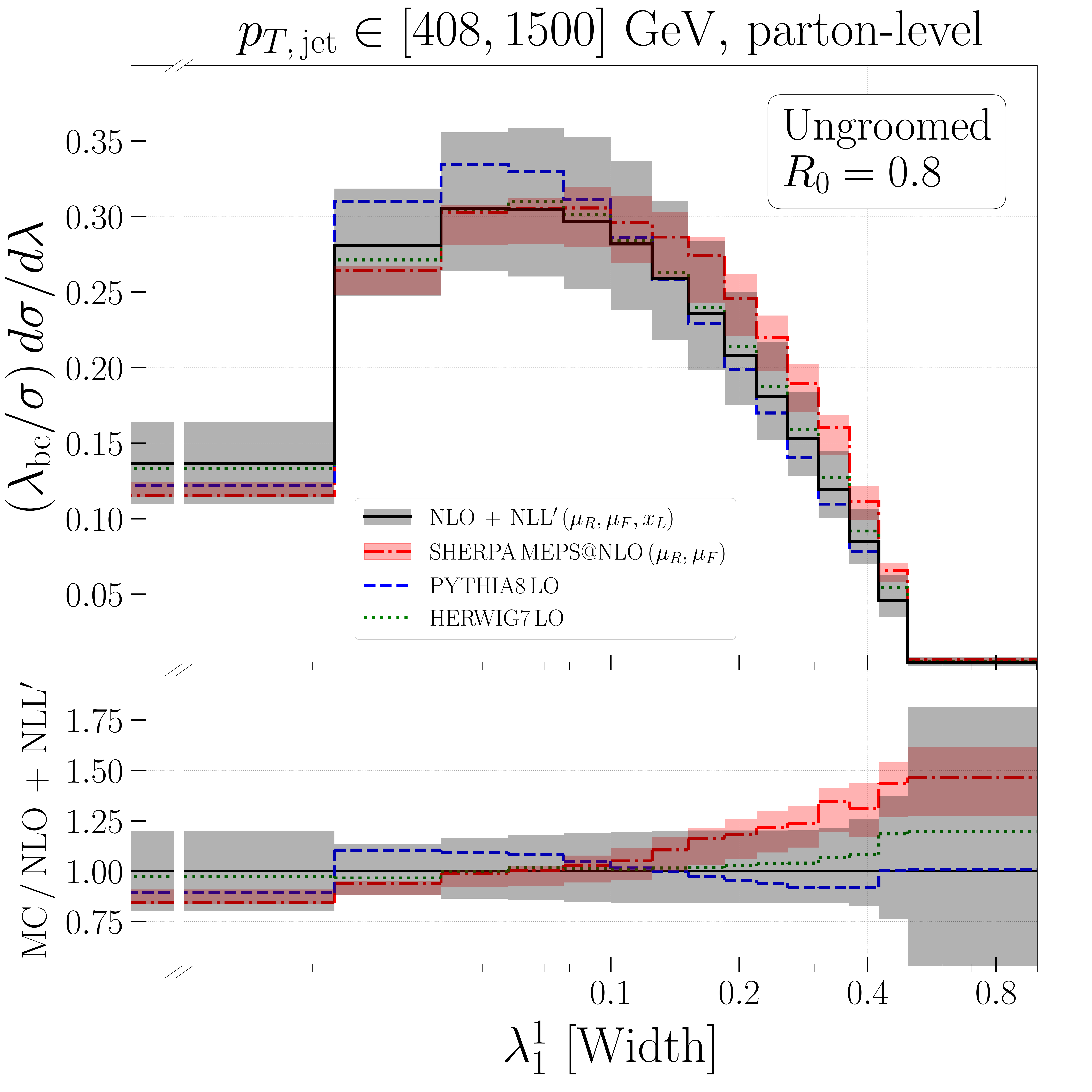}
  \hspace{1em}
  \includegraphics[width=0.44\linewidth]{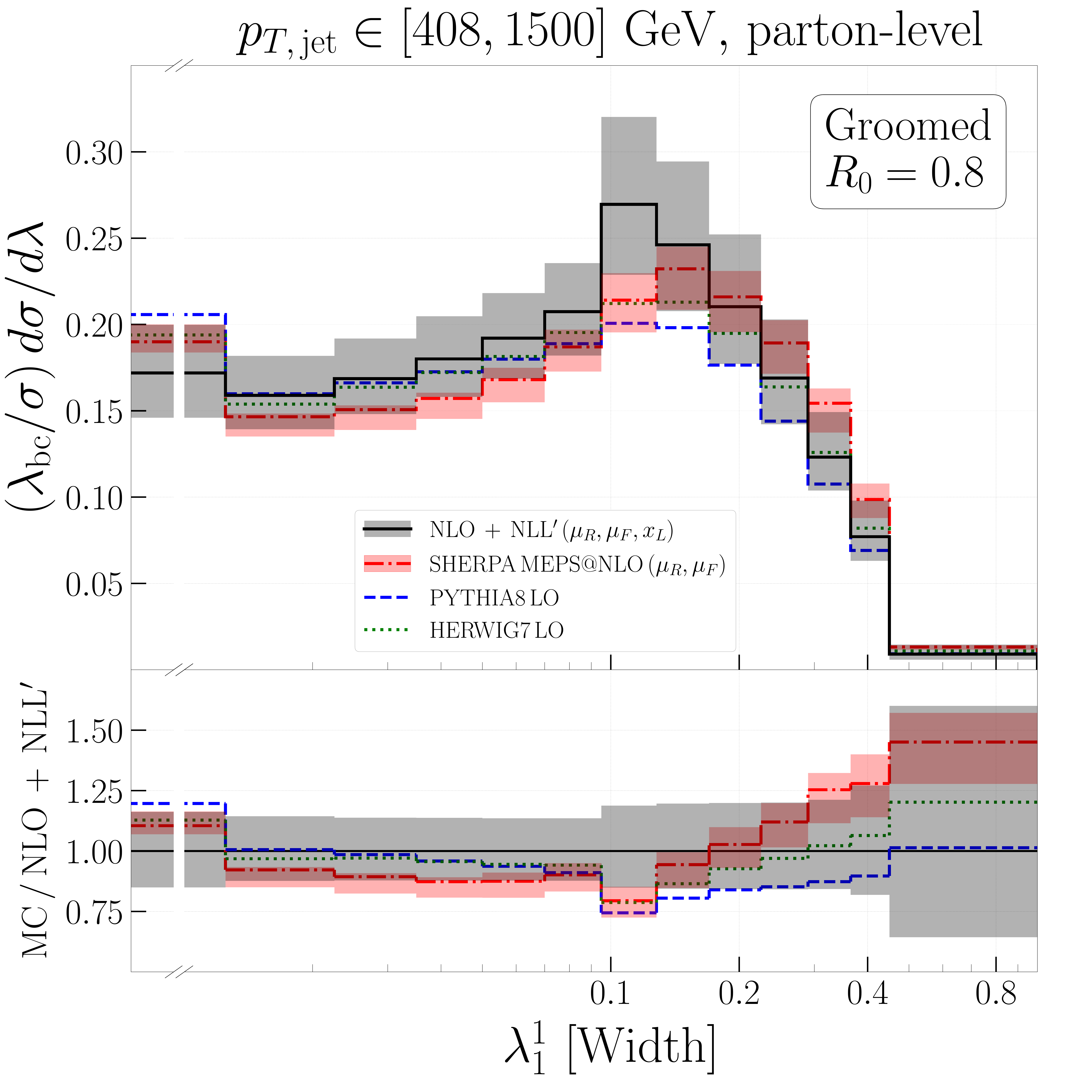}
  \centering
  \includegraphics[width=0.44\linewidth]{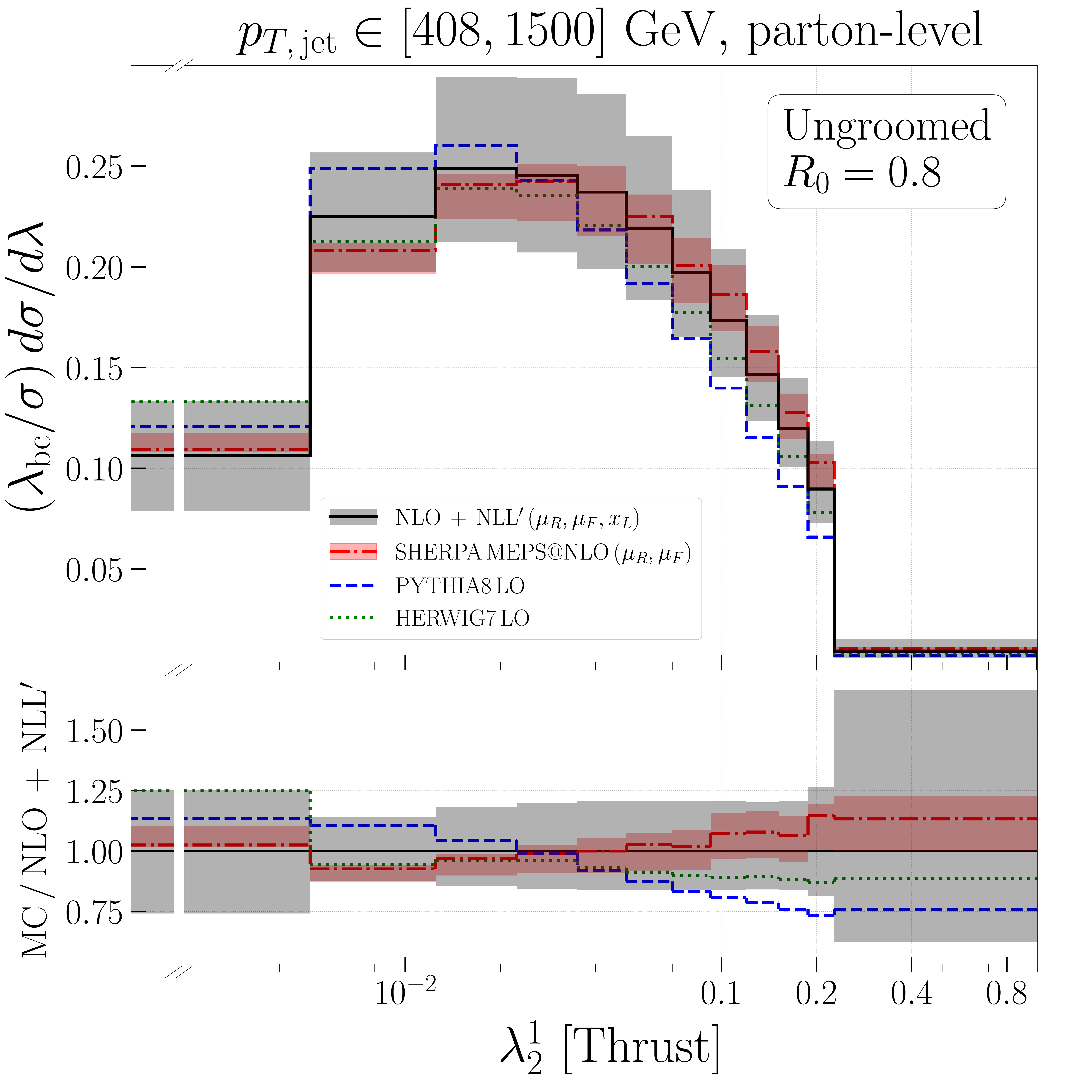}
  \hspace{1em}
  \includegraphics[width=0.44\linewidth]{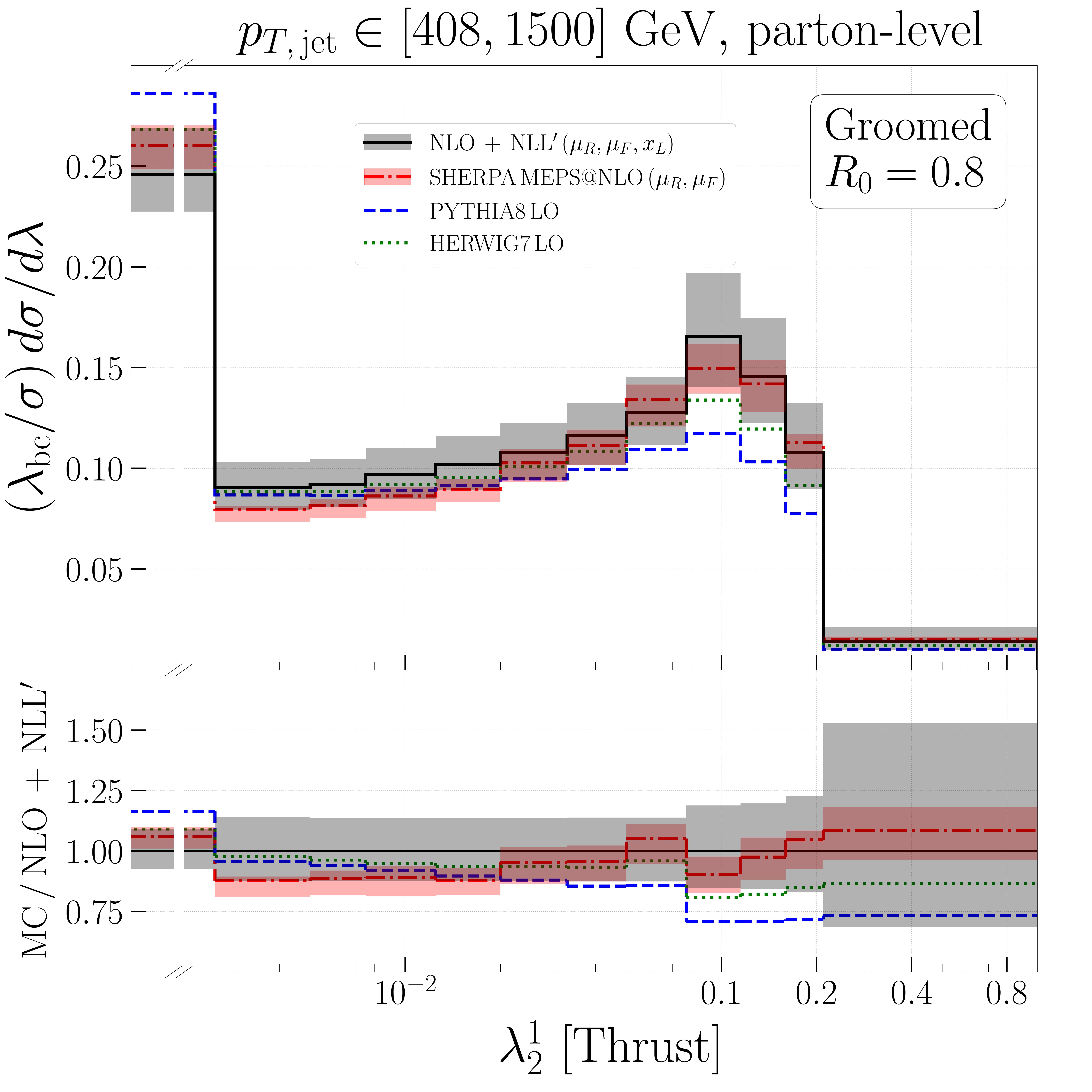}
  \caption{Comparison of parton-level predictions for ungroomed and groomed jet-angularities in $Zj$
    production from \pythia, \herwig (both LO), and \sherpa (MEPS@NLO) with \NLOpNLLp results for
    $p_{T,\text{jet}}\in[408,1500]\;\text{GeV}$. Here $\lambda_\text{bc}$ stands for the bin centre. }
\label{fig:all_mc_vs_res_ps_pT408} 
\end{figure}

Let us start by comparing our \NLOpNLLp results for the jet
angularities against different parton-level MC predictions.
As in Section~\ref{sec:def-mc} we consider the LHA ($\lambda^1_{1/2}$), Width ($\lambda^1_{1}$), and Thrust ($\lambda^1_{2}$)
observables.  In Figs.~\ref{fig:all_mc_vs_res_ps_pT120}
and~\ref{fig:all_mc_vs_res_ps_pT408}  we compare the \NLOpNLLp
distributions against the parton-level MEPS@NLO
predictions of \sherpa and the LO results from \pythia and \herwig. As before
we consider two $p_{T,\text{jet}}$ slices, namely \mbox{$p_{T,\text{jet}} \in [120, 150]$ GeV}  in
Fig.~\ref{fig:all_mc_vs_res_ps_pT120} and $p_{T,\text{jet}} \in [408, 1500]$ GeV in  Fig.~\ref{fig:all_mc_vs_res_ps_pT408}. 
The theoretical uncertainty for the resummed and matched predictions is obtained by varying renormalisation ($\mu_R$) and factorisation
($\mu_F$) scales, as well as the parameter $x_L$ that determines the resummation scale, as detailed in Sec.~\ref{sec:resummation}, while
the uncertainty band of the \sherpa MEPS@NLO predictions corresponds
to variations of $\mu_R$ and $\mu_F$ only, however, both in the
matrix element and parton-shower component. For this reason, the former yields a more conservative
assessment of the uncertainty than the latter. 

A useful viewpoint to understand our results is to consider the different types of contributions that affect the distributions. Roughly speaking, even at parton level, the behaviour of the distributions at low values of the observables strongly
depends on the details of the treatment of radiation in the infrared region, \emph{e.g.} the parton-shower cutoff or the Landau pole
in the resummation. This region is characterised by large non-perturbative corrections, mostly due to hadronisation. The size of the region depends on the angularity exponent $\alpha$, the presence of grooming and the considered transverse momentum region.
Following the analysis of, \emph{e.g.},
Refs.~\citep{Dasgupta:2013ihk,Larkoski:2014wba} we can estimate the
region of large non-perturbative corrections to be
\begin{align} 
\text{ungroomed jets or \softdrop jets with $\alpha\le 1$:} & \quad \lambda^1_{\alpha} \lesssim \left(\frac{\mu_\text{NP}}{p_{T,\text{jet}} R_0} \right)^{\min(\alpha,1)}\,,\label{nonpert-trans_UG}\\
\text{\softdrop jets with $\alpha>1$:} & \quad \lambda^1_{\alpha} \lesssim \left(\frac{\mu_\text{NP}}{p_{T,\text{jet}} R_0} \right)\left(\frac{\mu_\text{NP}}{\zc p_{T,\text{jet}} R_0} \right)^\frac{\alpha-1}{1+\beta}\,,\label{nonpert-trans_G}
\end{align}
where $\mu_\text{NP}$ is a non-perturbative scale that we take to be
$1$~GeV.
For $\alpha\le 1$, the non-perturbative corrections first come from
the region of hard-collinear radiation which is mostly unaffected by
the grooming procedure.
Conversely, when applying \softdrop for angularities with $\alpha>1$,
the factor in the second bracket of Eq.~\eqref{nonpert-trans_G} is
smaller than 1 and the boundary of the non-perturbative region is
pushed towards smaller observable values.\footnote{Applying \softdrop
  for angularities with $\alpha\le 1$ would still have the effect of
  reducing the impact of other non-perturbative corrections such as
  multi-parton interactions and pileup, as well as reducing the
  perturbative contributions from soft-large-angle radiation,
  including non-global logarithms.}
Above the thresholds identified by Eqs.~\eqref{nonpert-trans_UG}, \eqref{nonpert-trans_G}, we expect (resummed) perturbation theory to provide a
good description of the underlying physical processes. In this second region we can expect our \NLOpNLLp predictions and the MC ones
to agree best. Finally, we can identify a third region, which is characterised by very large values of the observables, close to the kinematic
endpoint of the distribution. 
In our resummed and matched prediction this region is under the jurisdiction of the NLO calculation, however the actual value of the kinematic
endpoint is strongly affected by the presence of additional emissions, as captured by the parton-shower simulations. For the same reason, we also
expect this very last bin to be rather sensitive to non-perturbative contributions, especially the UE, as we shall discuss below.

We can now move to analyse in more detail the various cases. 
First of all we consider the ungroomed $\lambda^1_{1/2}$ distribution for the lower transverse momentum bin in  Fig.~\ref{fig:all_mc_vs_res_ps_pT120}.
From the qualitative argument of Eq.~\eqref{nonpert-trans_UG}, we expect the boundary between the first two regions to be located
around $\lambda^1_{1/2} \simeq 0.1$. Indeed, in the region to the left of this value the agreement between the \NLOpNLLp and the three
parton-level MC predictions is rather poor, with the \NLOpNLLp and \sherpa MEPS@NLO uncertainty bands not overlapping. Above this value,
the agreement improves, as expected, until we end up in the third region, close to the endpoint, where the predictions from the two approaches
are in strong disagreement. 
The situation does not improve with \softdrop\!\!, as we have anticipated earlier. If anything the agreement between the two predictions
deteriorates at the boundary between the first and second region. We
attribute this to the presence of transition-point effects (remember
that we have chosen $\zc=0.1$). However, as these are formally NNLL effects,
they are not very-well modelled by our calculation. 
In the case of the higher transverse momentum bin, the boundary between the first two regions is pushed to smaller values of the observable
($\lambda^1_{1/2} \simeq 0.05$) and, consequently, the region where we find agreement between resummation and MC simulations should widen.
This appears clearly in the top-left plot of Fig.~\ref{fig:all_mc_vs_res_ps_pT408}, which represents the ungroomed case. In fact, the
\NLOpNLLp result is in very good agreement with the \pythia prediction. Changes in behaviour of the groomed distribution with respect to the
lower-$p_T$ case are instead minimal. 
In both cases, we still see large deviations in the endpoint region.
Therefore, we must conclude that the distribution of the LHA observable is not
well theoretically controlled in our \NLOpNLLp calculation. Since the different
parton-shower simulations agree among themselves much better, a more
systematic study of the differences with respect to the analytic approach
seems well motivated. In this context we note that Ref.~\cite{Hoeche:2017jsi}
presented such comparison for the closely related BKS observables
\cite{Berger:2002ig, Berger:2003iw} and fractional-energy correlations
\cite{Banfi:2004yd} in $e^+e^-$ collisions, in an approach that could be
extended to groomed observables, also in hadronic collision. In any case,
as we shall see shortly, the other angularities considered in this study do
not suffer from this problem to the same extend.

Now let us examine the remaining two cases, \emph{i.e.}\ the $\lambda^1_1$ and $\lambda^1_2$ distributions.\footnote{Note that, unlike
  in the $\lambda^1_{1/2}$ case, the binning for ungroomed and groomed distributions is chosen in a different way. More precisely, the groomed
  distributions have finer bin-spacing than the ungroomed variants. Also note that we use different binning for the $\lambda^1_1$ and the $\lambda^1_2$
  distributions.}
Similarly to the $\lambda^1_{1/2}$ case it is convenient to consider the three different $\lambda$ regions.
 Let us concentrate on the $p_{T,\text{jet}} \in [120, 150]$~GeV slice in Fig.~\ref{fig:all_mc_vs_res_ps_pT120} first.
 We observe differences between our \NLOpNLLp results and MC predictions in the first region for the
 ungroomed distributions, \emph{i.e.}\ $\lambda^1_{\alpha} \lesssim 0.01$. 

 However, we notice that grooming now significantly improves the overall agreement between the different predictions, especially for
 the $\lambda^1_2$ case, as expected from Eq.~\eqref{nonpert-trans_G}.
Differences between \NLOpNLLp and the parton-shower simulations in the
endpoint region are also seen in this case, but this time
 all predictions remain consistent within the estimated uncertainties.
 We also note the abrupt change of behaviour in the \softdrop distributions around $\lambda^1_\alpha \simeq \zcut=0.1$, which marks the
 boundary between the groomed and ungroomed regions. As already explained, this discontinuity is beyond the accuracy of our resummation and
 it is therefore reassuring that the theoretical uncertainty in this region is rather large. 
 Finally, we note that the overall agreement between our \NLOpNLLp calculation and the parton showers further improves for
 high transverse momentum jets, evidenced by Fig.~\ref{fig:all_mc_vs_res_ps_pT408}.
 Thus, unlike the case of the LHA observable, the \NLOpNLLp predictions for jet width and jet thrust in general agree with the
 corresponding results from parton-shower simulations. This allows us to conclude that the $\lambda^1_1$ and $\lambda^1_2$ distributions
 are well under theoretical control. 
It is interesting to point out that, in the context of quark--gluon discrimination, the
ATLAS collaboration has also observed~\cite{Aad:2014gea} a worsening of the
agreement between the experimental data and Monte Carlo simulations,
for smaller values of $\alpha$ although with large uncertainties.

\subsection{Impact of non-perturbative corrections}
\label{sec:nlo_nll_final_np}

Both the \NLOpNLLp and the MC predictions in Section \ref{sec:nlo_nll_final_pl} were given without
including non-perturbative corrections due to the UE and hadronisation. In this section we provide
an estimate of these contributions and demonstrate how one can introduce corresponding corrections
to our \NLOpNLLp results.

Let us consider the parton-to-hadron transition first. One way to add hadronisation effects to our
\NLOpNLLp results is to perform a field-theoretical analysis, based on non-perturbative matrix elements.
However, the jet structure may not only be significantly affected by hadronisation but also by the
UE. The underlying event, unlike hadronisation, should \emph{not} be seen as a non-perturbative
correction to our $Zj$ matrix elements, but rather as independent semi-hard processes, contributing
with additional final-state partons that ultimately hadronise. Therefore, the UE cannot be
straightforwardly included into the resummation framework beyond the simple model of uniform
radiation~\citep{Dasgupta:2007wa,Marzani:2017kqd}. Moreover, hadronisation may lead to non-trivial
clustering of partons from different processes, \emph{i.e.}\ the hard scattering and the UE,
into final-state hadrons. Therefore, we prefer to rely
on dynamical non-perturbative models as implemented in the \pythia, \herwig and \sherpa MC event
generators~\citep{Buckley:2011ms}. Each of these programs is using its own default combination of
non-perturbative models for hadronisation and UE. More precisely, in both the \pythia
and \sherpa frameworks the UE is described by the Poisson-based Sj\"ostrand--Zijl model presented
in~\citep{Sjostrand:1985vv, Sjostrand:1987su,Sjostrand:2004pf}  whereas in \herwig the UE is
simulated based on the eikonal-model for soft emissions~\citep{Bahr:2008dy,Gieseke:2016fpz}. The hadronisation
effects are simulated according to the Lund string model~\citep{Andersson:1983ia, Sjostrand:1984ic} in \pythia
and according to cluster-fragmentation models both in \herwig \citep{Webber:1983if, Kupco:1998fx} and
\sherpa \citep{Winter:2003tt}. 

\begin{figure}
  \centering
  \includegraphics[width=0.42\linewidth]{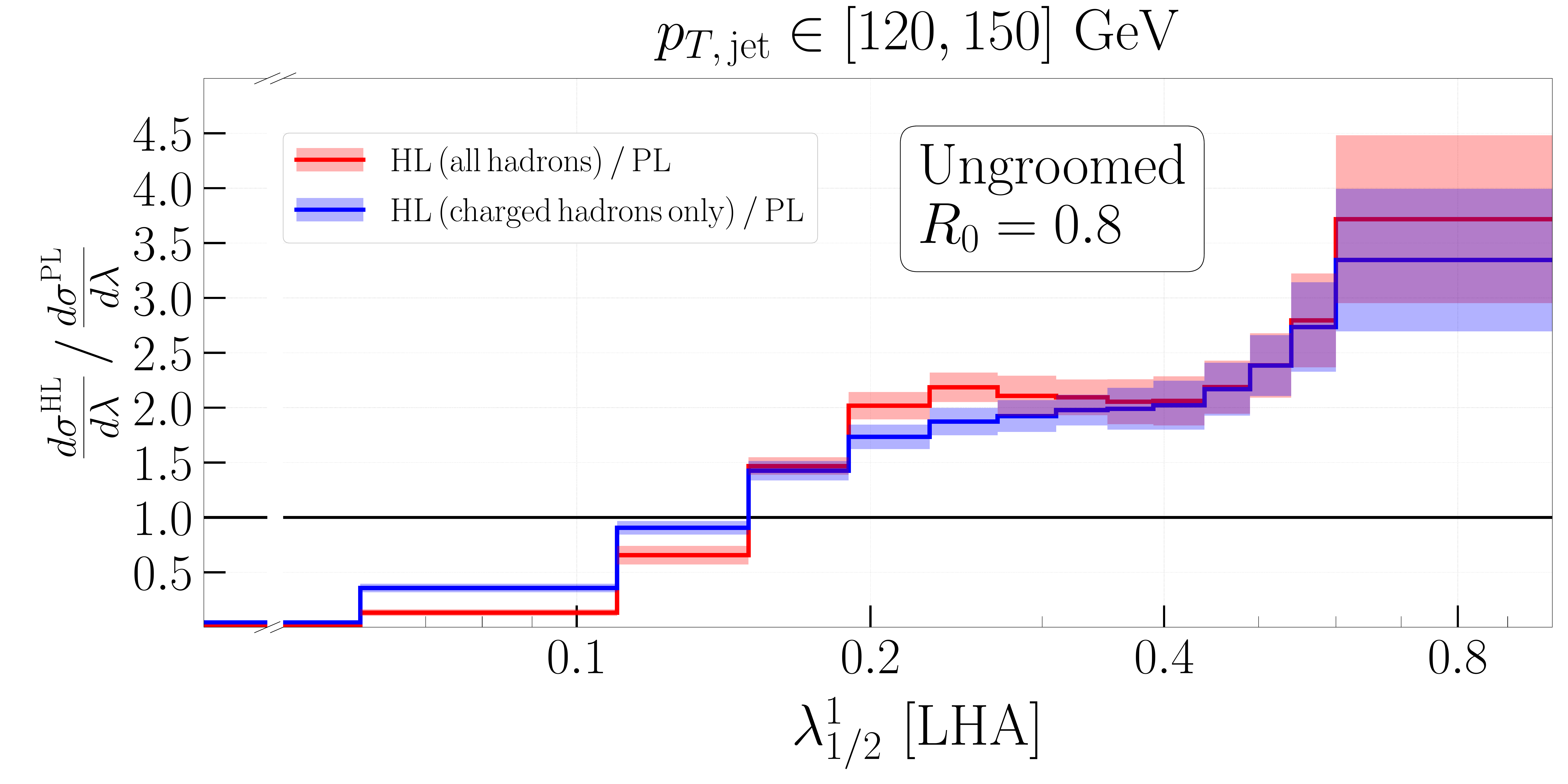}
  \hspace{0.1em}
  \includegraphics[width=0.42\linewidth]{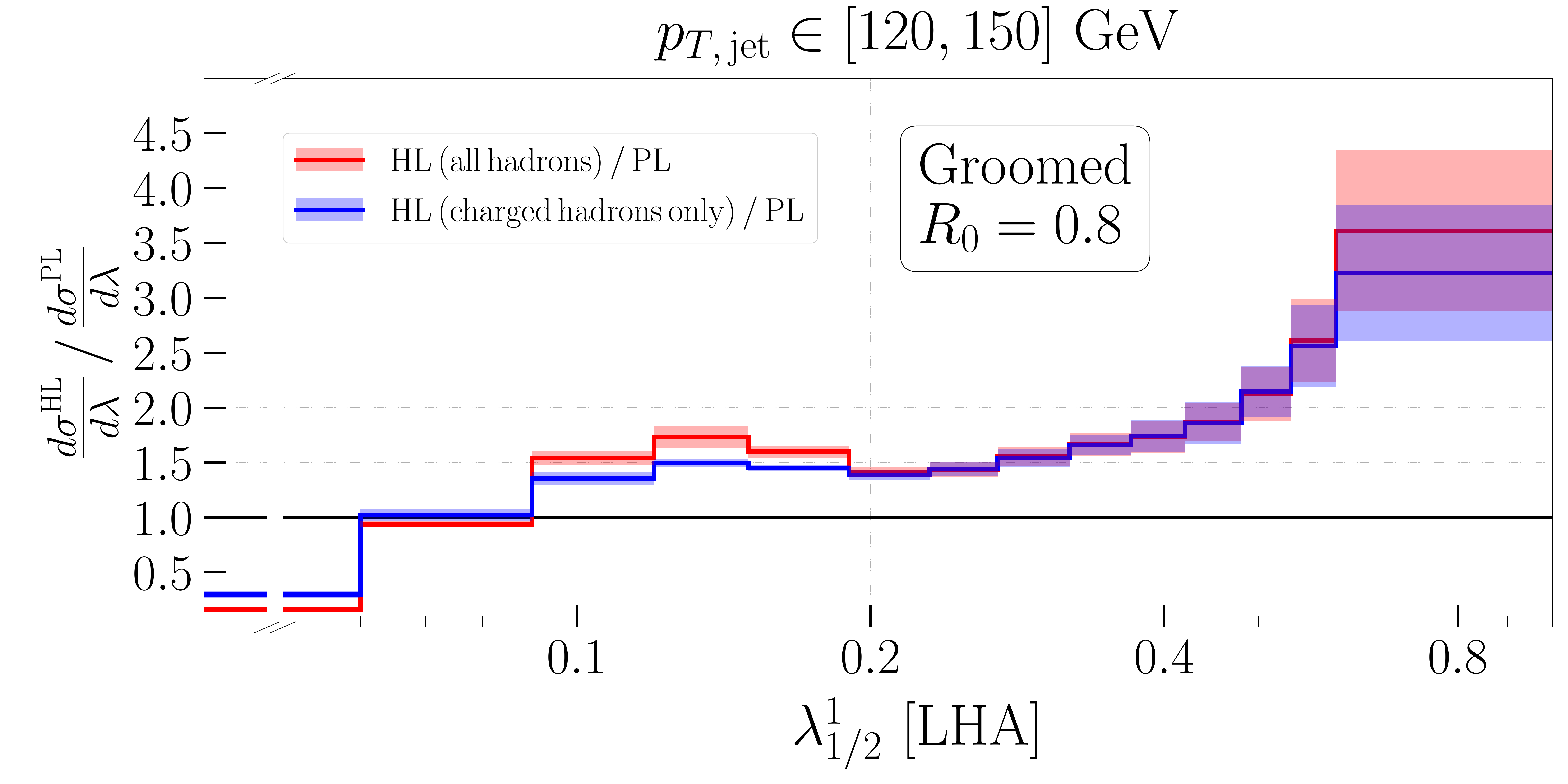}
  \centering
  \includegraphics[width=0.42\linewidth]{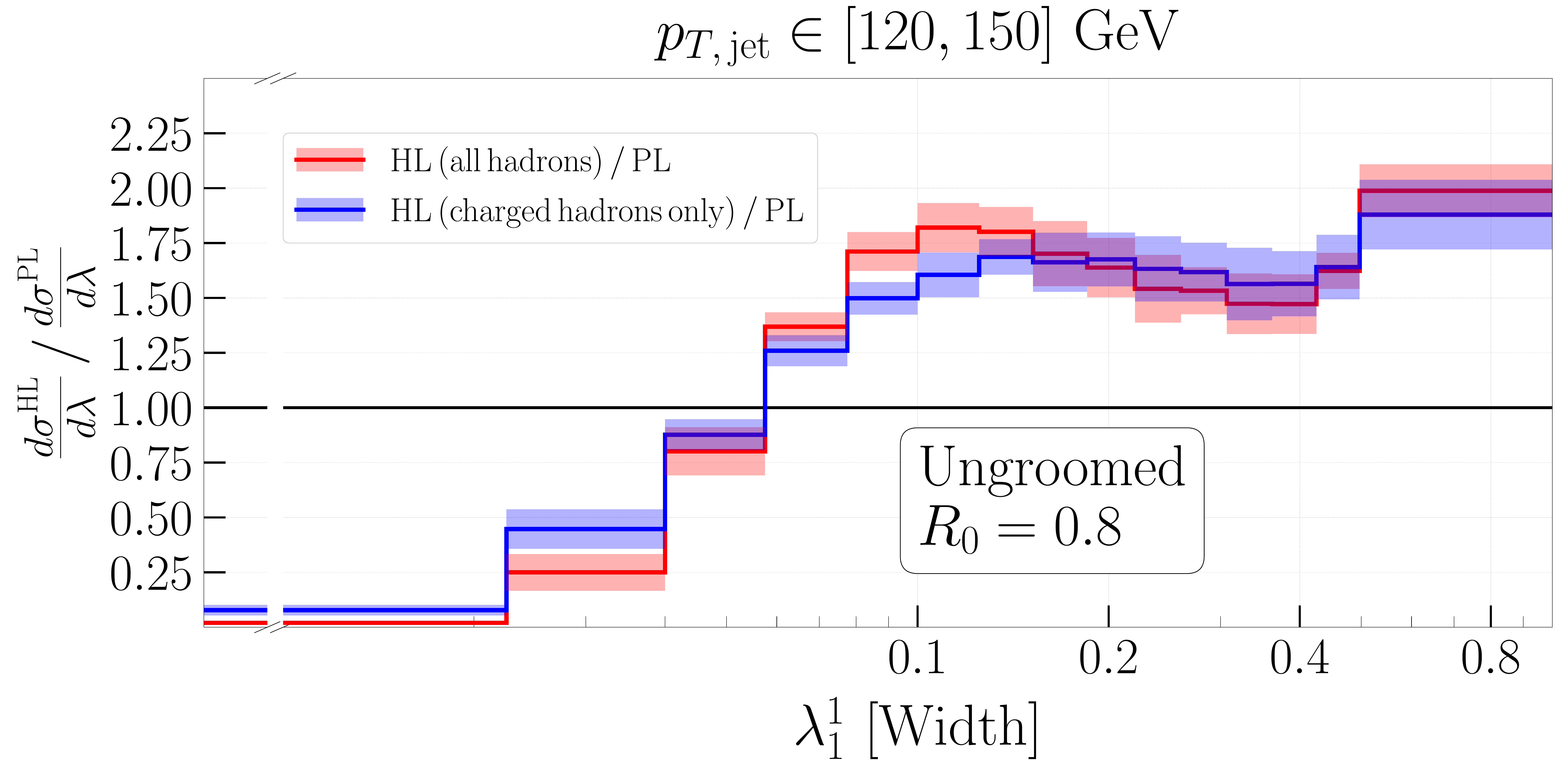}
  \hspace{0.1em}
  \includegraphics[width=0.42\linewidth]{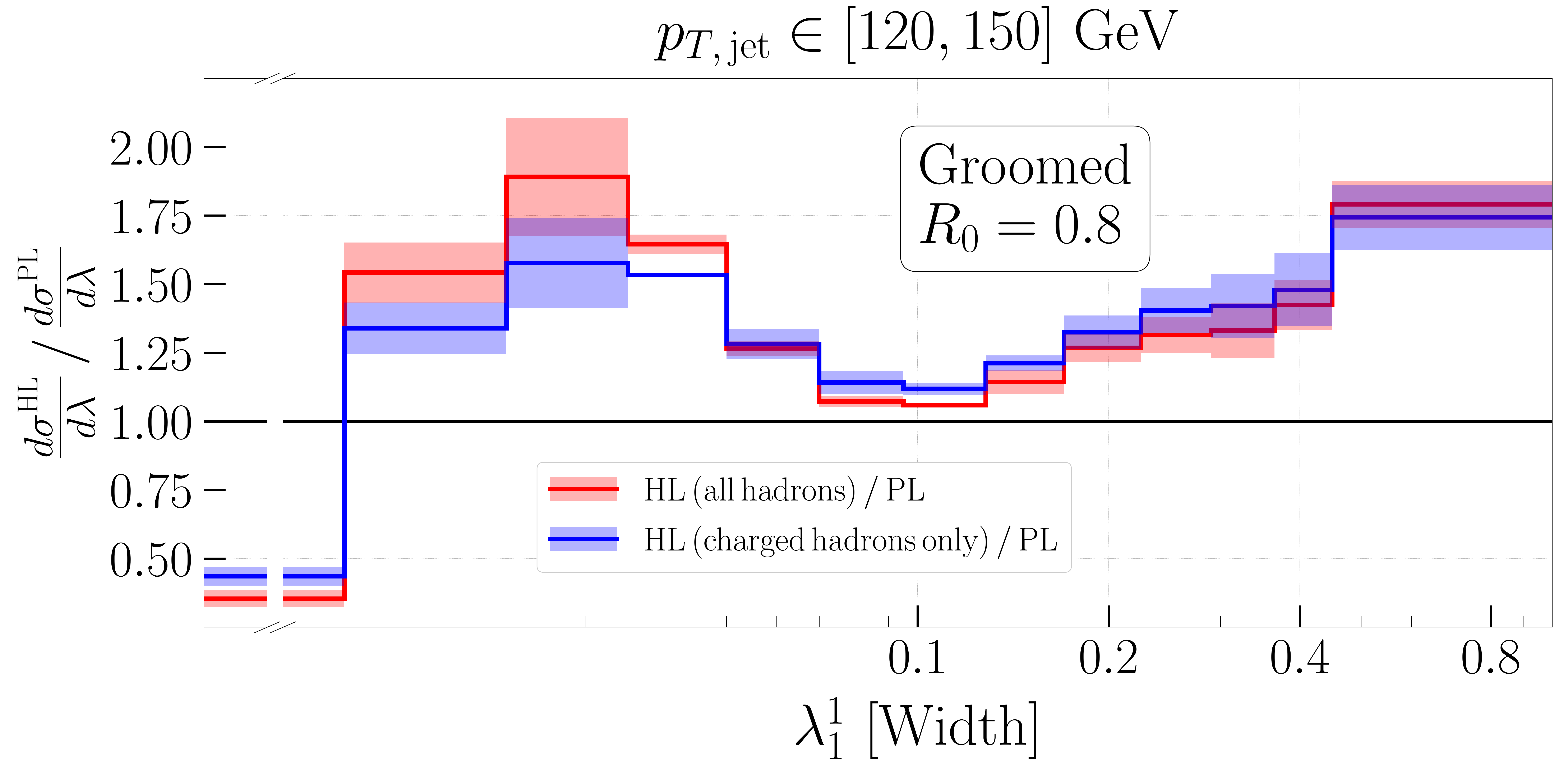}
  \centering
  \includegraphics[width=0.42\linewidth]{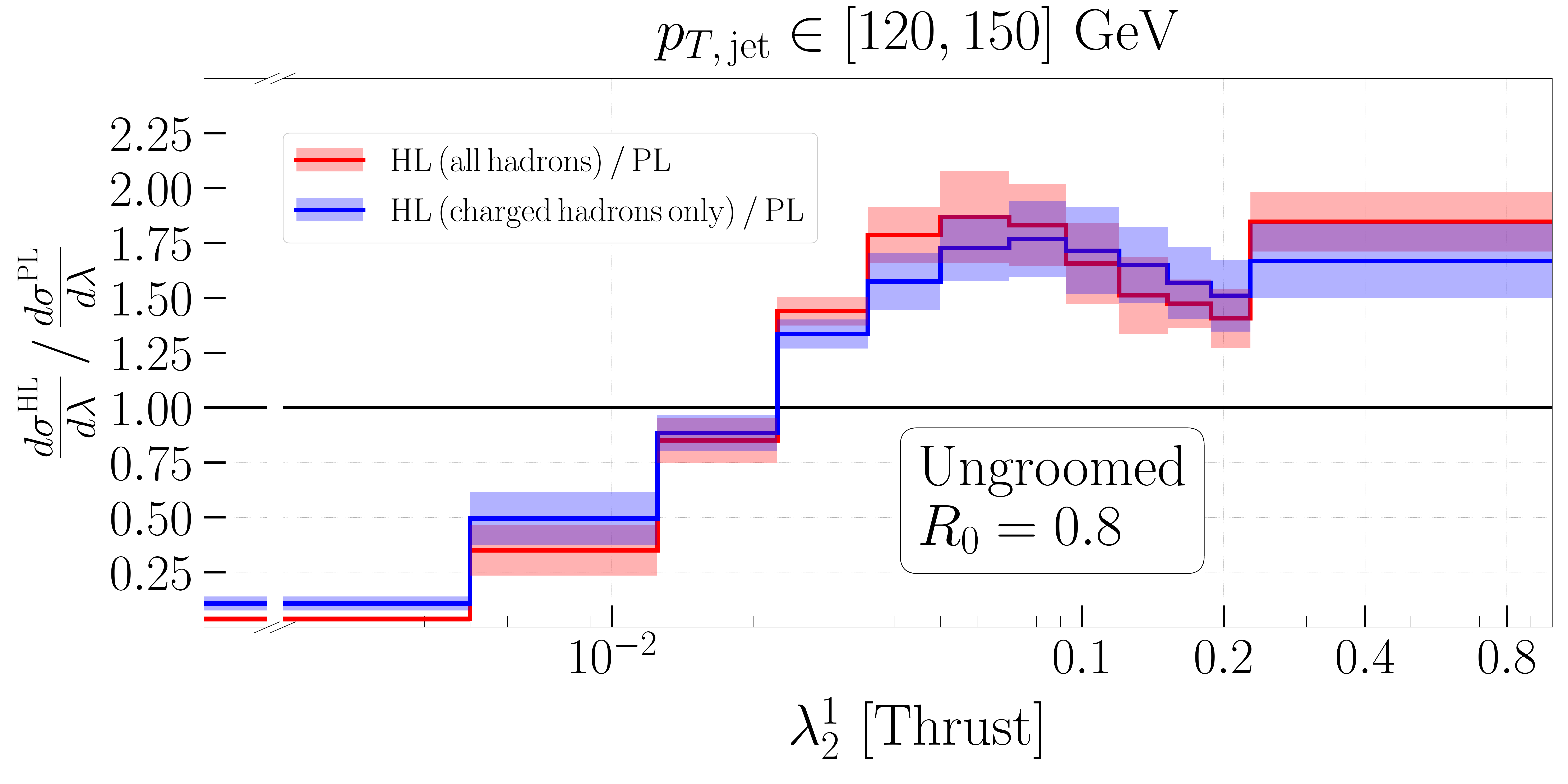}
  \hspace{0.1em}
  \includegraphics[width=0.42\linewidth]{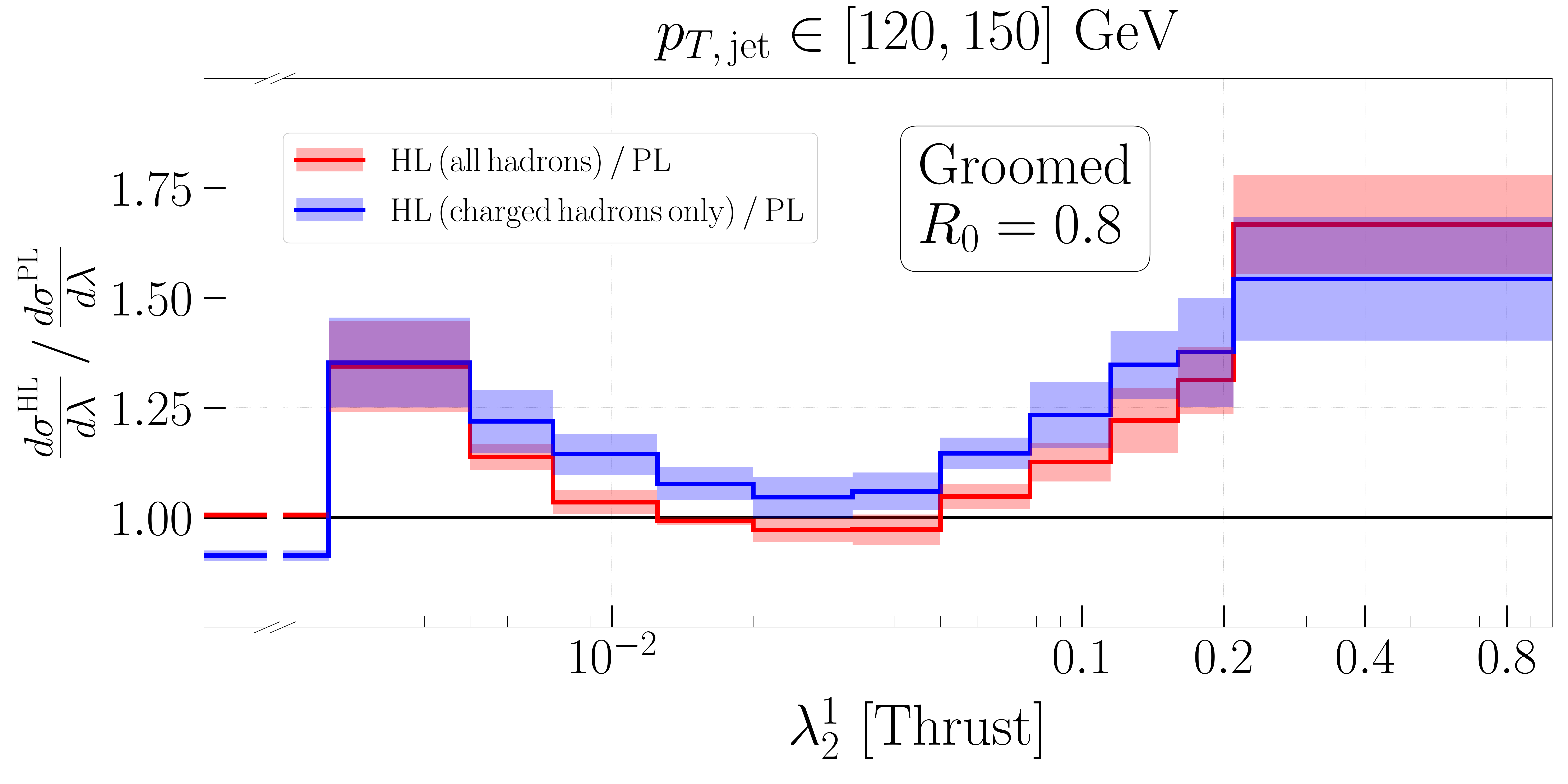}
  \centerline{(a)}
 \centering
  \includegraphics[width=0.42\linewidth]{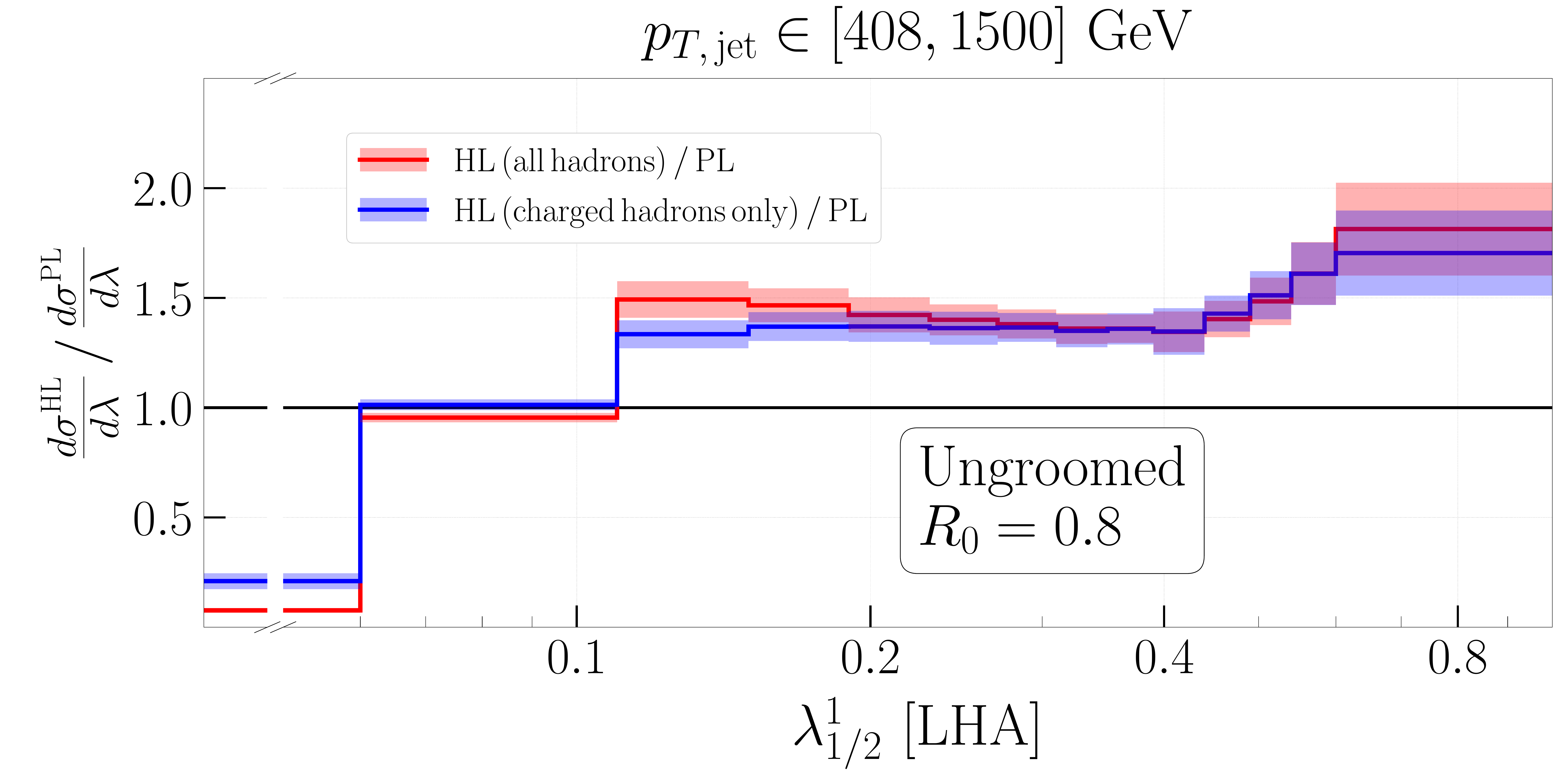}
  \hspace{0.1em}
  \includegraphics[width=0.42\linewidth]{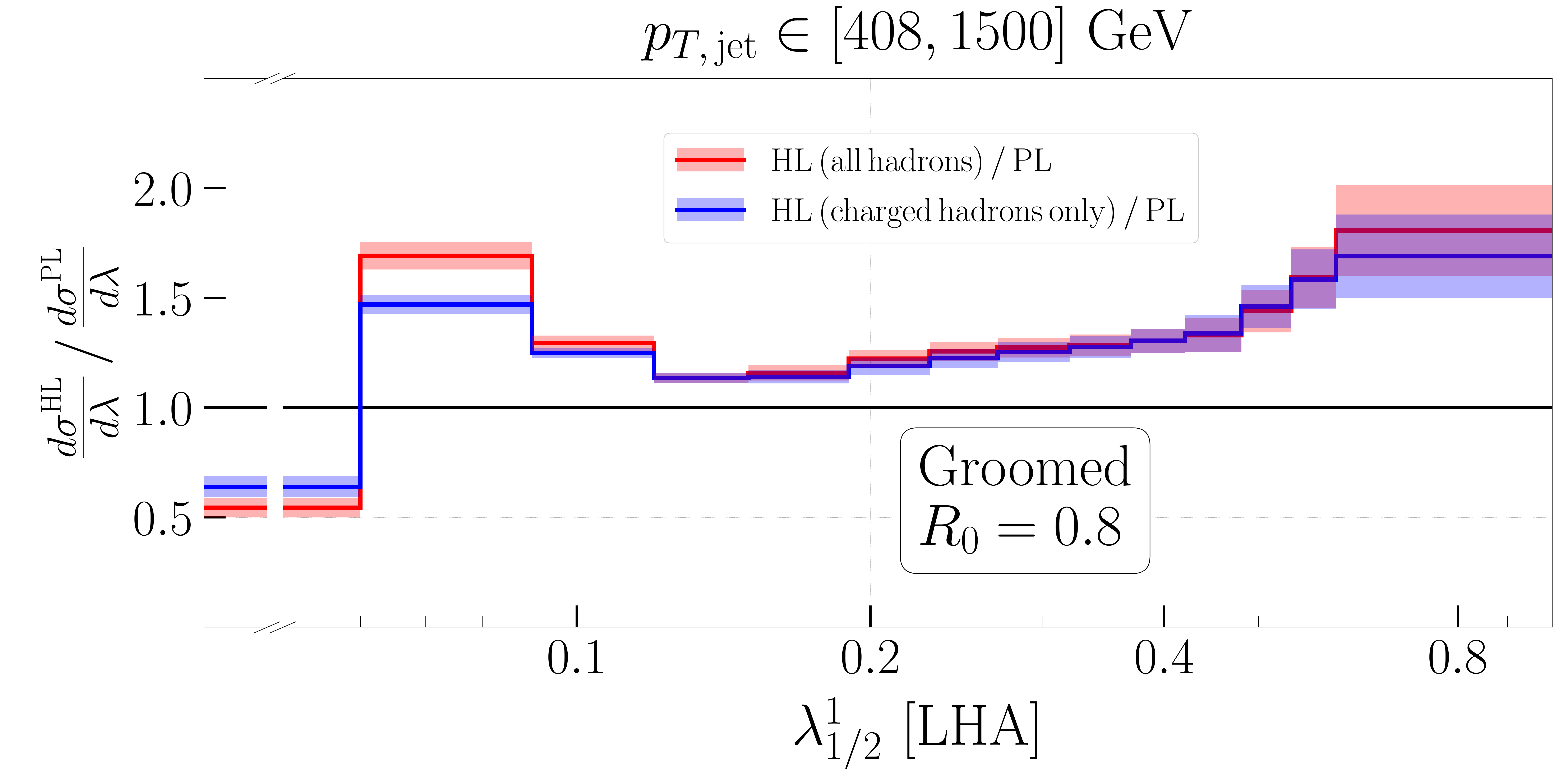}
  \centering
  \includegraphics[width=0.42\linewidth]{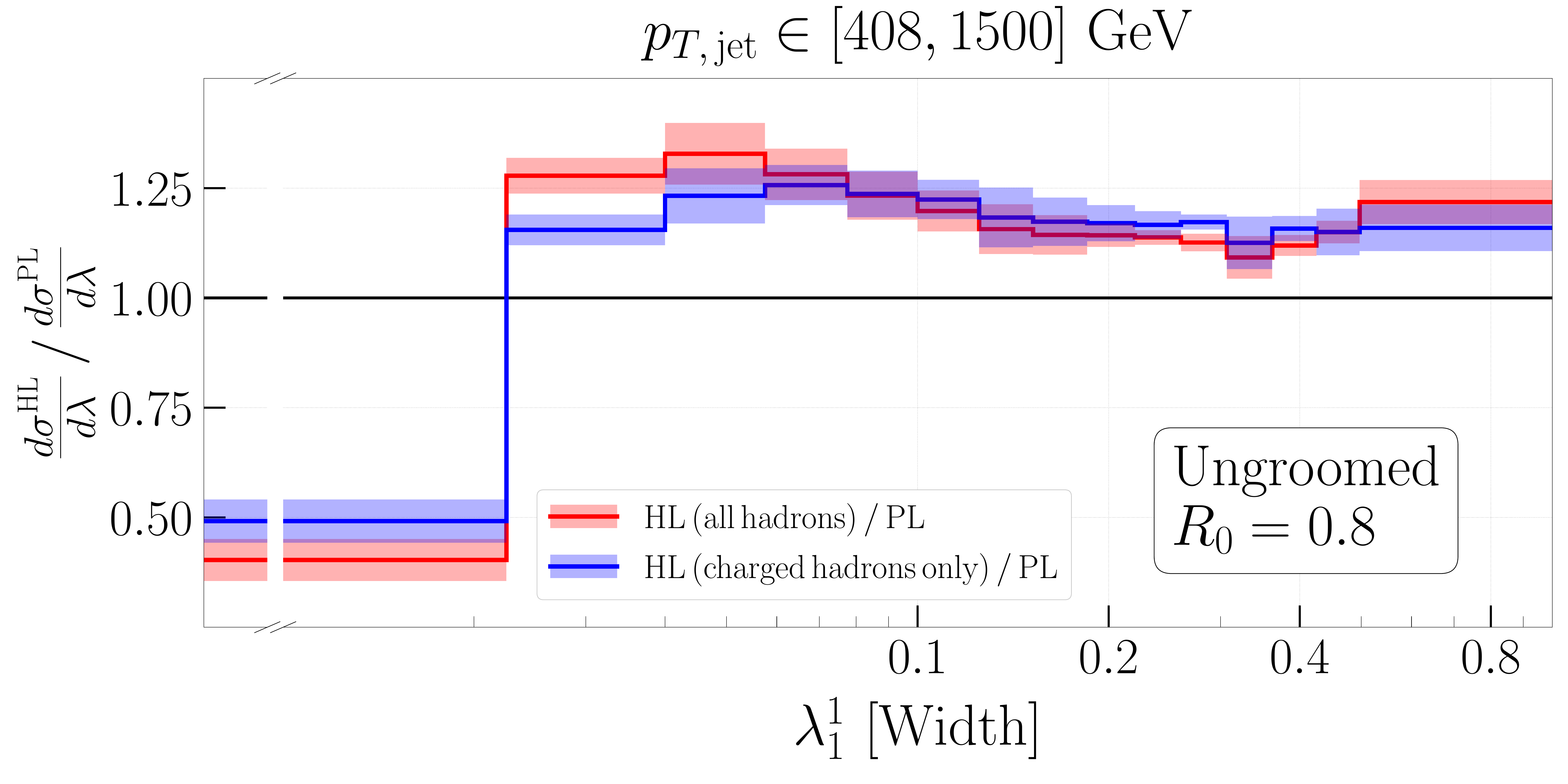}
  \hspace{0.1em}
  \includegraphics[width=0.42\linewidth]{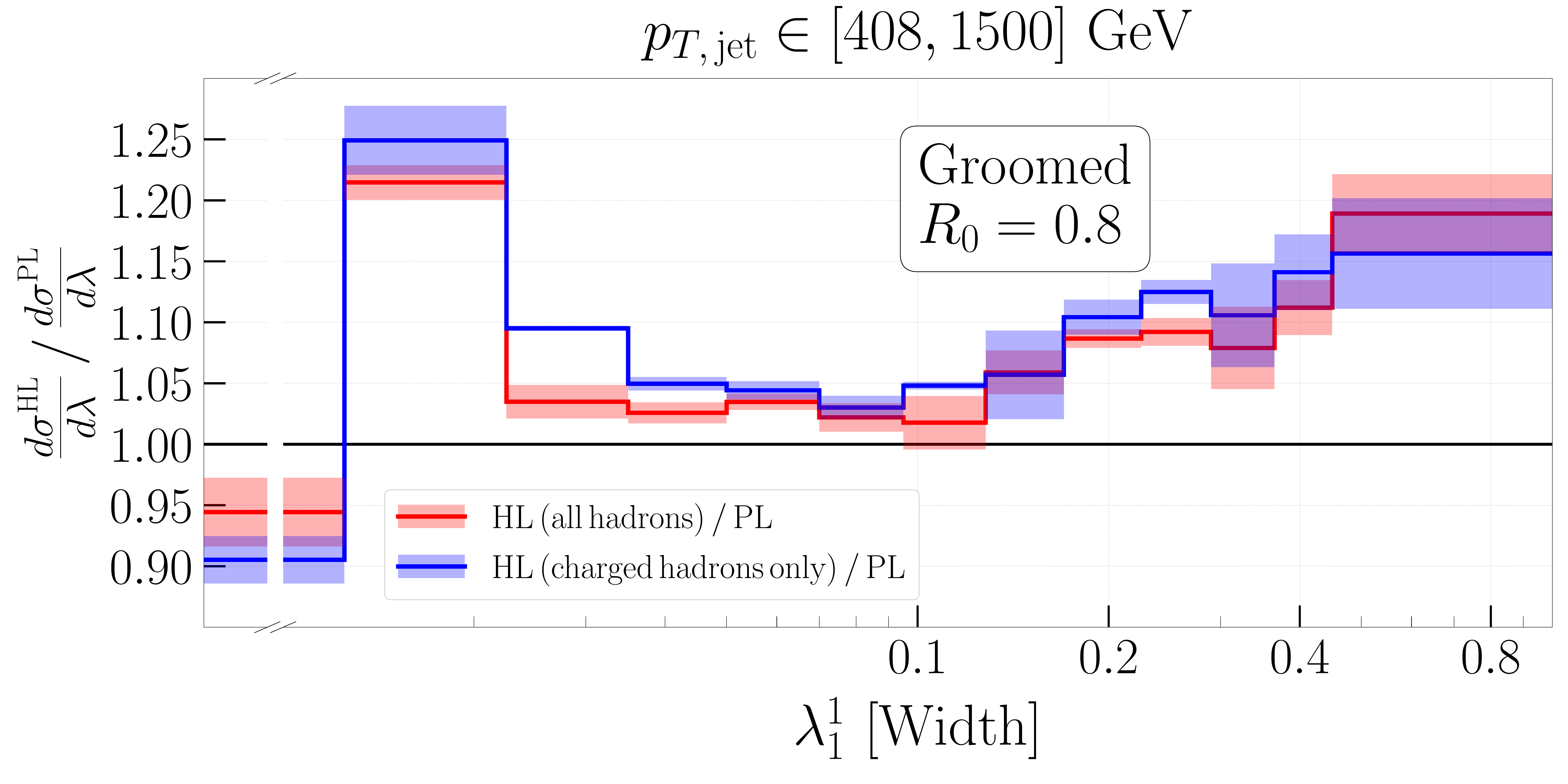}
  \centering
  \includegraphics[width=0.42\linewidth]{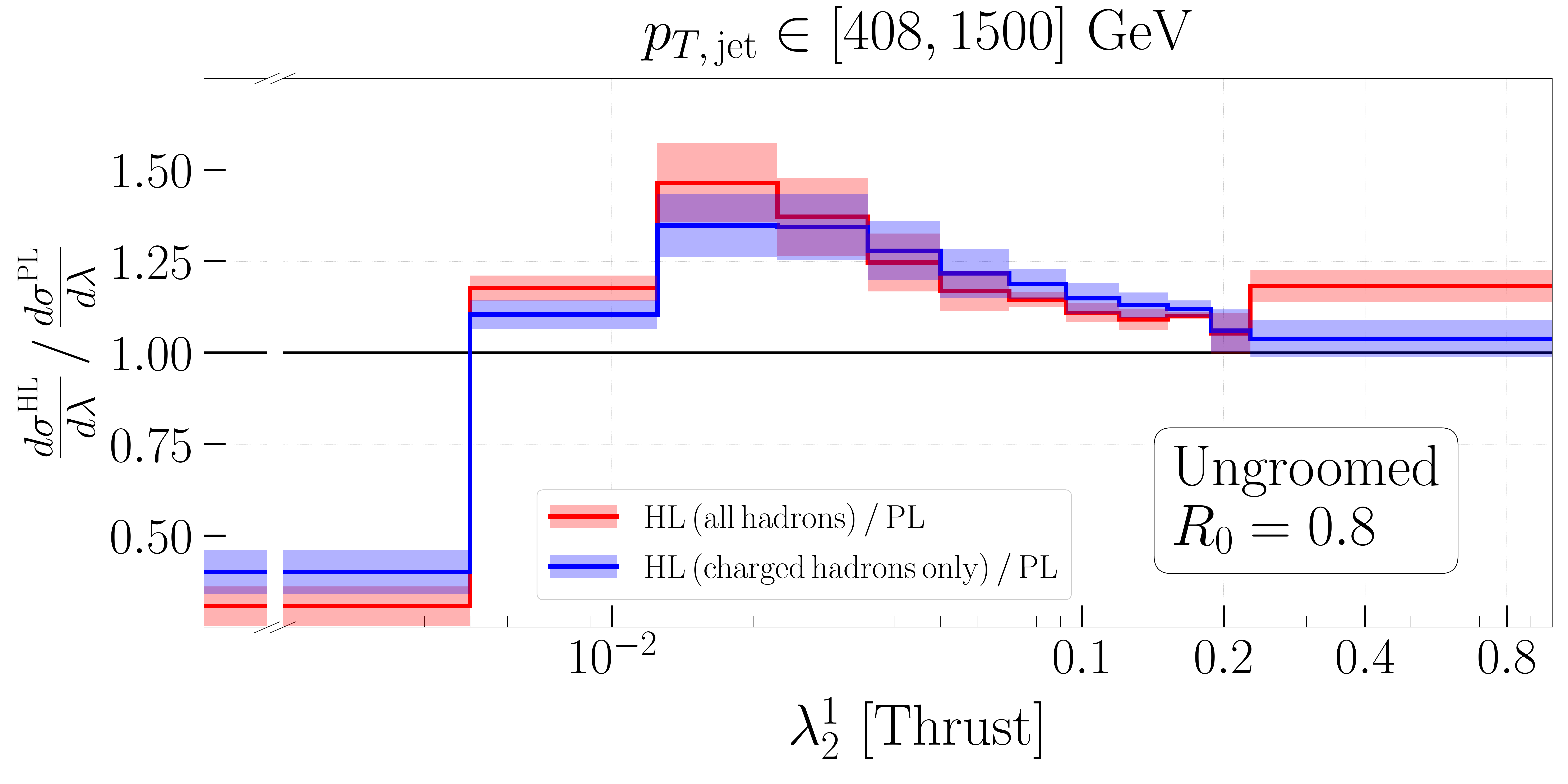}
  \hspace{0.1em}
  \includegraphics[width=0.42\linewidth]{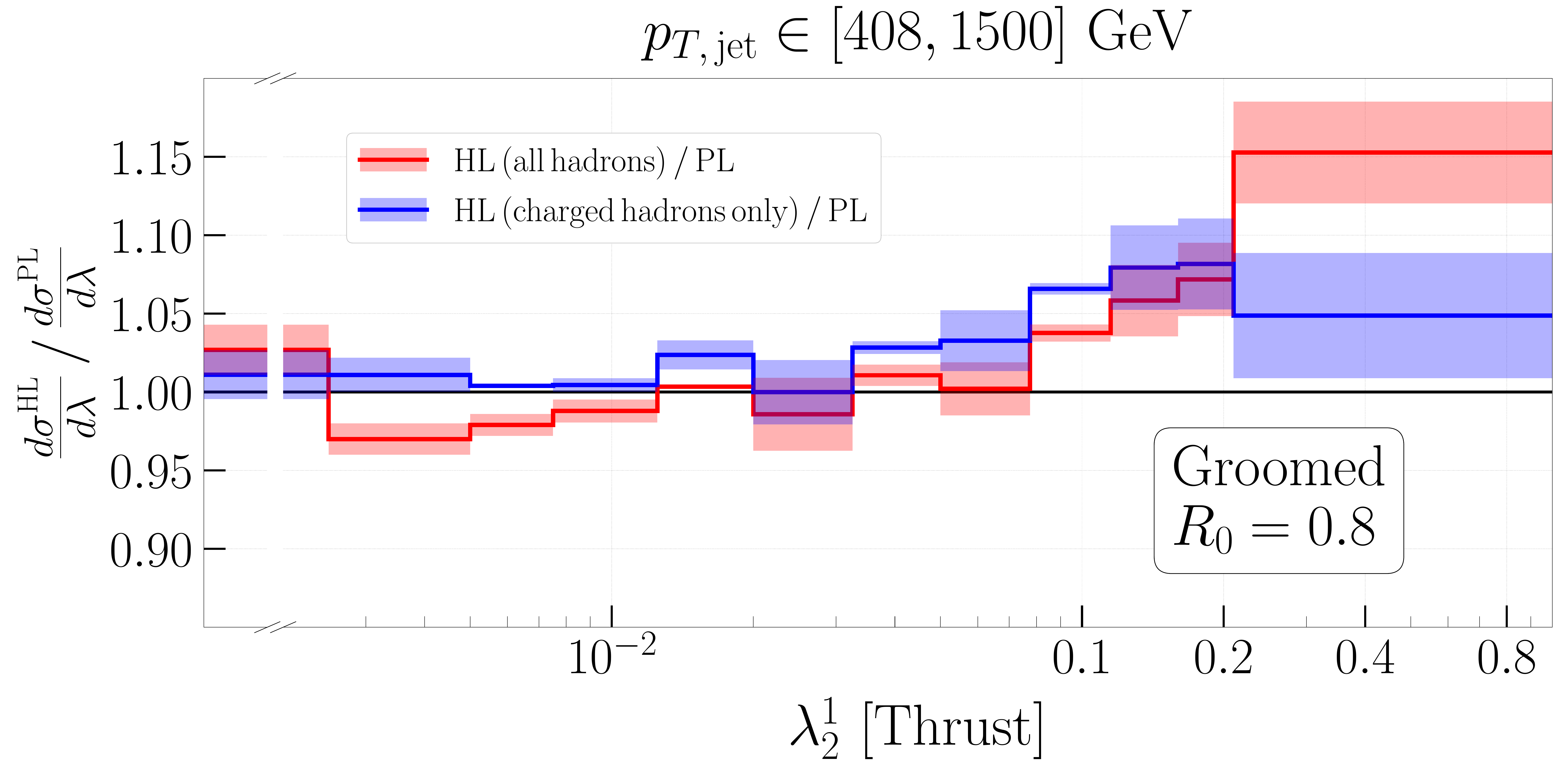}
  \centerline{(b)}
  \caption{Hadron-to-parton-level ratios with associated uncertainties extracted from
    MC simulations. Results are provided for jet angularities based on \emph{all} hadrons,
    and \emph{charged} hadrons only, for $p_{T,\text{jet}}\in[120,150]\;\text{GeV}$ (a) and $p_{T,\text{jet}}\in[408,1500]\;\text{GeV}$ (b).}
    \label{fig:all_mc_np_band} 
\end{figure}

In order to estimate the overall impact of non-perturbative effects we consider the ratios of hadron-to-parton-level
predictions. To this end, each MC simulation with \pythia, \herwig and \sherpa is performed twice: first at parton level
(without hadronisation and UE) and second at hadron level (with both UE and hadronisation switched on).
This approach allows us to consider the ratio \NPPS evaluated per $\lambda$-bin for each MC generator.
The \NPPS ratios obtained with \pythia, \herwig and \sherpa are then combined into an envelope, with extreme
points given by the maximal and minimal predictions per given angularity bin. The central value of the envelope
is determined by the respective arithmetic mean value. The resulting ratios, together with their uncertainties are
shown in Figs.~\ref{fig:all_mc_np_band} (a) and (b), differential in $\lambda^1_\alpha$, for the two representative
$p_{T,\text{jet}}$ bins, respectively.  
Experimental studies of jet structure may be based on all final-state particles, \emph{i.e.}\ jet constituents, or
purely on the tracks left by charged hadrons only. Therefore, it is convenient to consider two types of \NPPS ratios:
the first one where all final-state hadrons in the jet are considered (red solid areas in Fig.~\ref{fig:all_mc_np_band})
and the second one where only the charged jet constituents are used in the evaluation
of the angularities (blue hatched areas). We note that the shape and the size of non-perturbative corrections in the two cases
are rather similar.
By comparing Figs.~\ref{fig:all_mc_np_band} (a) and (b), we observe that the behaviour
of the non-perturbative correction factors obtained here feature all the expected properties. The overall size of the non-perturbative
corrections decreases as the transverse momentum increases, in line with our expectation for IRC safe observables. This remains true for the
charged-hadrons only case, despite IRC unsafety. Applying \softdrop typically reduces the size and the onset of non-perturbative
corrections, although this feature is less prominent in the LHA case, for which non-perturbative corrections remain rather large
also for groomed jets. 

Finally, we use the results in Fig.~\ref{fig:all_mc_np_band} to correct our \NLOpNLLp
predictions for non-perturbative effects. In order to do that we take our \NLOpNLLp distributions as in Figs.~\ref{fig:all_mc_vs_res_ps_pT120}
and \ref{fig:all_mc_vs_res_ps_pT408} and multiply them on a per-bin basis by the corresponding central value of the \NPPS ratios shown in
Fig.~\ref{fig:all_mc_np_band}. The final uncertainty bands are obtained by summing in quadrature the
perturbative and non-perturbative uncertainties. 

\begin{figure}
  \centering
  \includegraphics[width=0.44\linewidth]{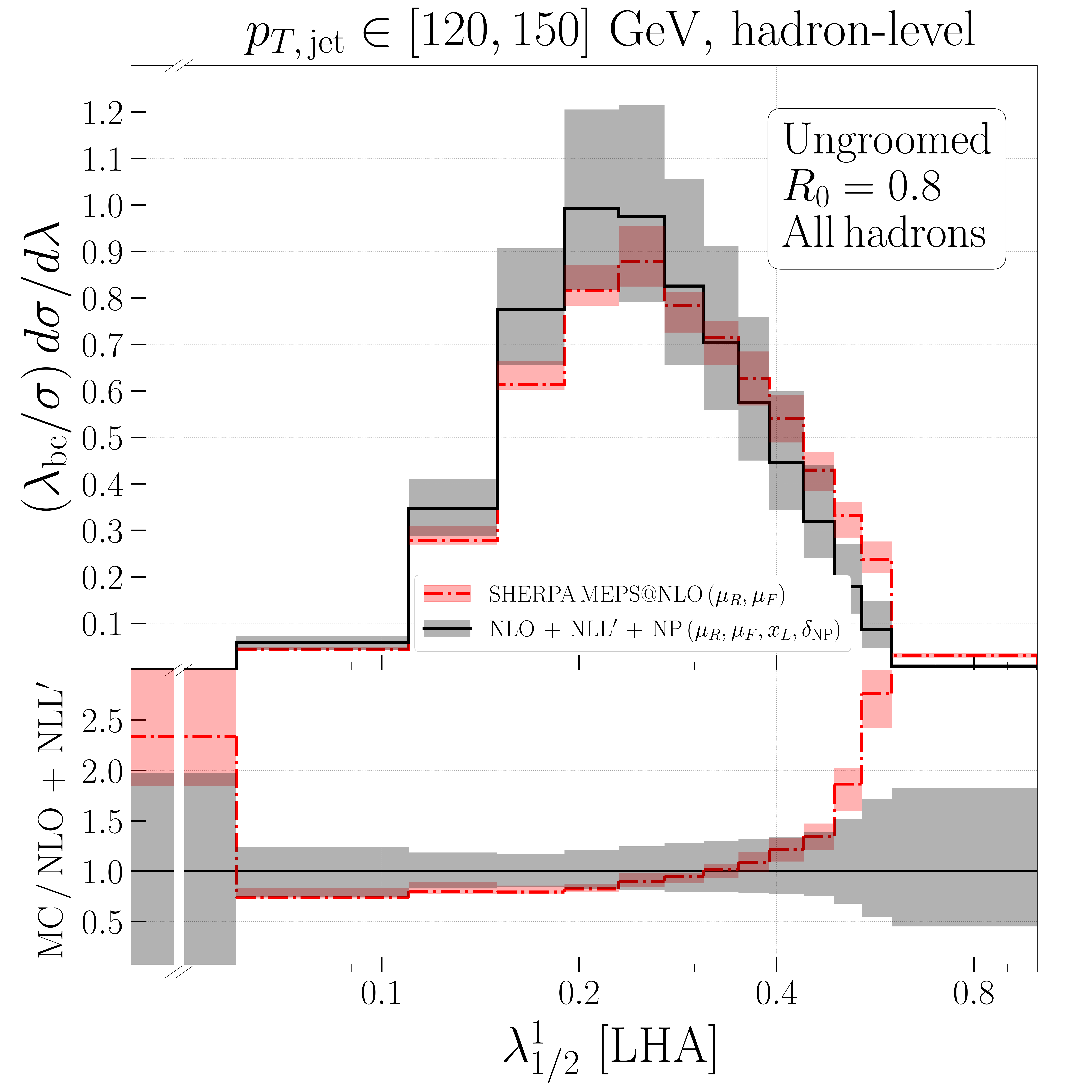}
  \hspace{1em}
  \includegraphics[width=0.44\linewidth]{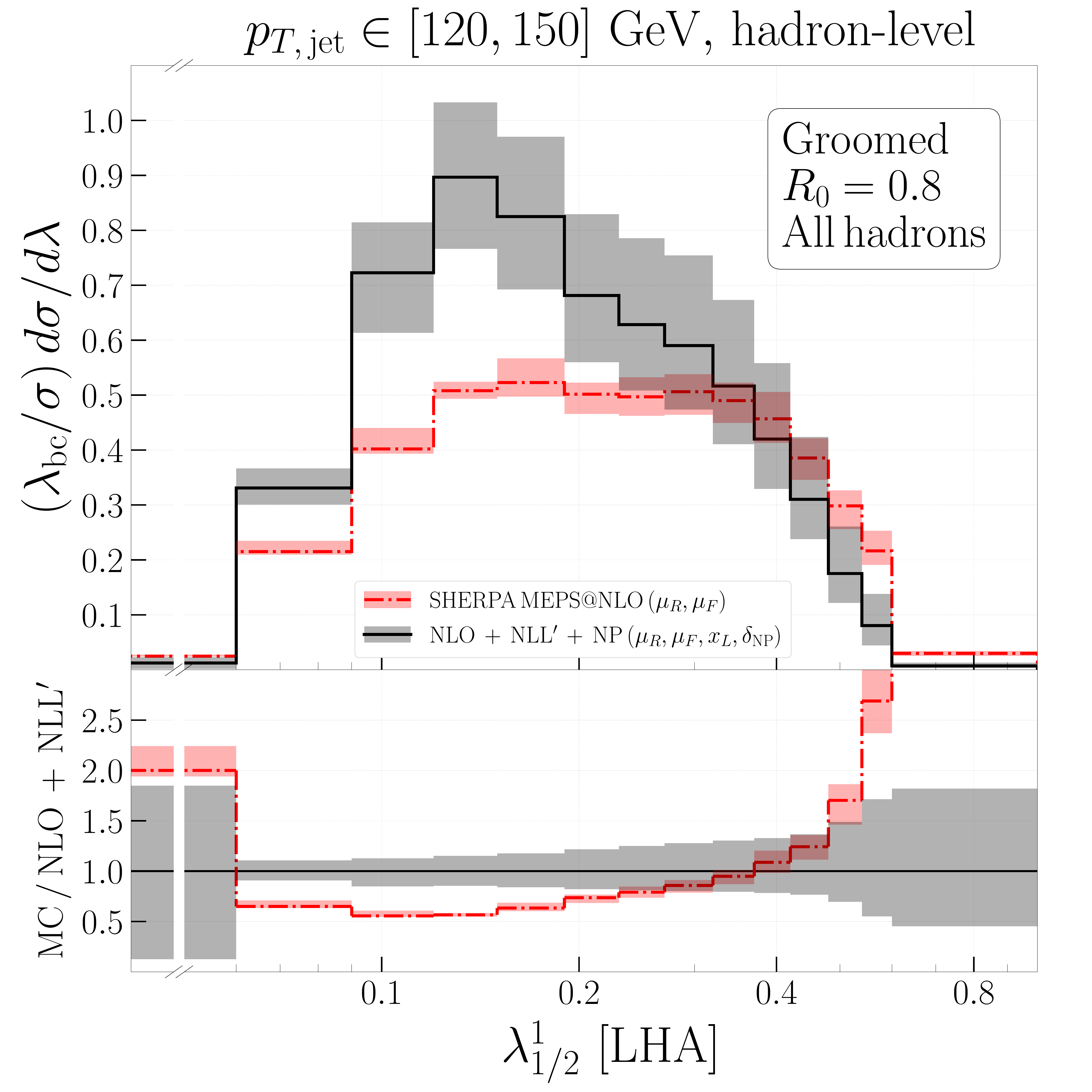}
  \centering
  \includegraphics[width=0.44\linewidth]{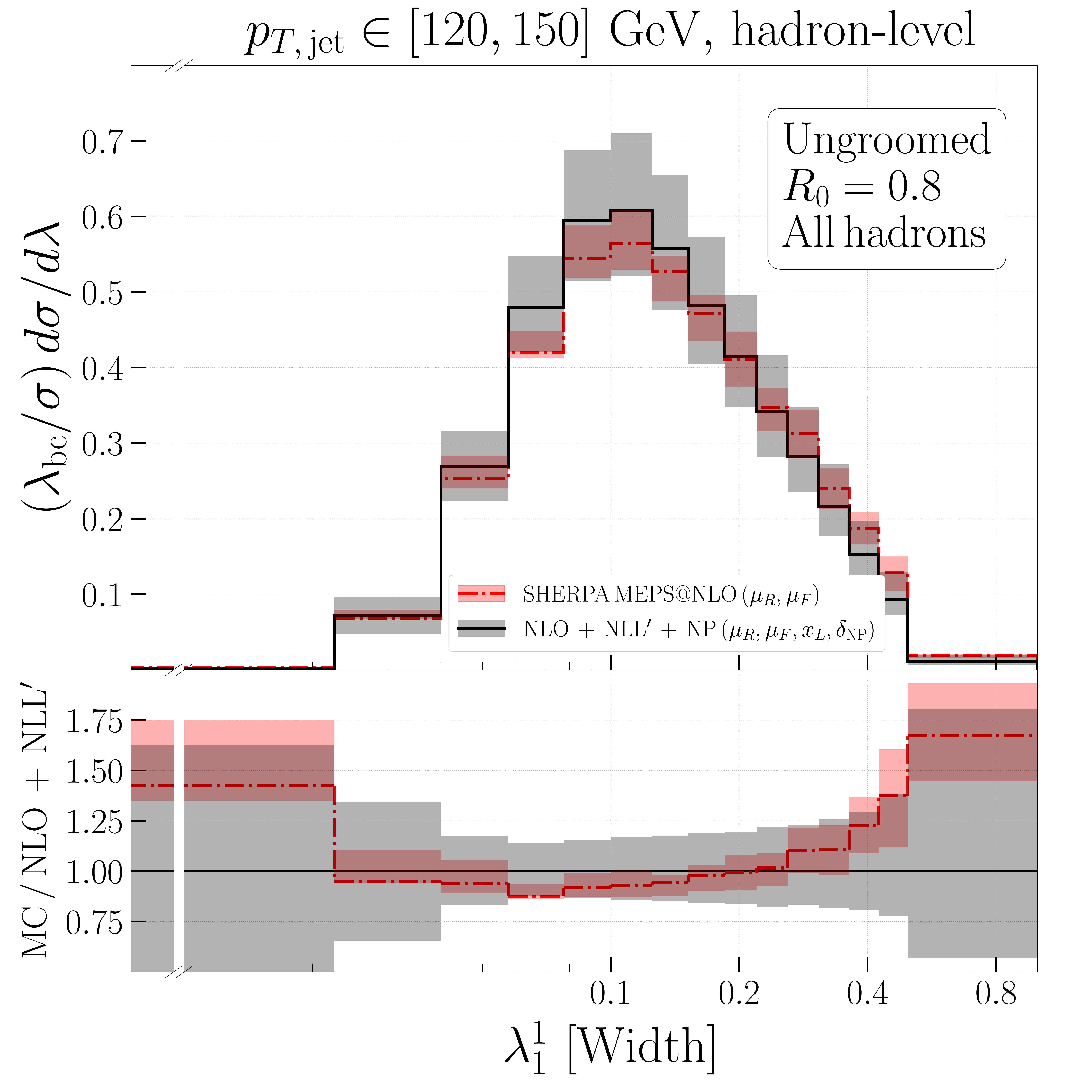}
  \hspace{1em}
  \includegraphics[width=0.44\linewidth]{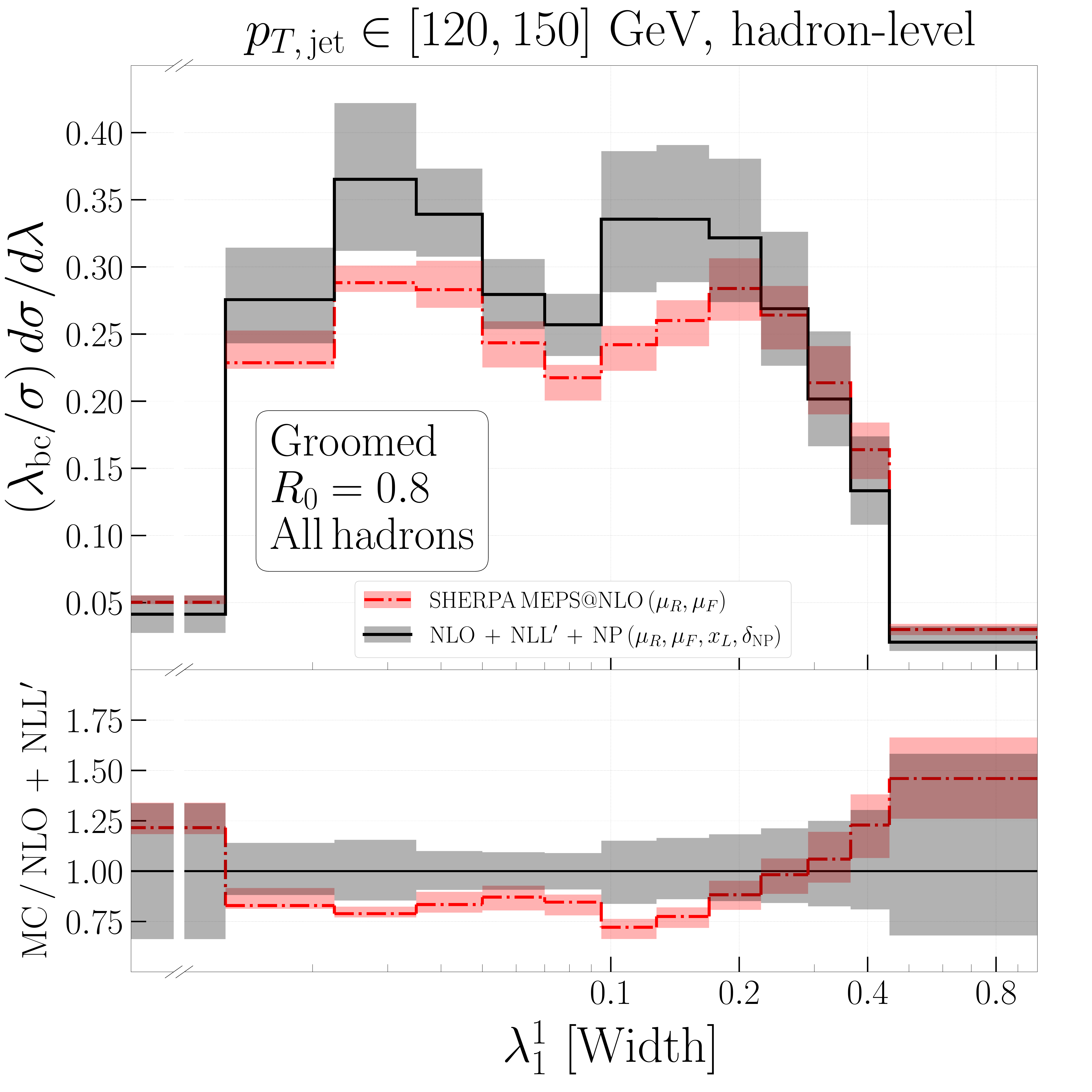}
  \centering
  \includegraphics[width=0.44\linewidth]{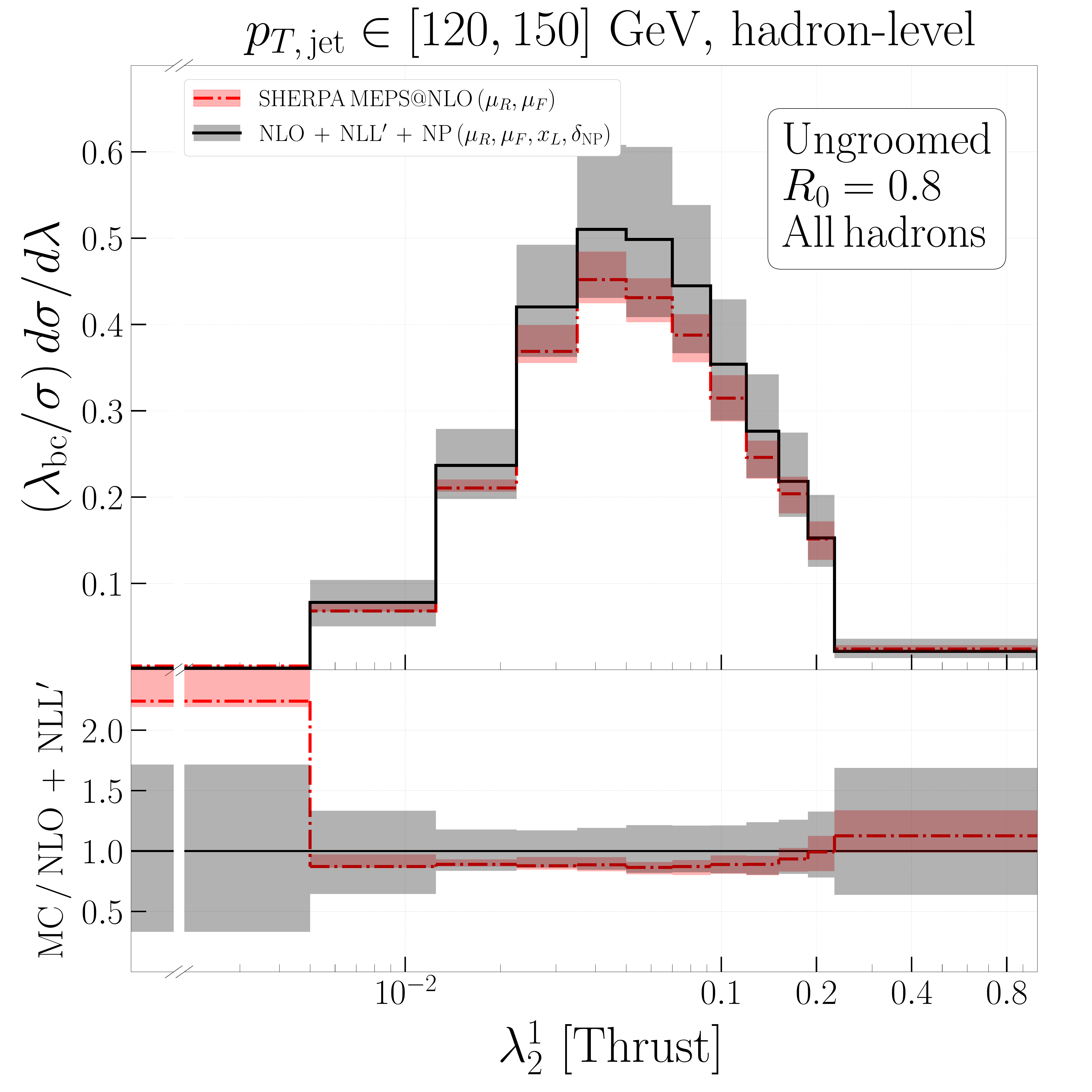}
  \hspace{1em}
  \includegraphics[width=0.44\linewidth]{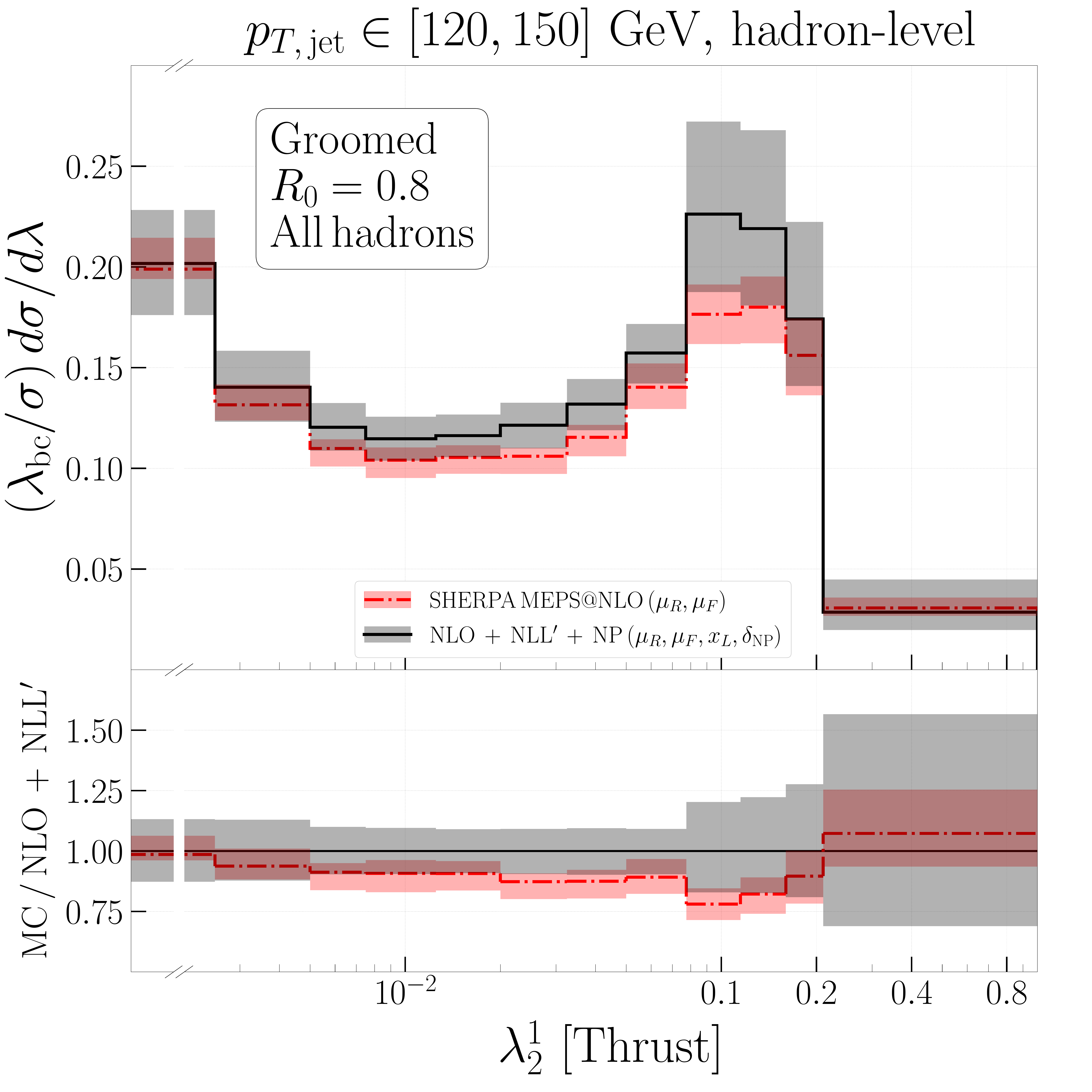}
  \caption{Comparison of hadron-level predictions from \sherpa (MEPS@NLO), based
    on all final state hadrons, for ungroomed and groomed jet-angularities in $Zj$ production, with
    $p_{T,\text{jet}}\in[120,150]\;\text{GeV}$, with \NLOpNLLp results corrected for non-perturbative effects.  Here $\lambda_\text{bc}$ stands for the bin centre.}
  \label{fig:res_plus_np_pT120_all}
\end{figure}

\begin{figure}
  \centering
  \includegraphics[width=0.44\linewidth]{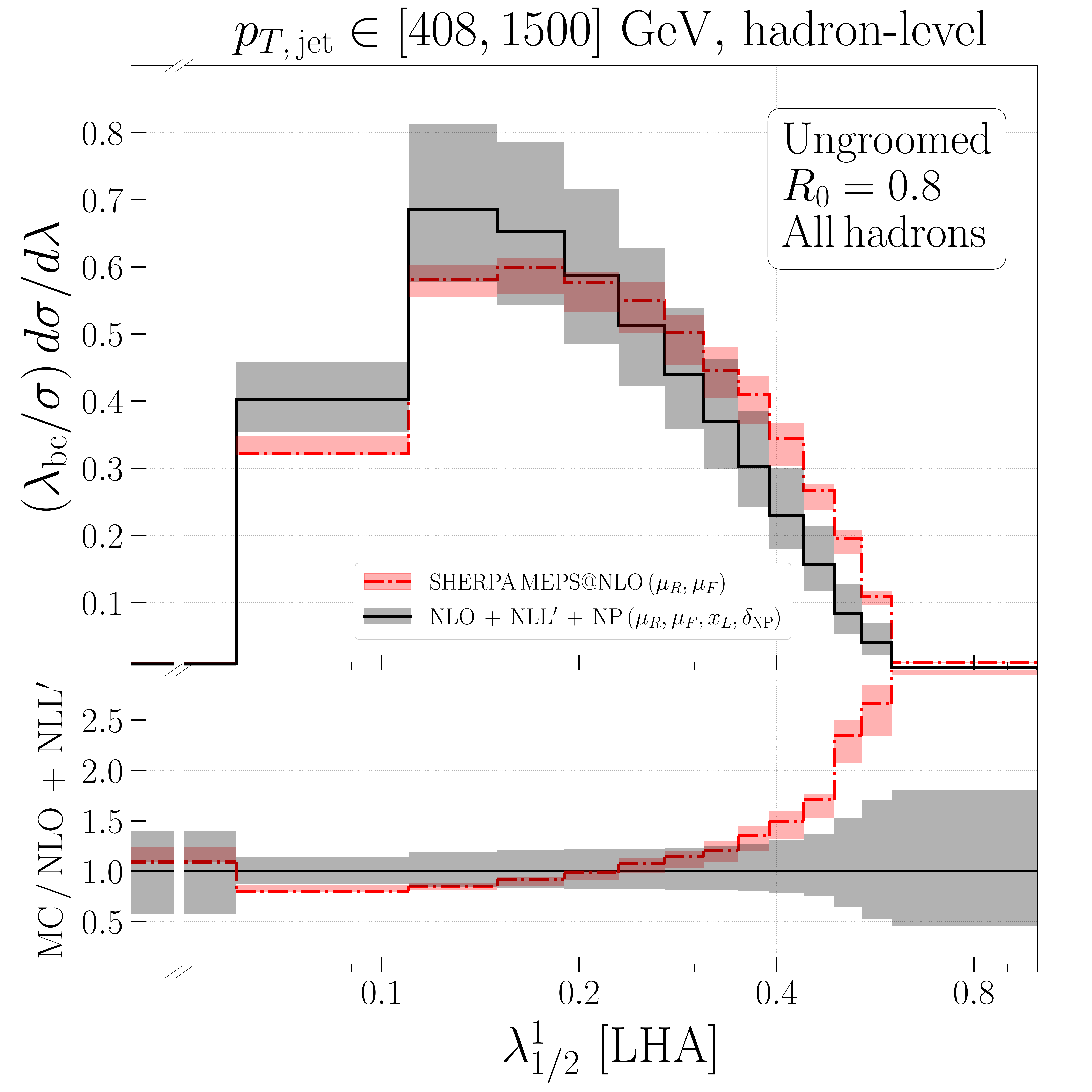}
  \hspace{1em}
  \includegraphics[width=0.44\linewidth]{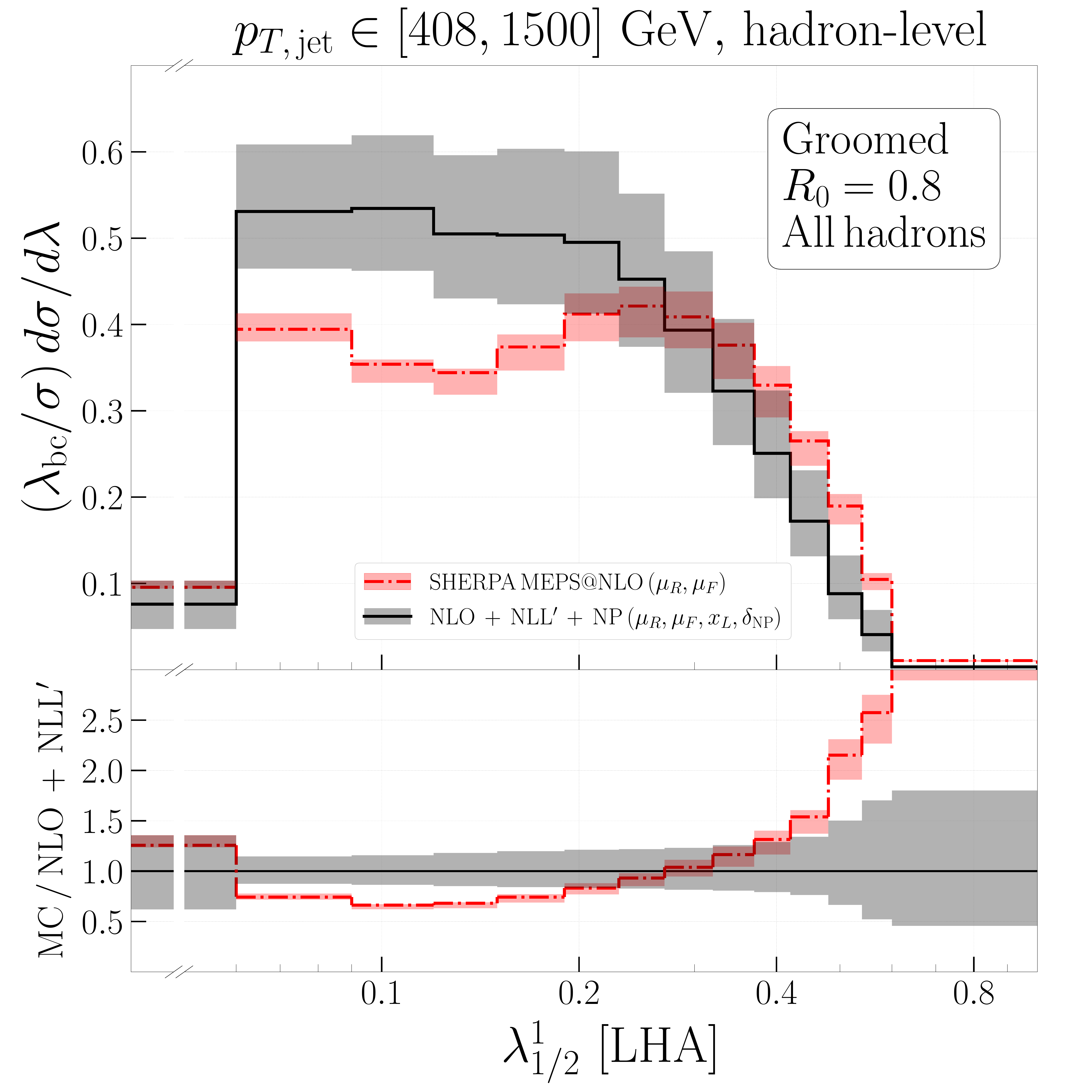}
  \centering
  \includegraphics[width=0.44\linewidth]{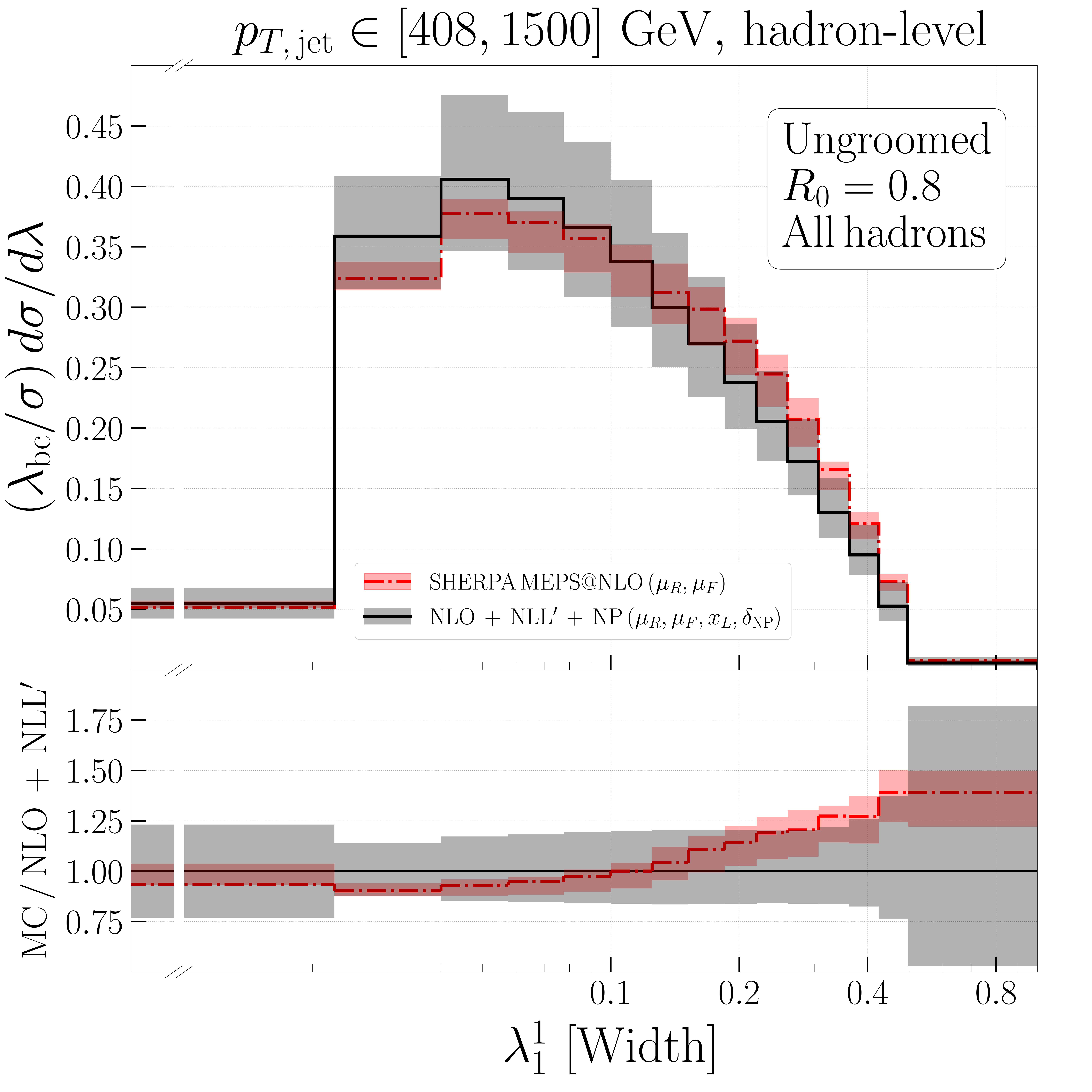}
  \hspace{1em}
  \includegraphics[width=0.44\linewidth]{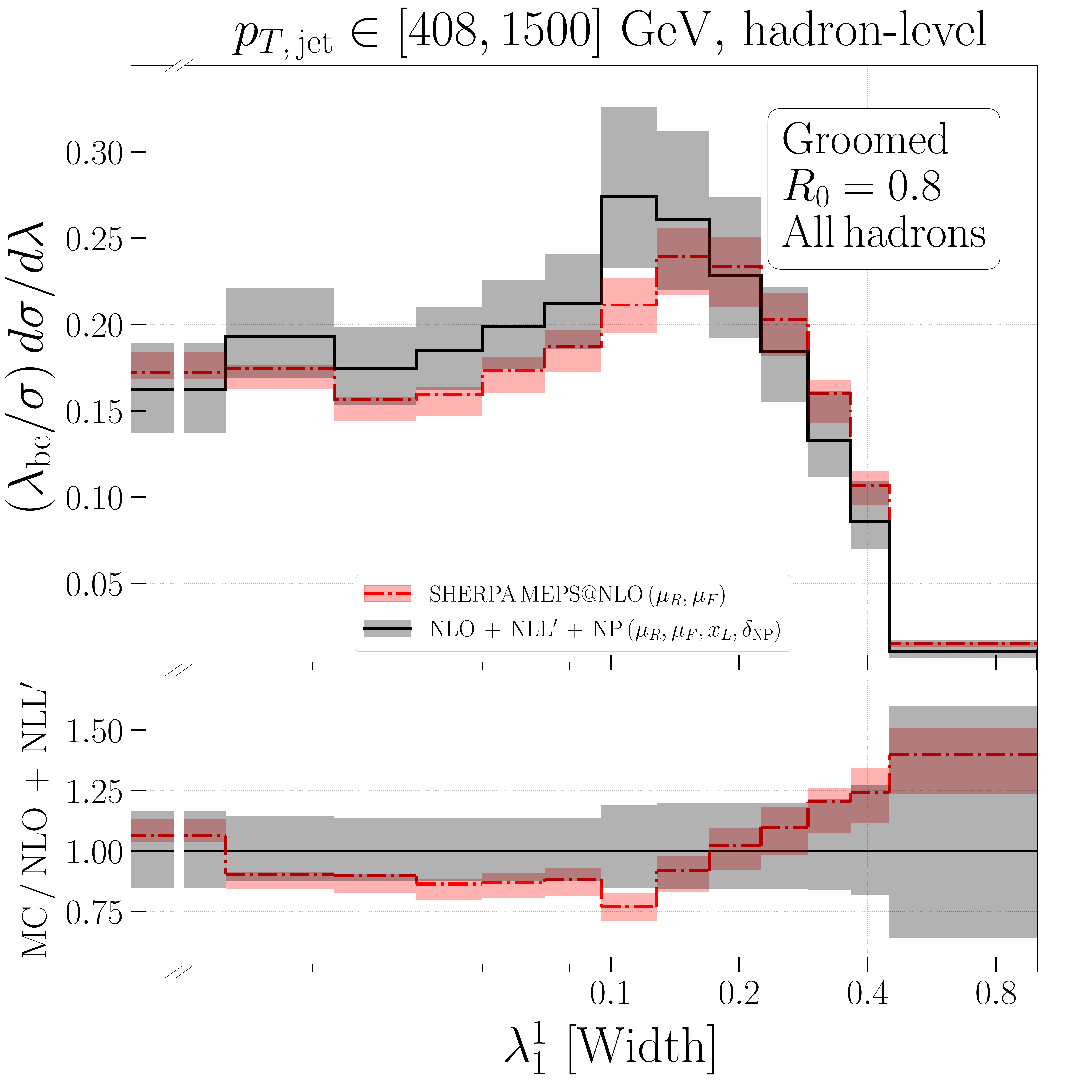}
  \centering
  \includegraphics[width=0.44\linewidth]{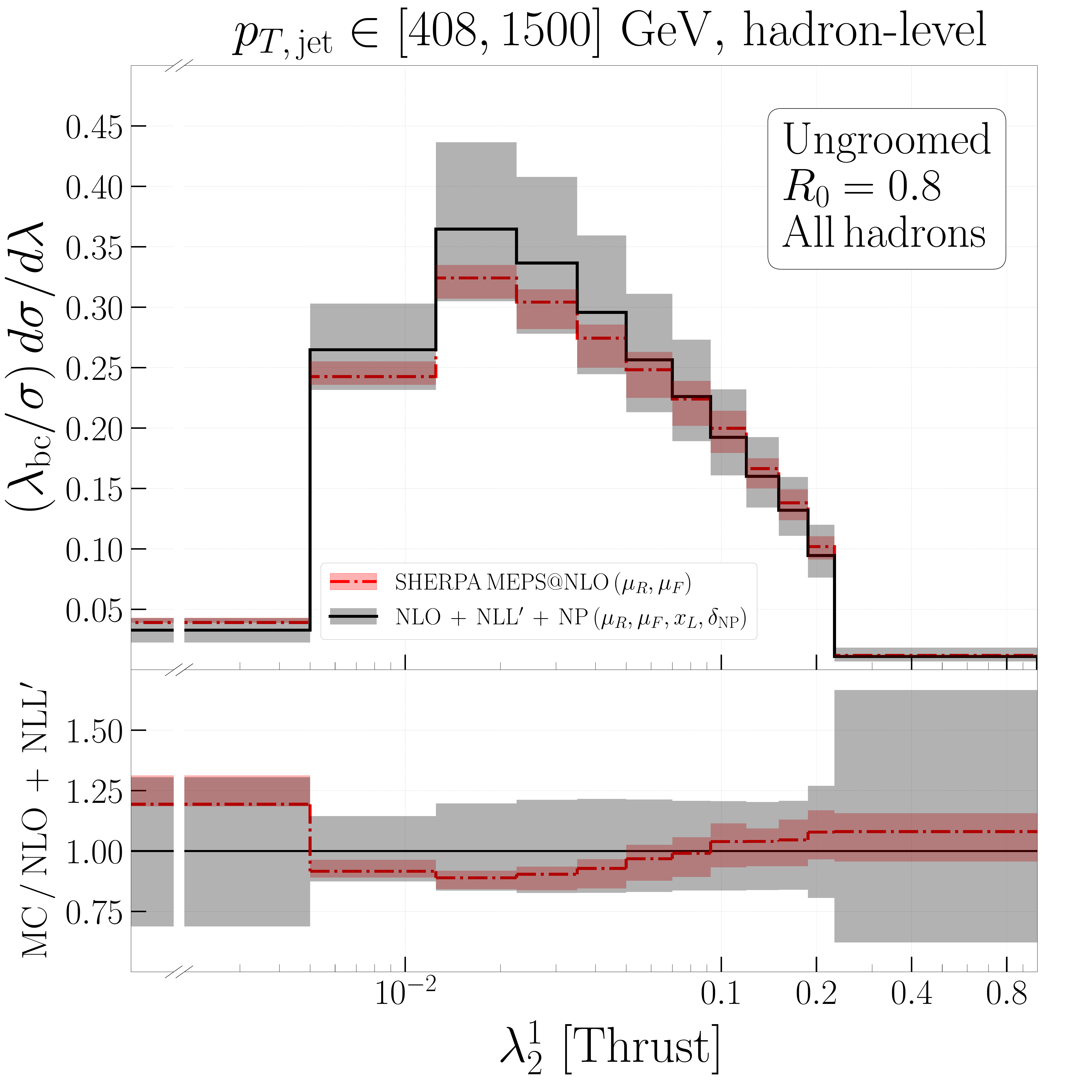}
  \hspace{1em}
  \includegraphics[width=0.44\linewidth]{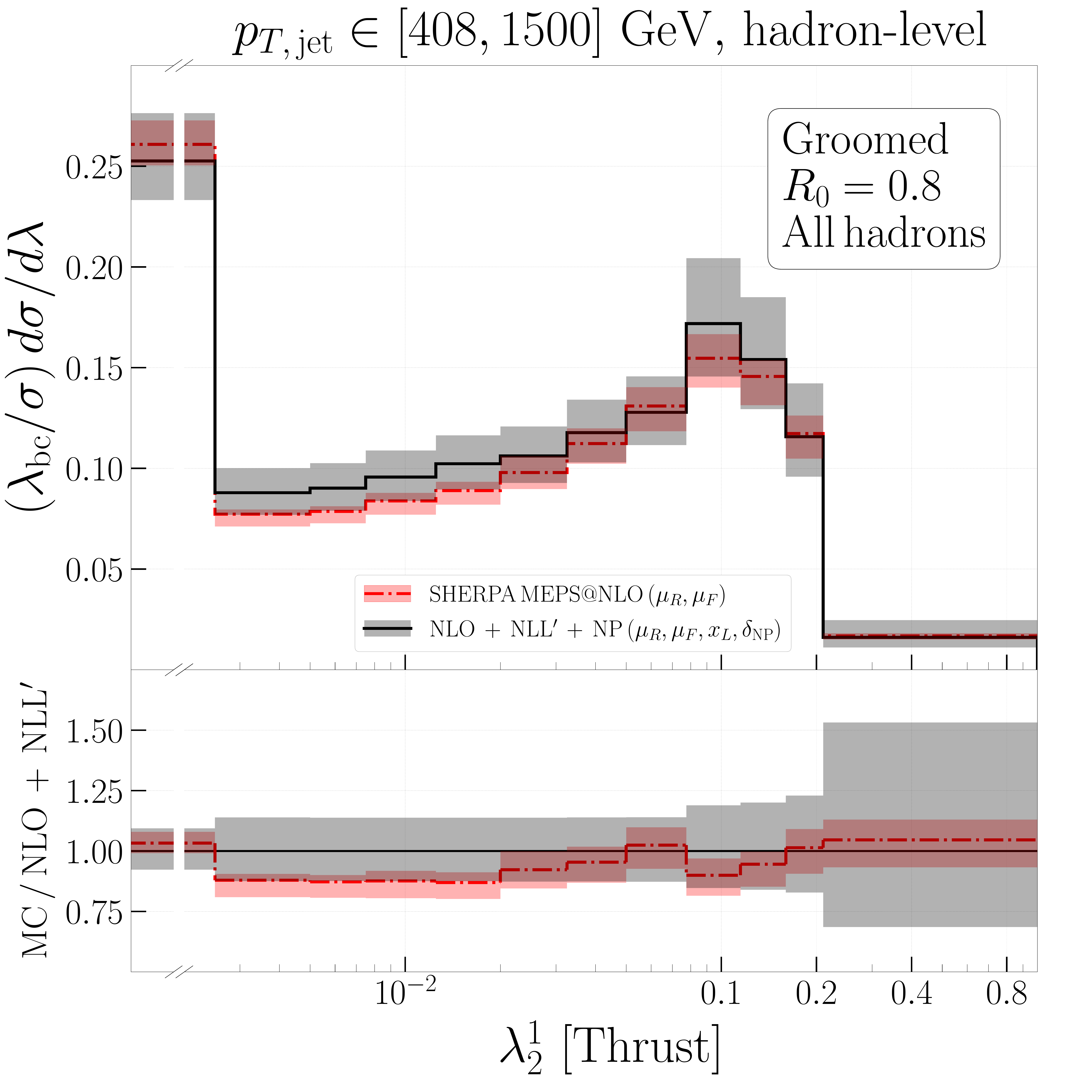}
 \caption{Comparison of hadron-level predictions from \sherpa (MEPS@NLO), based
    on all final state hadrons, for ungroomed and groomed jet-angularities in $Zj$ production, with
    $p_{T,\text{jet}}\in[408,1500]\;\text{GeV}$, with \NLOpNLLp results corrected for non-perturbative effects.  Here $\lambda_\text{bc}$ stands for the bin centre.}
\label{fig:res_plus_np_pT408_all}
\end{figure}

The \NLOpNLLp+ NP distributions, represented by black solid lines, in Figs.~\ref{fig:res_plus_np_pT120_all} and \ref{fig:res_plus_np_pT408_all}
present the main result of this paper. They feature our \NLOpNLLp perturbative predictions and include an MC-based estimate of
non-perturbative corrections, here for the case where all hadrons in the jet are considered, for the three angularities
$\lambda^1_\alpha$, for the two representative transverse momentum bins. Corresponding results for the case where only charged
tracks are considered in the observable calculation are presented in Appendix~\ref{app:NLL_NP_ch} in Figs.~\ref{fig:res_plus_np_pT120_ch}
and \ref{fig:res_plus_np_pT408_ch}. The gray uncertainty bands represent the perturbative, $(\mu_R,\mu_F, x_L)$ variations, and
non-perturbative, $\delta_{\text{NP}}$, systematics added in quadrature. For comparison, we also report
the \sherpa MEPS@NLO hadron-level predictions, red dashed-dotted lines, including systematic variations of the factorisation
and renormalisation scale, red hatched band. We see a good agreement between the two predictions for the jet Width and Thrust.
We expect that a comparison to upcoming experimental data would find agreement with these distributions.
However, the size of the
experimental uncertainties will tell us whether the theoretical calculation needs to be improved, in order to perform precision phenomenology.
In this context, these observables could be exploited to extract jet properties, such as the flavour, or, even more ambitiously, parameters
of the Standard Model such as couplings and masses.
The situation for the LHA distribution is instead in stark contrast. Here the resummed calculation and event-generator predictions are
not in agreement, especially in the groomed case. This is true both at parton- and hadron-level. In this case upcoming LHC data may help
us to shed light on the discrepancy. The data may favour the MC prediction, indicating that modelling used in the resummed
calculation, which for instance neglects recoil, is not appropriate. On the other hand, the data may favour the resummation, indicating
the need for a better understanding of the logarithmic structure that is achieved by the parton-shower models.

We conclude this section by reporting that most of the conclusions reached above, also hold for smaller jet radii. Explicit resummed and matched results for $R_0=0.4$ , as well as their comparison to \sherpa MEPS@NLO predictions, are collected in Appendix~\ref{app:NLL_NP_R4}.

\FloatBarrier

\section{Conclusions and Outlook}\label{sec:conclusions}

In this paper we have performed a detailed phenomenological analysis of variables that describe the
internal structure of jets produced in association with a $Z$ boson decaying into muons. Our study
has focussed on IRC safe jet angularities that are characterised by an angular exponent $\alpha$,
the variation of which allows us to probe QCD radiation in different ways. In particular, inspired
by an upcoming measurement by the CMS collaboration, we have derived results for $\alpha=1/2, 1,2$. 
Furthermore, we have considered \emph{standard} (ungroomed) and \emph{groomed} jets. Our grooming algorithm of
choice has been \softdrop\!\!, which is characterised by the momentum fraction parameter $\zc$ and
the angular exponent $\beta$. We have shown explicit results for the combination $\zc=0.1$, $\beta=0$,
corresponding to the default choice of the CMS collaboration in their ongoing study.

Comparisons between unfolded experimental measurements and
accurate first-principle theoretical predictions can allow us to
test, and improve, our description of jet angularities. These
observables are interesting for many reasons. For instance, they can
be employed to distinguish quark-like from gluon-like jets, in a
theoretically well-defined
way~\citep{Badger:2016bpw,Gras:2017jty,Amoroso:2020lgh}. They can
also be used to extract Standard Model parameters, such as the
strong coupling constant~\citep{Bendavid:2018nar}. Furthermore,
jet-angularity measurements can form an important input to constrain
Monte Carlo event generators in general and the tuning of
non-perturbative model parameters in particular~\citep{Amoroso:2020lgh}.
  
The first part of the paper has been devoted to a detailed comparison of different MC event
generators that provide an increasing sophistication in the way they include fixed-order matrix
elements. In the region of small angularities, where soft radiation, both of perturbative and
non-perturbative origin, dominates, we have confirmed  that the spread in the predictions provided
by widely used general-purpose event generators is largely reduced when grooming is considered. 
In our study we have also investigated the impact of fixed-order corrections to the parton shower.
In particular, we have compared LO predictions from \pythia and \herwig, to the ones obtained
with merged samples from \sherpa, both at LO and NLO. The inclusion of full NLO QCD corrections
in the \sherpa MEPS@NLO simulations leads to a significant reduction of systematic uncertainties,
estimated by consistent variations of the factorisation and renormalisation scales in both the
matrix element and parton-shower components. When comparing to the LO simulations, we have observed
noticeable effects at large values of the angularities, ranging from $\sim 10\%$ at
moderate transverse momentum to $\sim 30\%$ in the high-$p_{T,\text{jet}}$ region. This applies to both,
standard and groomed jets.
This behaviour is perhaps related to an analogous trend observed in the inclusive $K$-factors, which are increasing functions of the jet transverse momentum. 
Our findings underline the importance of using state-of-the-art generators for jet phenomenology, especially
when considering processes characterised by fairly large radiative corrections, as the one considered
here, \emph{i.e.}\ hadronic $Zj$ production.

The second part of this study has instead focussed on all-order predictions obtained with resummed
perturbation theory at single-logarithmic accuracy. Thanks to a flavour-dependent matching to fixed-order
predictions, we have been able to achieve \NLOpNLLp accuracy. This includes, for the ungroomed case,
the resummation of non-global logarithms, which we perform in the large-$N_c$ limit. The
resummed and matched calculations are implemented in the resummation plugin to \sherpa. This allows us
to exploit the main features of the event generation framework, including phase-space integration,
matrix-element evaluations and matching, obtaining \NLOpNLLp predictions that include fiducial cuts and
can be faithfully compared to unfolded measurements. To this end, we complemented our perturbative
predictions with non-perturbative corrections, derived from MC simulations, both for angularity
measurements based on \emph{all} or \emph{charged-only} jet constituents. As expected, groomed jets
benefit from smaller corrections, which however remain rather large for the $\lambda^1_{1/2}$ angularity. 
Although we have focussed on one particular combination of \softdrop parameters, our resummed calculation
is more general. It applies for all $\beta \ge0$ values and can be extended to negative $\beta$ if \softdrop is
used in tagging rather than grooming mode. If one wanted to consider much smaller values of $\zc$, in
addition the systematic resummation of related logarithmic corrections should be
included. Similarly, for large values of $\zc$, power corrections can become important and should be included.

The main deliverable of this paper are \NLOpNLLp predictions for jet angularities in $Z$+jets events in
proton--proton collisions that can be immediately exploited for data--theory comparisons. However, this
work has already shown us the path towards higher precision that we need to pursue. In this context, we
can identify four complementary lines of research. First of all, although we do not expect large corrections,
one should improve the accuracy of the resummation, in order to reduce the theoretical uncertainty. The
framework for NNLL calculations has been worked out using SCET, although
explicit results only exist, to the best of our knowledge, for the case $\alpha=2$~\citep{Frye:2016okc,Frye:2016aiz}.
Second, it would be desirable to match the resummation to NNLO distributions. However, we recognise that this endeavour
is much more formidable as it requires calculating $Z$ plus two
partons at NNLO accuracy.
Some first steps have however already been taken in this direction~\cite{Hartanto:2019uvl,Abreu:2020jxa,Canko:2020ylt,Badger:2021nhg}.
Next, given that the size of non-perturbative corrections can remain
large even for groomed angularities, \emph{e.g.}\ for $\alpha< 1$, it
would be interesting to study additional grooming strategies, such as
complementing the \softdrop procedure with a filtering
step~\cite{Butterworth:2008iy} or using the \textsf{Recursive
  SoftDrop} algorithm~\cite{Dreyer:2018tjj}, although these tools would most
likely further complicate the structure of the analytic resummation.
Finally, a SCET framework
to deal with non-perturbative effects has been recently developed~\citep{Hoang:2019ceu, Pathak:2020iue}
and it would be interesting to quantitatively compare its predictions to the phenomenological model employed here. 

\begin{acknowledgments}
We acknowledge many useful discussions with Robin Aggleton, Kaustuv Datta, Alejandro Gomez Espinosa, Andreas Hinzmann, Christine Angela McLean, Ashely Marie Parker, and Salvatore Rappoccio.

  The work leading to this publication was supported by the German Academic Exchange Service (DAAD) with funds from the German
  Federal Ministry of Education and Research (BMBF) and the People Programme (Marie Curie Actions) of the European Union Seventh
  Framework Programme (FP7/2007-2013) under REA grant agreement n.\ 605728 (P.R.I.M.E.\ Postdoctoral Researchers International
  Mobility Experience). The work of SC, SM and OF is supported by Universit\`a di Genova under the curiosity-driven grant ``Using jets to challenge the Standard Model of particle physics'' and by the Italian Ministry of Research (MUR) under grant PRIN 20172LNEEZ.
  SS acknowledges funding from the European Union's Horizon 2020 research
  and innovation programme as part of the Marie Sk\l{}odowska-Curie Innovative Training Network MCnetITN3
  (grant agreement no. 722104), the Fulbright-Cottrell Award and from BMBF (contract 05H18MGCA1).
  GS\ is supported in part by the French Agence Nationale de la
  Recherche, under grant ANR-15-CE31-0016.

All figures in this paper  were
created with the Matplotlib \citep{Hunter:2007ouj} and NumPy \citep{NumPy} libraries.

\end{acknowledgments}

\clearpage

\appendix

\section{Additional results}\label{app:appendix}
In this Appendix, we collect supplementary results. 
\subsection{Numerical effects of the global and non-global soft function for groomed
  observables}\label{app:ngl}
\begin{figure}[hb!]
  \begin{center}
  \includegraphics[width=0.36\textwidth]{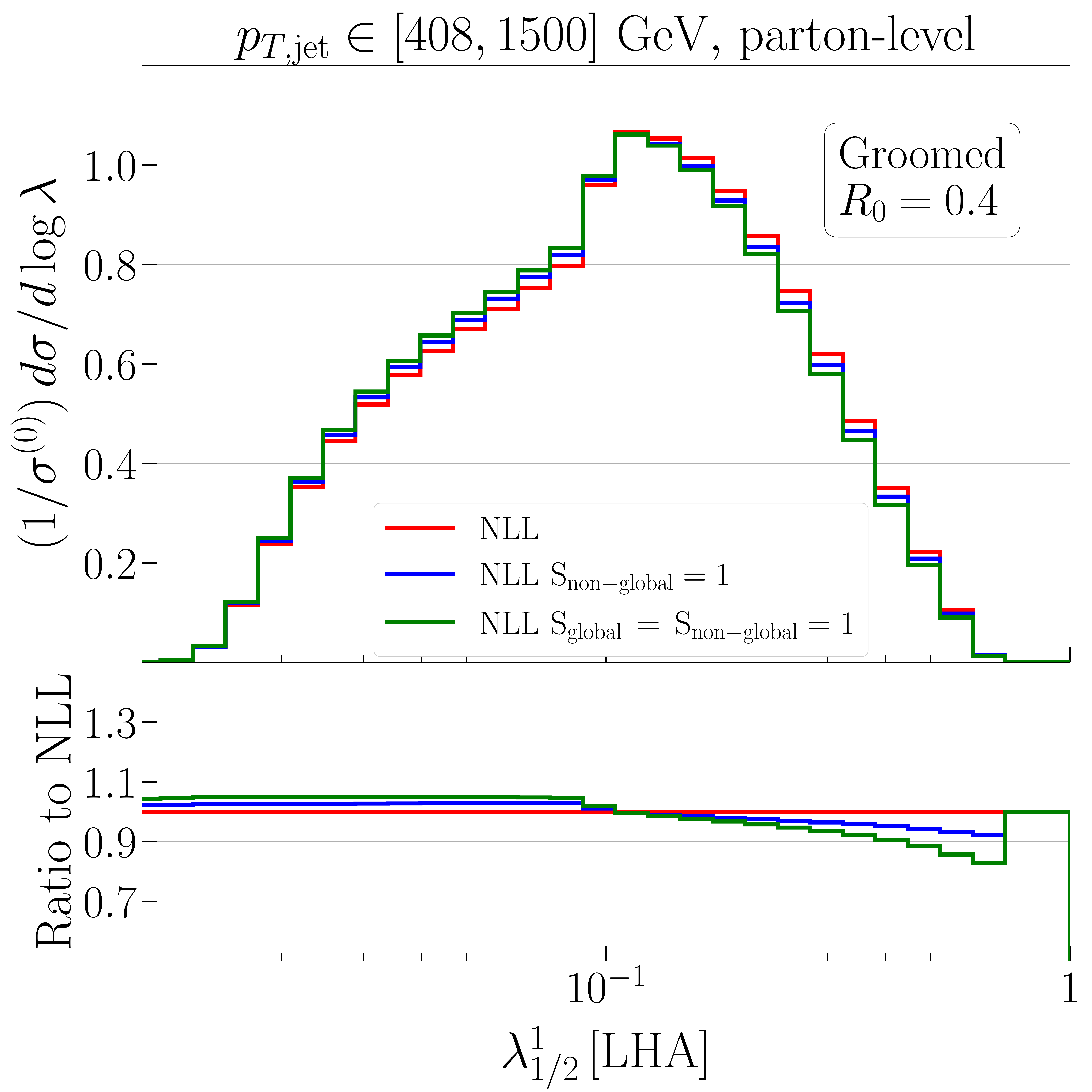}\hspace*{10mm}
  \includegraphics[width=0.36\textwidth]{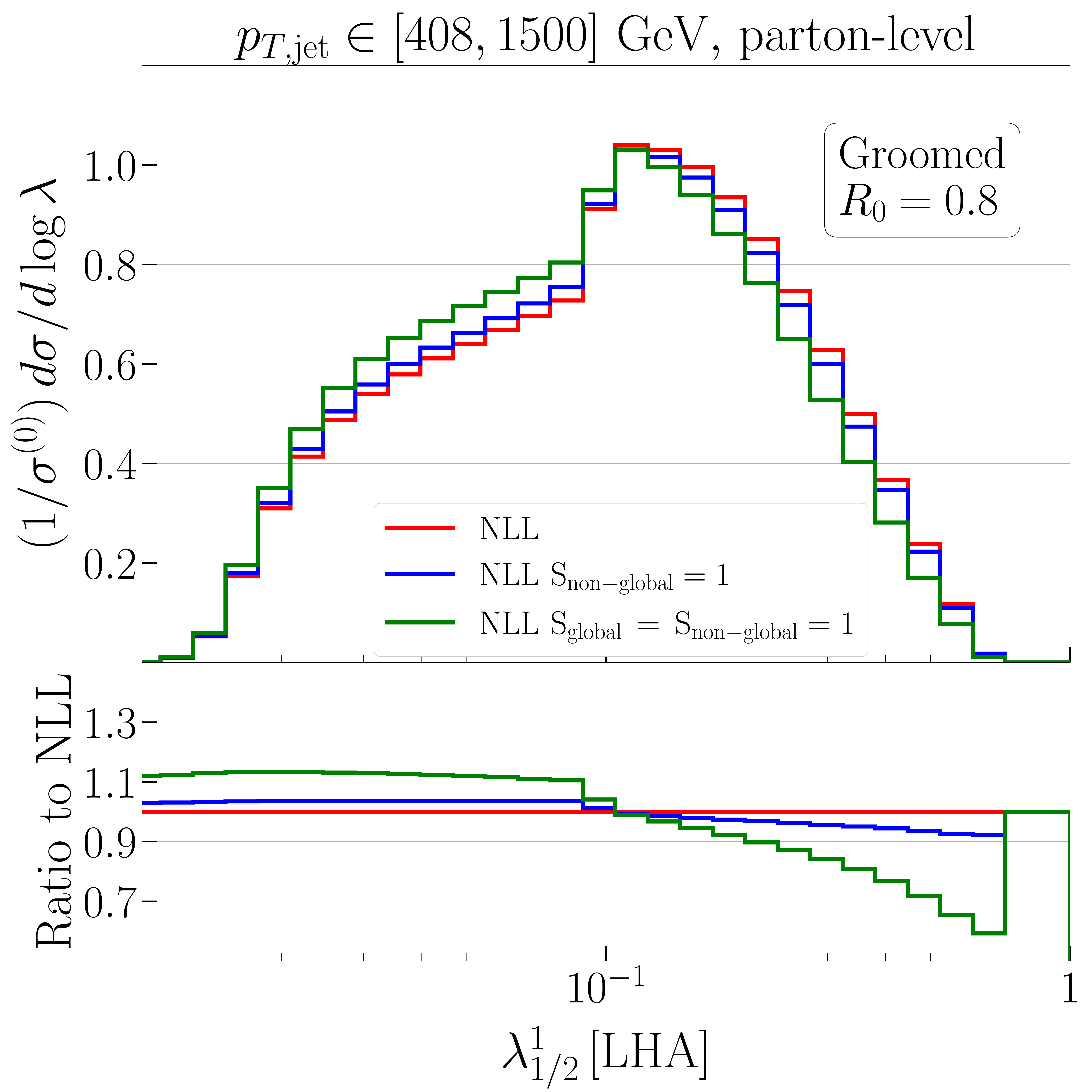}

  \includegraphics[width=0.36\textwidth]{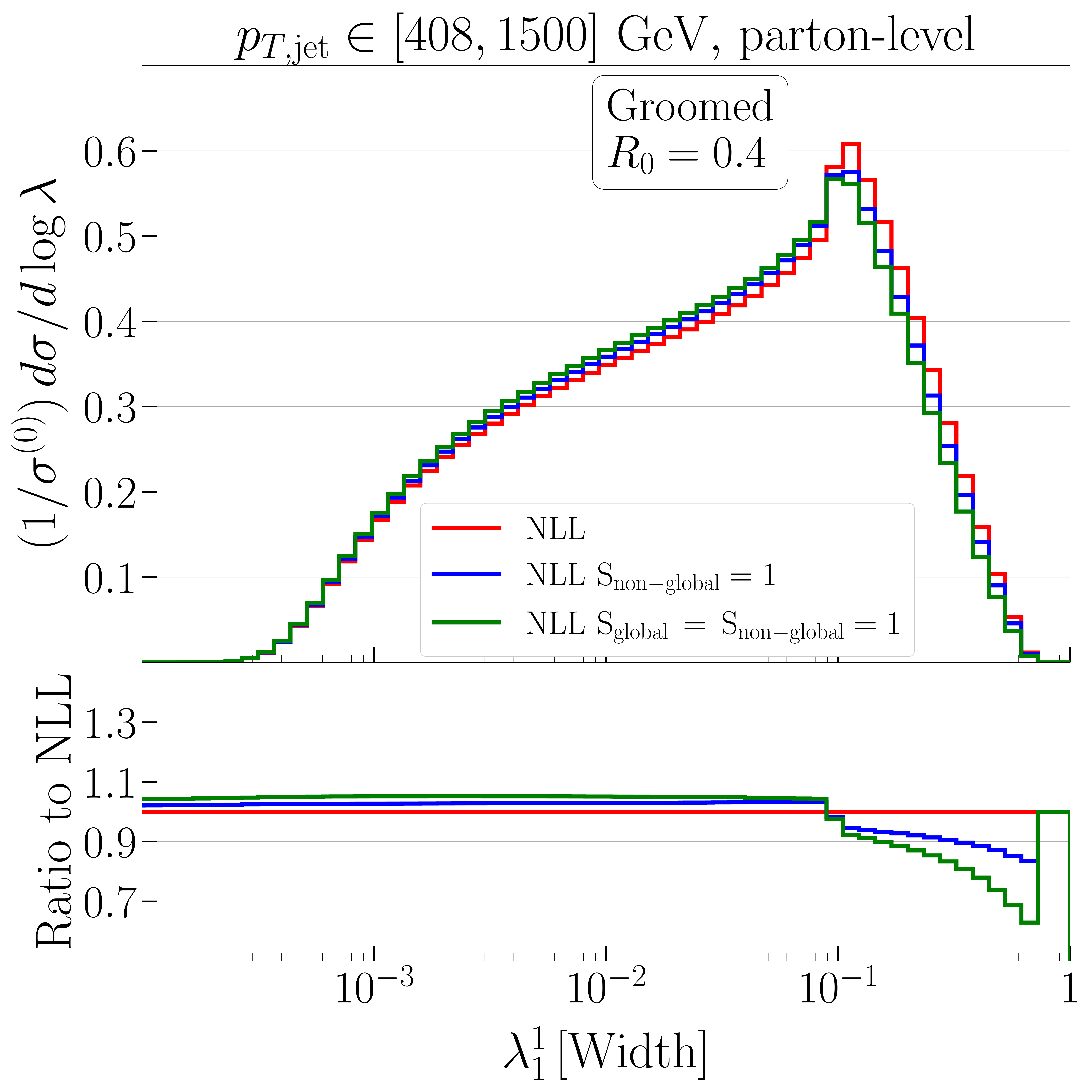}\hspace*{10mm}
  \includegraphics[width=0.36\textwidth]{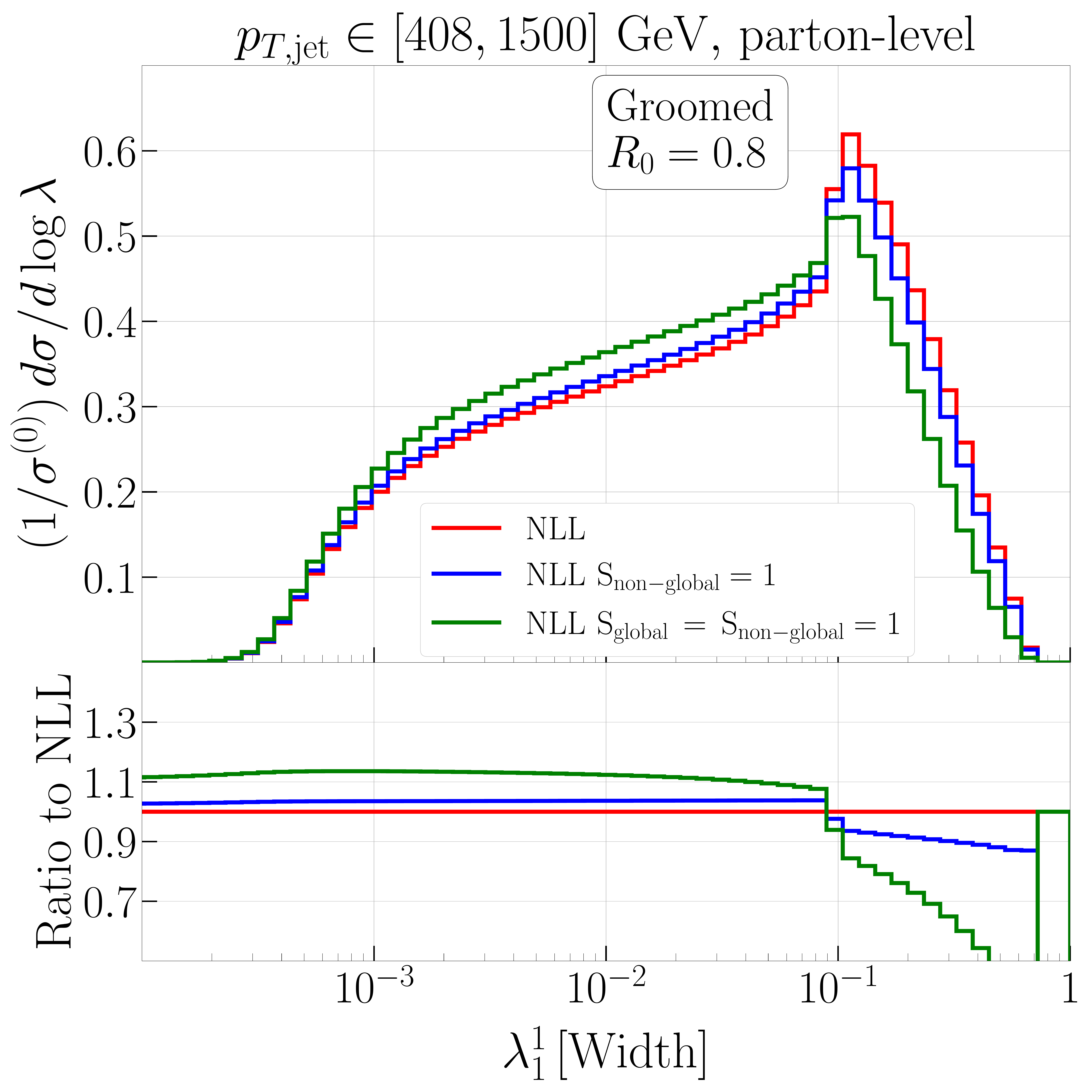}

  \includegraphics[width=0.36\textwidth]{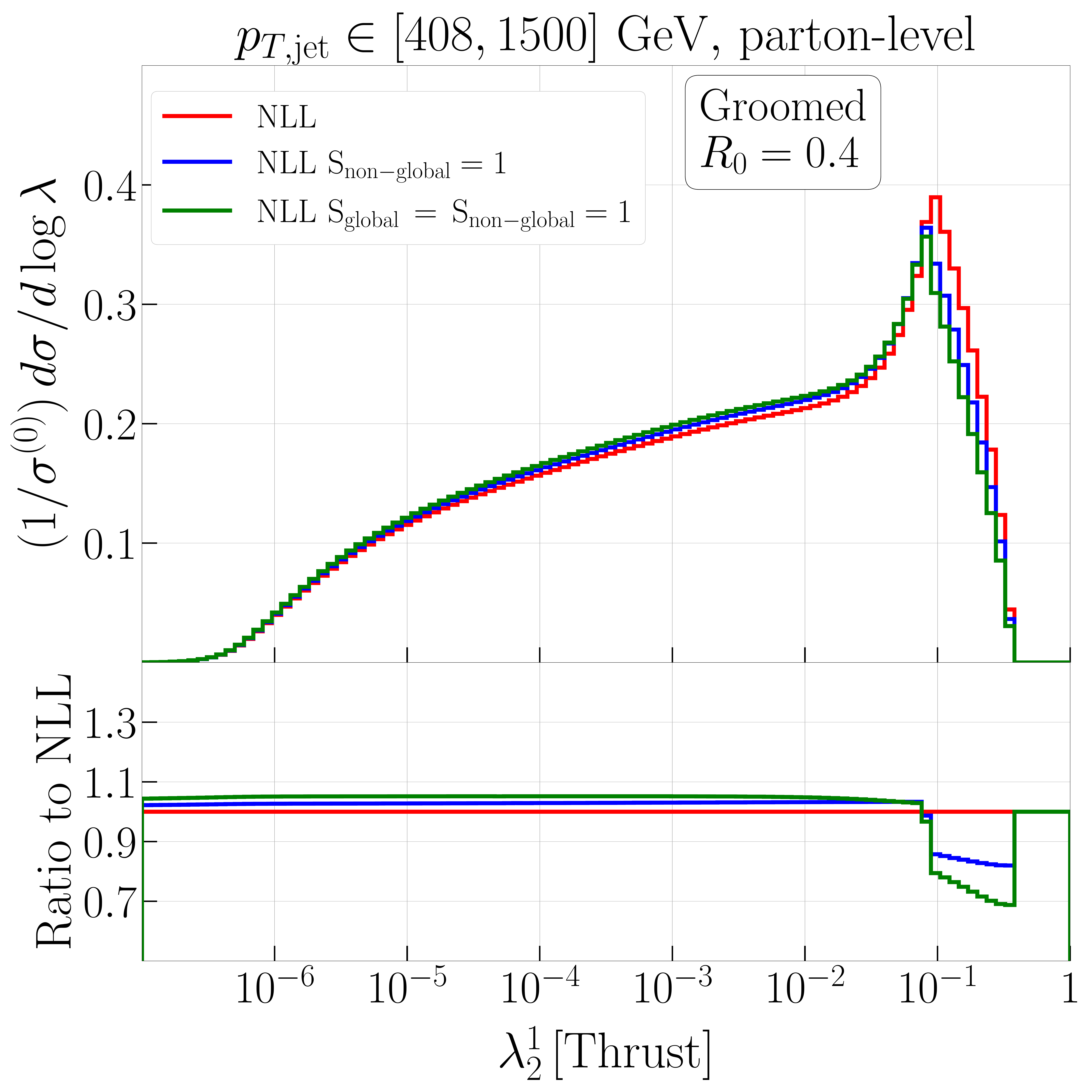}\hspace*{10mm}
  \includegraphics[width=0.36\textwidth]{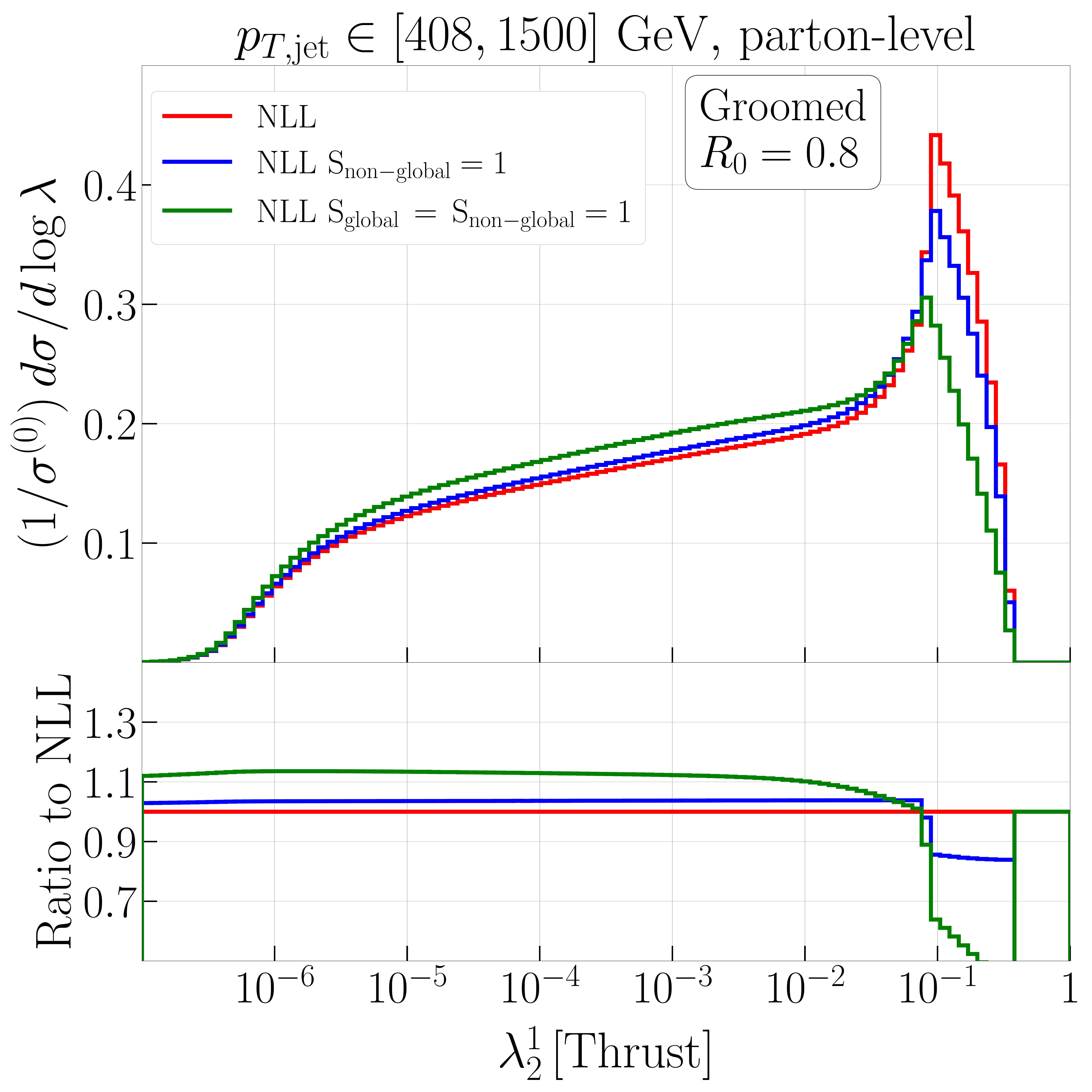}
  \end{center}
  \caption{Same as Fig.~\ref{fig:ngl_numerics} but with soft-drop grooming using
    $\zcut=0.1$, $\beta=0$.}\label{fig:ngl_numerics_groomed}
\end{figure}

In Fig.~\ref{fig:ngl_numerics_groomed} we show the same results as in
Fig.~\ref{fig:ngl_numerics} in the main text, but with grooming.

\clearpage

\subsection{Hadron-level predictions based on charged particles only}
\label{app:NLL_NP_ch}

\begin{figure}
  \centering
  \includegraphics[width=0.44\linewidth]{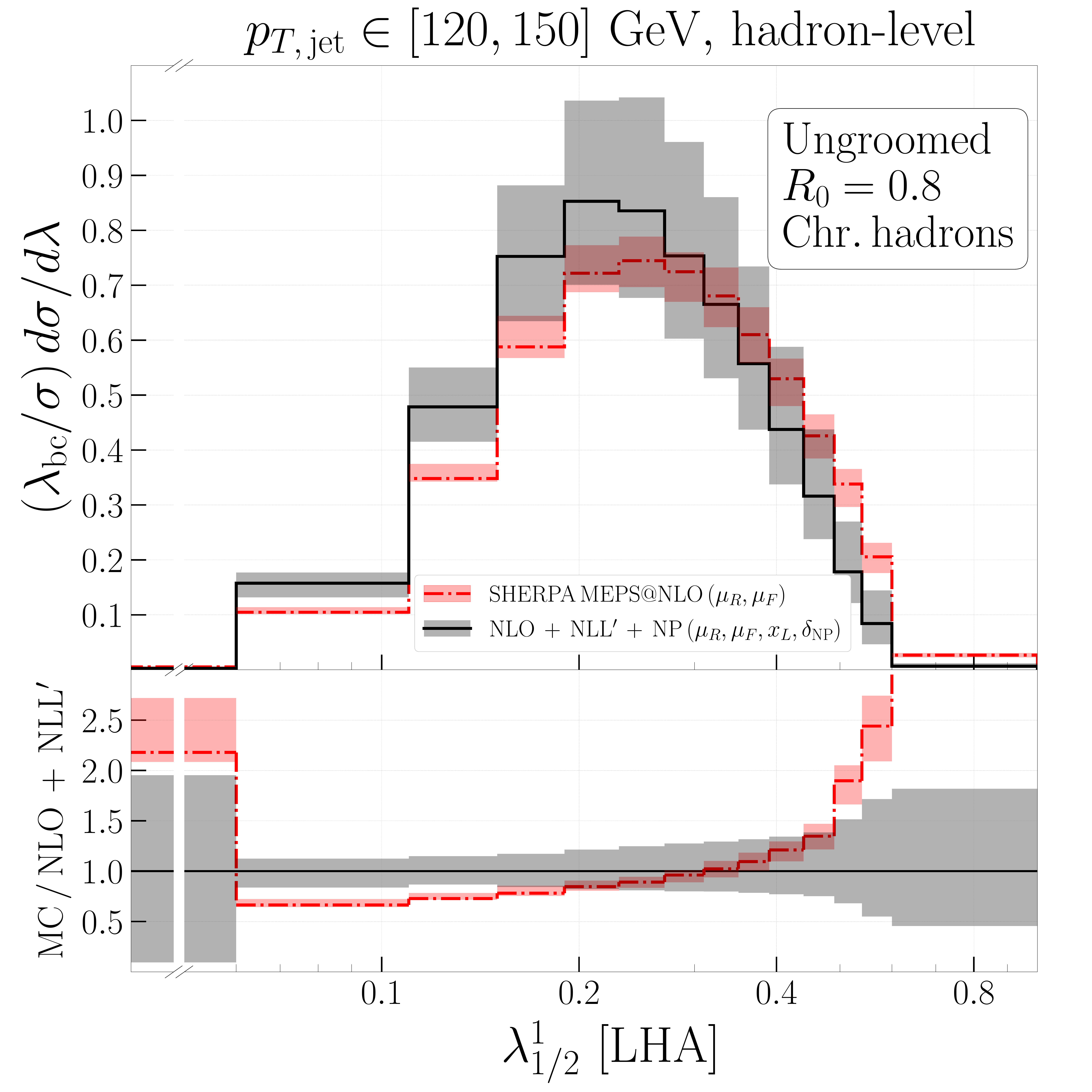}
  \hspace{1em}
  \includegraphics[width=0.44\linewidth]{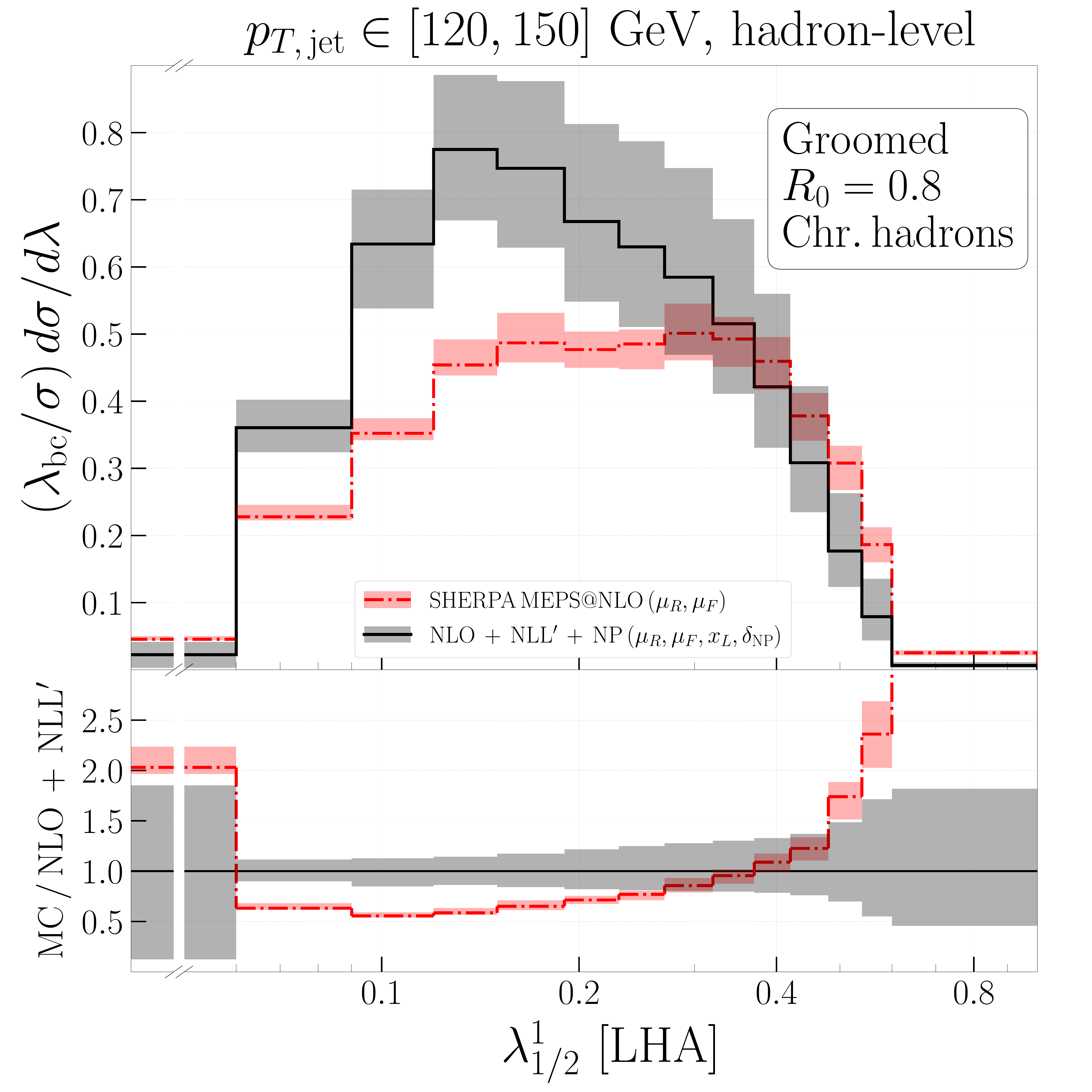}
  \centering
  \includegraphics[width=0.44\linewidth]{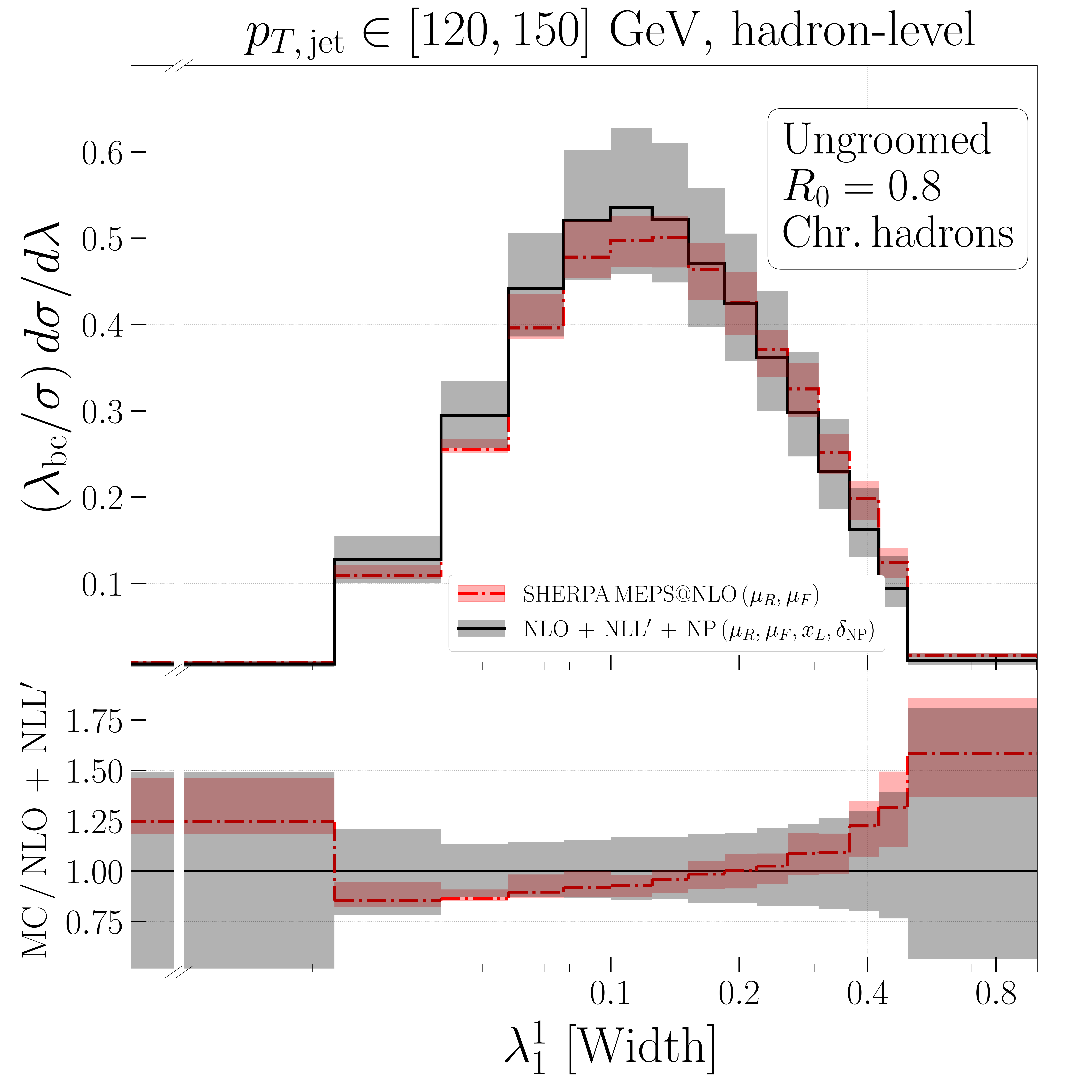}
  \hspace{1em}
  \includegraphics[width=0.44\linewidth]{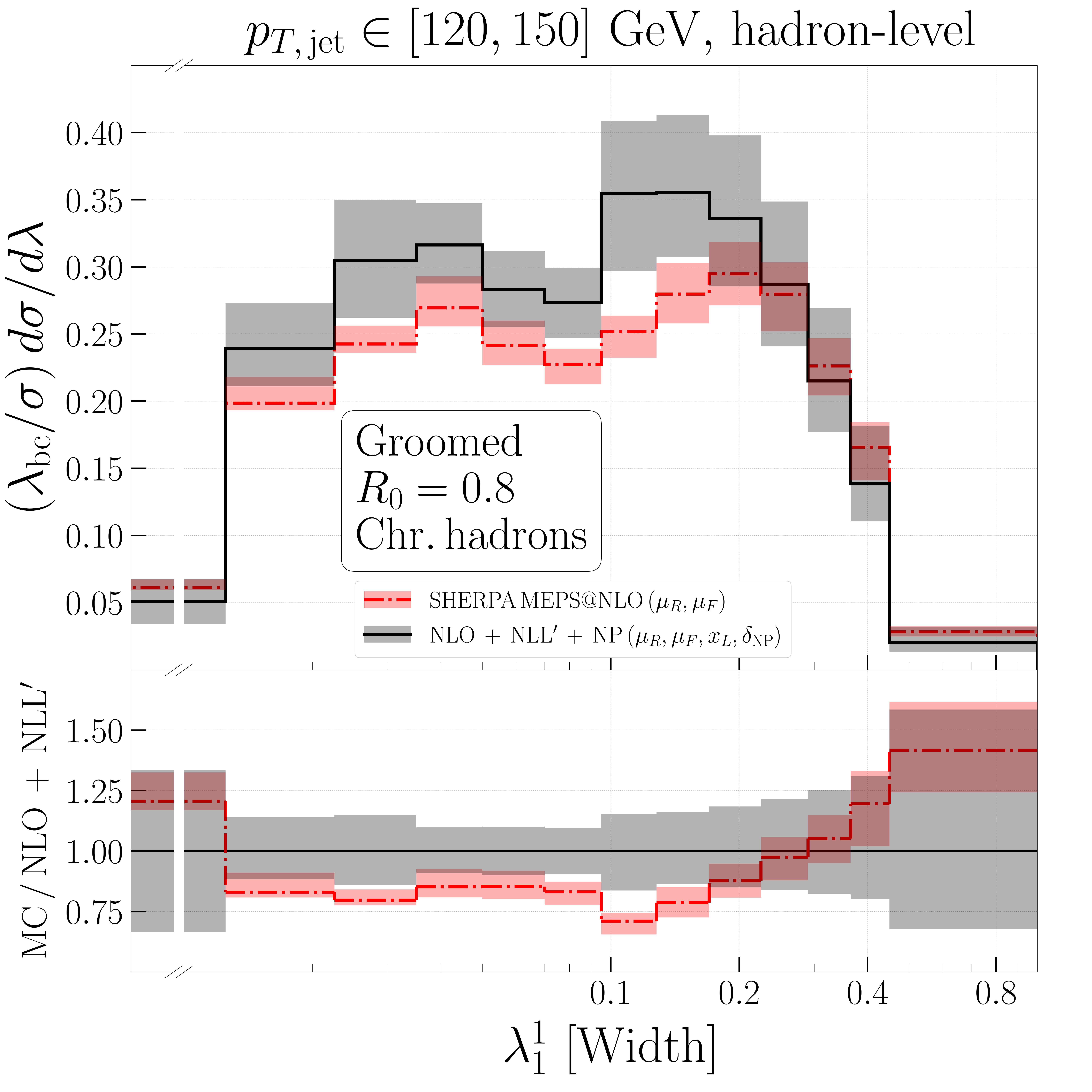}
  \centering
  \includegraphics[width=0.44\linewidth]{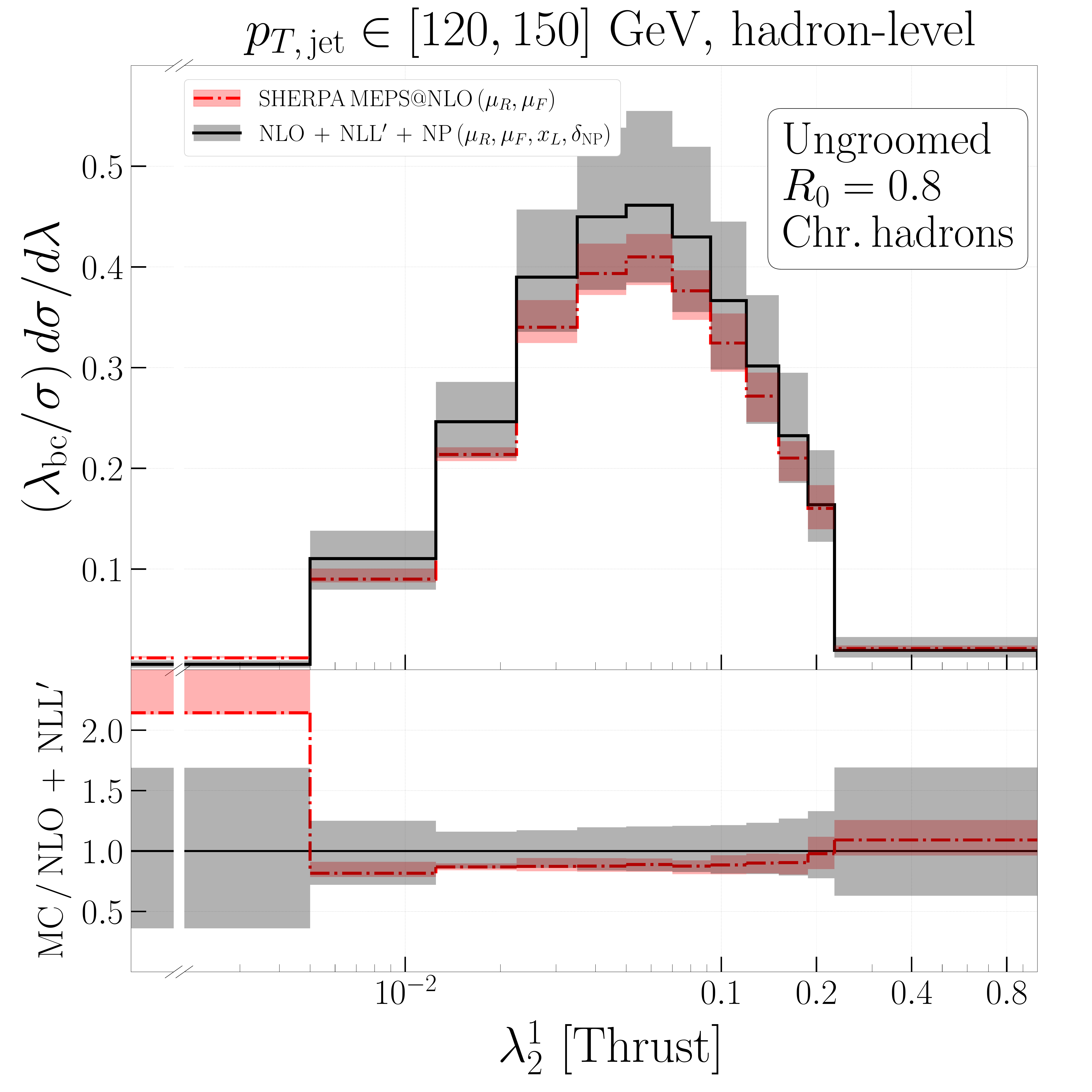}
  \hspace{1em}
  \includegraphics[width=0.44\linewidth]{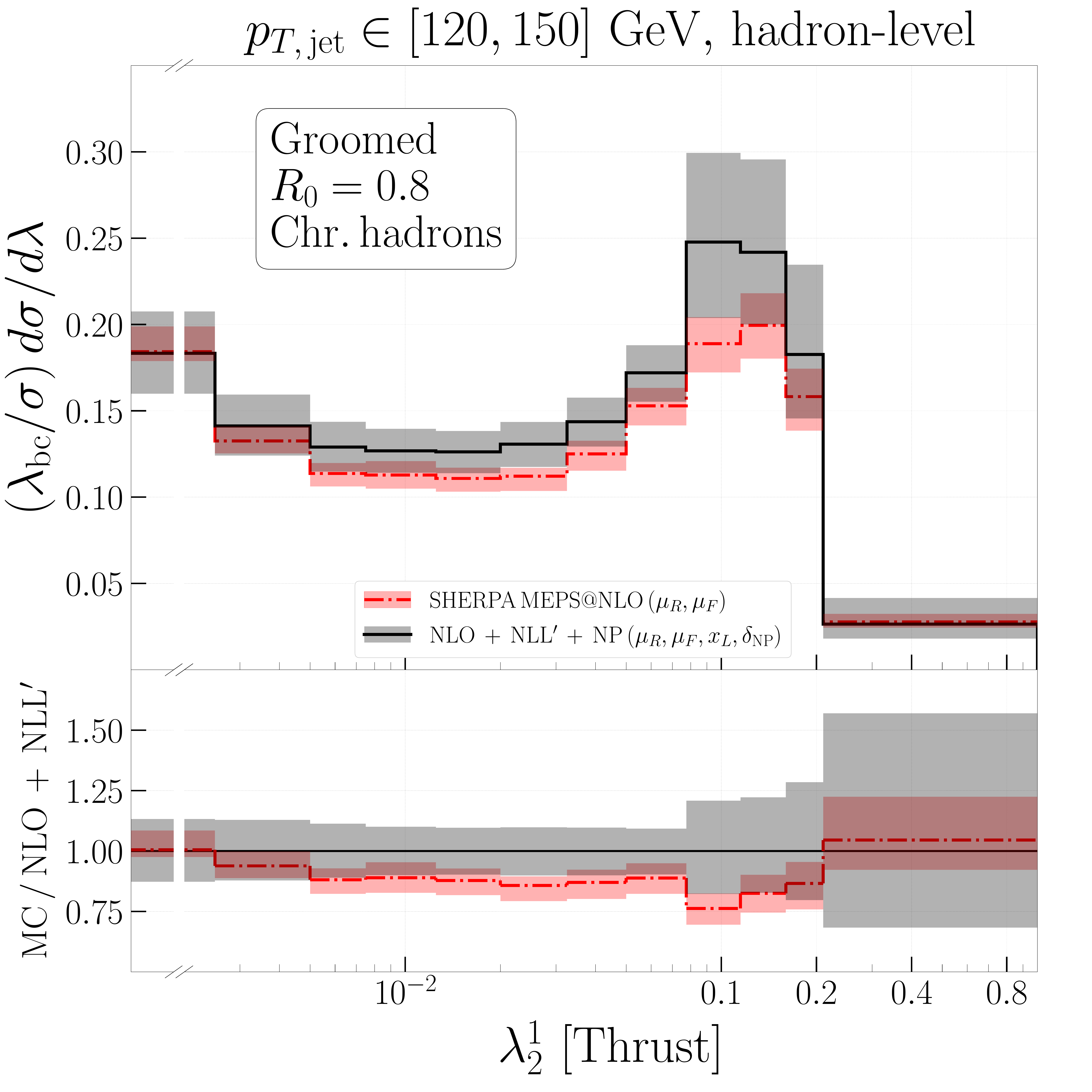}
  \caption{Comparison of hadron-level predictions from \sherpa (MEPS@NLO), based
    on charged hadrons only, for ungroomed and groomed jet-angularities in $Zj$ production, with
    $p_{T,\text{jet}}\in[120,150]\;\text{GeV}$, with \NLOpNLLp results corrected for non-perturbative effects.  Here $\lambda_\text{bc}$ stands for the bin centre.}
\label{fig:res_plus_np_pT120_ch}
\end{figure}

\begin{figure}
  \centering
  \includegraphics[width=0.44\linewidth]{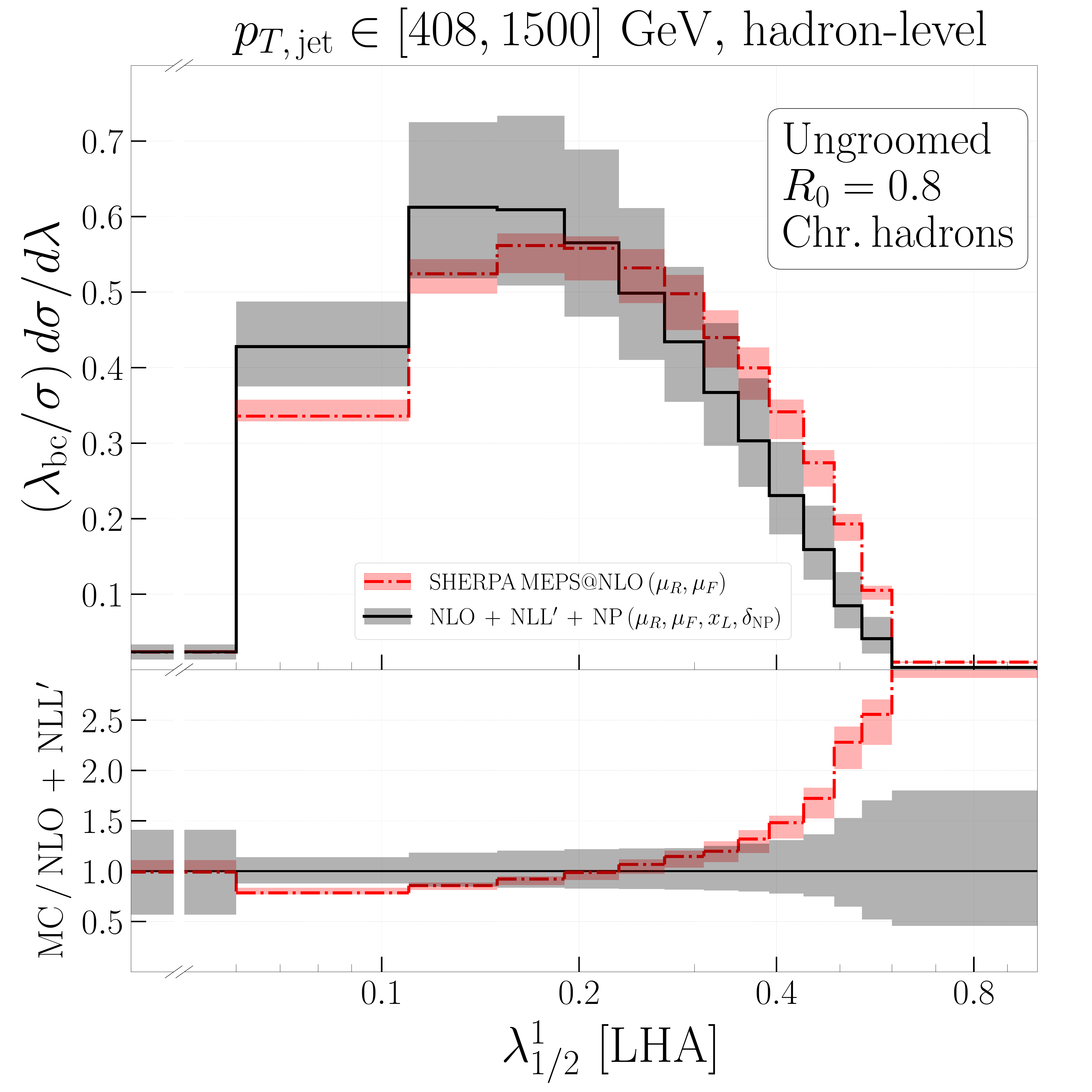}
  \hspace{1em}
  \includegraphics[width=0.44\linewidth]{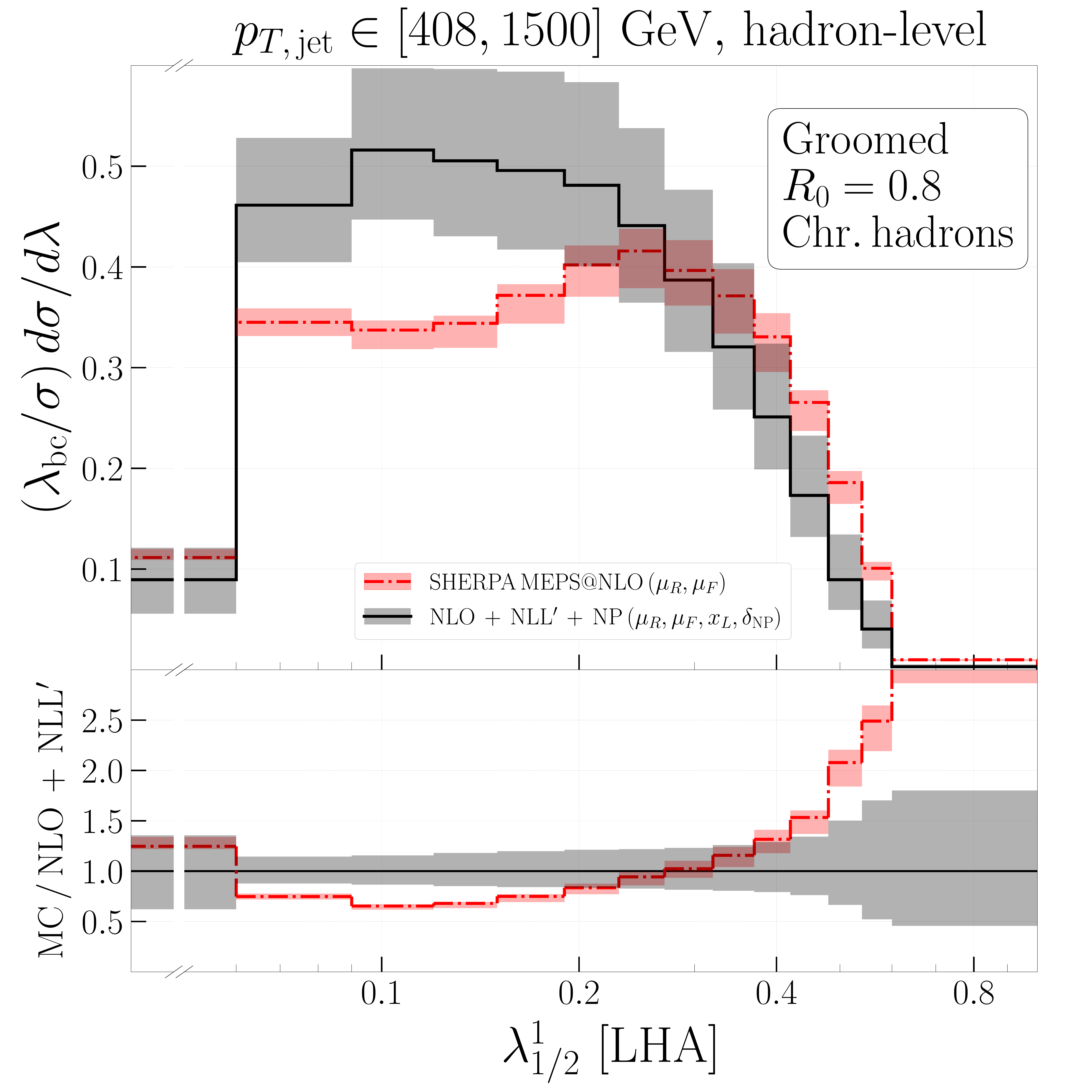}
  \centering
  \includegraphics[width=0.44\linewidth]{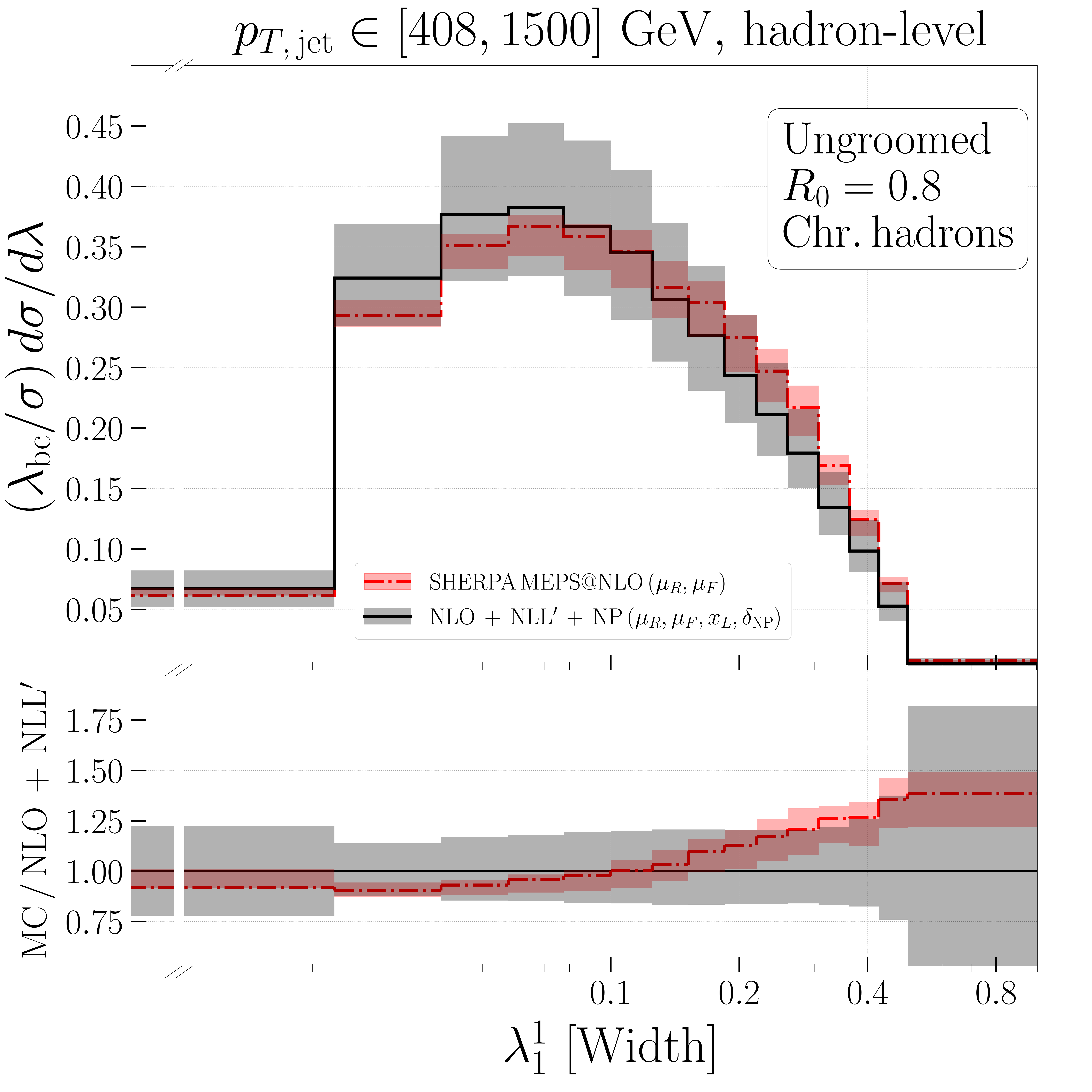}
  \hspace{1em}
  \includegraphics[width=0.44\linewidth]{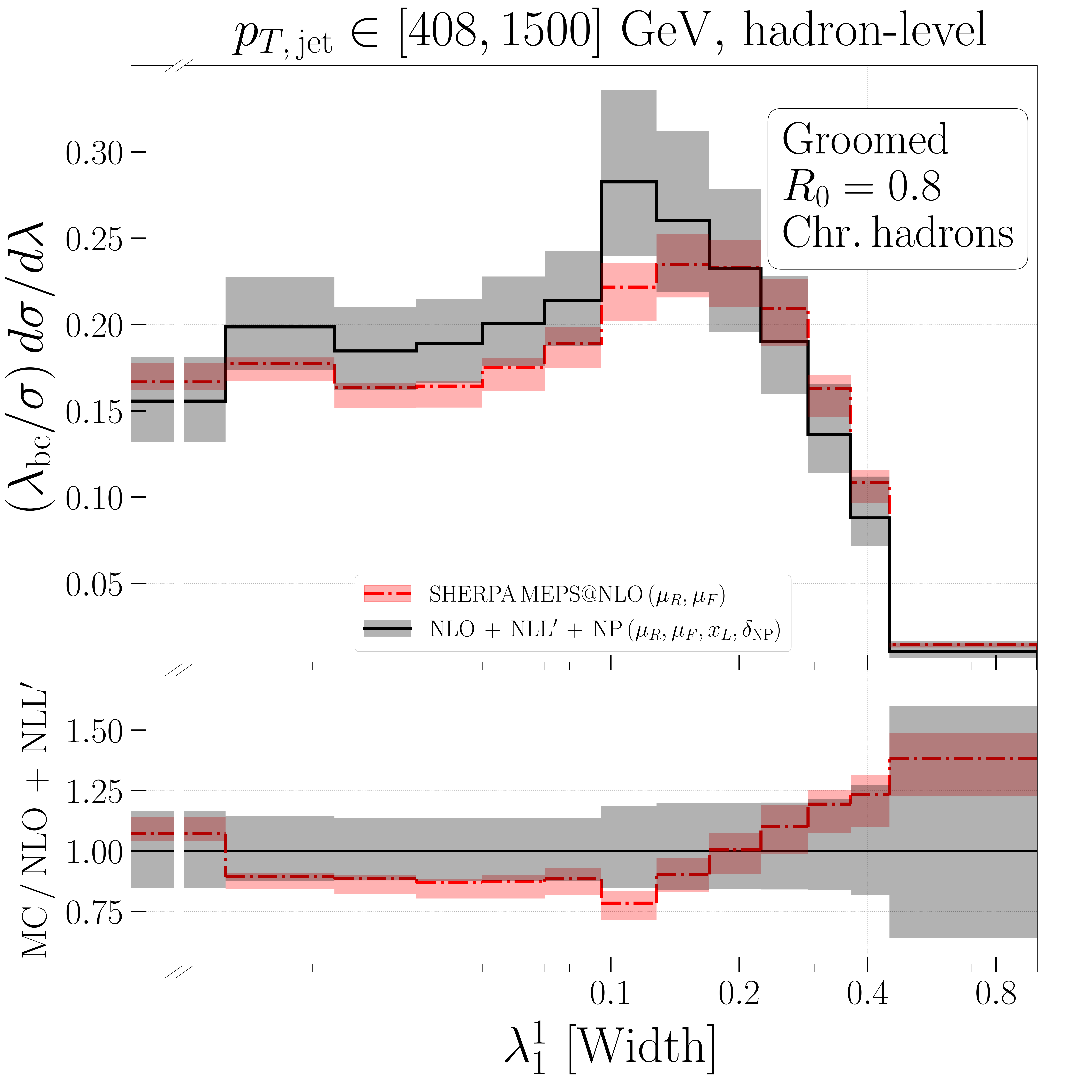}
  \centering
  \includegraphics[width=0.44\linewidth]{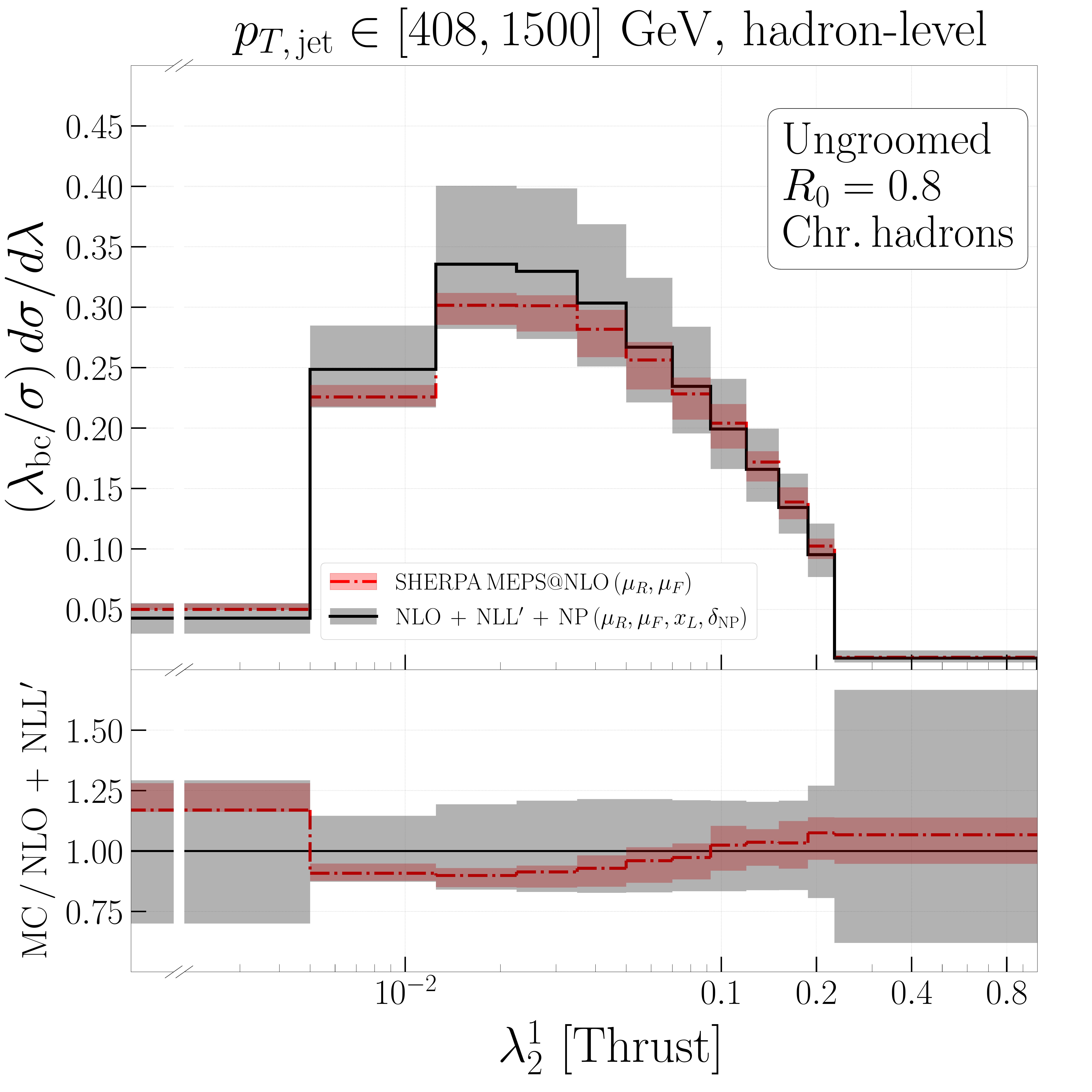}
  \hspace{1em}
  \includegraphics[width=0.44\linewidth]{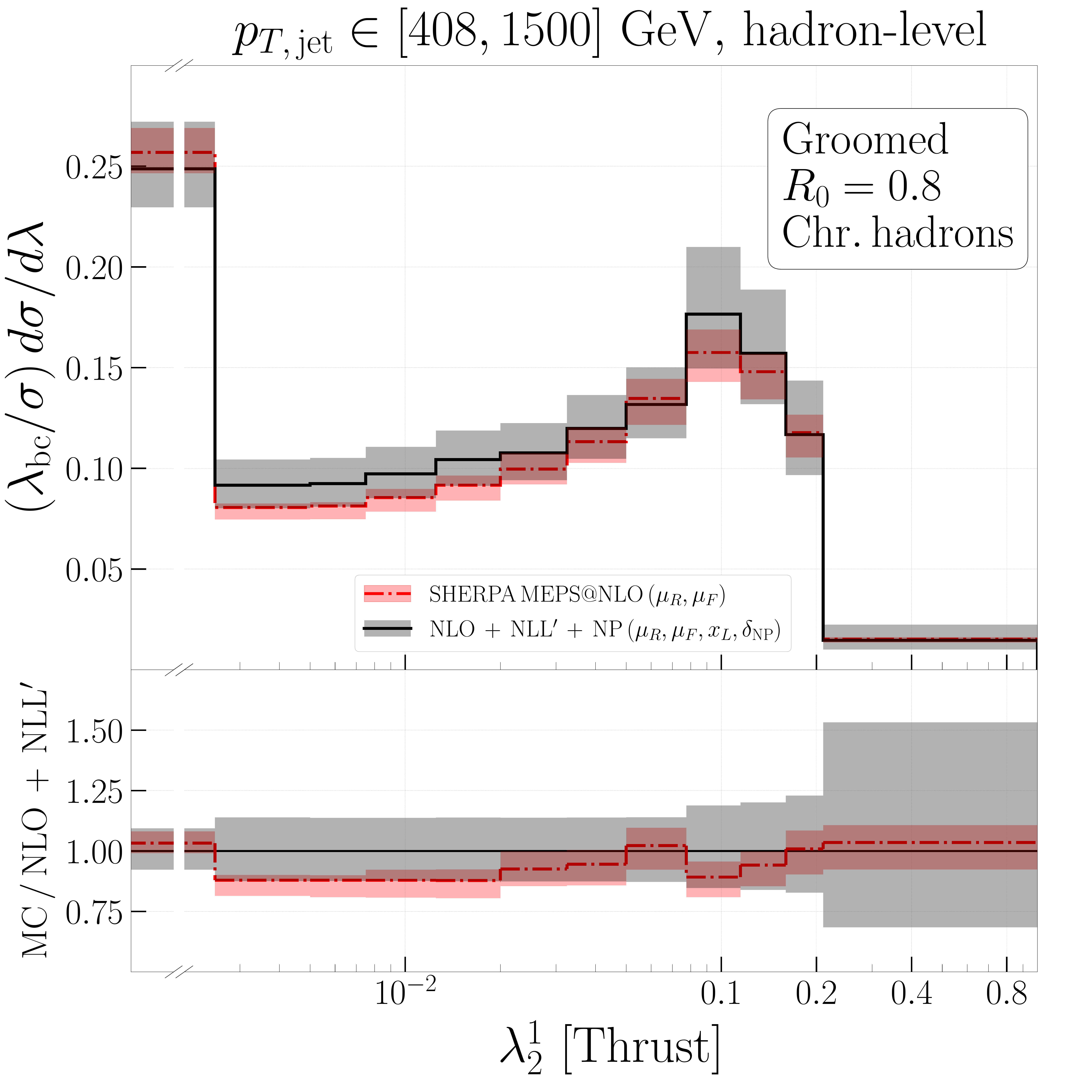}
 \caption{Comparison of hadron-level predictions from \sherpa (MEPS@NLO), based
    on charged hadrons only, for ungroomed and groomed jet-angularities in $Zj$ production, with
    $p_{T,\text{jet}}\in[408,1500]\;\text{GeV}$, with \NLOpNLLp results corrected for non-perturbative effects.  Here $\lambda_\text{bc}$ stands for the bin centre.}
\label{fig:res_plus_np_pT408_ch}
\end{figure}

In Figs.~\ref{fig:res_plus_np_pT120_ch} and \ref{fig:res_plus_np_pT408_ch}
we provide \NLOpNLLp+ NP predictions for the jet angularities based on the
charged jet constituents only for the two considered $p_{T,\text{jet}}$ slices,
\emph{i.e.}\ $p_{T,\text{jet}} \in [120,150]\;\text{GeV}$ and
$p_{T,\text{jet}} \in [408,1500]\;\text{GeV}$. The corresponding correction
factors have been presented in Fig.~\ref{fig:all_mc_np_band} in
Section~\ref{sec:nlo_nll_final_np}. The predictions are compared to MEPS@NLO
results from \sherpa . 

\subsection[Hadron-level results for $R_0 = 0.4$ jets]{Hadron-level results for \boldmath{$R_0 = 0.4$} jets}
\label{app:NLL_NP_R4}
\begin{figure}
  \centering
  \includegraphics[width=0.44\linewidth]{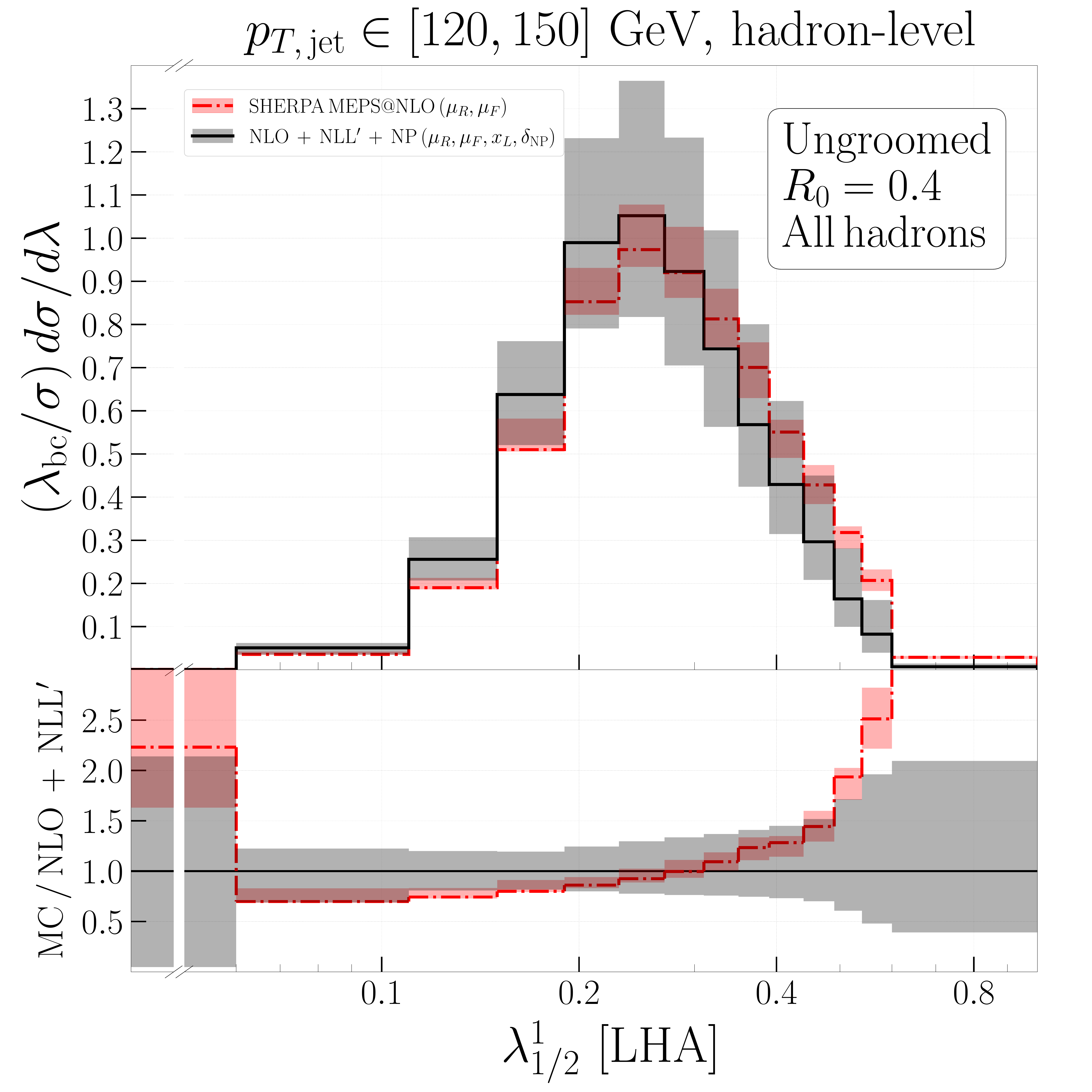}
  \hspace{1em}
  \includegraphics[width=0.44\linewidth]{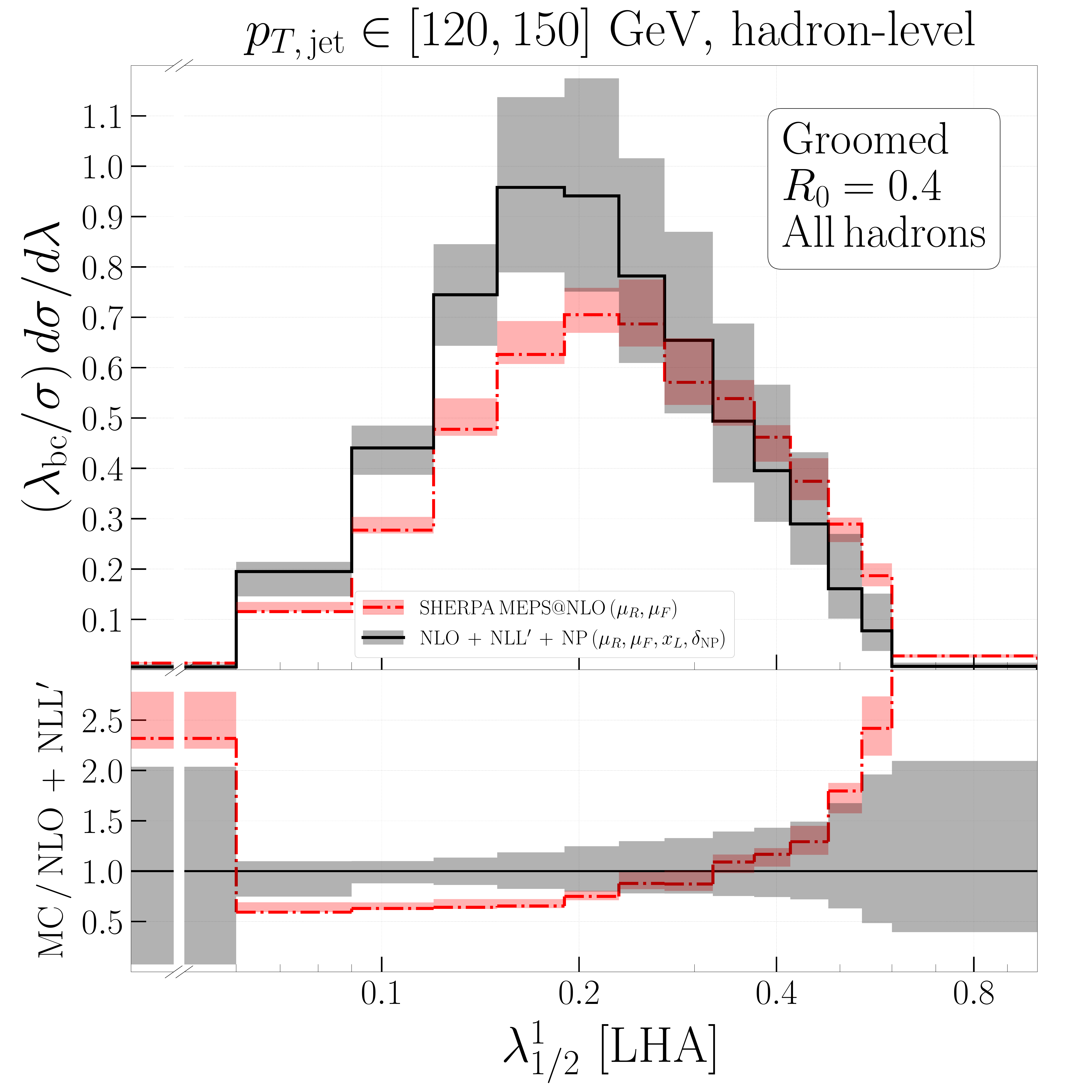}
  \centering
  \includegraphics[width=0.44\linewidth]{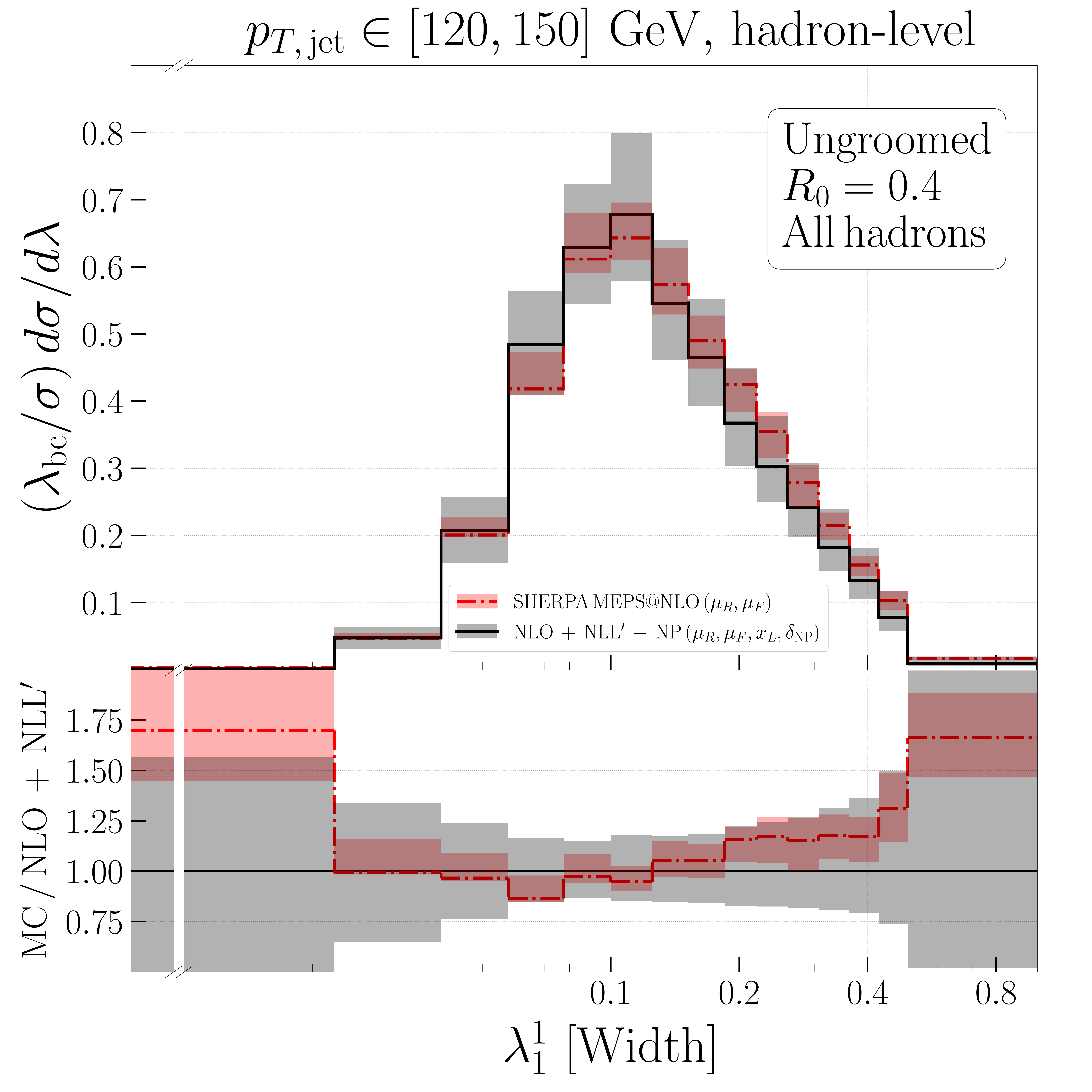}
  \hspace{1em}
  \includegraphics[width=0.44\linewidth]{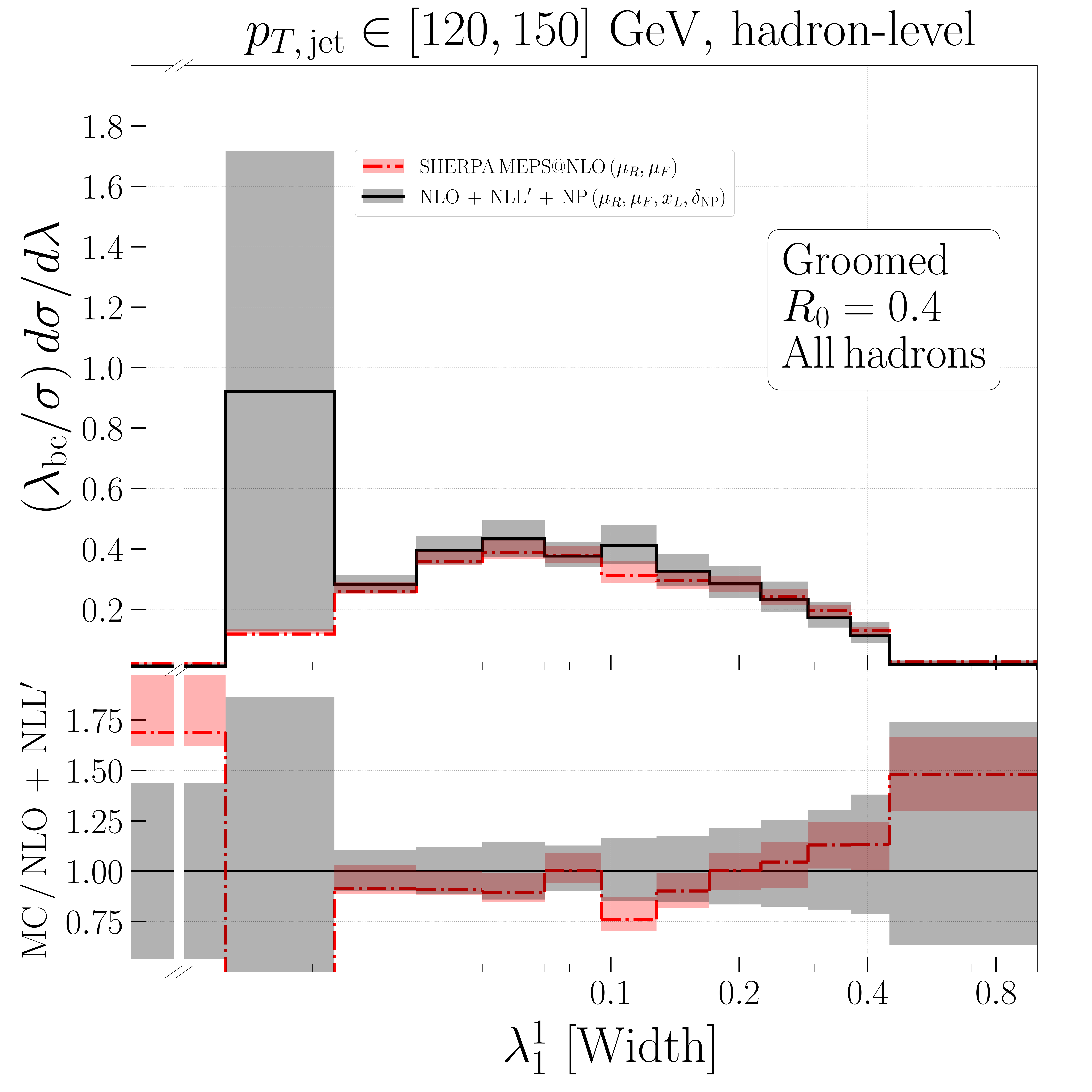}
  \centering
  \includegraphics[width=0.44\linewidth]{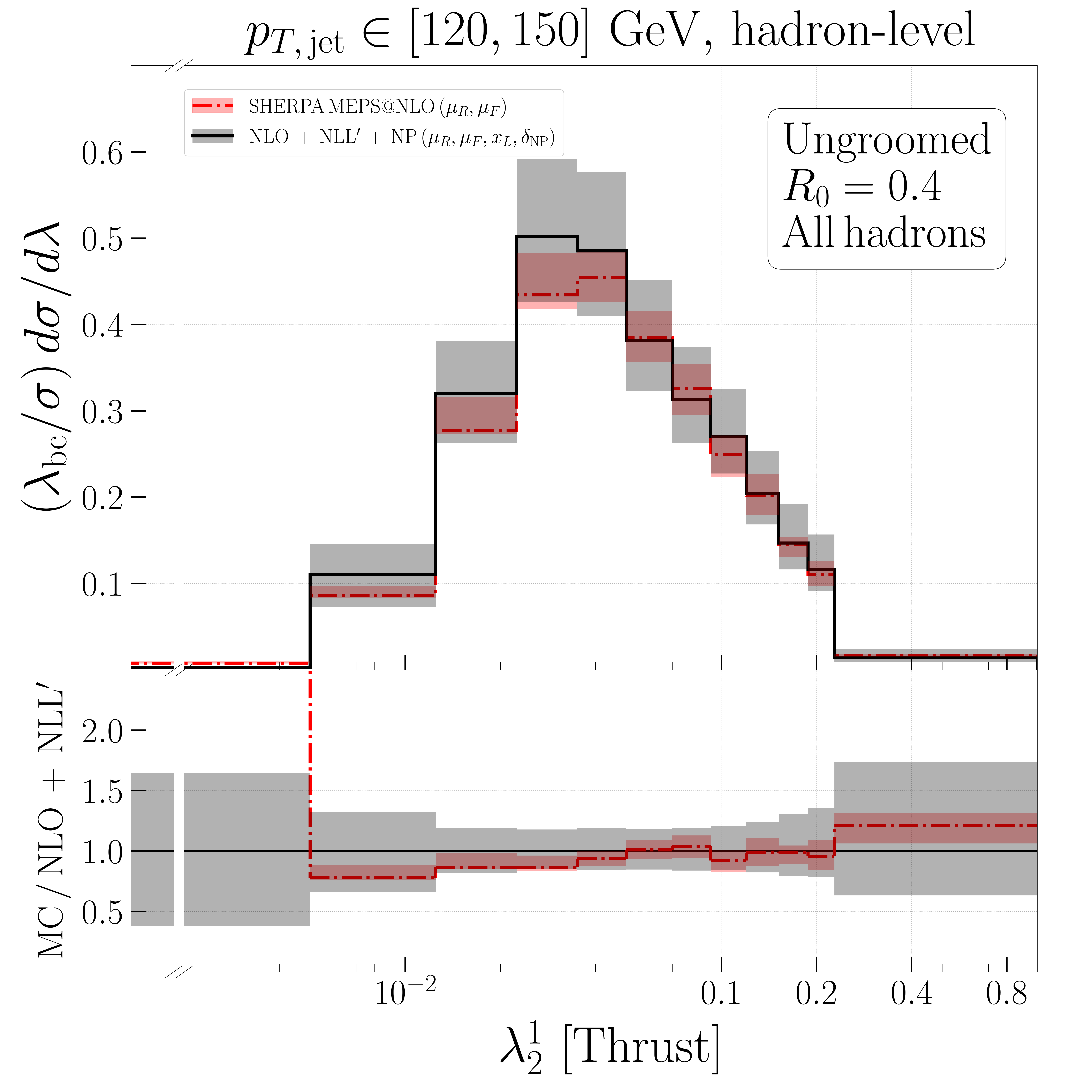}
  \hspace{1em}
  \includegraphics[width=0.44\linewidth]{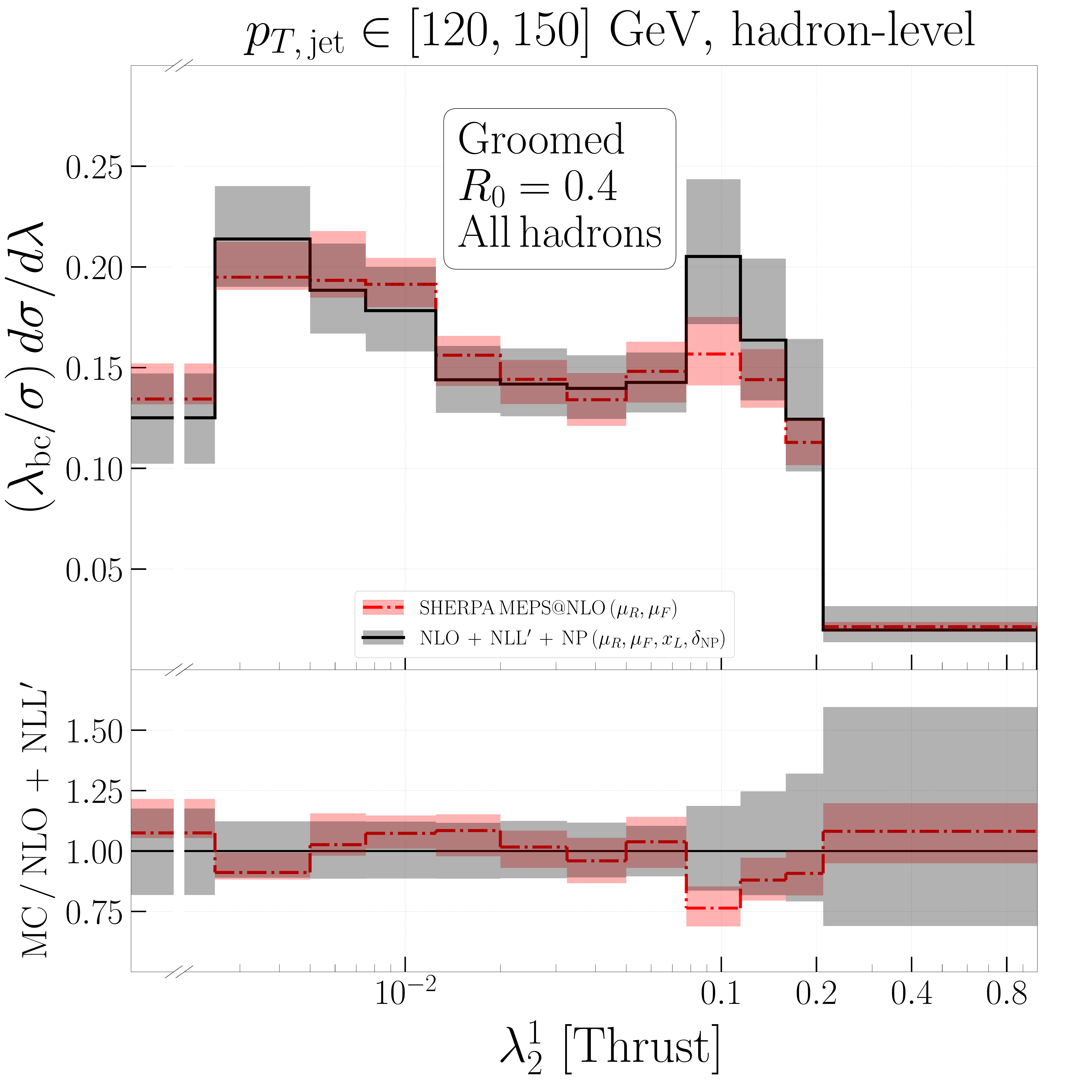}
  \caption{Same as Fig.~\ref{fig:res_plus_np_pT120_all} but for $R_0=0.4$ jets. }
  \label{fig:res_plus_np_pT120_all_R4}
\end{figure}

\begin{figure}
  \centering
  \includegraphics[width=0.44\linewidth]{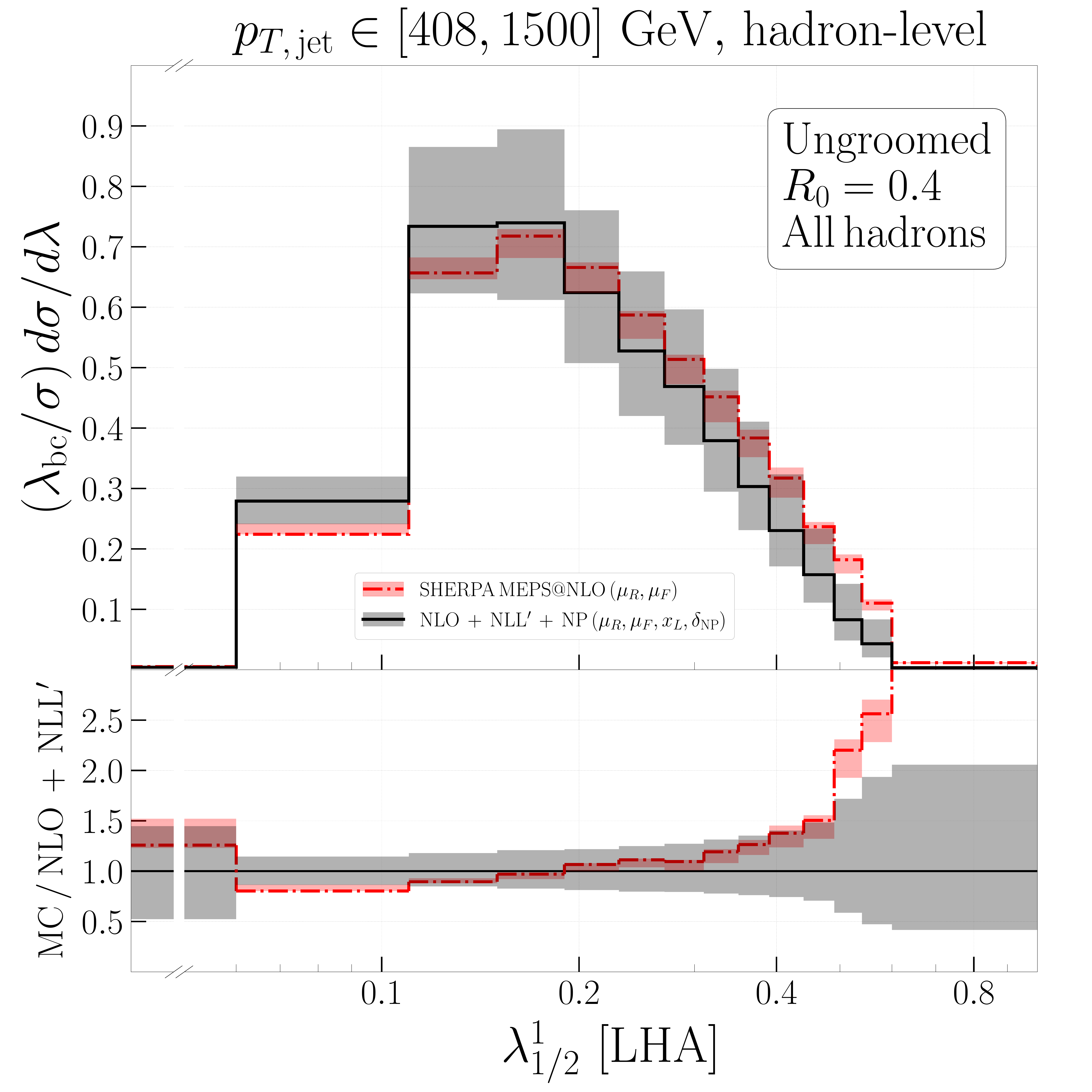}
  \hspace{1em}
  \includegraphics[width=0.44\linewidth]{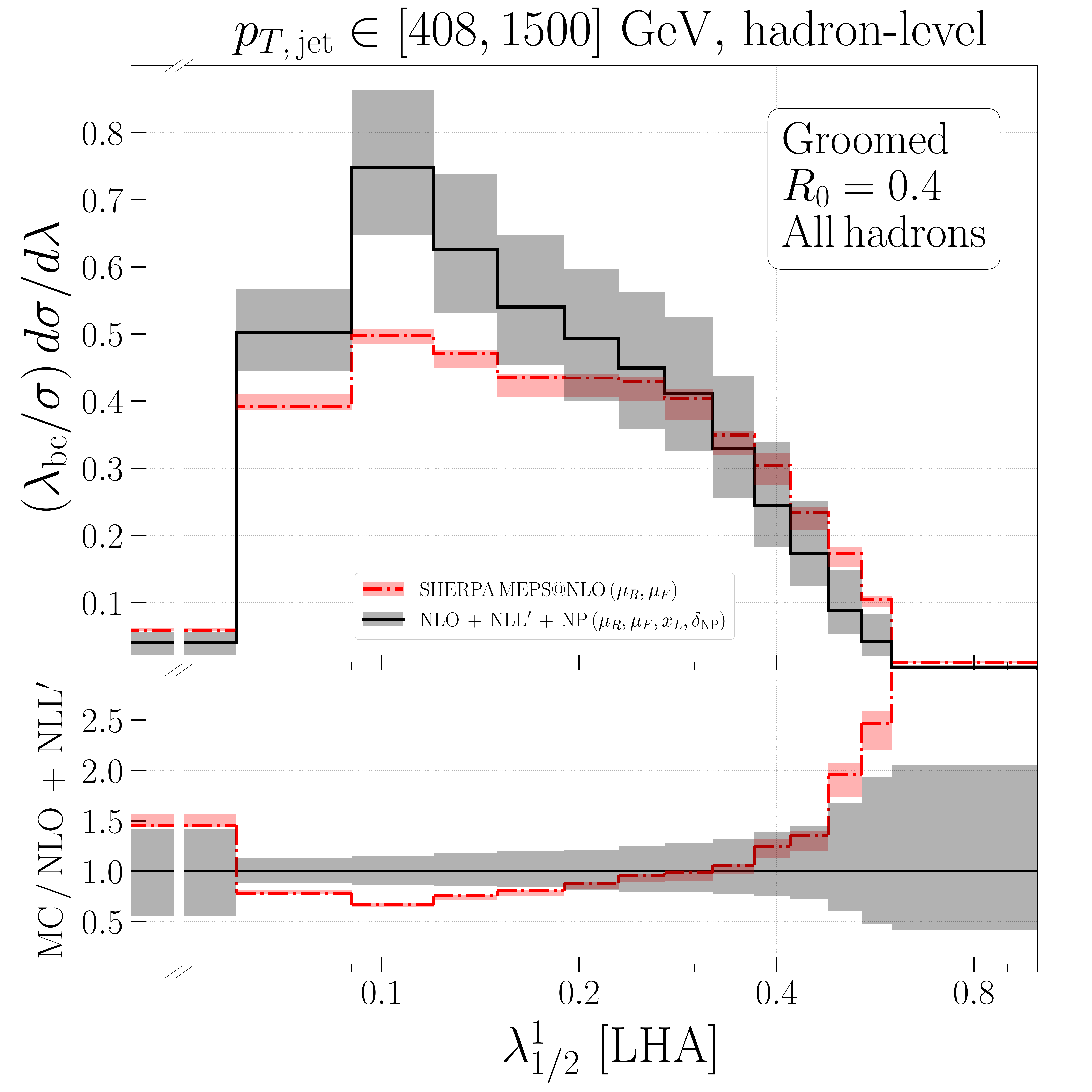}
  \centering
  \includegraphics[width=0.44\linewidth]{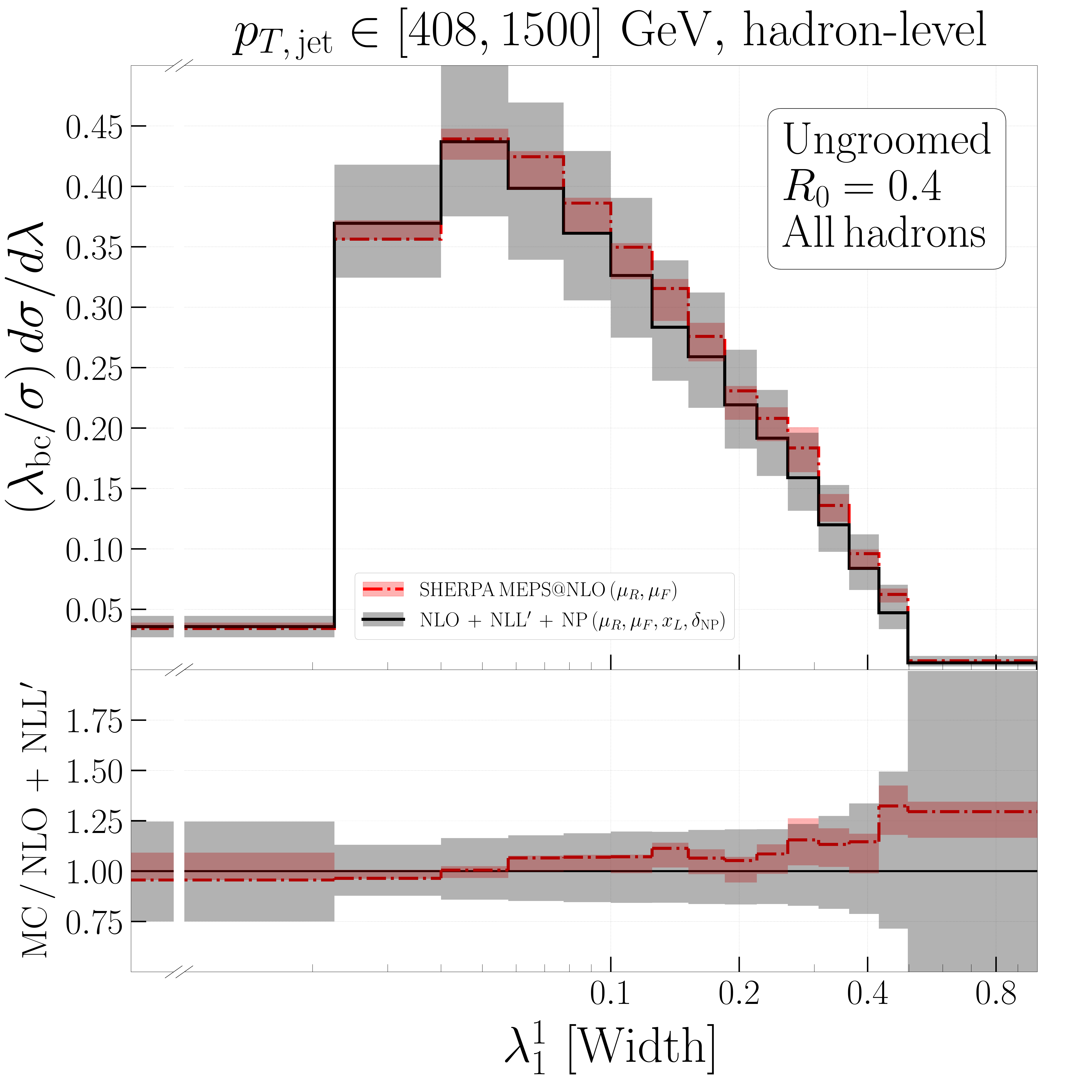}
  \hspace{1em}
  \includegraphics[width=0.44\linewidth]{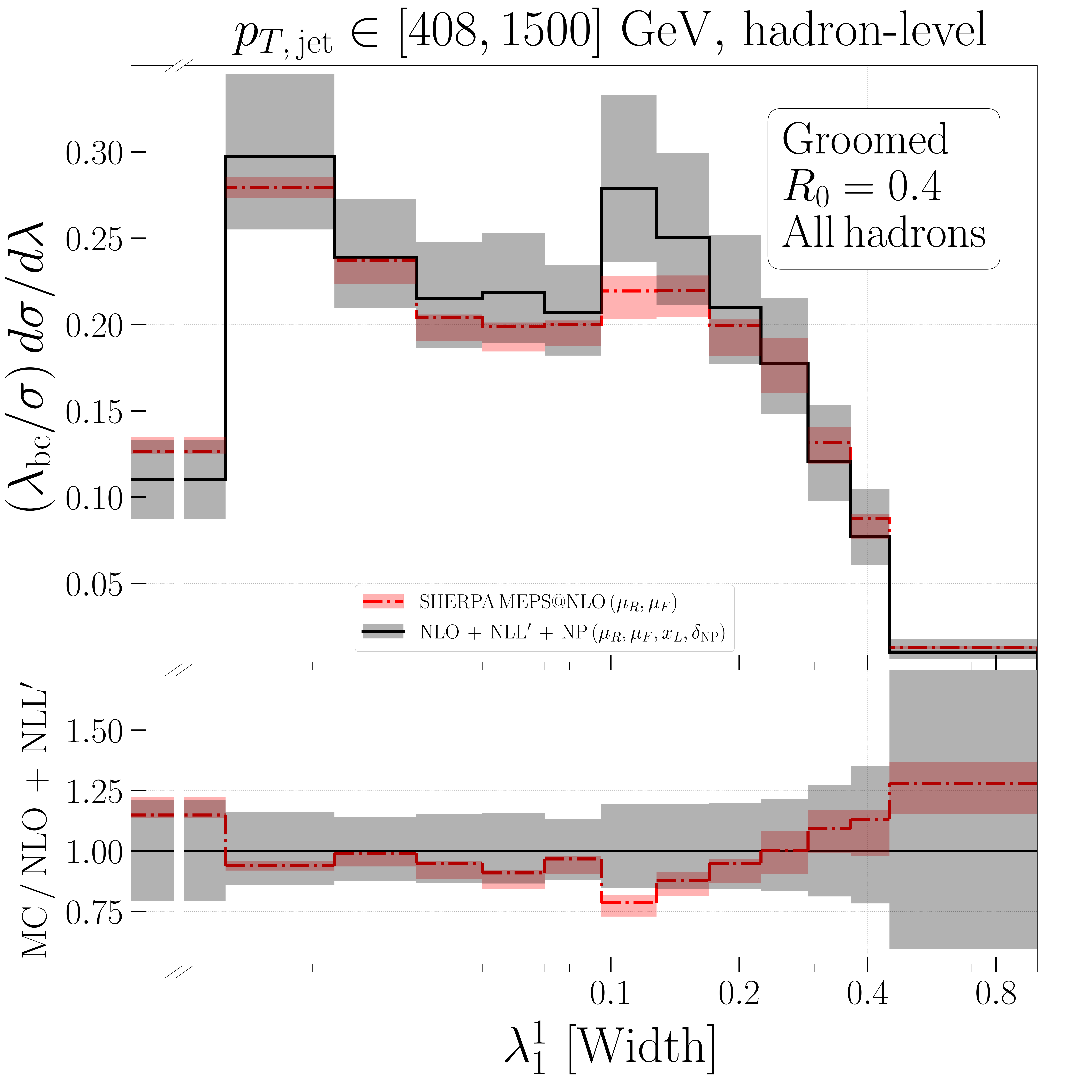}
  \centering
  \includegraphics[width=0.44\linewidth]{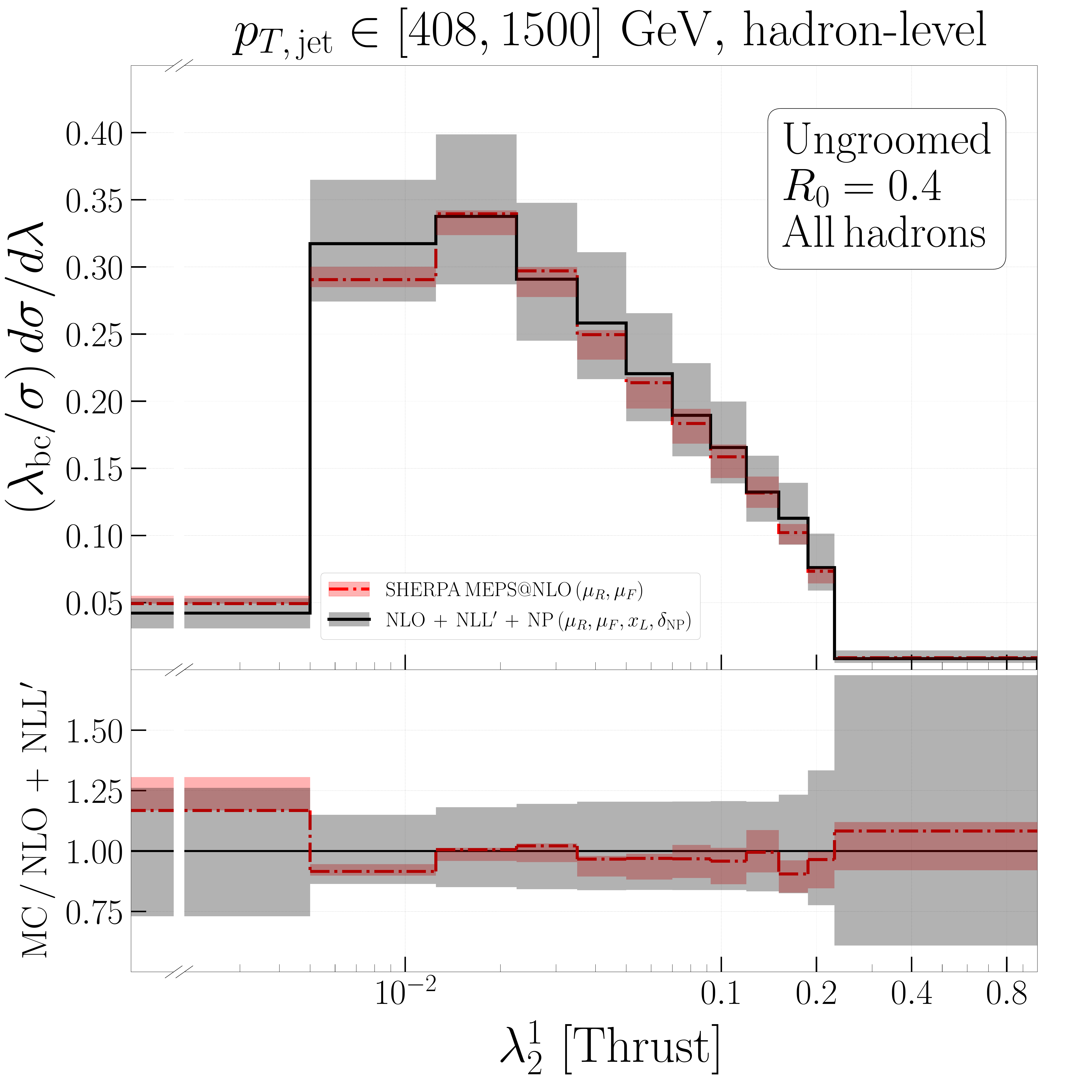}
  \hspace{1em}
  \includegraphics[width=0.44\linewidth]{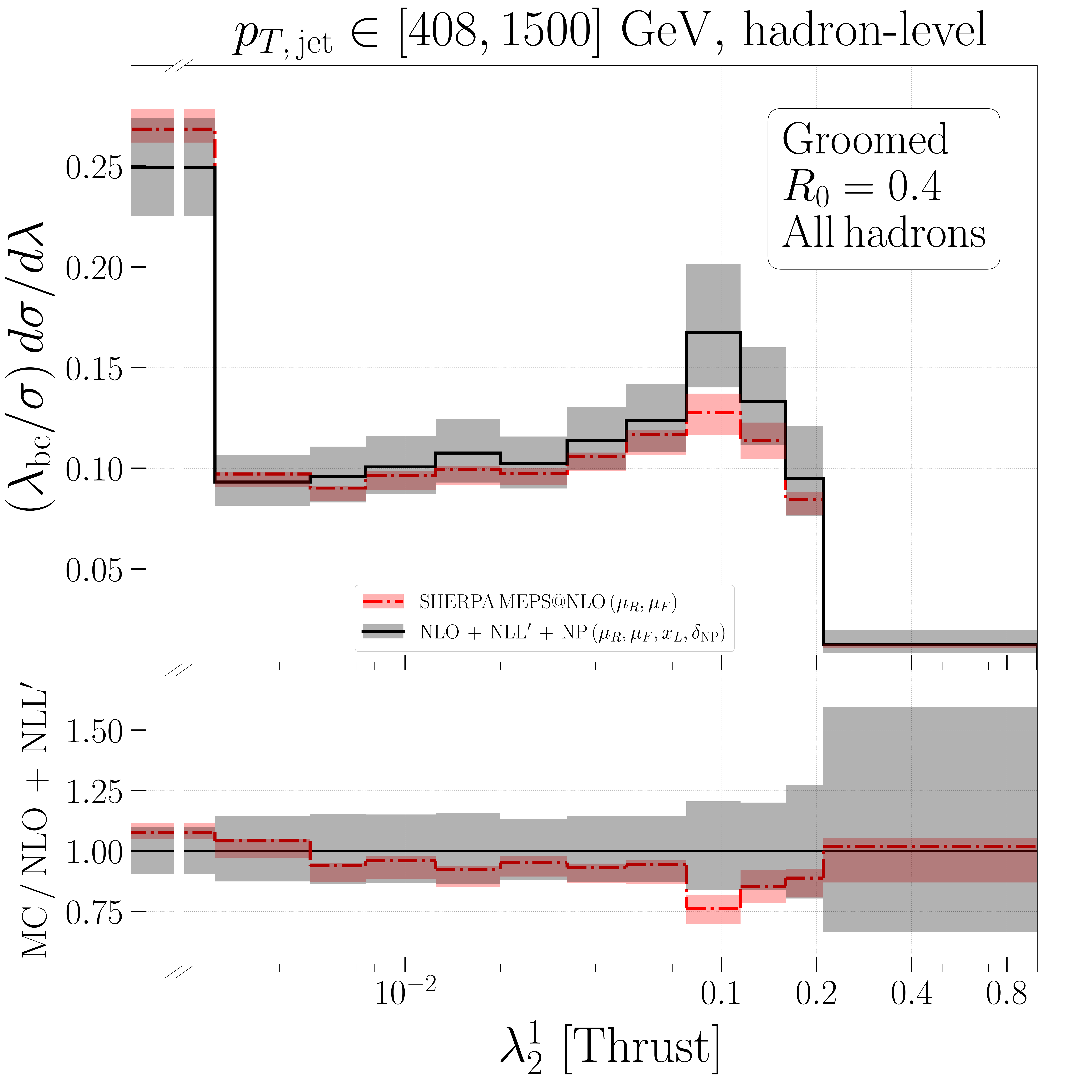}
 \caption{Same as Fig.~\ref{fig:res_plus_np_pT408_all} but for $R_0=0.4$ jets.}
\label{fig:res_plus_np_pT408_all_R4}
\end{figure}

\begin{figure}
  \centering
  \includegraphics[width=0.44\linewidth]{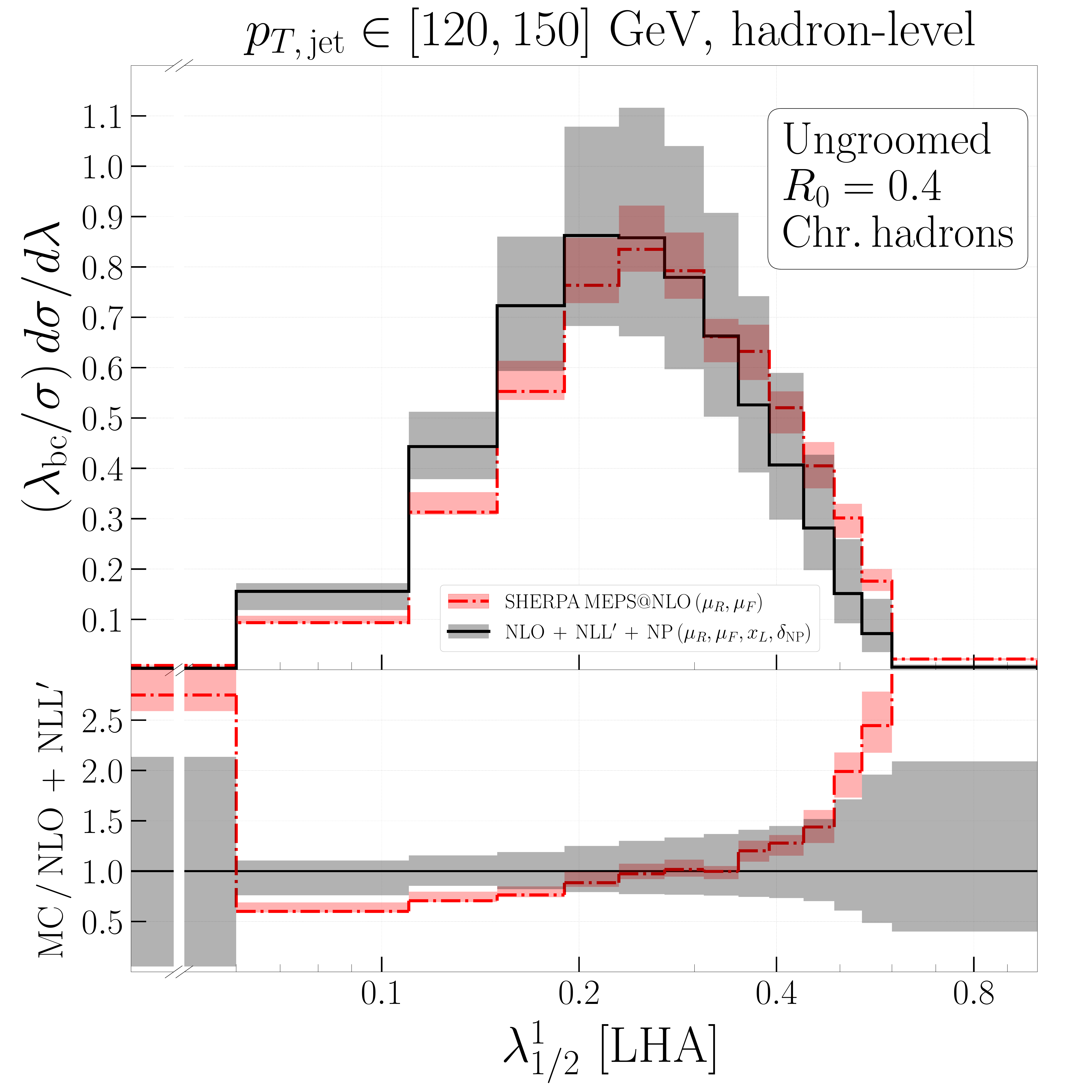}
  \hspace{1em}
  \includegraphics[width=0.44\linewidth]{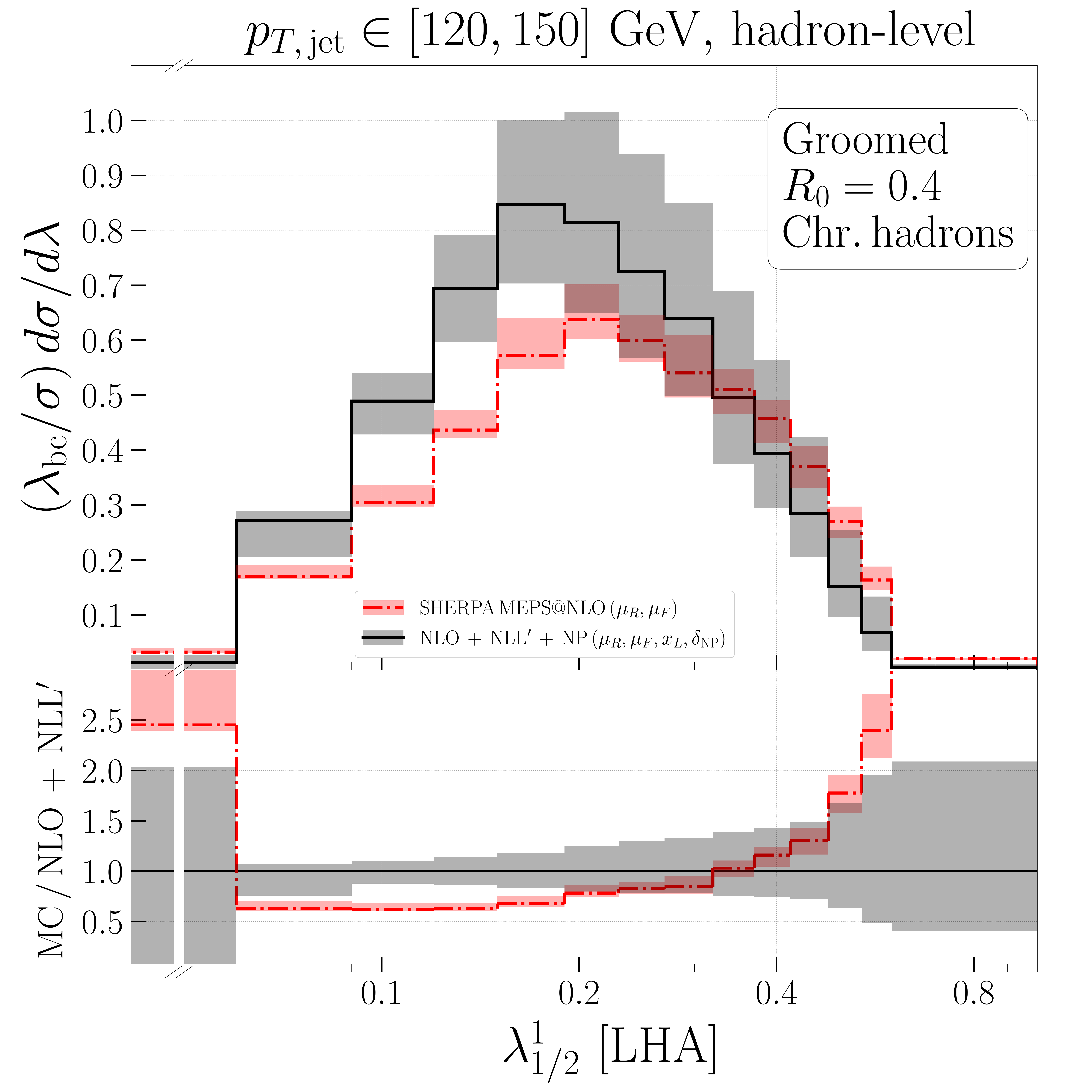}
  \centering
  \includegraphics[width=0.44\linewidth]{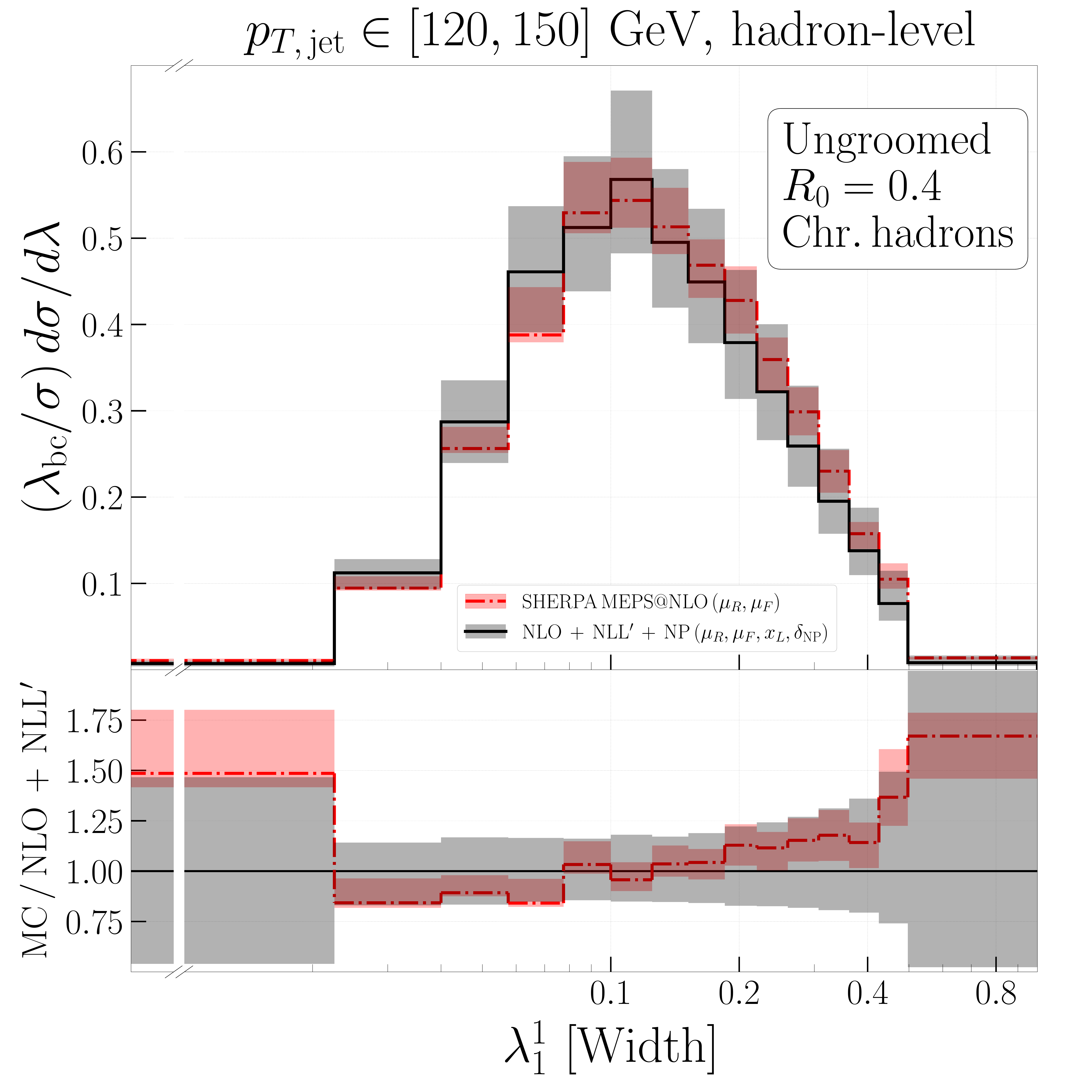}
  \hspace{1em}
  \includegraphics[width=0.44\linewidth]{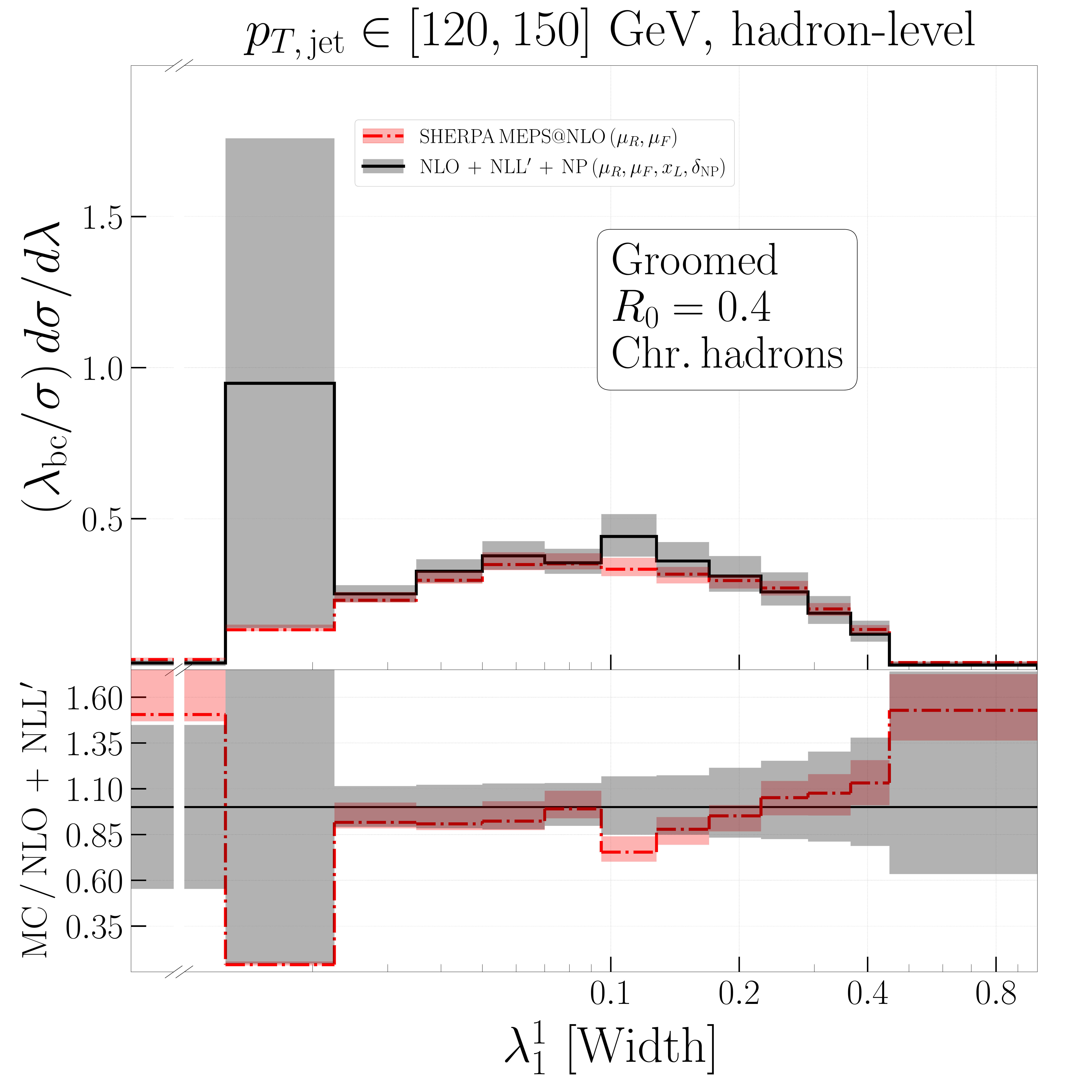}
  \centering
  \includegraphics[width=0.44\linewidth]{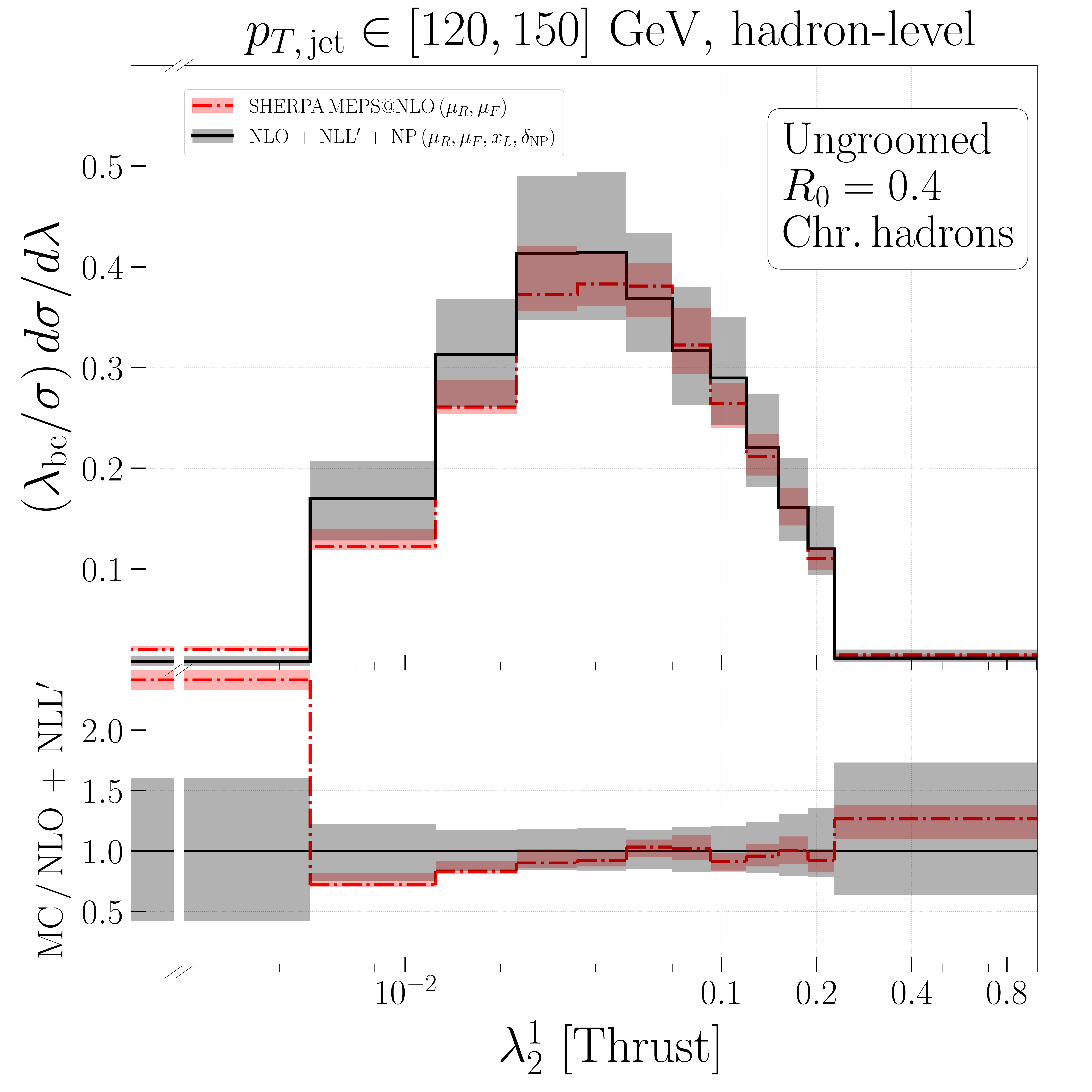}
  \hspace{1em}
  \includegraphics[width=0.44\linewidth]{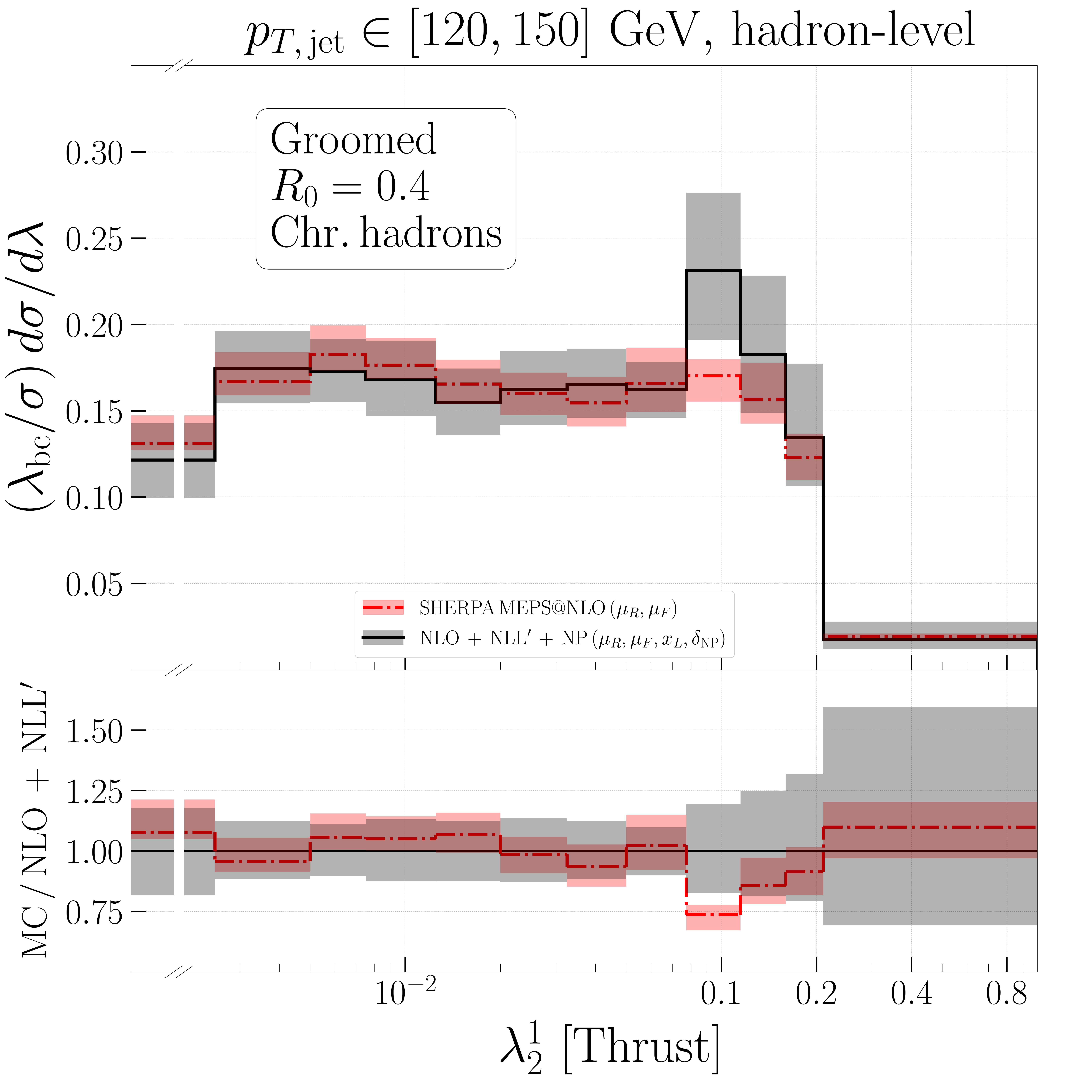}
  \caption{Same as Fig.~\ref{fig:res_plus_np_pT120_ch} but for $R_0=0.4$ jets. }
\label{fig:res_plus_np_pT120_ch_R4}
\end{figure}

\begin{figure}
  \centering
  \includegraphics[width=0.44\linewidth]{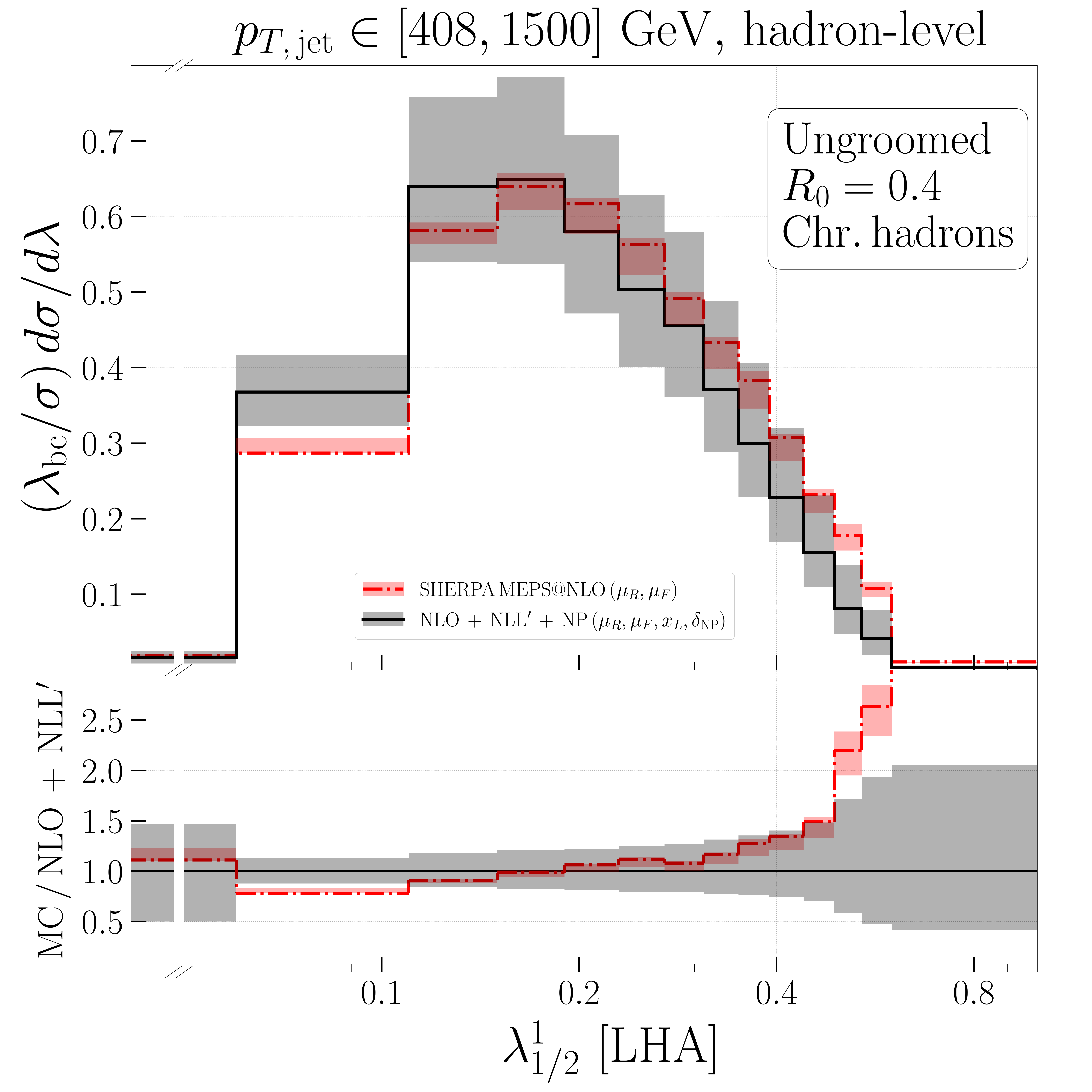}
  \hspace{1em}
  \includegraphics[width=0.44\linewidth]{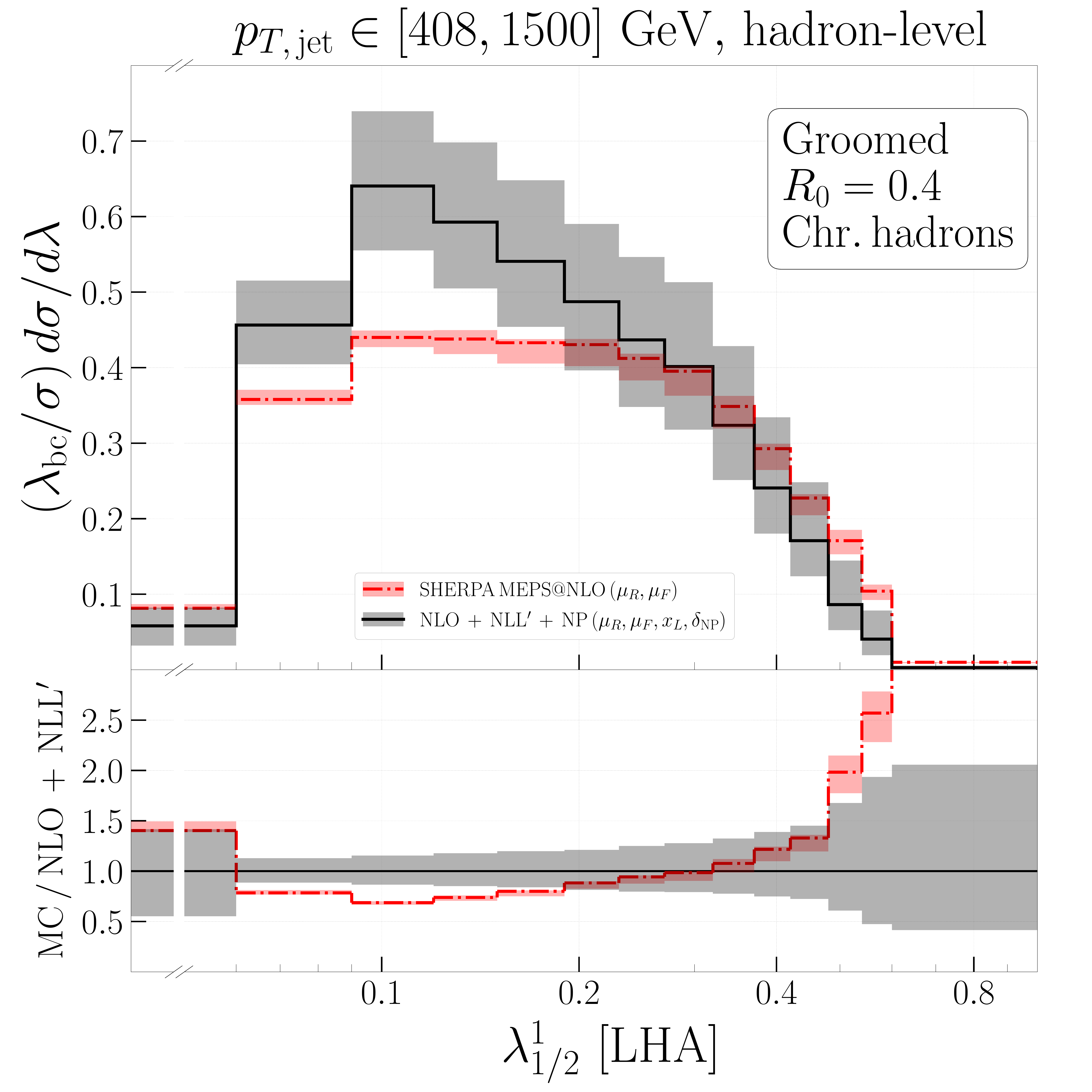}
  \centering
  \includegraphics[width=0.44\linewidth]{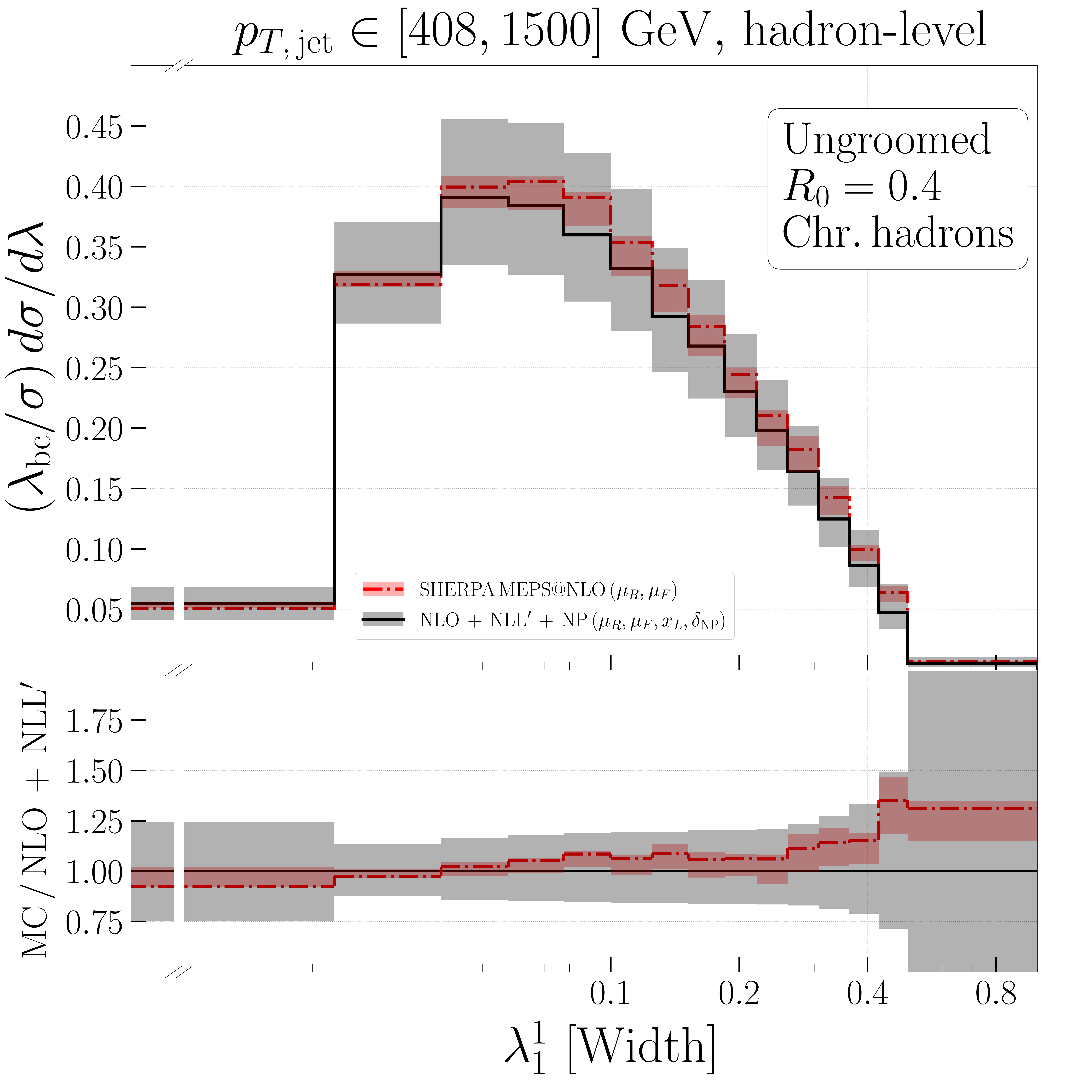}
  \hspace{1em}
  \includegraphics[width=0.44\linewidth]{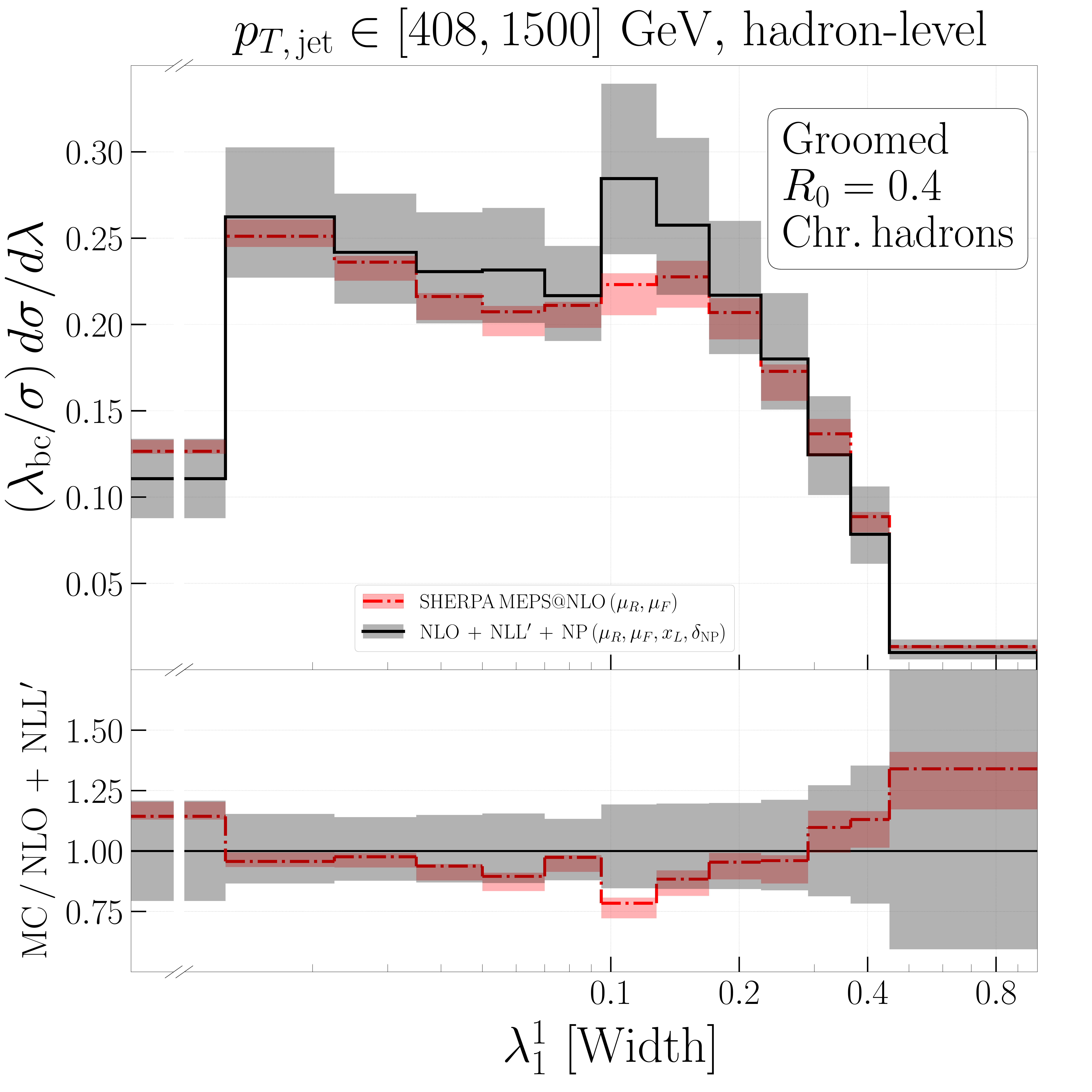}
  \centering
  \includegraphics[width=0.44\linewidth]{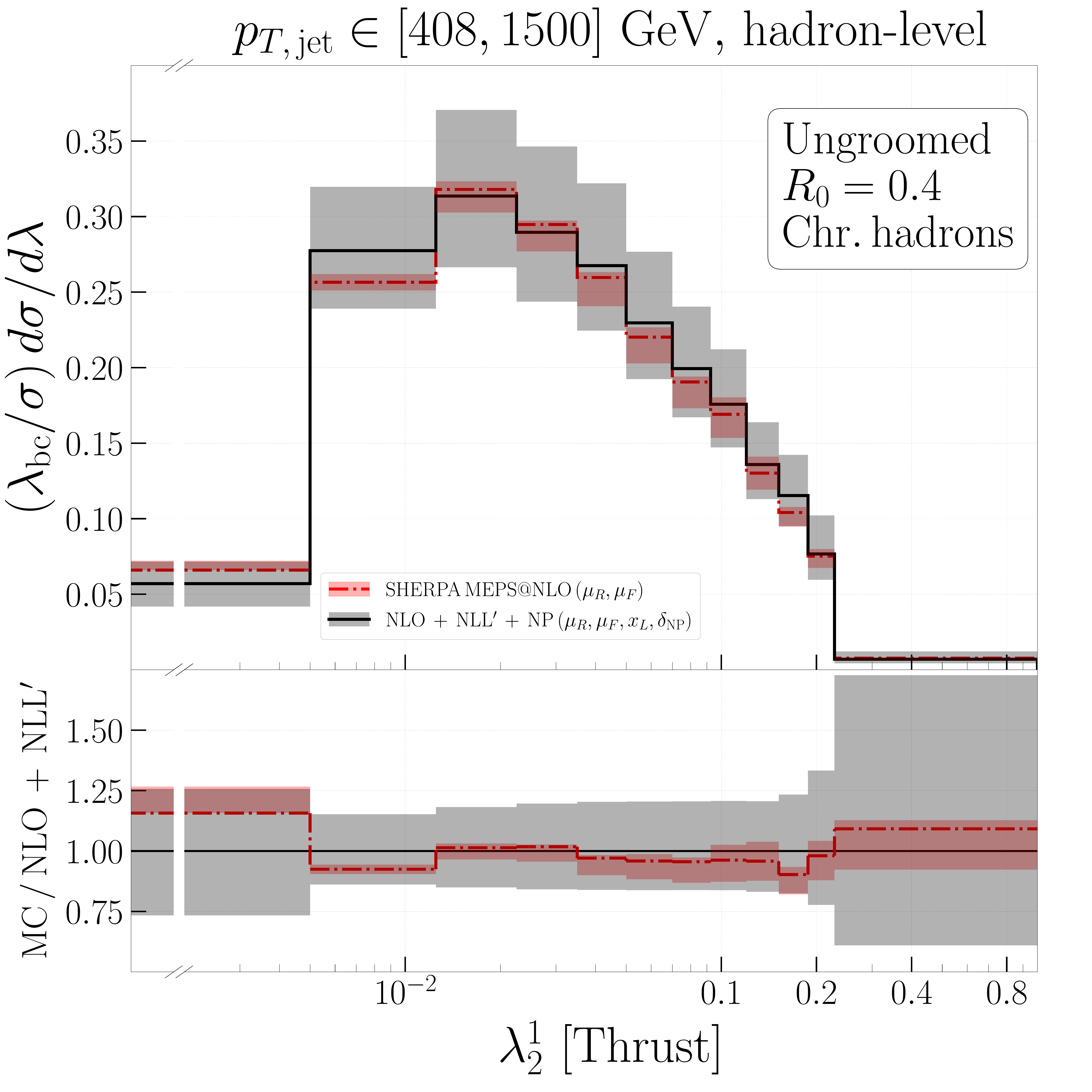}
  \hspace{1em}
  \includegraphics[width=0.44\linewidth]{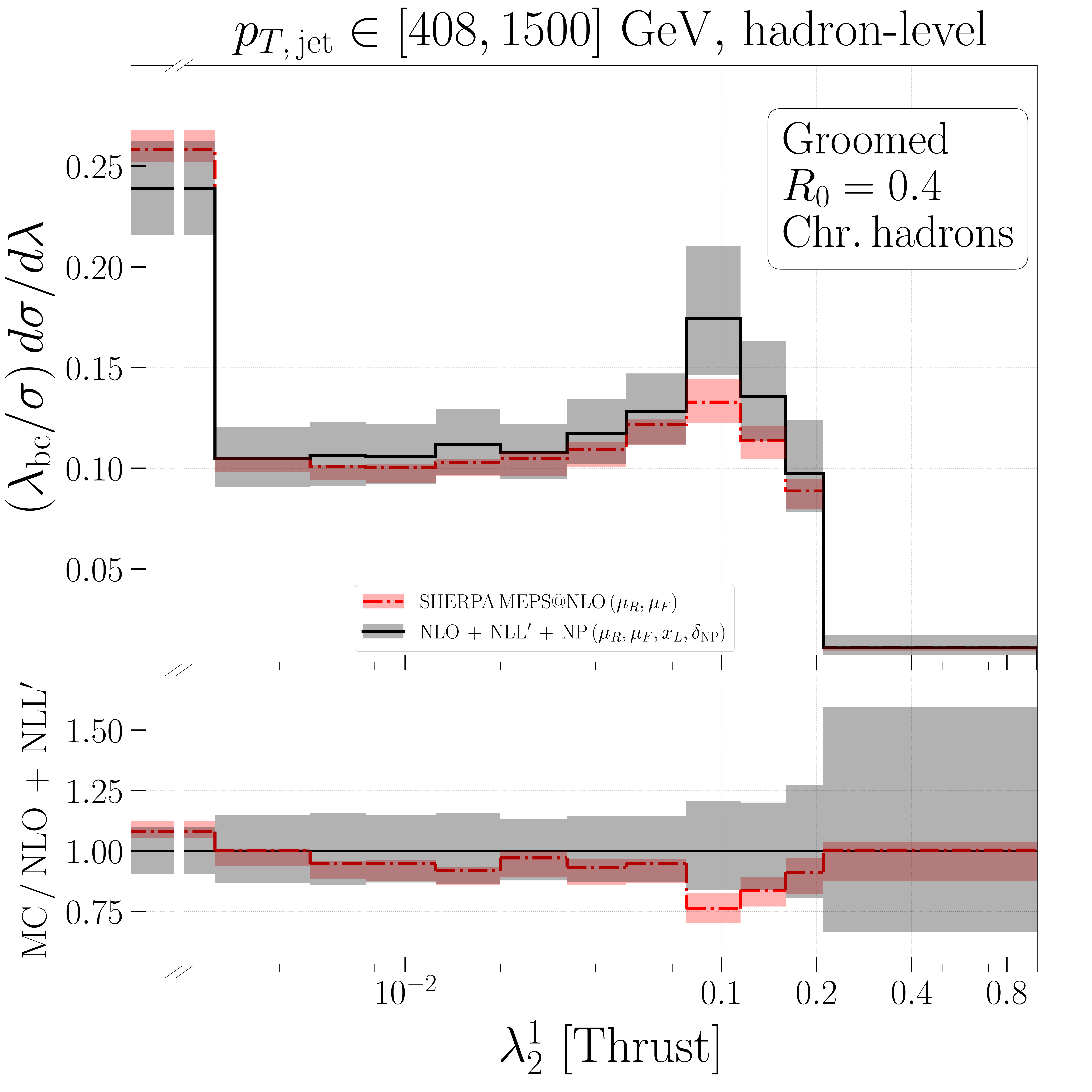}
 \caption{Same as Fig.~\ref{fig:res_plus_np_pT408_ch} but for $R_0=0.4$ jets.}
\label{fig:res_plus_np_pT408_ch_R4}
\end{figure}

Here we provide results analogous to the ones presented in Section~\ref{sec:nlo_nll_final_np} but
for a smaller jet radius, namely $R_0 = 0.4$ jets.
In particular, we show \NLOpNLLp predictions, including non-perturbative corrections, and compare them to
\sherpa MEPS@NLO results. This is done for the case of all-hadrons in Figs.~\ref{fig:res_plus_np_pT120_all_R4}
and~\ref{fig:res_plus_np_pT408_all_R4} for the two considered transverse momentum slices, respectively, while
in Figs.~\ref{fig:res_plus_np_pT120_ch_R4} and~\ref{fig:res_plus_np_pT408_ch_R4} we present corresponding results
based on charged hadrons only.

The comparison yields very similar findings to the $R_0=0.8$ case reported in the main text. However,
for the groomed width in the lower transverse momentum slice we here observe very large non-perturbative
corrections (with corresponding large uncertainties). This signals the breakdown of our approximate treatment
for including non-perturbative corrections. Here we probe a region of phase space close to the parton-shower
cutoffs, below which the Monte Carlo simulations are purely determined by non-perturbative effects and,
consequently,  the hadron-to-parton-level ratios get large. This effect becomes more important as we
lower $p_{T,\text{jet}}$ and/or the jet radius $R_0$. Its onset can be estimated using
Eqs.~\eqref{nonpert-trans_UG} and \eqref{nonpert-trans_G}.

\clearpage

\clearpage

\phantomsection
\addcontentsline{toc}{section}{References}
\bibliographystyle{jhep}
\bibliography{references}
\end{document}